\documentstyle[12pt,epsf,rotating]{article}

\input axodraw.sty

\catcode`@=11
\def\citer{\@ifnextchar [{\@tempswatrue\@citexr}{\@tempswafalse\@citexr[]}}
 
\def\@citexr[#1]#2{\if@filesw\immediate\write\@auxout{\string\citation{#2}}\fi
  \def\@citea{}\@cite{\@for\@citeb:=#2\do
    {\@citea\def\@citea{--\penalty\@m}\@ifundefined
       {b@\@citeb}{{\bf ?}\@warning
       {Citation `\@citeb' on page \thepage \space undefined}}%
\hbox{\csname b@\@citeb\endcsname}}}{#1}}
\catcode`@=12
 
\reversemarginpar

\newcommand{\lsim}{\raisebox{-0.13cm}{~\shortstack{$<$ \\[-0.07cm] $\sim$}}~}
\newcommand{\gsim}{\raisebox{-0.13cm}{~\shortstack{$>$ \\[-0.07cm] $\sim$}}~}

\newcommand{\ov}[1]{\overline{#1}}

\newcommand{\ra}{\rightarrow}

\newcommand{\GeV}{\mbox{GeV}}

\newcommand{\nn}{\noindent}
\newcommand{\non}{\nonumber}

\newcommand{\beq}{\begin{equation}}
\newcommand{\eeq}{\end{equation}}
\newcommand{\bea}{\begin{eqnarray}}
\newcommand{\eea}{\end{eqnarray}}

\newcommand{\tg}{\mbox{tg}}
\newcommand{\tb}{\mbox{tg$\beta$}}

\newcommand{\eps}{\epsilon}

\def\msb{\overline{\rm MS}}
\def\ep{\epsilon}

\newbox\mycount
\newcommand{\ctowidth}[2]{ \setbox\mycount=\hbox{$#2$}
                          \hbox to \wd\mycount{$ \hss #1 \hss $} }
\newcommand{\ltowidth}[2]{ \setbox\mycount=\hbox{$#2$}
                          \hbox to \wd\mycount{$\hskip0pt plus0pt minus1fil
                           #1 \hfill $} }
\newcommand{\rtowidth}[2]{ \setbox\mycount=\hbox{$#2$}
                          \hbox to \wd\mycount{$\hfill #1
                          \hskip0pt plus0pt minus1fil$} }

\begin{document}

\begin{titlepage}

\begin{flushright}
CERN-TH/97-68 \\
hep-ph/9705337
\end{flushright}

\vspace{1cm}

\begin{center}

{\Large\sc QCD Effects in Higgs Physics}

\vspace{1cm}

{\large \sc Michael Spira} \\
\vspace{1cm}

{\it Theoretical Physics Division, CERN, CH-1211 Geneva 23, Switzerland}

\end{center}

\vspace{2cm}

\begin{abstract}
\normalsize
\noindent
Higgs boson production at the LHC within the Standard Model and its minimal
supersymmetric extension is reviewed. The predictions for decay rates and
production cross sections are updated by choosing the present value of the top
quark mass and recent parton density sets. Moreover, all relevant higher order
corrections, some of which have been obtained only recently, are included in a
consistent way.
\end{abstract}

\vspace*{\fill}

\begin{flushleft}
CERN-TH/97-68 \\
hep-ph/9705337 \\
April 1997
\end{flushleft}

\end{titlepage}

\setcounter{page}{2}

\tableofcontents

\section{Introduction}
\subsection{Standard Model}
The Higgs mechanism is a cornerstone of the Standard Model (SM). 
To formulate the standard electroweak theory consistently, the 
introduction of the fundamental Higgs field is necessary \cite{higgs}. 
It allows the particles of the Standard Model to be weakly 
interacting up to high energies without violating the unitarity 
bounds of scattering amplitudes. The unitarity requirement 
determines the couplings of the Higgs particle to all the 
other particles. These basic ideas can be cast into an elegant and
physically deep theory by formulating the electroweak theory
as a spontaneously broken gauge theory. Due to the fact that 
the gauge symmetry, though hidden, is still preserved, the 
theory is renormalizable \cite{thooftren}. The massive gauge bosons and 
the fermions acquire their masses through the interaction with
the Higgs field \cite{higgs}. The minimal model requires the introduction
of one weak isospin doublet leading, after the spontaneous
symmetry breaking, to the existence of one 
elementary scalar Higgs boson. Since all the couplings are 
predetermined, the properties of this particle are fixed by
its mass, which is the only unknown parameter of the Standard 
Model Higgs sector. Once the Higgs mass will be known, all 
decay widths and production processes of the Higgs particle
will be uniquely determined \cite{hunter}. The discovery of the Higgs 
particle will be the {\it experimentum crucis} for the standard
formulation of the electroweak theory.

Although the Higgs mass cannot be predicted in the Standard 
Model, there are several constraints that can be deduced from
consistency conditions on the model \citer{trivia,lattice}. Upper bounds can
be derived from the requirement that the Standard Model can be extended up to
a scale $\Lambda$, before perturbation theory breaks down and new
non-perturbative phenomena dominate the predictions of the theory. If the SM
is required to be weakly interacting up to the scale of grand unified theories
(GUTs), which is of ${\cal O}(10^{16}$ GeV), the Higgs mass has to be less than
$\sim 200$ GeV. For a minimal cut-off $\Lambda\sim 1$ TeV and the condition
$M_H<\Lambda$, a universal upper bound of $\sim 700$ GeV can be obtained from
renormalization group analyses \cite{trivia,meta} and
lattice simulations of the SM Higgs sector \cite{lattice}.

If the top quark mass is large, the Higgs potential may become unbounded from
below, rendering the SM vacuum unstable and thus inconsistent. The negative
contribution of the top quark, however, can be compensated by a positive
contribution due to the Higgs self-interaction, which is proportional to the
Higgs mass. Thus for a given top mass $M_t= 175$ GeV \cite{leptop,topmass}
a lower bound of $\sim 55$ GeV can be obtained for the Higgs
mass, if the SM remains weakly interacting up to scales $\Lambda\sim 1$ TeV.
For $\Lambda\sim M_{GUT}$ this lower bound is enhanced to $M_H\gsim 130$ GeV.
However, the assumption that the vacuum is metastable, with a lifetime larger
than the age of the Universe, decreases these lower bounds significantly for
$\Lambda \sim 1$~TeV, but only slightly for $\Lambda \sim M_{GUT}$
\cite{meta}.

The direct search in the LEP experiments via the process $e^+e^-\to Z^*H$
yields a lower bound of $\sim 77$ GeV on the Higgs mass \cite{lephiggs}. This
search is being extended at the present LEP2 experiments, which probe Higgs
masses up to about 95 GeV via the Higgs-strahlung process $Z^*\to ZH$
\citer{prohiggs,ztozh}. After LEP2 the search for the SM Higgs particle
will be
continued at the LHC for Higgs masses up to the theoretical upper limit
\cite{higgslhc,atlascms}. The dominant Higgs production mechanism at the LHC
will be the gluon-fusion process \cite{glufus}
\begin{displaymath}
pp\to gg\to H \, ,
\end{displaymath}
which provides the largest production cross section for the whole Higgs mass
range of interest. For large Higgs masses the $W$ and $Z$ boson-fusion processes
\cite{vvh,vvhqcd}
\begin{displaymath}
pp\to qq\to qq+WW/ZZ\to Hqq
\end{displaymath}
become competitive. In the intermediate mass range $M_Z<M_H<2M_Z$
Higgs-strahlung off top quarks \cite{htt} and $W,Z$ gauge bosons
\cite{vhv,vhvqcd} provide alternative signatures for the Higgs boson search.

The detection of the Higgs boson at the LHC will be divided into two mass
regions:
\begin{description}
\item[(i)~] For $M_W\lsim M_H\lsim 140$ GeV the only promising decay mode is
the rare photonic one, $H\to \gamma\gamma$, which will be discriminated
against the large QCD continuum background by means of excellent energy and
angular resolutions of the detectors \cite{atlascms}. Alternatively excellent
$\mu$-vertex detectors might allow the detection of the dominant $b\bar b$ decay
mode \cite{btag}, although the overwhelming QCD background remains very
difficult to reject \cite{antibtag}. In order to reduce the background it may
be helpful to tag the additional $W$ boson in the Higgs-strahlung process
$pp\to HW$ \cite{vhv,vhvqcd} or the $t\bar t$ pair in Higgs bremsstrahlung
off top quarks, $pp\to Ht\bar t$ \cite{htt}.

\item[(ii)] In the mass range 140 GeV$ \lsim M_H \lsim 800$ GeV the search for
the Higgs particle can be performed by looking for final states containing 4
charged leptons, which originate from the Higgs decay $H\to ZZ^{(*)}$
\cite{atlascms}. The QCD
background will be small so that the signal can be extracted quite easily.
For the Higgs mass region 155 GeV $\lsim M_H \lsim 180$ GeV another possibility
arises from the Higgs decay $H\to WW^{(*)}\to l^+l^-\nu\bar \nu$
\cite{dreiditt}, because
the $W$ boson decay mode is dominating by more than one order of magnitude in
this mass range,
while the $Z$ pair decay mode may be difficult to detect due to a strong dip
in the branching ratio BR($H\to ZZ^*$) for Higgs masses around the $W$ pair
threshold. For Higgs masses above $\sim 800$ GeV the search may be extended by
looking for the decay chains $H\to ZZ,WW\to ll\nu\nu$. A Higgs boson search
up to $\sim 1$ TeV seems to be feasible at the LHC \cite{atlascms}.

\end{description}

In order to investigate the Higgs search potential of the LHC, it is of vital
importance to have reliable predictions for the production cross sections and
decay widths of the Higgs boson. In the past higher order corrections have
been evaluated for the most important processes. They are in general
dominated by QCD corrections. The present level leads to a significantly
improved and reliable determination of the signal processes involved in the
Higgs boson search at the LHC.

\subsection{Supersymmetric Extension}
Supersymmetric extensions of the SM \cite{susyref,mssmbase} are strongly
motivated by the idea of providing a
solution of the hierarchy problem in the SM Higgs sector. They allow for a
light Higgs particle in the context of GUTs \cite{lighth}, in contrast with the
SM, where the extrapolation requires an unsatisfactory fine-tuning of the
SM parameters. Supersymmetry is a symmetry between fermionic and bosonic
degrees of freedom and thus the most general symmetry of the $S$-matrix. The
minimal supersymmetric extension of the SM (MSSM) yields a prediction of the
Weinberg angle in agreement with present experimental measurements in the
context of GUTs \cite{sw}. Moreover, it does not exhibit any quadratic
divergences, in contrast with the SM Higgs sector. Throughout this review we
will concentrate on the MSSM only, although most of the results will also be
qualitatively valid for non-minimal supersymmetric extensions \cite{nmssm}.

In the MSSM two isospin Higgs doublets have to be introduced in order to
preserve supersymmetry \cite{twoiso}. After the electroweak symmetry-breaking
mechanism, three of the eight degrees of freedom are absorbed by the $Z$ and
$W$ gauge bosons,
leading to the existence of five elementary Higgs particles. These consist of
two CP-even neutral (scalar) particles $h,H$, one CP-odd neutral (pseudoscalar)
particle $A$, and two charged particles $H^\pm$.
In order to describe the MSSM Higgs sector one has to introduce
four masses $M_h$, $M_H$, $M_A$ and $M_{H^\pm}$ and
two additional parameters, which define the properties of the scalar particles
and their interactions with gauge bosons and fermions: the mixing angle
$\beta$, related to the ratio of the two vacuum expectation values,
$\tb = v_2/v_1$, and the mixing angle $\alpha$ in the
neutral CP-even sector. Due to supersymmetry there are several
relations among these parameters, and only two of them are
independent.  These relations lead to a hierarchical structure of the Higgs
mass spectrum [in lowest order: $M_h<M_Z , M_A < M_H$ and $M_W <M_{H^\pm}]$.
This is, however, broken by radiative corrections, which are dominated by
top-quark-induced contributions \cite{mssmrad1,mssmrad2}. The parameter $\tb$
will in general be assumed to be in the
range $1 < \tg \beta < m_t/m_b$ $[ \pi/4 < \beta < \pi/2] $, consistent with
the assumption that the MSSM is the low-energy limit of a supergravity model.

The input parameters of the MSSM Higgs sector are generally chosen to be the
mass $M_A$ of the pseudoscalar Higgs boson and $\tb$.
All other masses and the mixing angle $\alpha$ can be derived from these basic
parameters [and the top and squark masses, which enter through radiative
corrections]. In the following qualitative discussion of the radiative
corrections we shall neglect, for the sake of simplicity, non-leading effects
due to non-zero values of the supersymmetric Higgs mass parameter $\mu$ and
of the mixing parameters $A_t$ and $A_b$ in the soft symmetry-breaking
interaction. The radiative corrections are then determined by the parameter
$\epsilon$, which grows with the fourth power of the top quark mass $M_t$ and
logarithmically with the squark mass $M_S$,
\begin{equation}
\epsilon = \frac{3G_F}{\sqrt{2}\pi^2}\frac{M_t^4}{\sin^2\beta}\log\left(1+
\frac{M_S^2}{M_t^2} \right) \, .
\label{eq:epsusy}
\end{equation}
These corrections are positive and they increase the mass of the light neutral
Higgs boson $h$. The dependence of the upper limit of $M_h$ on the top
quark mass $M_t$ can be expressed as
\begin{equation}
M^2_h \le M_Z^2 \cos^2 2\beta + \epsilon \sin^2\beta \, .
\label{eq:hbound}
\end{equation}
In this approximation, the upper bound on $M_h$ is shifted from the tree level
value $M_Z$ up to $\sim$ 140 GeV for $M_t = 175$ GeV.
Taking $M_A$ and $\tb$ as the basic input parameters, the mass of the lightest
scalar state $h$ is given by
\begin{eqnarray}
M^2_h & = & \frac{1}{2} \left[ M_A^2 + M_Z^2 + \epsilon \right.
\non \\
& & \left. - \sqrt{(M_A^2+M_Z^2+\epsilon)^2
-4 M_A^2M_Z^2 \cos^2 2\beta
-4\epsilon (M_A^2 \sin^2\beta + M_Z^2 \cos^2\beta)} \right] \, .
\label{eq:hmass}
\end{eqnarray}
\nn The masses of the heavy neutral and charged Higgs bosons
are determined by the sum rules
\begin{eqnarray}
M_H^2 & = & M_A^2 + M_Z^2 - M_h^2 + \epsilon \non \\
M_{H^\pm}^2 & = & M_A^2 + M_W^2 \, .
\end{eqnarray}
The mixing parameter $\alpha$ is fixed by $\tb$ and
the Higgs mass $M_A$,
\begin{equation}
\tg 2 \alpha = \tg 2\beta \frac{M_A^2 + M_Z^2}{M_A^2 - M_Z^2 +
\epsilon/\cos 2\beta} \ \ \ \ \mbox{with} \ \ \  -\frac{\pi}{2}<\alpha <0 \, .
\end{equation}
The couplings of the various neutral Higgs bosons to fermions and gauge
bosons depend on the angles $\alpha$ and $\beta$. Normalized
to the SM Higgs couplings, they are listed in Table~\ref{tb:hcoup}.
The pseudoscalar particle $A$ does not couple to gauge bosons at tree level,
and its couplings to down (up)-type fermions are (inversely) proportional
to $\tb$.
\begin{table}[hbt]
\renewcommand{\arraystretch}{1.5}
\begin{center}
\begin{tabular}{|lc||ccc|} \hline
\multicolumn{2}{|c||}{$\Phi$} & $g^\Phi_u$ & $g^\Phi_d$ &  $g^\Phi_V$ \\
\hline \hline
SM~ & $H$ & 1 & 1 & 1 \\ \hline
MSSM~ & $h$ & $\cos\alpha/\sin\beta$ & $-\sin\alpha/\cos\beta$ &
$\sin(\beta-\alpha)$ \\
& $H$ & $\sin\alpha/\sin\beta$ & $\cos\alpha/\cos\beta$ &
$\cos(\beta-\alpha)$ \\
& $A$ & $ 1/\tg\beta$ & $\tg\beta$ & 0 \\ \hline
\end{tabular} 
\renewcommand{\arraystretch}{1.2}
\caption[]{\label{tb:hcoup}
\it Higgs couplings in the MSSM to fermions and gauge bosons [$V=W,Z$]
relative to SM couplings.}
\end{center}
\end{table}

\begin{figure}[hbtp]

\vspace*{-0.9cm}
\hspace*{-5.5cm}
\begin{turn}{-90}%
\epsfxsize=12cm \epsfbox{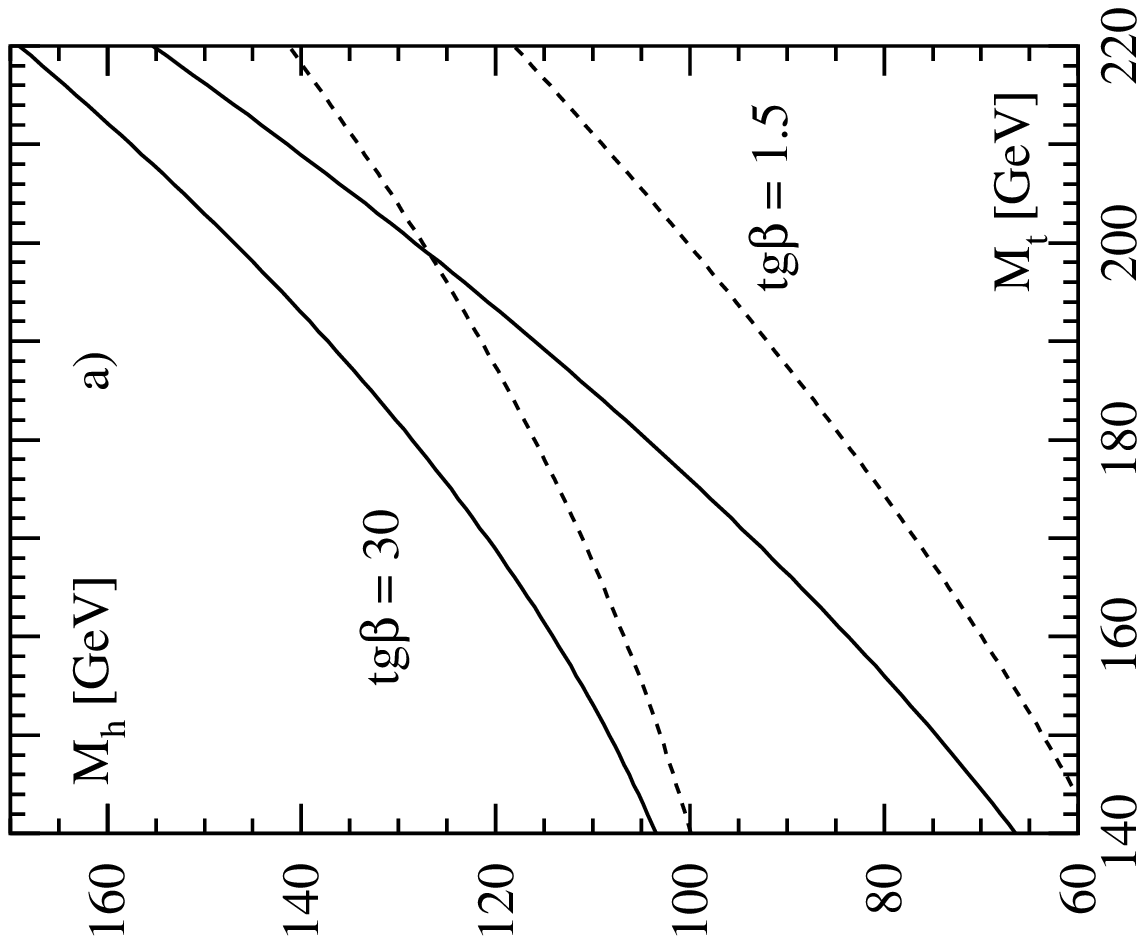}
\end{turn}
\vspace*{-12.05cm}

\hspace*{2.5cm}
\begin{turn}{-90}%
\epsfxsize=12cm \epsfbox{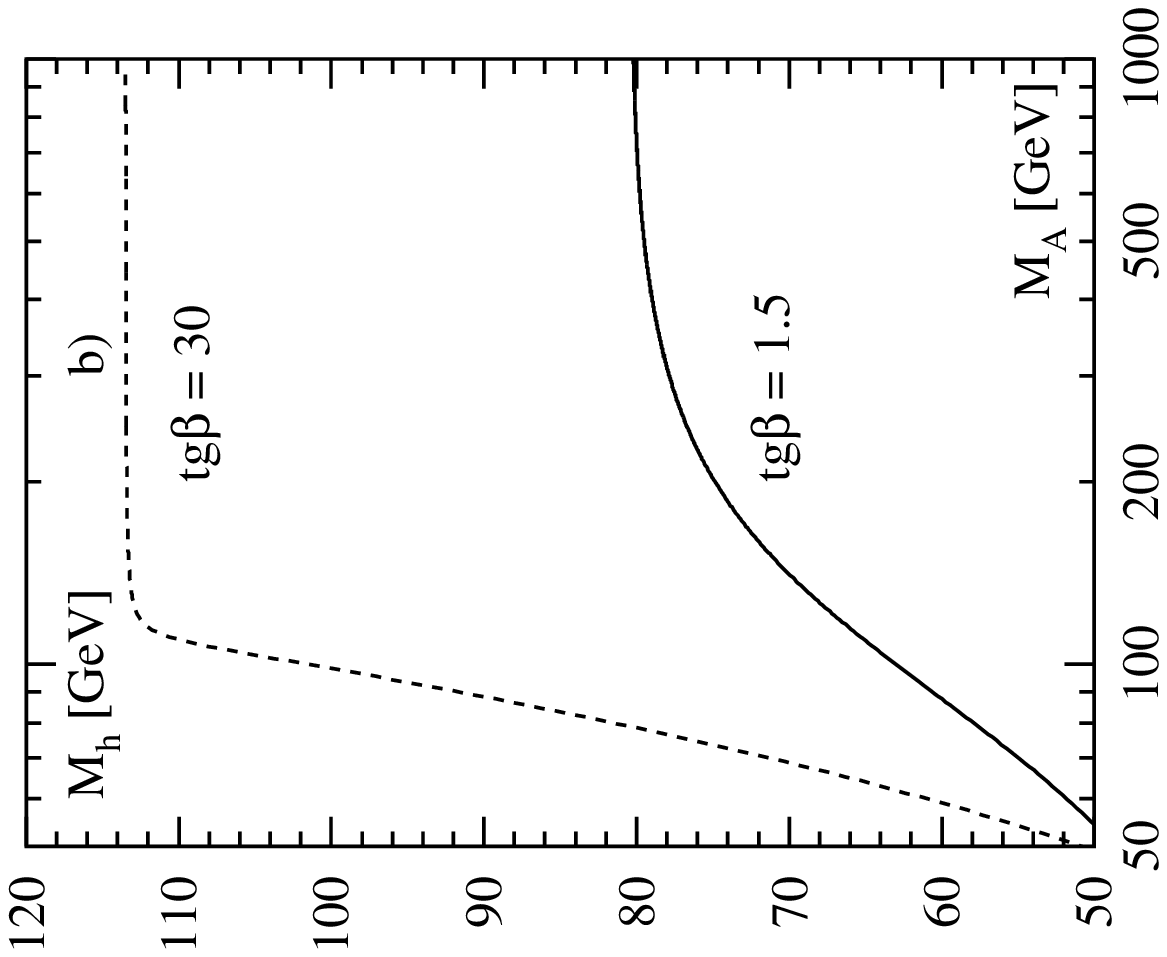}
\end{turn}
\vspace*{-3.5cm}

\hspace*{-5.5cm}
\begin{turn}{-90}%
\epsfxsize=12cm \epsfbox{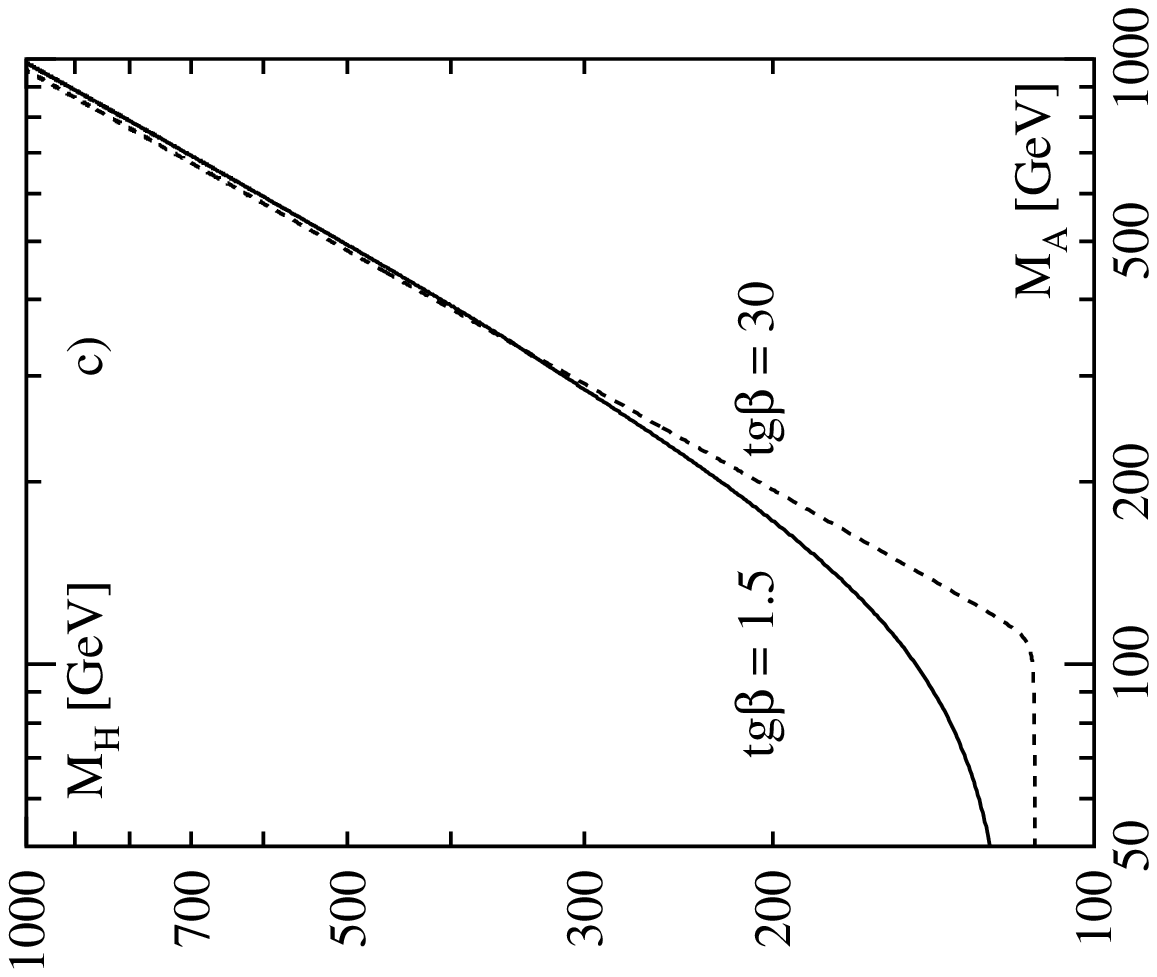}
\end{turn}
\vspace*{-12.05cm}

\hspace*{2.5cm}
\begin{turn}{-90}%
\epsfxsize=12cm \epsfbox{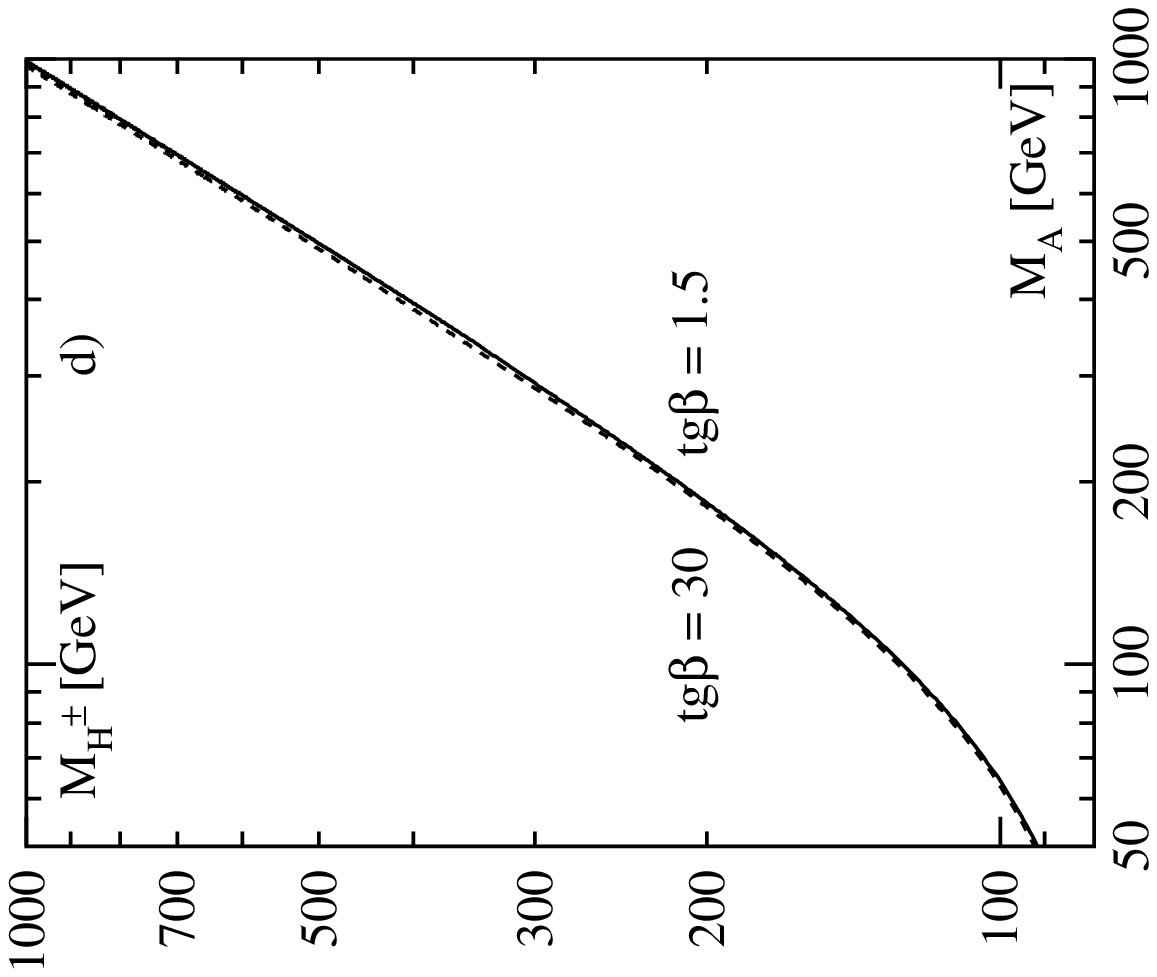}
\end{turn}
\vspace*{-2.5cm}

\caption[ ]{\label{fg:mssmhiggs} \it (a) The upper limit on the light scalar
Higgs pole mass in the MSSM as a function of the top quark mass for two values
of $\tb=1.5,30$. The top quark mass has been chosen as $M_t=175$ GeV and the
common squark mass as $M_S=1$ TeV. The full lines correspond to the maximal
mixing case [$A_t=\sqrt{6} M_S$, $A_b=\mu=0$] and the dashed lines to vanishing
mixing. The pole masses of the other Higgs bosons, $H,A,H^\pm$, are shown as a
function of the pseudoscalar mass in (b--d) for two values of $\tb=1.5, 30$ and
vanishing mixing.}
\end{figure}
Recently the radiative corrections to the MSSM Higgs sector have been calculated
up to the two-loop level in the effective potential approach \cite{mssmrad2}.
The two-loop
corrections are dominated by the QCD corrections to the top-quark-induced
contributions. They decrease the upper bound on the light scalar Higgs mass
$M_h$ by about 10 GeV. The variation of $M_h$ with the top quark mass is shown
in Fig.~\ref{fg:mssmhiggs}a for $M_S=1$ TeV and two representative values of
$\tb =1.5$ and 30. While the dashed curves correspond to the case of vanishing
mixing parameters $\mu=A_t=A_b=0$, the solid lines correspond to the maximal
mixing case, defined by the Higgs mass
parameter $\mu = 0$ and the Yukawa parameters $A_b=0$, $A_t = \sqrt{6}M_S$.
The upper bound on $M_h$ amounts to $\sim 130$ GeV for $M_t=175$ GeV. For the
two values of $\tb$ introduced above, the Higgs masses $M_h, M_H$ and
$M_{H^\pm}$ are presented in Figs.~\ref{fg:mssmhiggs}b-d as a function of the
pseudoscalar mass $M_A$ for vanishing mixing parameters.
The dependence on the mixing parameters
$\mu, A_t, A_b$ is rather weak and the effects on the masses are limited by a
few GeV \cite{mssmhiggs}.

\begin{figure}[hbtp]
\vspace*{1.5cm}
\hspace*{-7.0cm}
\begin{turn}{-90}%
\epsfxsize=19cm \epsfbox{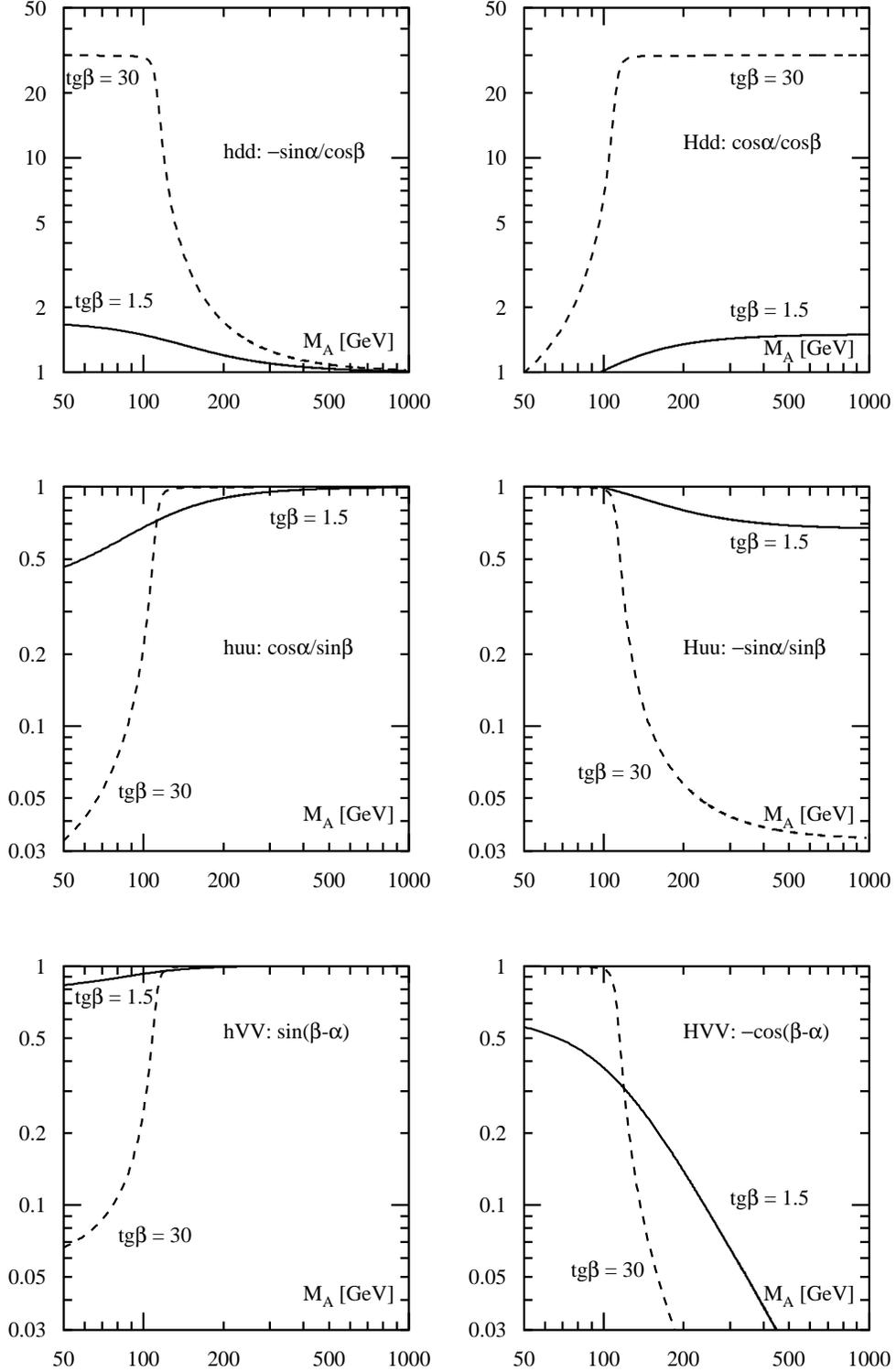}
\end{turn}
\vspace*{-0.5cm}
\caption[ ]{\label{fg:mssmcoup} \it The coupling parameters of the neutral MSSM
Higgs bosons as a function of the pseudoscalar mass $M_A$ for two values of $\tb
=1.5,30$ and vanishing mixing. They are defined in Table \ref{tb:hcoup}.}
\end{figure}
The MSSM couplings of Table \ref{tb:hcoup} are shown in Fig.~\ref{fg:mssmcoup}
as functions of the pseudoscalar mass $M_A$ for two values of $\tb=1.5$ and
30 and vanishing mixing parameters. The mixing effects are weak and thus
phenomenologically unimportant. For large values of $\tb$ the Yukawa couplings
to (up) down-type quarks are (suppressed) enhanced and vice
versa. Moreover, it can be inferred from Fig.~\ref{fg:mssmcoup} that the
couplings of the light scalar Higgs particle approach the SM values for
large pseudoscalar masses, i.e.~in the decoupling regime. Thus it will be
difficult to distinguish the light scalar MSSM Higgs boson from the SM Higgs
particle, in the region where all Higgs particles except the light scalar one
are very heavy.

\subsection{Organization of the Paper}
In this work we will review and update all Higgs decay widths and branching
ratios as well as all relevant Higgs boson production cross sections at the
LHC within the SM and MSSM. Previous reviews can be found in
Refs.~\cite{abdel, kniehl}. However, this work contains substantial
improvements due to our use of new results. Moreover, we will use recent
parametrizations of parton densities for the production cross sections at the
LHC.

This paper is organized as follows. In Section 2 we will review the decay rates
and production processes of the SM Higgs particle at the LHC. Section 3 will
present the corresponding decay rates and production cross sections for the
Higgs bosons of the minimal supersymmetric extension.  A summary will be given
in Section 4.

\section{Standard Model}
\subsection{Decay Modes}
The strength of the Higgs-boson interaction with SM particles grows with their
masses. Thus the Higgs boson predominantly couples to the heaviest
particles of the SM, i.e.\ $W,Z$ gauge bosons, top and bottom quark. The
decays into these particles will be dominant, if they are kinematically allowed.
All decay modes discussed in this section are obtained by means of the FORTRAN
program HDECAY \cite{hdecay,QCD}\footnote{The program can be obtained from
http://wwwcn.cern.ch/$\sim$mspira/.}.

\subsubsection{Lepton and heavy quark pair decays of the SM Higgs particle}
In lowest order the leptonic decay width of the SM Higgs boson is given by
\cite{prohiggs,hll}
\begin{equation}
\Gamma [H\to l^+ l^-] = \frac{G_F M_H } {4\sqrt{2}\pi}
\ m_l^2 \beta^3
\label{eq:hll}
\end{equation}
with $\beta = (1-4m_l^2/M_H^2)^{1/2}$ being the velocity of the leptons. The
branching ratio of decays into $\tau$ leptons amounts to about 10\% in the
intermediate mass range. Muonic decays can reach a level of a few $10^{-4}$,
and all other leptonic decay modes are phenomenologically unimportant.

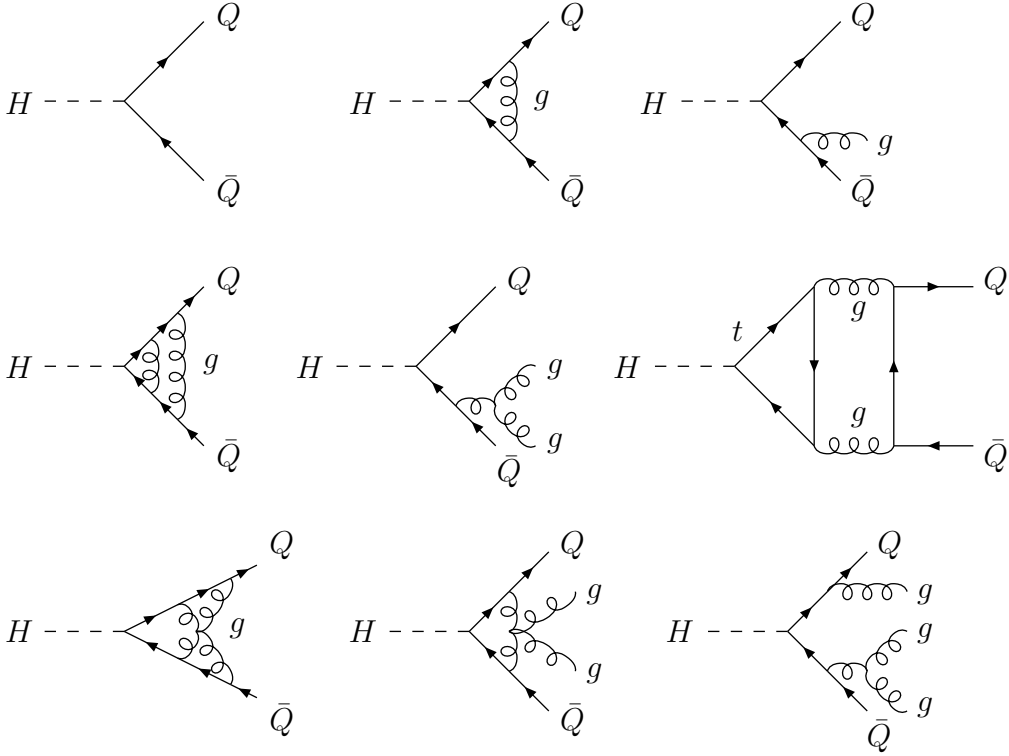
\begin{figure}[hbt]
\begin{center}
\setlength{\unitlength}{1pt}
\begin{picture}(350,300)(0,0)


\ArrowLine(30,250)(60,280)
\ArrowLine(60,220)(30,250)
\DashLine(0,250)(30,250){5}
\put(-15,246){$H$}
\put(65,280){$Q$}
\put(65,213){$\bar Q$}


\Gluon(175,235)(175,265){-3}{3}
\ArrowLine(160,250)(175,265)
\ArrowLine(175,265)(190,280)
\ArrowLine(190,220)(175,235)
\ArrowLine(175,235)(160,250)
\DashLine(130,250)(160,250){5}
\put(115,246){$H$}
\put(195,280){$Q$}
\put(195,213){$\bar Q$}
\put(185,250){$g$}

\Gluon(285,235)(310,235){3}{2}
\ArrowLine(270,250)(300,280)
\ArrowLine(300,220)(285,235)
\ArrowLine(285,235)(270,250)
\DashLine(240,250)(270,250){5}
\put(225,246){$H$}
\put(305,280){$Q$}
\put(305,213){$\bar Q$}
\put(315,233){$g$}


\Gluon(40,140)(40,160){-3}{2}
\Gluon(50,130)(50,170){-3}{4}
\ArrowLine(30,150)(40,160)
\ArrowLine(40,160)(50,170)
\ArrowLine(50,170)(60,180)
\ArrowLine(60,120)(50,130)
\ArrowLine(50,130)(40,140)
\ArrowLine(40,140)(30,150)
\DashLine(0,150)(30,150){5}
\put(-15,146){$H$}
\put(65,180){$Q$}
\put(65,113){$\bar Q$}
\put(60,150){$g$}

\Gluon(155,135)(170,135){3}{1}
\Gluon(170,135)(185,150){3}{2}
\Gluon(170,135)(185,120){-3}{2}
\ArrowLine(140,150)(170,180)
\ArrowLine(170,120)(155,135)
\ArrowLine(155,135)(140,150)
\DashLine(110,150)(140,150){5}
\put(95,146){$H$}
\put(175,180){$Q$}
\put(171,107){$\bar Q$}
\put(190,120){$g$}
\put(190,148){$g$}

\DashLine(230,150)(260,150){5}
\ArrowLine(260,150)(290,180)
\ArrowLine(290,120)(260,150)
\ArrowLine(290,180)(290,120)
\Gluon(290,180)(320,180){3}{3}
\Gluon(290,120)(320,120){-3}{3}
\ArrowLine(320,120)(320,180)
\ArrowLine(320,180)(350,180)
\ArrowLine(350,120)(320,120)
\put(215,146){$H$}
\put(355,180){$Q$}
\put(355,113){$\bar Q$}
\put(260,160){$t$}
\put(305,170){$g$}
\put(305,130){$g$}


\Gluon(50,40)(57,50){-3}{1}
\Gluon(57,50)(50,60){-3}{1}
\Gluon(70,30)(57,50){-3}{2}
\Gluon(57,50)(70,70){-3}{2}
\ArrowLine(30,50)(50,60)
\ArrowLine(50,60)(70,70)
\ArrowLine(70,70)(80,75)
\ArrowLine(80,25)(70,30)
\ArrowLine(70,30)(50,40)
\ArrowLine(50,40)(30,50)
\DashLine(0,50)(30,50){5}
\put(-15,46){$H$}
\put(85,80){$Q$}
\put(85,13){$\bar Q$}
\put(70,50){$g$}

\Gluon(175,35)(175,50){-3}{1}
\Gluon(175,50)(175,65){-3}{1}
\Gluon(175,50)(200,65){-3}{2}
\Gluon(175,50)(200,35){3}{2}
\ArrowLine(160,50)(175,65)
\ArrowLine(175,65)(190,80)
\ArrowLine(190,20)(175,35)
\ArrowLine(175,35)(160,50)
\DashLine(130,50)(160,50){5}
\put(115,46){$H$}
\put(195,80){$Q$}
\put(195,13){$\bar Q$}
\put(205,63){$g$}
\put(205,33){$g$}

\Gluon(295,65)(325,65){3}{3}
\Gluon(295,35)(310,35){3}{1}
\Gluon(310,35)(325,50){3}{2}
\Gluon(310,35)(325,20){-3}{2}
\ArrowLine(280,50)(295,65)
\ArrowLine(295,65)(310,80)
\ArrowLine(310,20)(295,35)
\ArrowLine(295,35)(280,50)
\DashLine(250,50)(280,50){5}
\put(235,46){$H$}
\put(315,80){$Q$}
\put(311,07){$\bar Q$}
\put(330,20){$g$}
\put(330,48){$g$}
\put(330,63){$g$}

\end{picture}  \\
\setlength{\unitlength}{1pt}
\caption[ ]{\label{fg:hqqdia} \it Typical diagrams contributing to $H\to
Q\bar Q$ at lowest order and one-, two- and three-loop QCD.}
\end{center}
\end{figure}
For large Higgs masses the particle width for decays to $b,c$ quarks [directly
coupling to the SM Higgs particle] is given up to three-loop
QCD corrections [typical diagrams are depicted in
Fig.~\ref{fg:hqqdia}] by the well-known expression \citer{drees,chet}
\begin{equation}
\Gamma [H\to Q{\overline{Q}}] = \frac{3 G_F M_H } {4\sqrt{2}\pi}
\ \overline{m}_Q^2(M_H)\ [\Delta_{\rm QCD}+\Delta_t]
\label{eq:hqq}
\end{equation}
with
\begin{eqnarray*}
\Delta_{\rm QCD} & = & 1 + 5.67 \frac{\alpha_s (M_H)}{\pi} + (35.94 - 1.36
N_F) \left( \frac{\alpha_s (M_H)}{\pi} \right)^2 \nonumber \\
& & + (164.14 - 25.77 N_F + 0.259 N_F^2) \left( \frac{\alpha_s(M_H)}{\pi}
\right)^3 \\
\Delta_t & = & \left(\frac{\alpha_s (M_H)}{\pi}\right)^2 \left[ 1.57 -
\frac{2}{3} \log \frac{M_H^2}{M_t^2} + \frac{1}{9} \log^2
\frac{\overline{m}_Q^2 (M_H)}{M_H^2} \right] \non
\end{eqnarray*}
in the ${\overline{\rm MS}}$ renormalization scheme; the running quark
mass and the QCD coupling are defined at the scale of the Higgs mass,
absorbing in this way large mass logarithms. The quark masses can be neglected
in general, except for heavy quark decays in the threshold region. The QCD
corrections in this case are given, in terms of the quark {\it pole} mass
$M_Q$, by \cite{drees}
\begin{eqnarray}
\Gamma [H \ra Q\bar{Q}\,]= \frac{3 G_F M_H}{4 \sqrt{2} \pi} \,
\, M_Q^2 \, \beta^3 \, \left[ 1 +\frac{4}{3} \frac{\alpha_s}{\pi}
\Delta^H \right]
\label{eq:qcdmass}
\end{eqnarray}
where $\beta = (1-4M_Q^2/M_H^2)^{1/2}$ denotes the velocity of the heavy quarks
$Q$. To leading order, the QCD correction factor reads as \cite{drees}
\begin{eqnarray}
\Delta^H = \frac{1}{\beta}A(\beta) + \frac{1}{16\beta^3}(3+34\beta^2-
13 \beta^4)\log \frac{1+\beta}{1-\beta} +\frac{3}{8\beta^2}(7 \beta^2-1) \, ,
\label{eq:dqcdmass}
\end{eqnarray}
with
\begin{eqnarray}
A(\beta) &= & (1+\beta^2) \left[ 4 {\rm Li}_2 \left( \frac{1-\beta}{1+\beta}
\right) +2 {\rm Li}_2 \left( -\frac{1-\beta}{1+\beta} \right) -3 \log
\frac{1+\beta}{1-\beta} \log \frac{2}{1+\beta} \right. \non \\
& & \left. -2 \log \frac{1+\beta}{1-\beta} \log \beta \right] -
3 \beta \log \frac{4}{1-\beta^2} -4 \beta \log \beta \, .
\non
\end{eqnarray}
[Li$_2$ denotes the Spence function, Li$_2(x)= -\int_0^x dy y^{-1} \log(1-y)$.]
Recently the full massive two-loop corrections of ${\cal O}(N_F
\alpha_s^2)$ have been computed; they are part of the full massive two-loop
result \cite{melnikov}.

The relation between the perturbative {\it pole} mass $M_Q$ of the heavy quarks
and the $\overline{\rm MS}$ mass $\overline{m}_Q(M_Q)$ at the scale of the pole
mass can be expressed as \cite{broad}
\begin{equation}
{\overline{m}}_{Q}(M_{Q})= \frac{M_{Q}}{\displaystyle 1+\frac{4}{3}
\frac{\alpha_{s}(M_Q)}{\pi} + K_Q \left(\frac{\alpha_s(M_Q)}{\pi}\right)^2}\, ,
\label{eq:mspole}
\end{equation}
where the numerical values of the NNLO coefficients are given by
$K_t\sim 10.9$, $K_b \sim 12.4$ and $K_c \sim 13.4$.
Since the relation between the pole mass $M_{c}$ of the charm quark and
the ${\overline{\rm MS}}$ mass ${\overline{m}}_{c}(M_{c})$ evaluated at the
pole mass is badly convergent \cite{broad}, the running
quark masses ${\overline{m}}_{Q}(M_{Q})$ have to be adopted as starting points.
[They have been extracted directly from QCD sum rules evaluated in a
consistent ${\cal O}(\alpha_{s})$ expansion \cite{narison}.]
In the following we will denote the pole mass corresponding to the full NNLO
relation in eq.~(\ref{eq:mspole}) by $M_Q^{pt3}$ and the pole
mass corresponding to the NLO relation [omitting the contributions of
$K_Q$] by $M_Q^{pt2}$ according to Ref.~\cite{narison}. Typical values of the
different mass definitions are presented in Table \ref{tb:qmass}. It
is apparent that the NNLO correction to the charm pole mass is comparable
to the NLO contribution starting from the $\overline{\rm MS}$ mass.
\begin{table}[hbt]
\renewcommand{\arraystretch}{1.5}
\begin{center}
\begin{tabular}{|c||c|c|c|c|} \hline
$Q$ & $\overline{m}_Q (M_Q)$ & $M_Q^{pt2}$ & $M_Q^{pt3}$ & $\overline{m}_Q
(100~\GeV)$ \\ \hline \hline
$c$ & 1.23 GeV               & 1.42 GeV    & 1.64 GeV  & 0.62 GeV  \\
$b$ & 4.23 GeV               & 4.62 GeV    & 4.87 GeV  & 2.92 GeV  \\
$t$ & 167.4 GeV              & 175.0 GeV   & 177.1 GeV & 175.1 GeV \\ \hline
\end{tabular}
\renewcommand{\arraystretch}{1.2}
\caption[]{\label{tb:qmass} \it Quark mass values for the $\overline{MS}$
mass
and the two different definitions of the pole masses. The strong coupling has
been chosen as $\alpha_s(M_Z)=0.118$, and the bottom and charm mass values are
taken from Ref.~\cite{narison}. The last column shows the values of the running
$\overline{MS}$ masses at a typical scale $\mu=100$ GeV.}
\end{center}
\end{table}

The evolution from $M_{Q}$ upwards to a renormalization scale $\mu$ can be
expressed as
\begin{eqnarray}
{\overline{m}}_{Q}\,(\mu )&=&{\overline{m}}_{Q}\,(M_{Q})
\,\frac{c\,[\alpha_{s}\,(\mu)/\pi ]}{c\, [\alpha_{s}\,(M_{Q})/\pi ]}
\label{eq:msbarev}
\end{eqnarray}
with the coefficient function \cite{runmass,runmass4}
\begin{eqnarray*}
c(x)&=&\left(\frac{9}{2}\,x\right)^{\frac{4}{9}} \,
[1+0.895x+1.371\,x^{2} + 1.952\, x^3]
\hspace{1.35cm} \mbox{for} \hspace{.2cm} M_{s}\,<\mu\,<M_{c}\\
c(x)&=&\left(\frac{25}{6}\,x\right)^{\frac{12}{25}} \,
[1+1.014x+1.389\,x^{2} + 1.091\, x^3]
\hspace{1.0cm} \mbox{for} \hspace{.2cm} M_{c}\,<\mu\,<M_{b}\\
c(x)&=&\left(\frac{23}{6}\,x\right)^{\frac{12}{23}} \,
[1+1.175x+1.501\,x^{2} + 0.1725\, x^3]
\hspace{1cm} \mbox{for} \hspace{.2cm} M_{b}\,<\mu \,< M_t \\
c(x)&=&\left(\frac{7}{2}\,x\right)^{\frac{4}{7}} \,
[1+1.398x+1.793\,x^{2} - 0.6834\, x^3]
\hspace{1.35cm} \mbox{for} \hspace{.2cm} M_{t}\,<\mu \, .
\end{eqnarray*}
For the charm quark mass the evolution is determined by eq.~(\ref{eq:msbarev})
up to the scale $\mu=M_b$, while for scales above the bottom mass the
evolution must be restarted at $M_Q = M_b$. The values of the
running $b,c$ masses at the scale $\mu = 100$ \GeV, characteristic of
the relevant Higgs masses, are typically 35\% (60\%) smaller than the bottom
(charm) pole
masses $M_b^{pt2}$ ($M_c^{pt2})$ as can be inferred from the last column in
Table \ref{tb:qmass}. Thus the QCD corrections turn out to be large in the
large Higgs mass regime reducing the lowest order expression [in terms of the
quark {\it pole} masses] by about 50\% (75\%) for bottom (charm) quarks. The
QCD corrections are moderate in the threshold regions apart from a Coulomb
singularity at threshold, which however is regularized by the finite heavy
quark decay width in the case of the top quark.

\begin{figure}[hbt]
\vspace*{0.5cm}
\hspace*{0.0cm}
\begin{turn}{-90}%
\epsfxsize=10cm \epsfbox{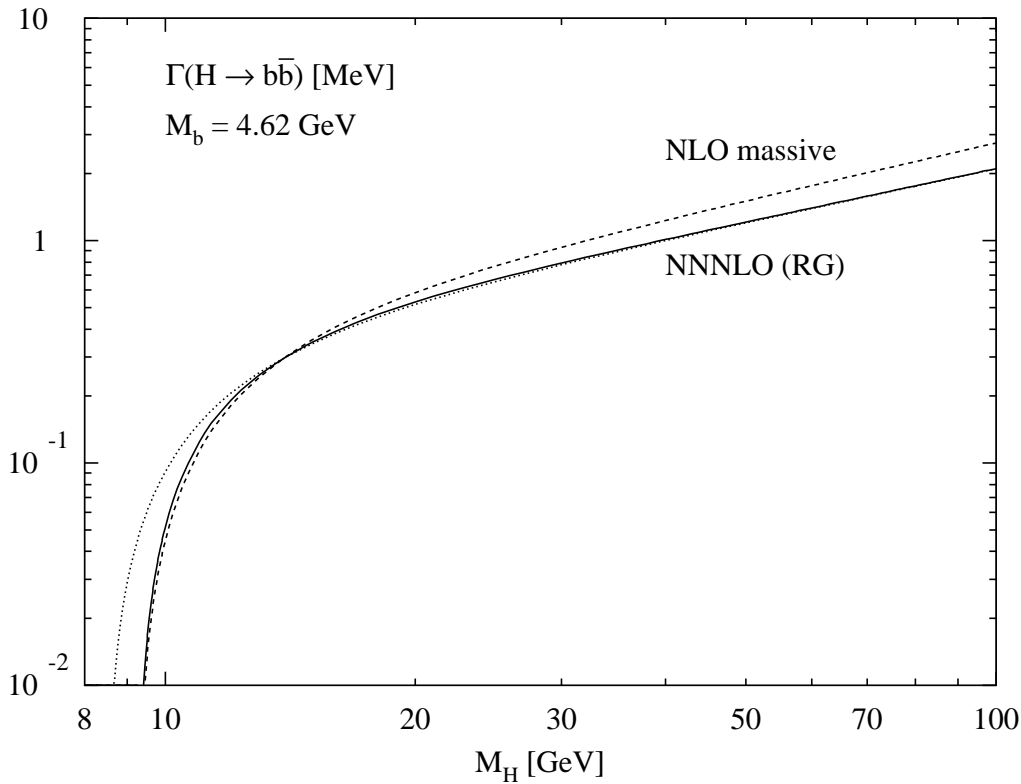}
\end{turn}
\vspace*{0.0cm}
\caption[ ]{\label{fg:qintpol} \it Interpolation between the full massive NLO
expression (dashed line) for the $b\bar b$ decay width of the Standard Higgs
boson and the renormalization group improved NNNLO result (dotted line). The
interpolated curve is presented by the full line.}
\end{figure}

In the threshold region mass effects are important so that the preferred
expression for the heavy quark decay width is given by eq.~(\ref{eq:qcdmass}).
Far above the threshold the massless ${\cal O}(\alpha_s^3)$ result of
eq.~(\ref{eq:hqq}) fixes the most improved result for this decay mode. The
transition between the two regions is performed by a linear interpolation as can
be inferred from Fig.~\ref{fg:qintpol}, thus yielding an optimized
description of the mass effects in the threshold region and the renormalization
group improved large Higgs mass regime.

Electroweak corrections to heavy quark and lepton decays are well under control
\cite{hqvelw,hqqelw}. In the intermediate mass range they can be approximated by
\cite{hqqelwapp}
\begin{equation}
\delta_{\rm elw} = \frac{3}{2} \frac{\alpha}{\pi}e_f^2 \left(\frac{3}{2} -
\log \frac{M_H^2}{M_f^2} \right) + \frac{G_F}{8 \pi^2 \sqrt{2}}
\left\{ k_f M_t^2 + M_W \left[ - 5 + \frac{3}{s_W^2} \log c_W^2 \right]
- M_Z^2 \frac{6 v_f^2 - a_f^2}{2} \right\}
\end{equation}
with $v_f = 2I_{3f} - 4e_f s_W^2$ and $a_f = 2I_{3f}$. $I_{3f}$ denotes the
third component of the electroweak isospin, $e_f$ the electric charge of the
fermion $f$ and $s_W = \sin\theta_W$ the Weinberg angle; $\alpha$ denotes the
QED coupling, $M_t$ the top quark mass and $M_W$ the $W$ boson mass.
The large logarithm $\log M_H^2/M_f^2$ can be absorbed in the running fermion
mass analogous to the QCD corrections.
The coefficient $k_f$ is equal to 7 for decays into leptons and light quarks;
for $b$ quarks it is reduced to 1 due to additional contributions involving top
quarks inside the vertex corrections. Recently the two- and three-loop QCD
corrections to the $k_f$ terms have been computed by means of low-energy
theorems \cite{hbb}. The results imply the replacements
\begin{eqnarray}
k_f & \to & k_f\times\left\{1-\frac{1}{7}\left(\frac{3}{2}+\zeta_2\right)
\frac{\alpha_s(M_t)}{\pi} \right\} \hspace{1cm}\mbox{for $f\neq b$}\nonumber \\
k_b & \to & k_b\times\left\{1-4\left(1+\zeta_2\right)\frac{\alpha_s(M_t)}{\pi}
\right\} \, .
\end{eqnarray}
The three-loop QCD corrections to the $k_f$ term can be found in \cite{hbb3}.
The electroweak corrections are small in the intermediate mass range and can
thus be neglected, but we have included them in the analysis. However, for large
Higgs masses the electroweak corrections may be important due to the enhanced
self-coupling of the Higgs bosons. In the large Higgs mass regime the leading
contributions can be expressed as \cite{hqqlam}
\begin{equation}
\Gamma (H\to f\bar f) = \Gamma_{LO}(H\to f\bar f) \left\{1 + 2.12 \hat \lambda
- 32.66 \hat \lambda^2 \right\}
\end{equation}
with the coupling constant
\begin{equation}
\hat \lambda = \frac{G_F M_H^2}{16 \sqrt{2} \pi^2} \, .
\label{eq:hatlam}
\end{equation}
For Higgs masses of about 1 TeV these corrections enhance the partial decay
widths by about 2\%.

In the case of $t\bar t$ decays of the Standard Higgs boson, below-threshold
decays $H\to t\bar t^* \to t\bar b W^-$ into off-shell top quarks may be
sizeable. Thus we have included them below the $t\bar t$ threshold. Their
Dalitz plot density reads as \cite{1OFF}
\begin{equation}
\frac{d\Gamma}{dx_1 dx_2} (H\to tt^*\to Wtb) = \frac{3G_F^2}{32\pi^3} M_t^2
M_H^3 \frac{\Gamma_0}{y_1^2 + \gamma_t \kappa_t}
\label{eq:httdalitz}
\end{equation}
with the reduced energies $x_{1,2}=2E_{t,b}/M_H$ and the scaling variables
$y_{1,2} = 1-x_{1,2}$, $\kappa_i = M_i^2/M_H^2$ and the reduced decay widths
of the virtual particles $\gamma_i=\Gamma_i^2/M_H^2$. The squared amplitude can
be written as
\begin{eqnarray}
\Gamma_0 & = & y_1^2(1-y_1-y_2+\kappa_W-5\kappa_t) + 2\kappa_W(y_1y_2-\kappa_W
-2\kappa_ty_1+4\kappa_t\kappa_W) \nonumber \\
& & -\kappa_ty_1y_2+\kappa_t(1-4\kappa_t)(2y_1+\kappa_W+\kappa_t) \, .
\end{eqnarray} 
The differential decay width in eq.~(\ref{eq:httdalitz}) has to be integrated
over the $x_1, x_2$ region, which is bounded by
\begin{equation}
\left| \frac{2(1-x_1-x_2+\kappa_t+\kappa_b-\kappa_W) + x_1x_2}
{\sqrt{x_1^2-4\kappa_t} \sqrt{x_2^2-4\kappa_b}} \right| \leq 1 \, .
\label{eq:dalitzbound}
\end{equation}
The transition from below to above the threshold is provided by a smooth cubic
interpolation. Below-threshold decays yield a $t\bar t$ branching ratio far
below the per cent level for Higgs masses $M_H\lsim 2M_t$.

\subsubsection{Higgs decay into gluons}
\begin{figure}[hbt]
\begin{center}
\setlength{\unitlength}{1pt}
\begin{picture}(180,100)(0,0)

\Gluon(100,20)(150,20){-3}{5}
\Gluon(100,80)(150,80){3}{5}
\ArrowLine(100,20)(100,80)
\ArrowLine(100,80)(50,50)
\ArrowLine(50,50)(100,20)
\DashLine(0,50)(50,50){5}
\put(-15,46){$H$}
\put(105,46){$t,b$}
\put(155,18){$g$}
\put(155,78){$g$}

\end{picture}  \\
\setlength{\unitlength}{1pt}
\caption[ ]{\label{fg:hgglodia} \it Diagrams contributing to $H\to gg$ at
lowest order.}
\end{center}
\end{figure}
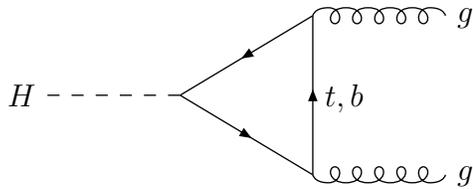
\begin{figure}[hbt]
\vspace*{0.5cm}
\hspace*{0.0cm}
\begin{turn}{-90}%
\epsfxsize=10cm \epsfbox{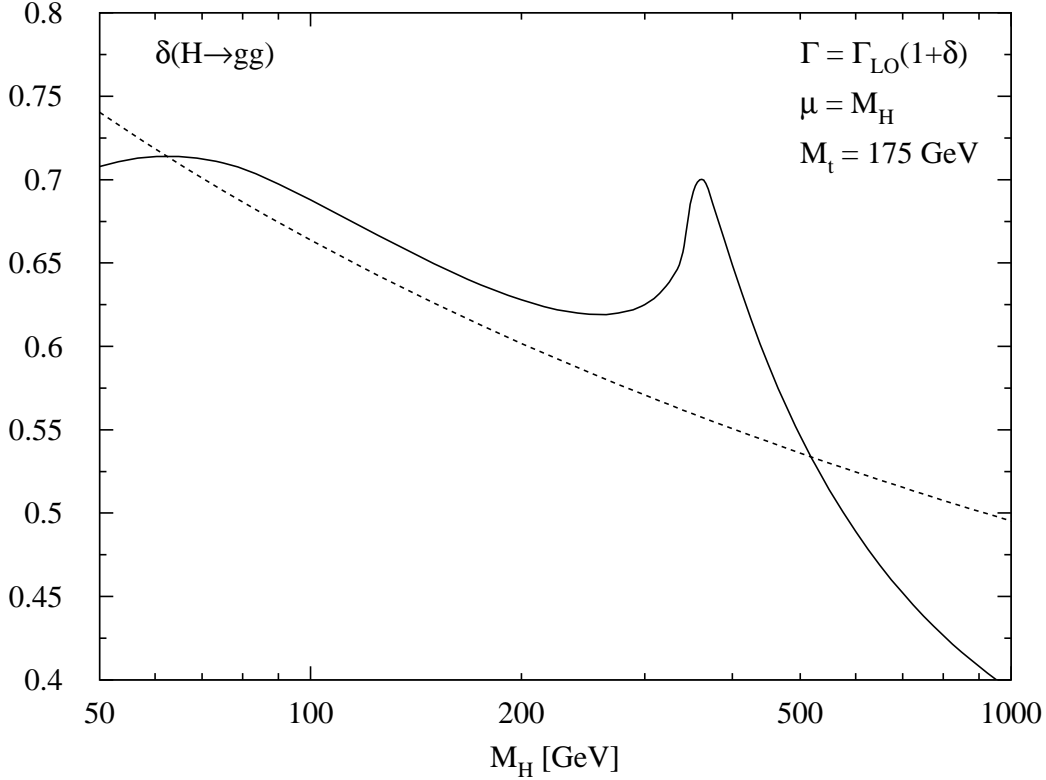}
\end{turn}
\vspace*{0.0cm}
\caption[ ]{\label{fg:hggqcd} \it The size of the QCD correction factor for
$H\to gg$, defined as $\Gamma = \Gamma_{LO} (1+\delta)$. The full line
corresponds to the full massive result, while the dashed line shows the heavy
top quark limit. The top and bottom masses have been chosen as $M_t=175$ GeV,
$M_b=5$~GeV and the NLO strong coupling constant is normalized as $\alpha_s(M_Z)
= 0.118$.}
\end{figure}

\noindent
The decay of the Higgs boson into gluons is mediated by heavy quark loops
in the Standard Model, see Fig.~\ref{fg:hgglodia}; at lowest order the partial
decay width \cite{prohiggs,higgsqcd,hgg0,hgg} is given by
\begin{eqnarray}
\Gamma_{LO}\, [H\ra gg] = \frac{G_{F}\, \alpha_{s}^{2}\,M_{H}^{3}}
{36\,\sqrt{2}\,\pi^{3}} \left| \sum_{Q} A_Q^H(\tau_Q) \right|^2
\label{eq:hgglo}
\end{eqnarray}
with the form factor
\begin{eqnarray}
A_Q^H (\tau) & = & \frac{3}{2} \tau \left[ 1+(1-\tau) f(\tau)
\right] \nonumber \\
f(\tau) & = & \left\{ \begin{array}{ll}
\displaystyle \arcsin^2 \frac{1}{\sqrt{\tau}} & \tau \ge 1 \\
\displaystyle - \frac{1}{4} \left[ \log \frac{1+\sqrt{1-\tau}}
{1-\sqrt{1-\tau}} - i\pi \right]^2 & \tau < 1
\end{array} \right.
\label{eq:ftau}
\end{eqnarray}
\begin{figure}[hbt]
\begin{center}
\setlength{\unitlength}{1pt}
\begin{picture}(400,100)(-20,0)

\Gluon(60,20)(90,20){-3}{3}
\Gluon(60,80)(90,80){3}{3}
\Gluon(45,35)(45,65){-3}{3}
\ArrowLine(60,20)(60,80)
\ArrowLine(60,80)(45,65)
\ArrowLine(45,65)(30,50)
\ArrowLine(30,50)(45,35)
\ArrowLine(45,35)(60,20)
\DashLine(0,50)(30,50){5}
\put(-15,46){$H$}
\put(65,46){$t,b$}
\put(95,18){$g$}
\put(95,78){$g$}
\put(50,48){$g$}

\Gluon(200,20)(230,20){-3}{3}
\Gluon(200,80)(230,80){3}{3}
\Gluon(200,50)(230,50){3}{3}
\ArrowLine(200,20)(200,50)
\ArrowLine(200,50)(200,80)
\ArrowLine(200,80)(170,50)
\ArrowLine(170,50)(200,20)
\DashLine(140,50)(170,50){5}
\put(125,46){$H$}
\put(165,60){$t,b$}
\put(235,18){$g$}
\put(235,48){$g$}
\put(235,78){$g$}

\Gluon(340,20)(370,20){-3}{3}
\Gluon(340,80)(355,80){3}{1}
\ArrowLine(355,80)(370,100)
\ArrowLine(370,60)(355,80)
\ArrowLine(340,20)(340,80)
\ArrowLine(340,80)(310,50)
\ArrowLine(310,50)(340,20)
\DashLine(280,50)(310,50){5}
\put(265,46){$H$}
\put(305,60){$t,b$}
\put(375,18){$g$}
\put(375,58){$\bar q$}
\put(375,98){$q$}

\end{picture}  \\
\setlength{\unitlength}{1pt}
\caption[ ]{\label{fg:hggdia} \it Typical diagrams contributing to the QCD
corrections to $H\to gg$.}
\end{center}
\end{figure}
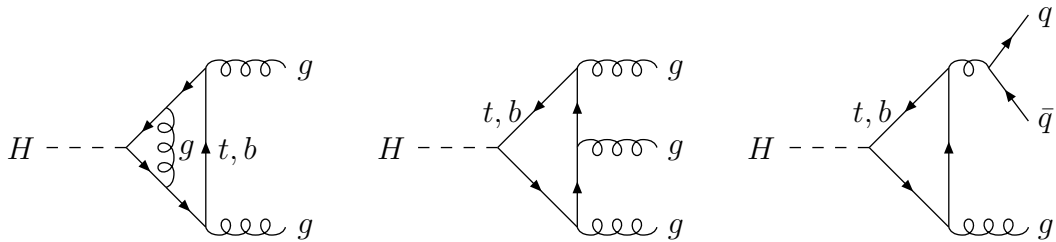
The parameter $\tau_Q= 4M_Q^2/M_H^2$ is defined by the pole mass $M_Q$ of the
heavy loop quark $Q$.
For large quark masses the form factor approaches unity. QCD radiative
corrections are built up by the exchange of virtual gluons, gluon radiation
from the quark triangle and the splitting of a gluon into two gluons or a
quark--antiquark pair, see Fig.~\ref{fg:hggdia}. If all quarks $u, \cdots,b$ are
treated as massless at the renormalization scale $\mu \sim M_{H} \sim 100$ \GeV,
the radiative corrections can be expressed as \citer{higgsqcd,hgg}
\begin{eqnarray}
&& \Gamma^{N_F}\,[H\ra gg\,(g),\,q{\overline{q}}g]= \Gamma_{LO}\,
\left[\alpha_{s}^{(N_{F})}(M_H )\right]\,\left\{1+E^{N_{F}}
\frac{\alpha_{s}^{(N_{F})}(M_H)}{\pi}\ \right\} \label{eq:hgg} \\
&& \hspace*{1.3cm}
E^{N_{F}} \to \frac{95}{4}-\frac{7}{6}N_{F} \hspace{1cm} \mbox{for $M_H^2\ll
4 M_Q^2$} \nonumber
\end{eqnarray}
with $N_{F}=5$ light quark flavors. The full massive result can be found in
\cite{higgsqcd}. The radiative corrections are plotted in Fig.~\ref{fg:hggqcd}
against the Higgs boson mass. They turn out to be very
large: the decay width is shifted by about 60--70\% upwards in the
intermediate mass range. The dashed line shows the approximated QCD corrections
defined by taking the coefficient $E^{N_F}$ in the limit of a heavy loop quark
$Q$ as presented in eq.~(\ref{eq:hgg}). It can be inferred from the figure that
the approximation is valid for the partial gluonic decay width within about
10\% for the whole relevant Higgs mass range up to 1 TeV. The reason for the
suppressed quark mass dependence of the relative QCD corrections is the
dominance of soft and collinear gluon contributions, which do not resolve
the Higgs coupling to gluons and are thus leading to a simple rescaling factor.

Recently the three-loop QCD corrections to the gluonic decay width have been
evaluated in the limit of a heavy top quark \cite{hgg3}. They contribute a
further amount of ${\cal O}(20\%)$ relative to the lowest order result and thus
increase the full NLO expression by ${\cal O}(10\%)$. The reduced size of
these corrections signals a significant stabilization of the perturbative
result and thus a reliable theoretical prediction.

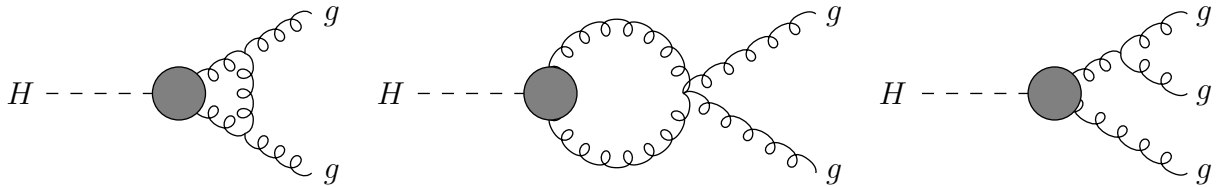
\begin{figure}[hbt]
\begin{center}
\setlength{\unitlength}{1pt}
\begin{picture}(500,100)(-15,0)

\DashLine(0,50)(50,50){5}
\Gluon(50,50)(75,65){3}{3}
\Gluon(75,65)(100,80){3}{3}
\Gluon(50,50)(75,35){-3}{3}
\Gluon(75,35)(100,20){-3}{3}
\Gluon(75,65)(75,35){3}{3}
\GCirc(50,50){10}{0.5}
\put(-15,46){$H$}
\put(105,18){$g$}
\put(105,78){$g$}

\DashLine(140,50)(190,50){5}
\GlueArc(215,50)(25,0,180){3}{8}
\GlueArc(215,50)(25,180,360){3}{8}
\Gluon(240,50)(290,80){3}{5}
\Gluon(240,50)(290,20){3}{5}
\GCirc(190,50){10}{0.5}
\put(125,46){$H$}
\put(295,18){$g$}
\put(295,78){$g$}

\DashLine(330,50)(380,50){5}
\Gluon(380,50)(405,65){3}{3}
\Gluon(405,65)(430,80){3}{2}
\Gluon(405,65)(430,50){-3}{2}
\Gluon(380,50)(430,20){-3}{5}
\GCirc(380,50){10}{0.5}
\put(315,46){$H$}
\put(435,18){$g$}
\put(435,78){$g$}
\put(435,48){$g$}

\end{picture}  \\
\setlength{\unitlength}{1pt}
\caption[ ]{\label{fg:hgglimdia} \it Typical effective diagrams contributing to
the QCD corrections to $H\to gg$ in the heavy quark limit.}
\end{center}
\end{figure}

The QCD corrections in the heavy quark limit can also be obtained by means of
a low-energy theorem \cite{prohiggs,hgagalo}. The starting point is that, for
vanishing Higgs momentum
$p_H\to 0$, the entire interaction of the Higgs particle with $W,Z$ bosons and
fermions can be generated by the substitution
\begin{equation}
M_i \to M_i \times \left[ 1 + \frac{H}{v} \right] \hspace*{1cm} (i=f,W,Z) \, ,
\end{equation}
where the Higgs field $H$ acts as a constant complex number. At higher orders
this substitution has to be expressed in terms of bare parameters
\cite{higgsqcd,let}. Thus there
is a relation between a bare matrix element with and without an external scalar
Higgs boson [$X$ denotes an arbitrary particle configuration]:
\begin{equation}
\lim_{p_H\to 0} {\cal M}(XH) = \frac{1}{v_0} m_0 \frac{\partial}{\partial m_0}
{\cal M}(X) \, .
\end{equation}
In most of the practical cases the external Higgs particle is defined as being
on-shell, so that $p_H^2 = M_H^2$ and the mathematical limit of vanishing
Higgs momentum coincides with the limit of small Higgs masses. In order to
calculate the Higgs coupling to two gluons one starts from the heavy quark $Q$
contribution to the bare gluon self-energy ${\cal M}(gg)$. The differentiation
with respect to the bare quark mass $m_0$ can be replaced by the
differentiation by the renormalized $\overline{\rm MS}$ quark mass
$\overline{m}_Q(\overline{m}_Q)$. In this way a finite
contribution from the quark anomalous mass dimension $\gamma_m(\alpha_s)$
arises:
\begin{equation}
m_0 \frac{\partial}{\partial m_0} = \frac{\overline{m}_Q(\overline{m}_Q)}
{1+\gamma_m
(\alpha_s)} \frac{\partial}{\partial \overline{m}_Q (\overline{m}_Q)} \, .
\end{equation}
The remaining mass differentiation of the gluon self-energy results in the
heavy quark contribution $\beta_Q(\alpha_s)$ to the QCD $\beta$ function at
vanishing momentum transfer and to an additional contribution of the
anomalous dimension of the gluon field operators, which can be expressed in
terms of the QCD $\beta$ function \cite{hgg0}. The final matrix element can be
converted into the effective Lagrangian
\cite{higgsqcd,hgg0,hgg3,let,dawson,gghresum}
\begin{equation}
{\cal L}_{eff} = \frac{\alpha_s}{4}~\frac{\beta(\alpha_s)/\alpha^2_s}
{\beta(\alpha^t_s)/[\alpha^t_s]^2}~
\frac{\beta_Q(\overline{\alpha}_s)/[\alpha_s^t]^2}{1+
\gamma_m(\overline{\alpha}_s)}~G^{a\mu\nu} G^a_{\mu\nu} \frac{H}{v}
\label{eq:hggleff}
\end{equation}
with $\overline{\alpha}_s = \alpha_s^{(6)}[\overline{m}_t(\overline{m}_t)]$ and
$\alpha^t_s = \alpha_s^{(5)}[\overline{m}_t(\overline{m}_t)]$. The strong
coupling $\alpha_s$ of the effective theory includes only $N_F = 5$ flavors.
The effective Lagrangian of eq.~(\ref{eq:hggleff}) is valid for the limiting
case $M_H^2 \ll 4M_Q^2$.
The anomalous mass dimension is given by \cite{anommass}
\begin{equation}
\gamma_m(\alpha_s) = 2\frac{\alpha_s}{\pi} + \left( \frac{101}{12} -
\frac{5}{18} [N_F+1] \right) \left(\frac{\alpha_s}{\pi}\right)^2
+ {\cal O}(\alpha_s^3) \, .
\label{eq:anommass}
\end{equation}
Up to NLO the heavy quark contribution to the QCD $\beta$ function coincides
with the corresponding part of the $\overline{\rm MS}$ result. But at NNLO an
additional piece arises from a threshold correction due to a mismatch between
the $\overline{\rm MS}$ scheme and the result for vanishing momentum transfer
\citer{betaqcd,larin}:
\begin{eqnarray}
\beta_Q(\alpha_s) & = & \beta_Q^{\overline{MS}}(\alpha_s) - \frac{11}{72}
~\frac{27-2N_F}{6}~\frac{\alpha_s^4}{\pi^3}
+ {\cal O}(\alpha_s^5) \nonumber \\
\beta_Q^{\overline{MS}}(\alpha_s) & = & \frac{\alpha_s^2}{3\pi}\left[ 1+
\frac{19}{4} \frac{\alpha_s}{\pi} + \frac{7387 - 325 N_F}{288}
\left(\frac{\alpha_s}{\pi} \right)^2 \right] + {\cal O}(\alpha_s^5)
\end{eqnarray}
The strong coupling constant $\overline{\alpha}_s$ of eq.~(\ref{eq:hggleff})
includes 6 flavors, and its scale is set by the top quark mass
$\overline{m}_t(\overline{m}_t)$. In order to decouple the top quark from the
couplings in the effective Lagrangian, the six-flavor
coupling $\alpha_s^{(6)}$ has to be replaced by the five-flavor expression
$\alpha_s^{(5)}$. They are related by \citer{bernwetz,bernpriv}\footnote{It
should be noted that eq.~(\ref{eq:matching}) differs from the result of
Ref.~\cite{bernwetz}. However, the difference can be traced back to the Abelian
part of the matching relation, which has been extracted by the author from the
analogous expression for the photon self-energy \cite{abgam}.}
\begin{equation}
\alpha_s^{(6)}[\overline{m}_t(\overline{m}_t)] = \alpha_s^{(5)}[\overline{m}_t
(\overline{m}_t)] \left\{ 1 - \frac{11}{72}
\left( \frac{\alpha_s^{(5)}[\overline{m}_t(\overline{m}_t)]}{\pi} \right)^2 +
{\cal O} (\alpha_s^3) \right\} \, .
\label{eq:matching}
\end{equation}
Finally the perturbative expansion of the effective Lagrangian can be cast
into the form \cite{hgg3,gghresum}
\begin{eqnarray}
{\cal L}_{eff} & = & \frac{\alpha_s^{(5)}}{12\pi}G^{a\mu\nu}G^a_{\mu\nu}
\frac{H}{v}
\left\{1 + \frac{\beta_1}{\beta_0} \frac{\alpha_s^{(5)}}{\pi}
+ \frac{\beta_2}{\beta_0}\left(\frac{\alpha_s^{(5)}}{\pi}\right)^2\right\}
\nonumber \\
& & \left\{1 + \left(\frac{11}{4}-\frac{\beta_1}{\beta_0}\right)
\frac{\alpha_s^{(5)}(M_t)}{\pi} \right. \nonumber \\
& & \left. \hspace*{1cm} + \left[\frac{2777-201 N_F}{288}
+\frac{\beta_1}{\beta_0}\left(
\frac{\beta_1}{\beta_0} - \frac{11}{4}\right) - \frac{\beta_2}{\beta_0} \right]
\left( \frac{\alpha_s^{(5)}(M_t)}{\pi} \right)^2 \right\} \, ,
\label{eq:hggeffnnlo}
\end{eqnarray}
where we have introduced the top quark pole mass $M_t$.
The coefficients of the QCD $\beta$ function in eq.~(\ref{eq:hggeffnnlo}) are
given by \cite{betaqcd}
\begin{eqnarray}
\beta_0 & = & \frac{33-2N_F}{12} \nonumber \\
\beta_1 & = & \frac{153-19N_F}{24} \nonumber \\
\beta_2 & = & \frac{1}{128} \left\{ 2857 - \frac{5033}{9} N_F + \frac{325}{27}
N_F^2 \right\} \, .
\end{eqnarray}
[The four-loop contribution has also been obtained recently \cite{betaqcd4}.]
$N_F$ denotes the number of light quark flavors and will be identified with 5.
For the calculation of the heavy quark limit given in eq.~(\ref{eq:hgg}) the
effective coupling has to be inserted into the blobs of the effective diagrams
shown in Fig.~\ref{fg:hgglimdia}. After evaluating these effective massless
one-loop contributions the result coincides with the explicit calculation of
the two-loop corrections in the heavy quark limit of eq.~(\ref{eq:hgg}) at NLO.

Using the discussed low-energy theorem, the electroweak corrections of
${\cal O}(G_F M_t^2)$ to the gluonic decay width, which are mediated by virtual
top quarks, can be obtained easily. For this purpose the leading top mass
corrections to the gluon self-energy have to be computed. The result has to be
differentiated by the bare top mass and the renormalization will be carried out
afterwards. The final result leads to a simple rescaling of the lowest order
decay width \cite{abdelgambino}
\begin{equation}
\Gamma(H\to gg) = \Gamma_{LO}(H\to gg) \left[ 1+
\frac{G_F M_t^2}{8\sqrt{2}\pi^2} \right] \, .
\label{eq:hggelw}
\end{equation}
They enhance the gluonic decay width by about 0.3\% and are thus negligible.

\begin{figure}[hbt]
\begin{center}
\setlength{\unitlength}{1pt}
\begin{picture}(120,95)(0,0)

\Gluon(75,35)(110,35){3}{4}
\ArrowLine(50,50)(100,80)
\ArrowLine(100,20)(75,35)
\ArrowLine(75,35)(50,50)
\DashLine(0,50)(50,50){5}
\put(-15,46){$H$}
\put(105,80){$Q$}
\put(105,13){$\bar Q$}
\put(115,33){$g$}

\end{picture}  \\
\setlength{\unitlength}{1pt}
\caption[ ]{\label{fg:hqqgdia} \it Typical diagram contributing to $H\to
Q\bar Qg$.}
\end{center}
\end{figure}
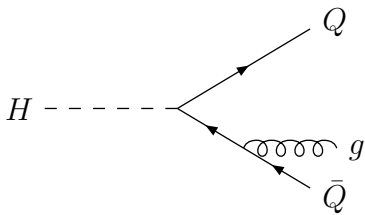
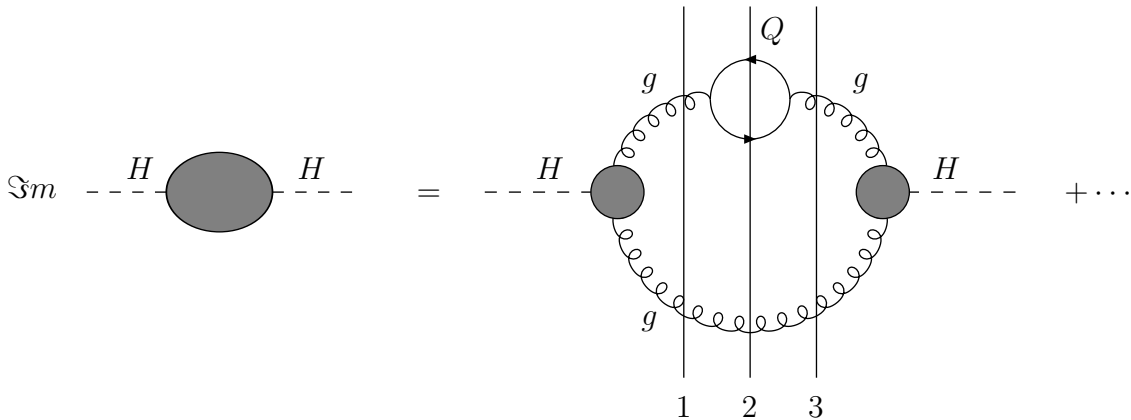
\begin{figure}[hbt]
\begin{center}
\setlength{\unitlength}{1pt}
\begin{picture}(450,150)(10,-40)

\DashLine(50,50)(80,50){5}
\DashLine(120,50)(150,50){5}
\GOval(100,50)(15,20)(0){0.5}
\put(20,48){$\Im m$}
\put(65,55){$H$}
\put(130,55){$H$}
\put(175,47){$=$}

\DashLine(200,50)(250,50){5}
\DashLine(350,50)(400,50){5}
\GlueArc(315,50)(35,0,90){3}{6}
\GlueArc(285,50)(35,90,180){3}{6}
\GlueArc(300,50)(50,180,360){3}{18}
\GCirc(250,50){10}{0.5}
\GCirc(350,50){10}{0.5}
\ArrowArc(300,85)(15,0,180)
\ArrowArc(300,85)(15,180,360)
\Line(275,-20)(275,120)
\Line(325,-20)(325,120)
\Line(300,-20)(300,120)
\put(220,55){$H$}
\put(370,55){$H$}
\put(305,108){$Q$}
\put(260,90){$g$}
\put(340,90){$g$}
\put(260,0){$g$}
\put(420,47){$+ \cdots$}
\put(273,-35){$1$}
\put(298,-35){$2$}
\put(323,-35){$3$}

\end{picture}  \\
\setlength{\unitlength}{1pt}
\caption[]{\label{fg:hggcut} \it Cut diagrams, involving heavy quark $Q$ loops,
contributing to the imaginary part of the Higgs self-energy at the two-loop
level.}
\end{center}
\end{figure}

The final states $H\,\ra\, b{\overline{b}}g$ and $c{\overline{c}}g$ are also
generated through processes in which the $b,c$ quarks directly couple to the
Higgs boson, see Fig.~\ref{fg:hqqgdia}. Gluon splitting $g\,\ra\,
b{\overline{b}}$ in $H\,\ra\,gg$ increases the inclusive decay probabilities
$\Gamma(H \ra b\bar{b}+ \dots)$ etc. Since $b$ quarks, and eventually $c$
quarks, can in principle be tagged
experimentally, it is physically meaningful to consider the particle width of
Higgs decays to gluon and light $u,d,s$ quark final jets separately.
If one naively subtracts the final state gluon splitting contributions for $b$
and $c$ quarks and keeps the quark masses finite to regulate the emerging mass
singularities, one ends up with large logarithms of the $b,c$ quark masses
[in the limit of heavy loop quark masses $M_Q$]
\begin{equation}
\delta E^{b,c} = -\frac{7}{3} + \frac{1}{3} \left[ \log\frac{M_H^2}{M_b^2}
+ \log\frac{M_H^2}{M_c^2} \right] \, ,
\label{eq:bcnaive}
\end{equation}
which have to be added to the $b\bar b$ and $c\bar c$ decay widths [the finite
part emerges from the non-singular phase-space integrations]. On the other hand
the KLN theorem \cite{kln} ensures that all final-state mass singularities of
the real
corrections cancel against a corresponding part of the virtual corrections
involving the same particle. Thus the mass-singular logarithms $\log M_H^2/
M_{b,c}^2$ in eq.~(\ref{eq:bcnaive}) have to cancel against the corresponding
heavy quark loops in the external gluons, i.e. the sum of the cuts $1,2,3$
in Fig.~\ref{fg:hggcut} has to be finite for small quark masses $M_Q$.
[The blobs at the $Hgg$ vertices in Fig.~\ref{fg:hggcut} represent the effective
couplings in the heavy top quark limit. In the general massive case they have
to be replaced by the top and bottom triangle loops\footnote{It should be
noted that the bottom quark triangle loop develops a logarithmic behaviour
$\propto M_b^2/M_H^2 \times \log^2 M_H^2/M_b^2$, which arises from the
integration region of the loop momentum, where the $b$ quark, exchanged between
the two gluons, becomes nearly on-shell. These mass logarithms do {\it not}
correspond to final-state mass singularities in pure QCD and are thus not
required to cancel by the KLN theorem.}.] Thus in order to resum these large
final-state mass logarithms in the gluonic decay width, the heavy quarks
$Q=b,c$ have to be decoupled from the running strong coupling constant, which
has to be defined with three light flavors, if $b,c$ quark final states are
subtracted,
\begin{eqnarray}
\alpha_s^{(5)} (M_H) & = & \alpha_s^{(4)} (M_H) \times \left\{ 1 +
\frac{\alpha_s^{(4)}(M_H)}{6\pi} \log \frac{M_H^2}{M_b^2}
+ {\cal O}(\alpha_s^2) \right\} \nonumber \\
\alpha_s^{(4)} (M_H) & = & \alpha_s^{(3)} (M_H) \times \left\{ 1 +
\frac{\alpha_s^{(3)}(M_H)}{6\pi} \log \frac{M_H^2}{M_c^2}
+ {\cal O}(\alpha_s^2) \right\}
\label{eq:als345}
\end{eqnarray}
Expressed in terms of three light flavors, the gluonic decay width is free of
explicit mass singularities in the bottom and charm quark masses.
The resummed contributions of $b,c$ quark final states are given by
the difference of the gluonic widths [eq.~(\ref{eq:hgg})] for the corresponding
number of flavors $N_F$ \cite{QCD},
\begin{eqnarray}
\delta \Gamma [H\to c\bar c+\dots] & = & \Gamma^{4} - \Gamma^{3}
\nonumber \\
\delta \Gamma [H\to b\bar b+\dots] & = & \Gamma^{5} - \Gamma^{4}
\end{eqnarray}
in the limit $M_{H}^{2}\,\gg\,M_{b,c}^{2}$. In this way large mass logarithms
$\log M_H^2/M_{c,b}^2$ in the remaining gluonic decay mode are absorbed into
the strong coupling by changing the
number of active flavors according to the number of contributing flavors in
the final states. It should be noted that by virtue of eqs.~(\ref{eq:als345})
the large logarithms are implicitly contained in the strong couplings for
different numbers of active flavors. The subtracted parts may be added to the
partial decay widths into $c$ and $b$ quarks. 
In $\alpha_s^{(4)}(M_Z)$ the contribution of the $b$ quark
is subtracted and in $\alpha_s^{(3)} (M_Z)$ the contributions of both
the $b$ and $c$ quarks are. The values for $\alpha_s^{(4)}(M_Z)$ are typically
5\% smaller and those of $\alpha_s^{(3)}(M_Z)$ about 15\% smaller than
$\alpha_s^{(5)}(M_Z)$, see Table \ref{tb:alphas345}.
\begin{table}[hbt]
\renewcommand{\arraystretch}{1.5}
\begin{center}
\begin{tabular}{|c|c|c|} \hline
$\alpha_s^{(5)} (M_Z)$ & $\alpha_s^{(4)} (M_Z)$ & $\alpha_s^{(3)} (M_Z)$ \\
\hline \hline
0.112 & 0.107 & 0.101 \\
0.118 & 0.113 & 0.105 \\
0.124 & 0.118 & 0.110 \\ \hline
\end{tabular}
\renewcommand{\arraystretch}{1.2}
\caption[]{\label{tb:alphas345} \it Strong coupling constants $\alpha_s (M_Z)$
for different numbers of flavors contributing to the scale dependence. In
$\alpha_s^{(4)}$ the $b$ quark contribution is subtracted and in
$\alpha_s^{(3)}$ the $b$ and $c$ quark contributions are.}
\end{center}
\end{table}

\subsubsection{Higgs decay to photon pairs}
\begin{figure}[hbt]
\begin{center}
\setlength{\unitlength}{1pt}
\begin{picture}(400,100)(0,0)

\Photon(60,20)(90,20){-3}{4}
\Photon(60,80)(90,80){3}{4}
\ArrowLine(60,20)(60,80)
\ArrowLine(60,80)(30,50)
\ArrowLine(30,50)(60,20)
\DashLine(0,50)(30,50){5}
\put(-15,46){$H$}
\put(70,46){$f$}
\put(95,18){$\gamma$}
\put(95,78){$\gamma$}

\Photon(210,20)(240,20){-3}{4}
\Photon(210,80)(240,80){3}{4}
\Photon(210,80)(210,20){3}{7}
\Photon(210,20)(180,50){3}{5}
\Photon(180,50)(210,80){3}{5}
\DashLine(150,50)(180,50){5}
\put(135,46){$H$}
\put(220,46){$W$}
\put(245,18){$\gamma$}
\put(245,78){$\gamma$}

\DashLine(300,50)(330,50){5}
\PhotonArc(345,50)(15,0,180){3}{6}
\PhotonArc(345,50)(15,180,360){3}{6}
\Photon(360,50)(390,80){3}{5}
\Photon(360,50)(390,20){3}{5}
\put(285,46){$H$}
\put(335,70){$W$}
\put(395,18){$\gamma$}
\put(395,78){$\gamma$}

\end{picture}  \\
\setlength{\unitlength}{1pt}
\caption[ ]{\label{fg:hgagalodia} \it Typical diagrams contributing to $H\to
\gamma \gamma$ at lowest order.}
\end{center}
\end{figure}
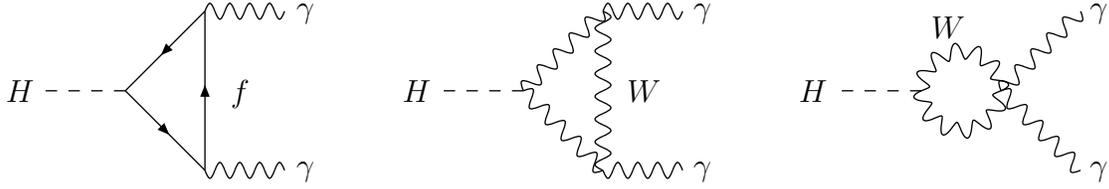
\noindent
The decay of the Higgs boson to photons is mediated by $W$ and heavy fermion
loops in the Standard Model, see Fig.~\ref{fg:hgagalodia}; the partial decay
width \cite{hgagalo} can be cast into the form
\begin{eqnarray}
\Gamma\, [H\ra \gamma\gamma] = \frac{G_{F}\, \alpha^{2}\,M_{H}^{3}}
{128\,\sqrt{2}\,\pi^{3}} \left| \sum_{f} N_{cf} e_f^2 A_f^H(\tau_f) +
A^H_W(\tau_W) \right|^2
\label{eq:hgaga}
\end{eqnarray}
with the form factors
\begin{eqnarray*}
A_f^H (\tau) & = & 2 \tau \left[ 1+(1-\tau) f(\tau)
\right] \\
A_W^H (\tau) & = & - \left[ 2+3\tau+3\tau (2-\tau) f(\tau) \right]
\end{eqnarray*}
and the function $f(\tau)$ defined in eq.~(\ref{eq:ftau}).
The parameters $\tau_i= 4M_i^2/M_H^2~~(i=f,W)$ are defined by the
corresponding masses of the heavy loop particles. For large loop masses the
form factors approach constant values:
\begin{eqnarray}
A_f^H & \to & \frac{4}{3} \hspace*{1cm} \mbox{for $M_H^2 \ll 4 M_Q^2$}
\nonumber \\
A_W^H & \to & -7 \hspace*{0.75cm} \mbox{for $M_H^2 \ll 4 M_W^2$}
\end{eqnarray}
The $W$ loop provides the dominant contribution in the intermediate Higgs mass
range, and the fermion loops interfere destructively. Only far above the
thresholds, for Higgs masses $M_H\sim 600$ GeV, does the top quark loop become
competitive, nearly cancelling the $W$ loop contribution.

\begin{figure}[hbt]
\begin{center}
\setlength{\unitlength}{1pt}
\begin{picture}(200,100)(0,0)

\DashLine(0,50)(50,50){5}
\ArrowLine(100,20)(100,80)
\ArrowLine(100,80)(75,65)
\ArrowLine(75,65)(50,50)
\ArrowLine(50,50)(75,35)
\ArrowLine(75,35)(100,20)
\Gluon(75,65)(75,35){3}{3}
\Photon(100,20)(150,20){-3}{6}
\Photon(100,80)(150,80){3}{6}
\put(-15,46){$H$}
\put(105,46){$t,b$}
\put(82,48){$g$}
\put(155,18){$\gamma$}
\put(155,78){$\gamma$}

\end{picture}  \\
\setlength{\unitlength}{1pt}
\caption[ ]{\label{fg:hgagaqcddia} \it Typical diagram contributing to the QCD
corrections to $H\to \gamma \gamma$.}
\end{center}
\end{figure}
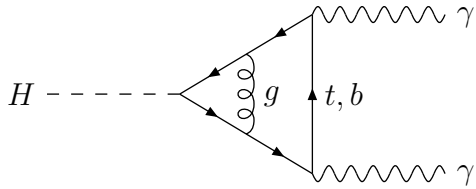
In the past the two-loop QCD corrections to the quark loops have been
calculated \cite{higgsqcd,hgaga}. They are built up by virtual gluon exchange
inside the quark triangle [see Fig.~\ref{fg:hgagaqcddia}]. Owing to charge
conjugation invariance and color conservation, radiation of a single gluon is
not possible. Hence the QCD corrections simply rescale the lowest order quark
amplitude by a factor that only depends on the ratios of the Higgs and quark
masses
\begin{eqnarray}
A_Q^H(\tau_Q) & \to & A_Q^H(\tau_Q) \times \left[1+ C_H(\tau_Q)
\frac{\alpha_s}{\pi} \right] \nonumber \\
C_H(\tau_Q) & \to & -1 \hspace*{1cm} \mbox{for $M_H^2\ll 4M_Q^2$}
\label{eq:hgagaqcd}
\end{eqnarray}
According to the low-energy theorem discussed before, the NLO QCD corrections in
the heavy quark limit can be obtained from the effective Lagrangian
\cite{higgsqcd,let}
\begin{equation}
{\cal L}_{eff} = \frac{e_Q^2}{4} \frac{\beta^Q_\alpha/\alpha}{1+
\gamma_m(\alpha_s)} F^{\mu\nu} F_{\mu\nu} \frac{H}{v} \, ,
\end{equation}
where $\beta_\alpha^Q/\alpha = 2 (\alpha/\pi) [1+\alpha_s/\pi+\cdots]$ denotes
the heavy quark $Q$ contribution to the QED $\beta$ function and $\gamma_m
(\alpha_s)$ the anomalous mass dimension given in eq.~(\ref{eq:anommass}). The
NLO expansion of the effective Lagrangian reads as \cite{higgsqcd,let}
\begin{equation}
{\cal L}_{eff} = e_Q^2 \frac{\alpha}{2\pi} F^{\mu\nu} F_{\mu\nu} \frac{H}{v}
\left[1 - \frac{\alpha_s}{\pi} + {\cal O}(\alpha_s^2) \right] \, ,
\label{eq:leffhgaga}
\end{equation}
which agrees with the $C$-value of eq.~(\ref{eq:hgagaqcd}) in the heavy quark
limit.

\begin{figure}[hbt]
\vspace*{0.5cm}
\hspace*{0.0cm}
\begin{turn}{-90}%
\epsfxsize=10cm \epsfbox{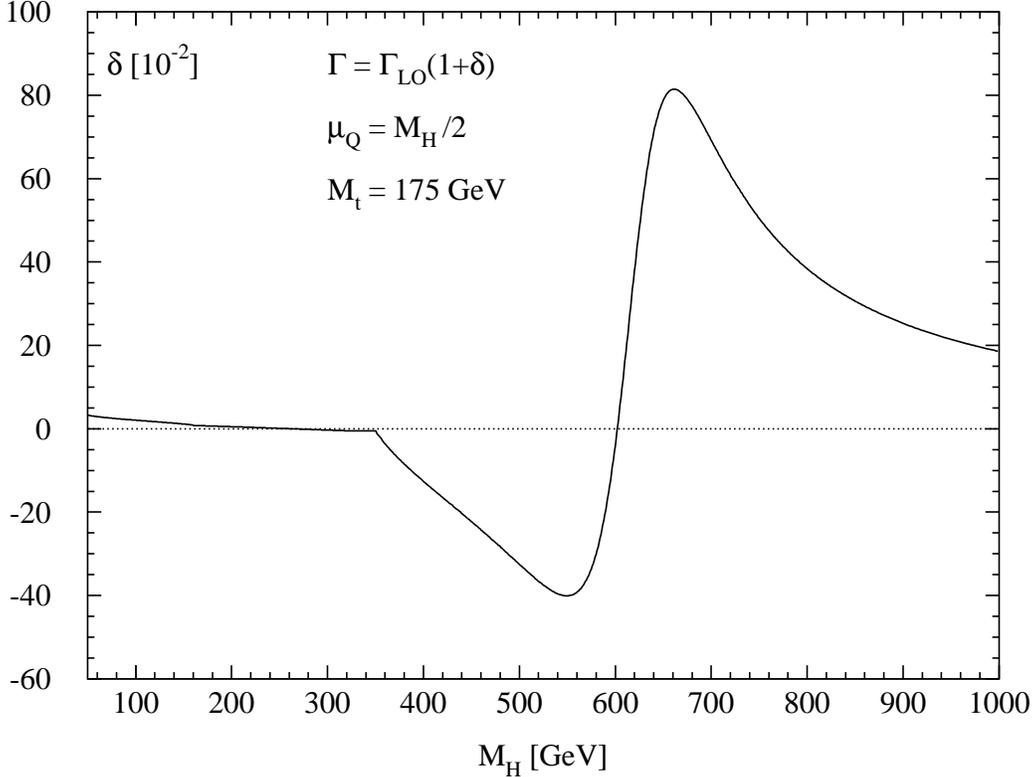}
\end{turn}
\vspace*{0.0cm}
\caption[ ]{\label{fg:hgagaqcd} \it The size of the QCD correction factor for
$H\to \gamma\gamma$, defined by $\Gamma = \Gamma_{LO} (1+\delta)$. The top and
bottom masses have been chosen as $M_t=175$ GeV, $M_b=5$ GeV and the strong
coupling constant has been normalized to $\alpha_s (M_Z)=0.118$ at NLO. The
quark masses are replaced by their running masses at the scale $\mu_Q=M_H/2$.}
\end{figure}
The QCD corrections for finite Higgs and quark masses are presented in
Fig.~\ref{fg:hgagaqcd} as a function of the Higgs mass. In order to improve the
perturbative behaviour of the quark loop contributions they should be
expressed preferably in terms of the running quark masses $m_Q(M_H/2)$, which
are normalized to the {\it pole} masses $M_Q$ via
\begin{equation}
m_Q(\mu_Q=M_Q) = M_Q \, ;
\end{equation}
their scale is identified with $\mu_Q=M_H/2$ within the photonic decay mode.
These definitions imply a proper definition of the $Q\bar Q$ thresholds $M_H =
2 M_Q$, without artificial displacements due to finite shifts between the
{\it pole} and running quark masses, as is the case for the running
$\overline{\rm MS}$ masses. It can be inferred from Fig.~\ref{fg:hgagaqcd} that
the residual QCD corrections are moderate, of ${\cal O}(10\%)$, apart from a
broad region around $M_H\sim 600$ GeV, where the $W$ loop nearly cancels the
top quark contributions in the lowest order decay width. Consequently the
relative QCD corrections are only artificially enhanced, and the perturbative
expansion is reliable in this mass region, too. Since the QCD
corrections are small in the intermediate mass range, where the photonic decay
mode is important, they are neglected in this analysis. Recently the
three-loop QCD corrections to the effective Lagrangian of
eq.~(\ref{eq:leffhgaga}) have been calculated \cite{hgaga3}. They lead to a
further contribution of a few per mille.

The electroweak corrections of ${\cal O}(G_F M_t^2)$ have been evaluated
recently. This part of the correction arises from all diagrams, which contain
a top quark coupling to a Higgs particle or would-be Goldstone boson. The
final expression results in a rescaling factor to the top quark loop amplitude,
given by \cite{hgagamt}
\begin{equation}
A_t^H(\tau_t) \to A_t^H(\tau_t) \times \left[1 -\frac{3}{4e_t^2}\left( 4e_te_b
+ 5 - \frac{14}{3} e_t^2 \right) \frac{G_F M_t^2}{8\sqrt{2}\pi^2}
\right] \, ,
\end{equation}
where $e_{t,b}$ are the electric charges of the top and bottom quarks. The
effect is an enhancement of the photonic decay width by less than 1\%,
so that these corrections are negligible.

In the large Higgs mass regime the leading electroweak corrections to the $W$
loop have been computed by means of the equivalence theorem \cite{ztozh,equiv}.
This ensures that for large Higgs masses the dominant contributions arise from
longitudinal would-be Goldstone interactions, whereas the contributions of the
transverse $W$ and $Z$ components are suppressed. The final result decreases
the $W$ form factor by a finite amount \cite{hgagaw},
\begin{equation}
A_W^H(\tau_W) \to A_W^H(\tau_W)\left[1 - 3.027 \frac{G_F M_H^2}{8\sqrt{2}\pi^2}
\right] \hspace*{1cm} \mbox{for $M_H^2 \gg 4M_W^2$} \, .
\end{equation}
These electroweak corrections are only sizeable in the region around $M_H\sim
600$ GeV, where the lowest order decay width develops a minimum due to the
strong cancellation of the $W$ and $t$ loops and for very large Higgs masses
$M_H\sim 1$ TeV. Since the photonic branching ratio is only
important in the intermediate mass range, where it reaches values of a few
$10^{-3}$, the electroweak corrections are neglected in the present analysis.

\subsubsection{Higgs decay to photon and $Z$ boson}
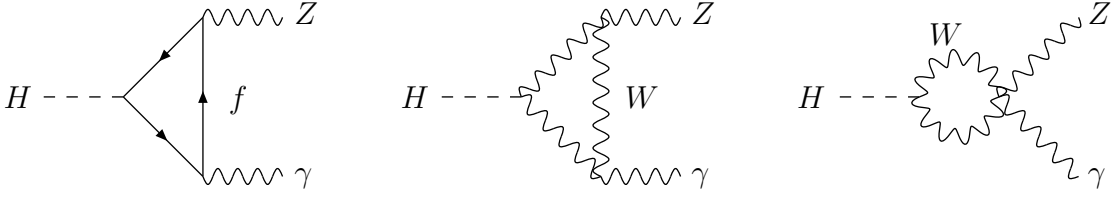
\begin{figure}[hbt]
\begin{center}
\setlength{\unitlength}{1pt}
\begin{picture}(400,100)(0,0)

\Photon(60,20)(90,20){-3}{4}
\Photon(60,80)(90,80){3}{4}
\ArrowLine(60,20)(60,80)
\ArrowLine(60,80)(30,50)
\ArrowLine(30,50)(60,20)
\DashLine(0,50)(30,50){5}
\put(-15,46){$H$}
\put(70,46){$f$}
\put(95,18){$\gamma$}
\put(95,78){$Z$}

\Photon(210,20)(240,20){-3}{4}
\Photon(210,80)(240,80){3}{4}
\Photon(210,80)(210,20){3}{7}
\Photon(210,20)(180,50){3}{5}
\Photon(180,50)(210,80){3}{5}
\DashLine(150,50)(180,50){5}
\put(135,46){$H$}
\put(220,46){$W$}
\put(245,18){$\gamma$}
\put(245,78){$Z$}

\DashLine(300,50)(330,50){5}
\PhotonArc(345,50)(15,0,180){3}{6}
\PhotonArc(345,50)(15,180,360){3}{6}
\Photon(360,50)(390,80){3}{5}
\Photon(360,50)(390,20){3}{5}
\put(285,46){$H$}
\put(335,70){$W$}
\put(395,18){$\gamma$}
\put(395,78){$Z$}

\end{picture}  \\
\setlength{\unitlength}{1pt}
\caption[ ]{\label{fg:hzgalodia} \it Typical diagrams contributing to $H\to
Z \gamma$ at lowest order.}
\end{center}
\end{figure}
\noindent
The decay of the Higgs boson to a photon and a $Z$ boson is mediated by $W$ and
heavy fermion loops, see Fig.~\ref{fg:hzgalodia}; the partial decay
width can be obtained as \cite{hunter,hzga}
\begin{eqnarray}
\Gamma\, [H\ra Z\gamma] = \frac{G^2_{F}M_W^2\, \alpha\,M_{H}^{3}}
{64\,\pi^{4}} \left( 1-\frac{M_Z^2}{M_H^2} \right)^3 \left|
\sum_{f} A_f^H(\tau_f,\lambda_f) + A^H_W(\tau_W,\lambda_W) \right|^2 \, ,
\label{eq:hzga}
\end{eqnarray}
with the form factors
\begin{eqnarray}
A_f^H (\tau,\lambda) & = & 2 N_{cf} \frac{e_f (I_{3f} - 2e_f\sin^2\theta_W )}
{\cos\theta_W} \left[I_1(\tau,\lambda) - I_2(\tau,\lambda)
\right] \nonumber \\
A_W^H (\tau,\lambda) & = & \cos \theta_W \left\{ 4(3-\tan^2\theta_W)
I_2(\tau,\lambda) \right. \nonumber \\
& & \left. + \left[ \left(1+\frac{2}{\tau}\right) \tan^2\theta_W
- \left(5+\frac{2}{\tau} \right) \right] I_1(\tau,\lambda) \right\} \, .
\label{eq:hzgaform}
\end{eqnarray}
The functions $I_1,I_2$ are given by
\begin{eqnarray*}
I_1(\tau,\lambda) & = & \frac{\tau\lambda}{2(\tau-\lambda)}
+ \frac{\tau^2\lambda^2}{2(\tau-\lambda)^2} \left[ f(\tau) - f(\lambda) \right]
+ \frac{\tau^2\lambda}{(\tau-\lambda)^2} \left[ g(\tau) - g(\lambda) \right] \\
I_2(\tau,\lambda) & = & - \frac{\tau\lambda}{2(\tau-\lambda)}\left[ f(\tau)
- f(\lambda) \right]
\end{eqnarray*}
where the function $g(\tau)$ can be expressed as
\begin{equation}
g(\tau) = \left\{ \begin{array}{ll}
\displaystyle \sqrt{\tau-1} \arcsin \frac{1}{\sqrt{\tau}} & \tau \ge 1 \\
\displaystyle \frac{\sqrt{1-\tau}}{2} \left[ \log \frac{1+\sqrt{1-\tau}}
{1-\sqrt{1-\tau}} - i\pi \right] & \tau < 1
\end{array} \right.
\label{eq:gtau}
\end{equation}
and the function $f(\tau)$ is defined in eq.~(\ref{eq:ftau}).
The parameters $\tau_i= 4M_i^2/M_H^2$ and $\lambda_i= 4M_i^2/M_Z^2~~(i=f,W)$
are defined in terms of the corresponding masses of the heavy loop particles.
Due to charge conjugation invariance, only the vectorial $Z$ coupling
contributes to the fermion loop so that problems with the axial $\gamma_5$
coupling do not arise.
The $W$ loop dominates in the intermediate Higgs mass range, and the heavy
fermion loops interfere destructively.

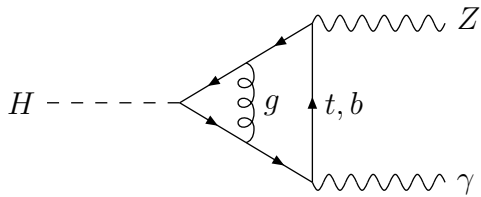
\begin{figure}[hbt]
\begin{center}
\setlength{\unitlength}{1pt}
\begin{picture}(200,100)(0,0)

\DashLine(0,50)(50,50){5}
\ArrowLine(100,20)(100,80)
\ArrowLine(100,80)(75,65)
\ArrowLine(75,65)(50,50)
\ArrowLine(50,50)(75,35)
\ArrowLine(75,35)(100,20)
\Gluon(75,65)(75,35){3}{3}
\Photon(100,20)(150,20){-3}{6}
\Photon(100,80)(150,80){3}{6}
\put(-15,46){$H$}
\put(105,46){$t,b$}
\put(82,48){$g$}
\put(155,18){$\gamma$}
\put(155,78){$Z$}

\end{picture}  \\
\setlength{\unitlength}{1pt}
\caption[ ]{\label{fg:hzgaqcddia} \it Typical diagram contributing to the QCD
corrections to $H\to Z \gamma$.}
\end{center}
\end{figure}
The two-loop QCD corrections to the top quark loops have been calculated
\cite{hzgaqcd} in complete analogy to the photonic case. They are generated by
virtual gluon exchange inside the quark triangle [see Fig.~\ref{fg:hzgaqcddia}].
Due to charge conjugation invariance and color conservation, radiation of a
single gluon is not possible. Hence the QCD corrections can simply be expressed
as a rescaling of the lowest order amplitude by a factor that only depends on
the ratios $\tau_i$ and $\lambda_i~~(i=f,W)$, defined above:
\begin{eqnarray}
A_Q^H(\tau_Q,\lambda_Q) & \to & A_Q^H(\tau_Q,\lambda_Q) \times \left[1+
D_H(\tau_Q,\lambda_Q) \frac{\alpha_s}{\pi} \right] \nonumber \\
D_H(\tau_Q,\lambda_Q) & \to & -1 \hspace*{1cm} \mbox{for $M_Z^2 \ll M_H^2\ll
4M_Q^2$} \, .
\label{eq:hzgaqcd}
\end{eqnarray}
In the limit $M_Z\to 0$ the quark amplitude approaches the corresponding form
factor of the photonic decay mode [{\it modulo} couplings], which has been
discussed before.
Hence the QCD correction in the heavy quark limit for small $Z$ masses has to
coincide with the heavy quark limit of the photonic decay mode of
eq.~(\ref{eq:hgagaqcd}). The QCD corrections for finite Higgs, $Z$ and quark
masses are presented in \cite{hzgaqcd} as a function of the Higgs mass. They
amount to less than 0.3\% in the intermediate mass range, where this decay
mode is relevant, and can thus be neglected.

\subsubsection{Intermediate gauge boson decays}
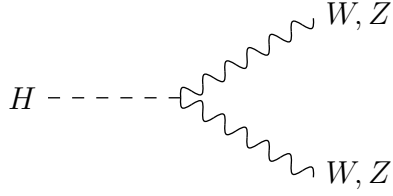
\begin{figure}[hbt]
\begin{center}
\setlength{\unitlength}{1pt}
\begin{picture}(200,100)(0,0)

\DashLine(0,50)(50,50){5}
\Photon(50,50)(100,80){3}{6}
\Photon(50,50)(100,20){-3}{6}
\put(-15,46){$H$}
\put(105,18){$W,Z$}
\put(105,78){$W,Z$}

\end{picture}  \\
\setlength{\unitlength}{1pt}
\caption[ ]{\label{fg:hvvdia} \it Diagram contributing to $H\to VV$ [$V=W,Z$].}
\end{center}
\end{figure}
\noindent
Above the $WW$ and $ZZ$ decay thresholds, the partial decay widths into pairs
of massive gauge bosons ($V = W,Z$) at lowest order [see Fig.~\ref{fg:hvvdia}]
are given by \cite{ztozh}
\begin{equation}
\Gamma(H\to VV) = \delta_V \frac{G_F M_H^3}{16\sqrt{2}\pi} \beta (1-4x+12x^2)
\, ,
\end{equation}
with $x = M_V^2/M_H^2$, $\beta = \sqrt{1-4x}$ and $\delta_V = 2\,(1)$ for
$V = W\, (Z)$.

The electroweak corrections have been computed in \cite{hqvelw,hvvelw} at the
one-loop
level. They are small and amount to less than about 5\% in the intermediate
mass range. Furthermore the QCD corrections to the leading top mass corrections
of ${\cal O}(G_F M_t^2)$ have been calculated up to three loops. They rescale
the $WW, ZZ$ decay widths by \cite{let,hzzmt}
\begin{eqnarray}
\Gamma(H\to ZZ) & = & \Gamma_{LO}(H\to ZZ) \left\{1 - x_t \left[ 5 -
(15-2\zeta_2)\frac{\alpha_s}{\pi} \right] \right\} \, , \\
\Gamma(H\to WW) & = & \Gamma_{LO}(H\to WW) \left\{1 - x_t \left[ 5 -
(9-2\zeta_2)\frac{\alpha_s}{\pi} \right] \right\} \, ,
\end{eqnarray}
with $x_t = G_F M_t^2 / (8\sqrt{2}\pi^2)$. The three-loop corrections can be
found in \cite{hvv3}. Since the electroweak corrections are
small in the intermediate mass regime, they are neglected in the analysis.
For large Higgs masses, higher order corrections due to the self-couplings of
the Higgs particles are relevant. They are given by \cite{hvvlam}
\begin{equation}
\Gamma (H\to VV) = \Gamma_{LO}(H\to VV) \left\{1 + 2.80 \hat \lambda
+ 62.03 \hat \lambda^2 \right\}
\end{equation}
with the coupling constant $\hat \lambda$ defined in eq.~(\ref{eq:hatlam}).
They start to be sizeable for $M_H \gsim 400$ GeV and increase the decay width
by about 20\% at Higgs masses of the order of $\sim 1$ TeV.

Below threshold the decays into off-shell gauge particles are
important. The partial decay widths into single off-shell gauge bosons can
be obtained in analytic form \cite{hvvp}
\begin{equation}
\Gamma(H\to VV^*) = \delta'_V \frac{3G^2_F M_V^4M_H}{16\pi^3} R\left(
\frac{M_V^2}{M_H^2} \right)
\label{eq:hvvp}
\end{equation}
with $\delta'_W = 1$, $\delta'_Z = 7/12 - 10\sin^2\theta_W/9 + 40
\sin^4\theta_W/27$ and
\begin{eqnarray}
R(x) & = & 3 \frac{1-8x+20x^2}{\sqrt{4x-1}} \arccos \left(\frac{3x-1}{2x^{3/2}}
\right) - \frac{1-x}{2x} (2-13x+47x^2) \\
& & \hspace*{7cm} - \frac{3}{2} (1-6x+4x^2) \log x \, . \nonumber
\end{eqnarray}
For Higgs masses slightly larger than the corresponding gauge boson mass the
decay widths into pairs of off-shell gauge bosons play a significant
role. Their contribution can be cast into the form \cite{2OFF}
\begin{equation}
\Gamma(H\to V^*V^*) = \frac{1}{\pi^2}\int_0^{M_H^2} \frac{dQ_1^2 M_V \Gamma_V}
{(Q_1^2 - M_V^2)^2 + M_V^2 \Gamma_V^2}
\int_0^{(M_H-Q_1)^2} \frac{dQ_2^2 M_V \Gamma_V} {(Q_2^2 - M_V^2)^2 + M_V^2
\Gamma_V^2} \Gamma_0
\end{equation}
with $Q_1^2, Q_2^2$ being the squared invariant masses of the virtual gauge
bosons, $M_V$ and $\Gamma_V$ their masses and total decay widths; $\Gamma_0$
is given by
\begin{equation}
\Gamma_0 = \delta_V \frac{G_F M_H^3}{16\sqrt{2}\pi}
\sqrt{\lambda(Q_1^2,Q_2^2;M_H^2)} \left[ \lambda(Q_1^2,Q_2^2;M_H^2) +
12\frac{Q_1^2Q_2^2}{M_H^4} \right] \, ,
\end{equation}
with the phase-space factor $\lambda(x,y;z)=(1-x/z-y/z)^2-4xy/z^2$.
The branching ratios of double off-shell decays reach the per cent level
for Higgs masses above about 100 (110) GeV for off-shell $W(Z)$ boson pairs.
They are therefore included in the analysis.

\subsubsection{Three-body decay modes}
The branching ratios of three-body decay modes may reach the per cent level for
large Higgs masses \cite{threebody}. The decays $H\to W^+W^- \gamma, t\bar t
\gamma (g)$ are already contained in the QED (QCD) corrections to the
corresponding decay widths $H\to W^+ W^-, t\bar t$.
However, the decay modes $H\to W^+W^- Z, t\bar t Z$ are not contained
in the electroweak corrections to the $WW, t\bar t$ decay widths. Their
branching ratios can reach values of up to about $10^{-2}$ for Higgs masses
$M_H\sim 1$ TeV. As they do not exceed the per cent level,
they are neglected in the present analysis. The analytical expressions are
rather involved and can be found in \cite{threebody}.

\subsubsection{Total decay width and branching ratios}
\begin{figure}[hbtp]

\vspace*{0.5cm}
\hspace*{0.0cm}
\begin{turn}{-90}%
\epsfxsize=9.5cm \epsfbox{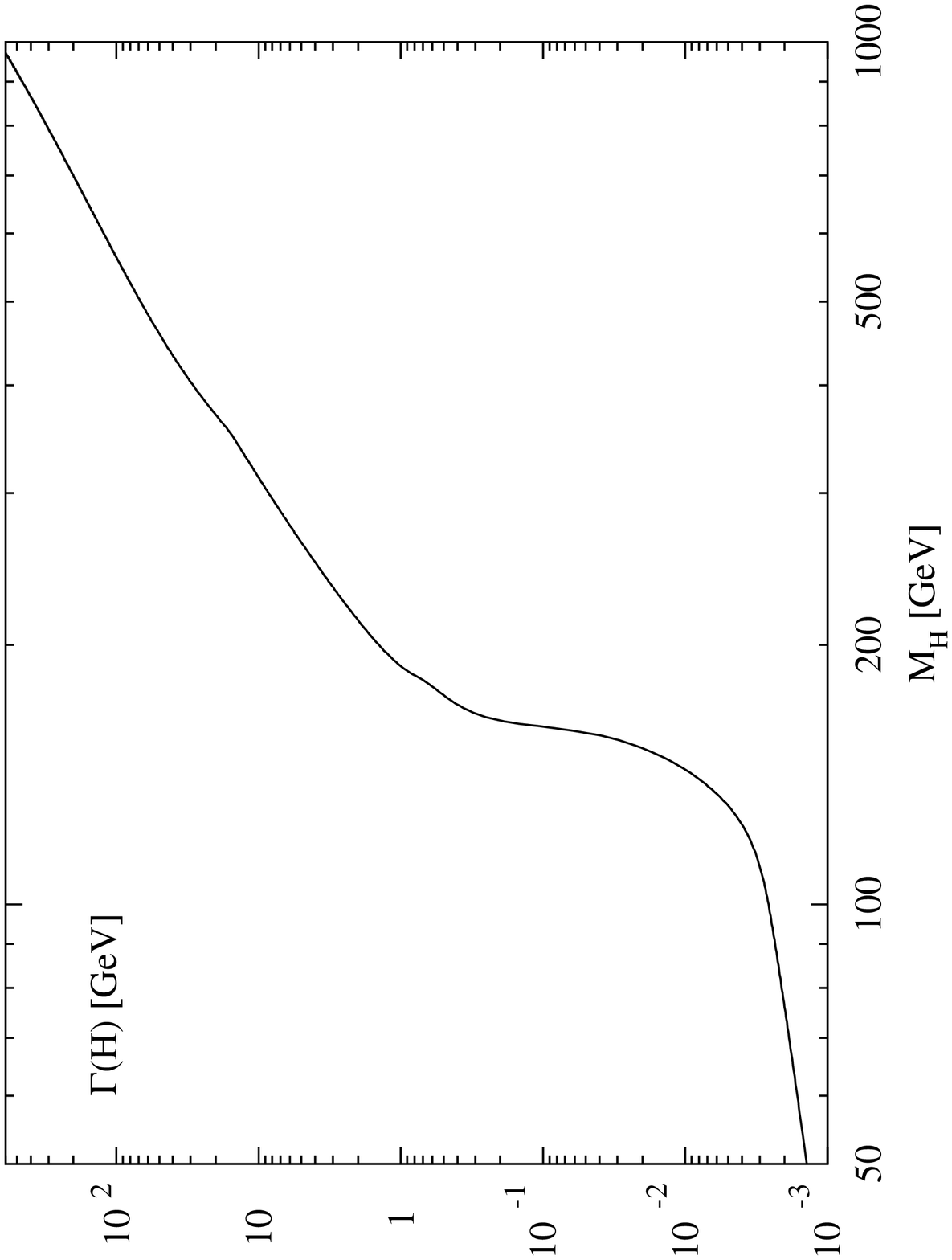}
\end{turn}

\vspace*{0.5cm}
\hspace*{0.0cm}
\begin{turn}{-90}%
\epsfxsize=9.5cm \epsfbox{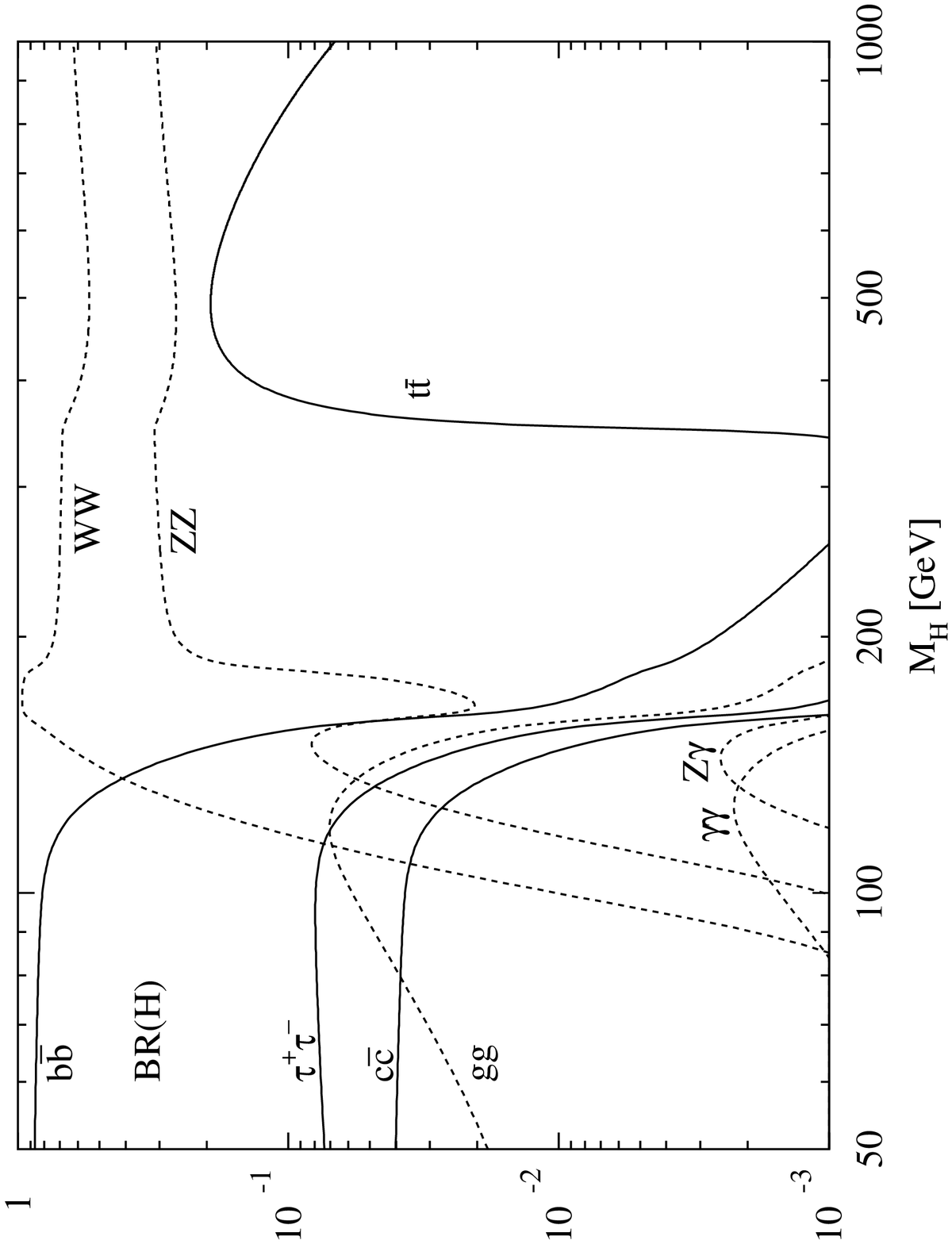}
\end{turn}
\vspace*{-0.0cm}

\caption[ ]{\label{fg:wtotbr} \it (a) Total decay width (in GeV) of the SM
Higgs boson as a function of its mass. (b) Branching ratios of the
dominant decay modes of the SM Higgs particle. All relevant higher order
corrections are taken into account.}
\end{figure}

In Fig.~\ref{fg:wtotbr}  the total decay width and branching ratios of the
Standard Model Higgs boson are shown as a function of the Higgs mass. For Higgs
masses below $\sim 140$ GeV, where the total width amounts to less than 10 MeV,
the dominant decay mode is the $b\bar b$ channel with a branching ratio up to
$\sim 85\%$. The remaining 10--20\% are supplemented by the $\tau^+\tau^-,
c\bar c$ and $gg$ decay modes, the branching ratios of which amount to 6.6\%,
4.6\% and 6\% respectively, for $M_H=120$ GeV [the $b\bar b$ branching ratio is
about 78\% for this Higgs mass]. The $\gamma\gamma$ ($Z\gamma$) branching
ratio turns out to be sizeable only for Higgs masses $80\, (120)~\GeV \lsim M_H
\lsim 150\, (160)$ GeV, where they exceed the $10^{-3}$ level.

Starting from $M_H\sim 140$ GeV the $WW$ decay takes over the dominant r\^ole
joined by the $ZZ$ decay mode. Around the $WW$ threshold of $150~\GeV \lsim M_H
\lsim 180$ GeV, where the $W$ pair of the dominant $WW$ channel becomes
on-shell, the $ZZ$ branching ratio drops down to a level of $\sim 2\%$ and
reaches again a branching ratio $\sim 30\%$ above the $ZZ$ threshold. Above the
$t\bar t$ threshold $M_H = 2M_t$, the $t\bar t$ decay mode opens up quickly,
but never exceeds a branching ratio of $\sim 20\%$. This is caused by the
fact that the leading $WW$ and $ZZ$ decay widths grow with the third power
of the Higgs mass [due to the longitudinal $W,Z$ components, which are
dominating for large Higgs masses], whereas the $t\bar t$ decay width
increases only with the first power. Consequently the total Higgs width grows
rapidly at large Higgs masses and reaches a level of $\sim 600$ GeV at $M_H=1$
TeV, rendering the Higgs width of the same order as its mass. At $M_H=1$
TeV the $WW$ ($ZZ$) branching ratio approximately reaches its asymptotic value
of 2/3 (1/3).

\subsection{Higgs Boson Production at the LHC}

\subsubsection{Gluon fusion: $gg\to H$}
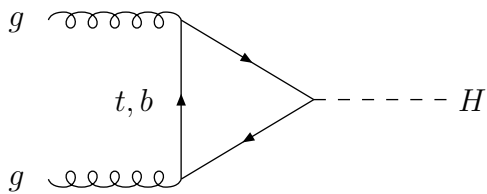
\begin{figure}[hbt]
\begin{center}
\setlength{\unitlength}{1pt}
\begin{picture}(180,100)(0,0)

\Gluon(0,20)(50,20){-3}{5}
\Gluon(0,80)(50,80){3}{5}
\ArrowLine(50,20)(50,80)
\ArrowLine(50,80)(100,50)
\ArrowLine(100,50)(50,20)
\DashLine(100,50)(150,50){5}
\put(155,46){$H$}
\put(25,46){$t,b$}
\put(-15,18){$g$}
\put(-15,78){$g$}

\end{picture}  \\
\setlength{\unitlength}{1pt}
\caption[ ]{\label{fg:gghlodia} \it Diagrams contributing to $gg\to H$
at lowest order.}
\end{center}
\end{figure}

The gluon-fusion mechanism \cite{glufus}
\begin{displaymath}
pp \to gg \to H
\end{displaymath}
provides the dominant production mechanism of Higgs bosons at the LHC
in the entire relevant Higgs mass range up to about 1 TeV. As in the case of the
gluonic decay mode, the gluon coupling to the Higgs boson in the SM is mediated
by triangular loops of top and bottom quarks, see Fig.~\ref{fg:gghlodia}. Since
the Yukawa coupling of the Higgs particle to heavy quarks
grows with the quark mass, thus balancing the decrease of the amplitude, the
form factor reaches a constant value for large loop quark masses. If the masses
of heavier quarks beyond the third generation are fully generated by the Higgs
mechanism, these particles would add the same amount to the form factor as the
top quark in the asymptotic heavy quark limit. Thus gluon fusion can serve as
a counter of the number of heavy quarks, the masses of which are generated
by the conventional Higgs mechanism. On the other hand, if these novel heavy
quarks will not be produced directly at the LHC, gluon fusion will allow to
measure the top quark Yukawa coupling. This, however, requires a precise
knowledge of the cross section within the SM with three generations of quarks.

The partonic cross section, Fig.~\ref{fg:gghlodia}, can be expressed by the
gluonic width of the Higgs boson at lowest order \cite{glufus},
\begin{eqnarray}
\hat\sigma_{LO} (gg\to H) & = & \sigma_0 \delta
(1 - z) \label{eq:gghlo} \\
\sigma_0 = \frac{\pi^2}{8M_H^3} \Gamma_{LO} (H\to gg)
& = & \frac{G_{F}\alpha_{s}^{2}(\mu)}{288 \sqrt{2}\pi} \
\left| \sum_{Q} A_Q^H (\tau_{Q}) \right|^{2}
\nonumber
\end{eqnarray}
where the scaling variables are defined as $z=M_H^2/\hat s$, $\tau_Q=4M_Q^2/
M_H^2$, and $\hat{s}$ denotes the partonic c.m.~energy squared. The amplitudes
$A_Q^H(\tau_Q)$ are presented in eq.~(\ref{eq:ftau}).

In the narrow-width approximation the hadronic cross section can be cast into
the form \cite{glufus}
\begin{equation}
\sigma_{LO}(pp\to H) = \sigma_0 \tau_H \frac{d{\cal L}^{gg}}{d\tau_H}
\end{equation}
with
\begin{equation}
\frac{d{\cal L}^{gg}}{d\tau} = \int_\tau^1 \frac{dx}{x}~g(x,M^2)
g(\tau /x,M^2)
\label{eq:gglum}
\end{equation}
denoting the gluon luminosity at the factorization scale $M$, and the scaling
variable is defined, in analogy
to the Drell--Yan process, as $\tau_H = M^2_H/s$, with $s$ specifying the total
hadronic c.m.~energy squared.

\begin{figure}[hbt]
\begin{center}
\setlength{\unitlength}{1pt}
\begin{picture}(450,100)(-10,0)

\Gluon(0,20)(30,20){-3}{3}
\Gluon(0,80)(30,80){3}{3}
\Gluon(30,20)(60,20){-3}{3}
\Gluon(30,80)(60,80){3}{3}
\Gluon(30,20)(30,80){3}{5}
\ArrowLine(60,20)(60,80)
\ArrowLine(60,80)(90,50)
\ArrowLine(90,50)(60,20)
\DashLine(90,50)(120,50){5}
\put(125,46){$H$}
\put(65,46){$t,b$}
\put(-10,18){$g$}
\put(-10,78){$g$}
\put(15,48){$g$}

\Gluon(180,100)(210,100){3}{3}
\Gluon(210,100)(270,100){3}{6}
\Gluon(180,0)(210,0){-3}{3}
\Gluon(210,100)(210,60){3}{4}
\ArrowLine(210,0)(210,60)
\ArrowLine(210,60)(240,30)
\ArrowLine(240,30)(210,0)
\DashLine(240,30)(270,30){5}
\put(275,26){$H$}
\put(215,26){$t,b$}
\put(165,-2){$g$}
\put(165,98){$g$}
\put(195,78){$g$}
\put(275,98){$g$}

\ArrowLine(330,100)(360,100)
\ArrowLine(360,100)(420,100)
\Gluon(330,0)(360,0){-3}{3}
\Gluon(360,100)(360,60){3}{4}
\ArrowLine(360,0)(360,60)
\ArrowLine(360,60)(390,30)
\ArrowLine(390,30)(360,0)
\DashLine(390,30)(420,30){5}
\put(425,26){$H$}
\put(365,26){$t,b$}
\put(315,-2){$g$}
\put(315,98){$q$}
\put(425,98){$q$}
\put(345,78){$g$}

\end{picture}  \\
\setlength{\unitlength}{1pt}
\caption[ ]{\label{fg:gghqcddia} \it Typical diagrams contributing to the
virtual and real QCD corrections to $gg\to H$.}
\end{center}
\end{figure}
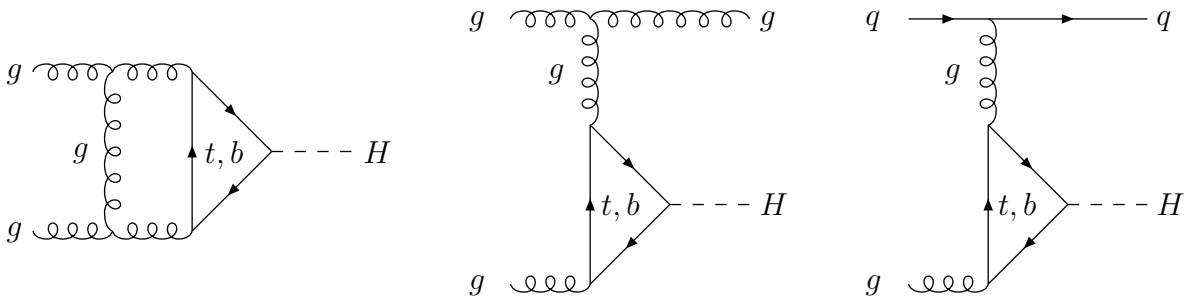
\paragraph{QCD corrections.}
In the past the two-loop QCD corrections to the gluon-fusion cross
section, Fig.~\ref{fg:gghqcddia}, have been calculated
\cite{higgsqcd,hgg,gghsm,phd}. They consist of
virtual corrections to the basic $gg\to H$ process and real corrections
due to the associated production of the Higgs boson with massless partons,
\begin{displaymath}
gg \rightarrow Hg \hspace{0.5cm} \mbox{and} \hspace{0.5cm}
gq \rightarrow Hq,~q\overline{q} \rightarrow Hg \, .
\end{displaymath}
These subprocesses contribute to the Higgs production at ${\cal O}
(\alpha_s^3)$. The virtual corrections rescale the lowest-order fusion cross
section with a coefficient depending only on the ratios of the Higgs and quark
masses. Gluon radiation leads to
two-parton final states with invariant energy $\hat{s}\ge m_H^2$ in the $gg,gq$
and $q\overline{q}$ channels. The final result for the hadronic cross section
can be split into five parts \cite{higgsqcd,hgg,gghsm,phd},
\begin{equation}
\sigma(pp \rightarrow H+X) = \sigma_{0} \left[ 1+ C
\frac{\alpha_{s}}{\pi} \right] \tau_{H} \frac{d{\cal L}^{gg}}{d\tau_{H}} +
\Delta \sigma_{gg} + \Delta \sigma_{gq} + \Delta \sigma_{q\bar{q}} \, ,
\end{equation}
with the renormalization scale in $\alpha_s$ and the factorization scale of
the parton densities to be fixed properly. The lengthy analytic expressions
for arbitrary Higgs boson and quark masses can be found in
Refs.~\cite{higgsqcd,phd}.
The quark-loop mass has been identified with the pole mass $M_Q$, while the
QCD coupling is defined in the $\overline{\rm MS}$ scheme. We have adopted
the $\overline{\rm MS}$ factorization scheme for the NLO parton densities. 

The coefficient $C(\tau_Q)$ denotes the finite part of the virtual two-loop
corrections. It splits into the infrared part $\pi^2$, a
logarithmic term depending on the renormalization scale $\mu$
and a finite quark-mass-dependent piece $c(\tau_Q)$,
\begin{equation}
C(\tau_Q) = \pi^{2}+ c(\tau_Q) + \frac{33-2N_{F}}{6} \log
\frac{\mu^{2}}{M_{H}^{2}} \, .
\label{eq:Cvirt}
\end{equation}
The term $c(\tau_Q)$ can be reduced analytically to a one-dimensional
Feynman-parameter integral, which has been evaluated numerically
\cite{higgsqcd,gghsm,phd}. In
the heavy-quark limit $\tau_Q = 4M^2_Q / M^2_H \gg 1$ and in the light-quark
limit $\tau_Q \ll 1$, the integrals could be solved analytically.

The finite parts of the hard contributions from gluon radiation in $gg$
scattering, $gq$ scattering and $q \overline{q}$ annihilation depend on the
renormalization scale $\mu$ and the factorization scale $M$ of the parton
densities:
\begin{eqnarray}
\Delta \sigma_{gg} & = & \int_{\tau_{H}}^{1} d\tau \frac{d{\cal
L}^{gg}}{d\tau} \times \frac{\alpha_{s}}{\pi} \sigma_{0} \left\{ - z
P_{gg} (z) \log \frac{M^{2}}{\hat{s}} + d_{gg} (z,\tau_Q) \right. \non \\
& & \left. \hspace{3.7cm} + 12 \left[ \left(\frac{\log
(1-z)}{1-z} \right)_+ - z[2-z(1-z)] \log (1-z) \right] \right\} \non \\ \non \\
\Delta \sigma_{gq} & = & \int_{\tau_{H}}^{1} d\tau \sum_{q,
\bar{q}} \frac{d{\cal L} ^{gq}}{d\tau} \times \frac{\alpha_{s}}{\pi}
\sigma_{0} \left\{ -\frac{z}{2} P_{gq}(z) \log\frac{M^{2}}{\hat{s}(1-z)^2}
+ d_{gq} (z,\tau_Q) \right\} \non \\ \non \\
\Delta \sigma_{q\bar{q}} & = & \int_{\tau_{H}}^{1} d\tau
\sum_{q} \frac{d{\cal L}^{q\bar{q}}}{d\tau} \times \frac{\alpha_{s}}{\pi}
\sigma_{0}~d_{q\bar q} (z,\tau_Q) \, ,
\label{eq:gghqcd}
\end{eqnarray}
with $z = \tau_H / \tau = M_H^2/\hat s$; $P_{gg}$ and $P_{gq}$ are the
standard Altarelli--Parisi splitting functions \cite{apsplit}:
\begin{eqnarray}
P_{gg}(z) & = & 6 \left\{ \left( \frac{1}{1-z} \right)_+ + \frac{1}{z} -2 +
z (1-z) \right\} + \frac{33-2N_F}{6} \delta(1-z) \nonumber \\
P_{gq}(z) & = & \frac{4}{3} \frac{1+ (1-z)^2}{z} \, ;
\label{eq:APKernel}
\end{eqnarray}
$F_+$ denotes the usual $+$ distribution: $F(z)_+ = F(z) - \delta
(1 - z) \int_0^1 dz' F(z')$. The
coefficients $d_{gg}, d_{gq}$ and $d_{q \ov{q}}$
can be reduced to one-dimensional integrals, which have
been evaluated numerically \cite{higgsqcd,gghsm,phd} for arbitrary quark masses.
They can be calculated analytically in the heavy- and
light-quark limits.

In the heavy-quark limit $\tau_Q \gg 1$ the coefficients $c(\tau_Q)$ and
$d_{ij} (z,\tau_Q)$ reduce to very simple expressions
\cite{higgsqcd,hgg,dawson},
\begin{eqnarray}
c(\tau_Q) & \displaystyle \to\frac{11}{2} \hspace{4.1cm} d_{gg}(z,\tau_Q) & \to
-\frac{11}{2} (1-z)^3 \non \\
d_{gq}(z,\tau_Q) & \displaystyle \to\frac{2}{3}z^2 - (1 - z)^2 \hspace{2cm}
d_{q\bar q}(z,\tau_Q) & \to \frac{32}{27} (1-z)^3 \, .
\label{eq:gghqcdlim}
\end{eqnarray}
The corrections of ${\cal O} (M_H^2/M_Q^2)$ in a systematic Taylor
expansion have been demonstrated to be very small \cite{gghsub}. In fact,
the leading term provides an excellent approximation
up to the quark threshold $M_H \sim 2 M_Q$.
In the opposite limit where the Higgs mass is much larger than the top mass,
the analytic result can be found in \cite{higgsqcd}.

\begin{figure}[hbt]

\vspace*{0.5cm}
\hspace*{0.0cm}
\begin{turn}{-90}%
\epsfxsize=10cm \epsfbox{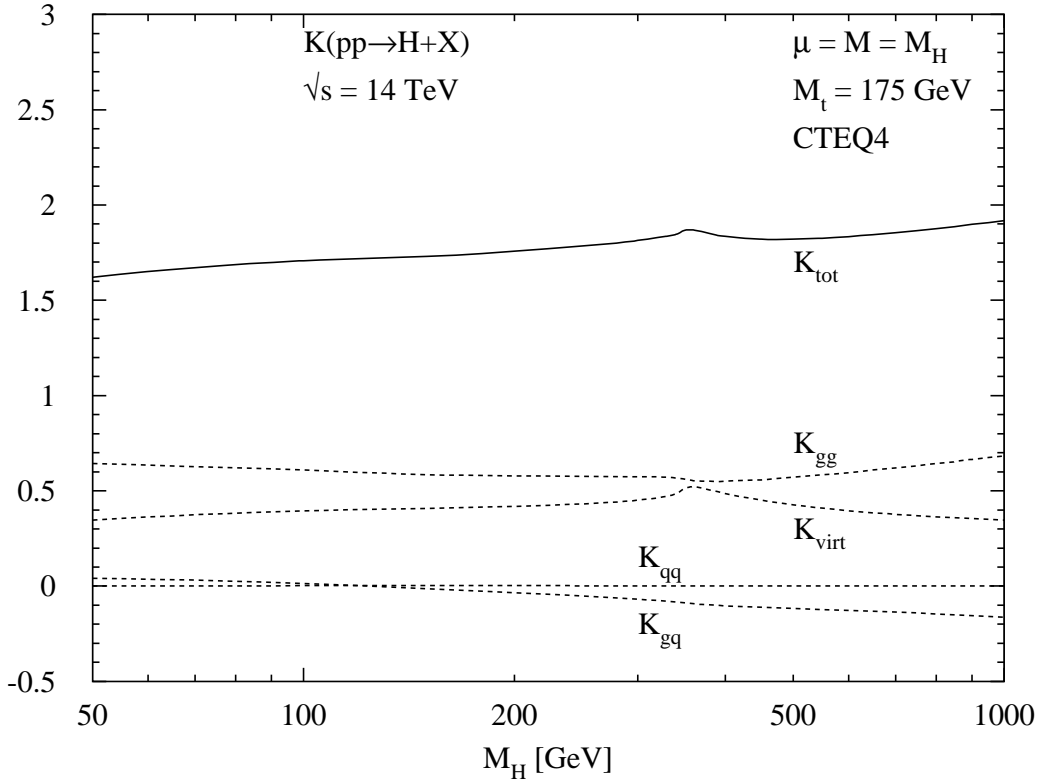}
\end{turn}
\vspace*{-0.0cm}

\caption[]{\label{fg:gghk} \it K factors of the QCD-corrected gluon-fusion
cross section $\sigma(pp \to H+X)$ at the LHC with c.m.~energy $\sqrt{s}=14$
TeV. The dashed lines show the individual contributions of the four terms of
the QCD corrections given in eq.~(\ref{eq:gghqcd}). The renormalization and
factorization scales have been identified with the Higgs mass, $\mu=M=M_H$
and the CTEQ4 parton densities have been adopted.}
\end{figure}
We define $K$ factors as the ratio
\begin{eqnarray}
K_{tot} = \frac{\sigma_{NLO}}{\sigma_{LO}} \, .
\end{eqnarray}
The cross sections $\sigma_{NLO}$ in next-to-leading order
are normalized to the leading-order cross sections $\sigma_{LO}$,
convoluted consistently with parton densities and $\alpha_s$
in leading order; the NLO and LO strong couplings are chosen from the CTEQ4
parametrizations \cite{cteq4} of the structure functions, $\alpha_s^{NLO}(M_Z)
=0.116,~\alpha_s^{LO}(M_Z)=0.132$. The $K$ factor can be
decomposed into several characteristic components: $K_{virt}$ accounts
for the regularized virtual corrections, corresponding to the coefficient
$C$; $K_{AB}$ [$A,B=g, q, \bar{q}$] for the real corrections as defined in
eqs.~(\ref{eq:gghqcd}). These $K$ factors are presented for LHC energies in
Fig.~\ref{fg:gghk} as a function of the Higgs boson mass. Both the
renormalization and the factorization scales have been identified with the
Higgs mass, $\mu = M = M_H$. Apparently $K_{virt}$
and $K_{gg}$ are of the same size, of order 50\%, while
$K_{gq}$ and $K_{q\overline{q}}$ turn out to be quite small.
Apart from the threshold region $M_H\sim 2M_t$, $K_{tot}$ is insensitive to
the Higgs mass.

\begin{figure}[hbt]

\vspace*{0.5cm}
\hspace*{-1.0cm}
\begin{turn}{-90}%
\epsfxsize=11cm \epsfbox{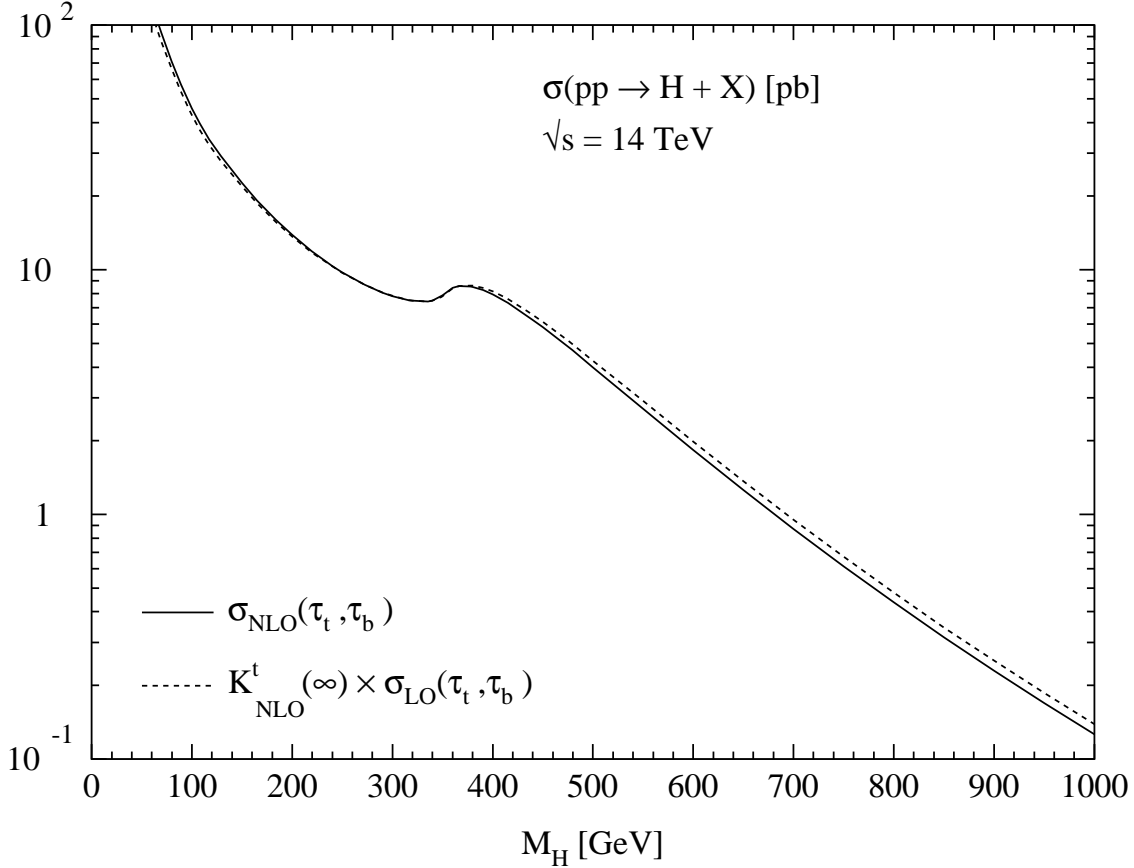}
\end{turn}
\vspace*{-0.1cm}

\caption[]{\label{fg:gghapprox} \it Comparison of the exact and approximate
NLO cross section $\sigma(pp\to H+X)$ at the LHC with c.m.~energy $\sqrt{s}=14$
TeV. The solid line shows the exact cross section including the full $t,b$
quark mass dependence and the dashed line corresponds to the approximation
defined in eq.~(\ref{eq:gghapprox}). The renormalization and factorization
scales have been identified with the Higgs mass, $\mu=M=M_H$ and the CTEQ4
parton densities \cite{cteq4} with NLO strong coupling [$\alpha_s(M_Z)=0.116$]
have been
adopted. The top mass has been chosen as $M_t=175$ GeV and the bottom mass as
$M_b=5$ GeV.}
\end{figure}
The corrections are positive and large, increasing the Higgs production cross
section at the LHC by about 60\% to 90\%. Comparing the exact numerical results
with the analytic expressions in the heavy-quark limit, it turns out that these
asymptotic $K$ factors provide an excellent approximation even for Higgs masses
above the top-decay threshold. We explicitly define the approximation by
\begin{eqnarray}
\sigma_{app} & = & K_{NLO}^t (\infty) \times \sigma_{LO}(\tau_t,\tau_b)
\label{eq:gghapprox} \\
K_{NLO}^t (\infty) & = & \lim_{M_t \to \infty} K_{tot} \nonumber
\end{eqnarray}
where we neglect the $b$ quark contribution in $K_{NLO}^t (\infty)$, while the
leading order cross section $\sigma_{LO}$ includes the full $t,b$ quark mass
dependence. The comparison with the full massive NLO result is presented in
Fig.~\ref{fg:gghapprox}. The solid line corresponds to the exact cross
section and the broken line to the approximate one. For Higgs masses below
$\sim$ 1 TeV, the deviations of the asymptotic approximation from the full
NLO result are less than 10\%.

Theoretical uncertainties in the prediction of the Higgs cross section
originate from two sources, the dependence of the cross section on different
parametrizations of the parton densities and the unknown NNLO corrections.

\begin{figure}[hbt]

\vspace*{0.5cm}
\hspace*{0.0cm}
\begin{turn}{-90}%
\epsfxsize=10cm \epsfbox{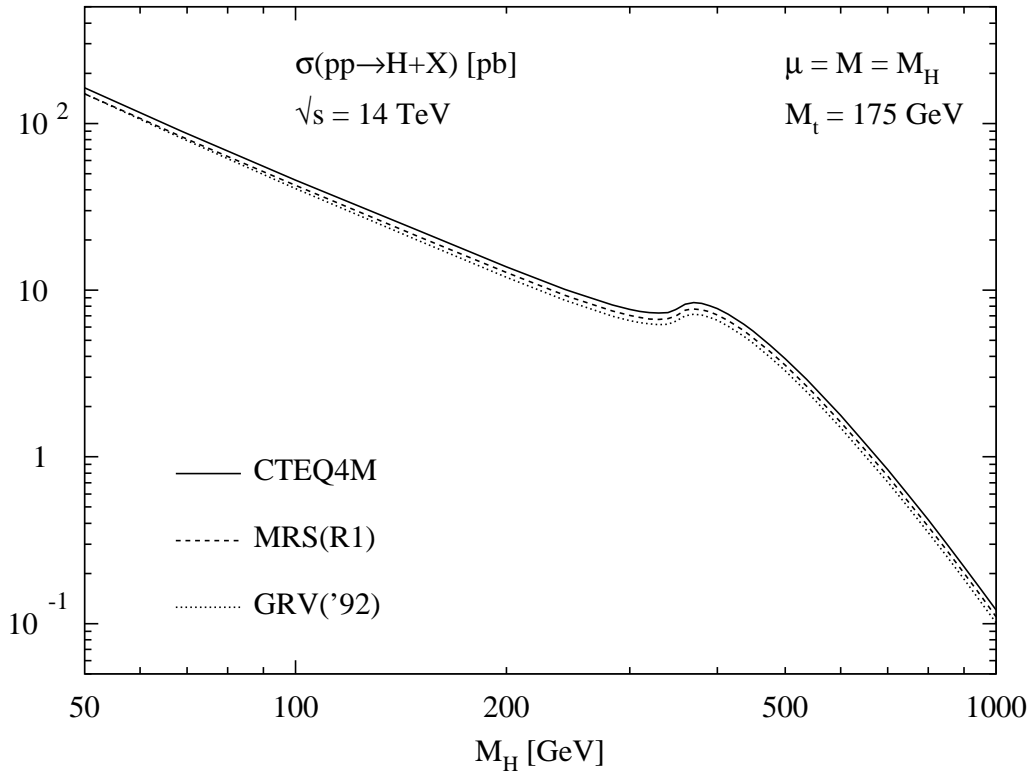}
\end{turn}
\vspace*{0.0cm}

\caption[]{\label{fg:gghparton} \it Higgs production cross section for three
different sets of parton densities [CTEQ4M, MRS(R1) and GRV('92)].}
\end{figure}
The uncertainty of the gluon density causes one of the main uncertainties in
the prediction of the Higgs production cross section.  This distribution can
only indirectly be extracted through order $\alpha_s$ effects from
deep-inelastic lepton--nucleon scattering, or by means of complicated analyses
of final states in lepton--nucleon and hadron--hadron scattering.  Adopting a
representative set of recent parton distributions \cite{cteq4,stfu}, we find
a variation of about $\pm 10\%$ of the cross section for the entire Higgs mass
range. The cross section for these different sets of parton
densities is presented in Fig.~\ref{fg:gghparton} as a function of the Higgs
mass. The uncertainty will be smaller in the near future, when the
deep-inelastic electron/positron--nucleon scattering experiments at HERA will
have reached the anticipated level of accuracy.

\begin{figure}[hbtp]

\vspace*{1.7cm}
\hspace*{-6.0cm}
\begin{turn}{-90}%
\epsfxsize=18cm \epsfbox{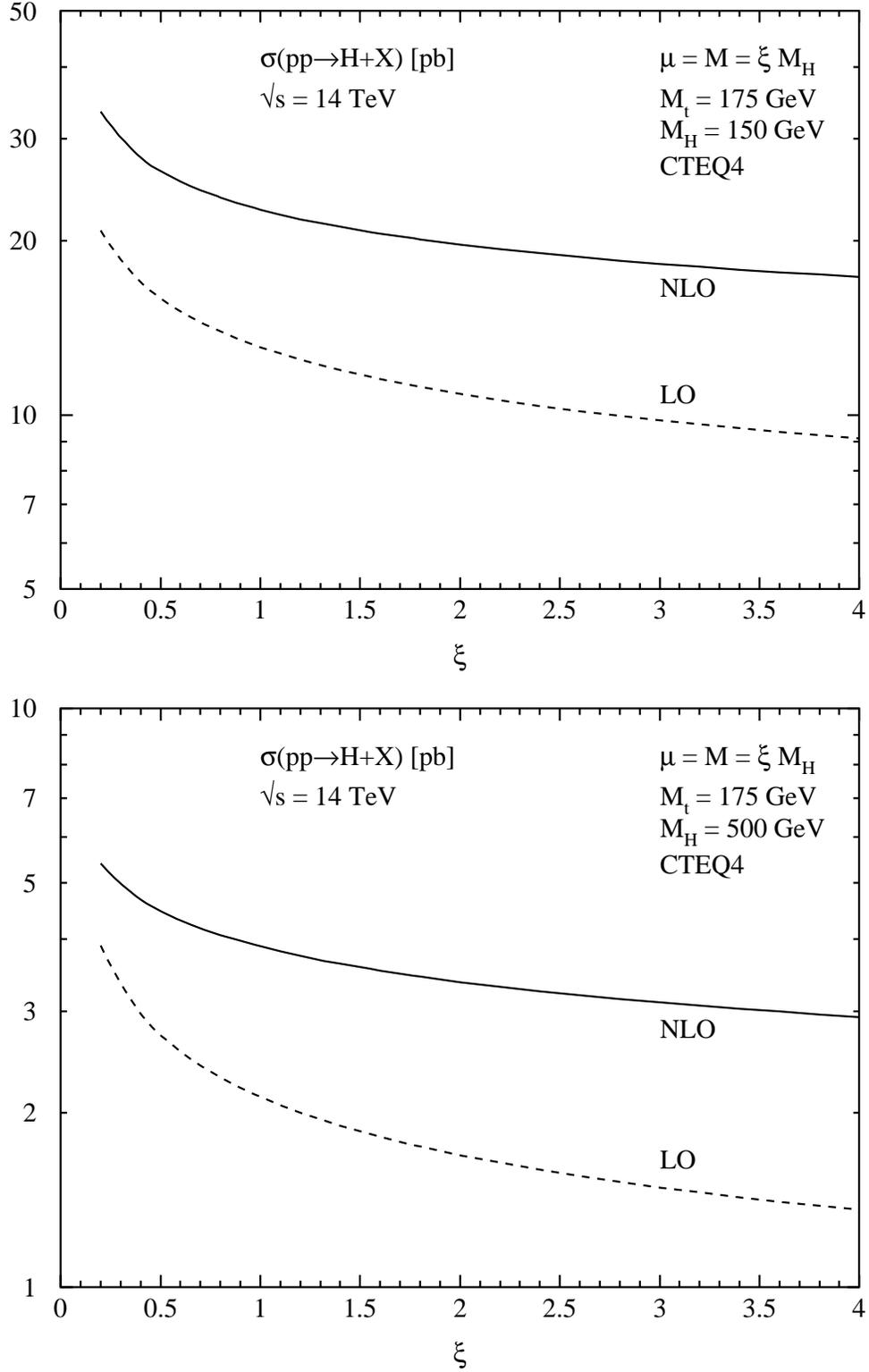}
\end{turn}
\vspace*{0.5cm}

\caption[]{\label{fg:gghscale} \it The renormalization and factorization scale
dependence of the Higgs production cross section at lowest and next-to-leading
order for two different Higgs masses $M_H = 150$ and $500$ GeV.}
\end{figure}
The [unphysical] variation of the cross section with the
renormalization and factorization scales is reduced by
including the next-to-leading order corrections. This is
shown in Fig.~\ref{fg:gghscale} for two typical values
of the Higgs mass, $M_H = 150, 500$ GeV. The renormalization/factorization
scale $\mu = M$ is varied in units of the Higgs mass, $\mu = \xi M_H$ for
$\xi$ between 1/2 and 2.
The ratio of the cross sections for $\xi=1/2$ and 2 is reduced from 1.62
in leading order to 1.32 in next-to-leading order for $M_H = 500~\GeV$.
Since, for small Higgs masses, the dependence on $\mu$ for
$\xi \sim 1$ is already small at the LO level, the improvement by the NLO
corrections is less significant for a Higgs mass $M_H=150$~GeV. However, the
figures indicate that further improvements are required, because the $\mu$
dependence of the cross section is still monotonic in the parameter range set
by the natural scale $\mu\sim M\sim M_H$. The uncertainties due to the scale
dependence appear to be less than about 15\%.

\paragraph{Soft gluon resummation.}
Recently soft and collinear gluon radiation effects for the total gluon-fusion
cross section have been resummed. The perturbative expansion of the resummed
result leads to an approximation of the three-loop NNLO corrections
of the partonic cross section in the heavy top mass limit, which approximates
the full massive NLO result with a reliable precision [see
Fig.~\ref{fg:gghapprox}]. Owing to the low-energy theorem discussed before [see
the gluonic decay mode $H\to gg$], the unrenormalized partonic cross section
factorizes, in $n=4-2\eps$ dimensions, as
\beq
\hat\sigma^0_{gg} = \sigma^\eps_0~\kappa~\rho_0\left(z,\frac{M_H^2}{\mu^2},
\alpha_s(\mu),\epsilon\right) \, ,
\label{eq:sigfac}
\eeq
where $\kappa$ originates from the effective Lagrangian of
eq.~(\ref{eq:hggeffnnlo}),
\bea
\kappa & = & 1 + \frac{11}{2} \frac{\alpha^{(5)}_s(M_t)}{\pi}
+ \frac{3866 - 201\, N_F}{144} \left(\frac{\alpha_s^{(5)}(M_t)}{\pi} \right)^2
\nonumber \\
& & \hspace*{1cm} + \frac{153-19 N_F}{33-2 N_F}
~\frac{\alpha^{(5)}_s(M_{H}) - \alpha^{(5)}_s(M_t)}{\pi}
+ {\cal O}(\alpha_s^3)
\label{eq:kappa}
\eea
[with $N_F=5$]
and the factor $\sigma_0^\eps$ reads as
\beq
\sigma_0^\eps = \frac{\Gamma^2(1+\eps)}{1-\eps} \left(\frac{4\pi}{M_t^2}
\right)^{2\eps} \sigma_0 \, ,
\eeq
where the coefficient $\sigma_0$ is defined in eq.~(\ref{eq:gghlo}) with the
strong coupling $\alpha_s(\mu)$ replaced by the bare one, $\alpha_{s0}$. The
bare correction factor $\rho_0(z,M_H^2/\mu^2,\eps)$ arises from the effective
diagrams analogous to Fig.~\ref{fg:hgglimdia} in higher orders. In the following
we shall neglect
the contributions from $gq$ and $q\bar q$ initial states, which contribute less
than $\sim 10\%$ to the gluon-fusion cross section at NLO. The hadronic cross
section can be obtained by convoluting eq.~(\ref{eq:sigfac}) with the bare gluon
densities,
\beq
\sigma (\tau_H,M_H^2,\mu^2) = \int_{\tau_H}^1 dx_1
\int_{\tau_H/x_1}^1 dx_2~g_0(x_1)~g_0(x_2)~\hat \sigma^0_{gg}
(z,M_H^2,\mu^2,\epsilon)
\eeq
with the scaling variables $z=\tau_H/(x_1 x_2)$ and $\tau_H=M_H^2/s$, where $s$
denotes the hadronic c.m.~energy squared. The moments of the hadronic cross
section factorize into three factors:
\beq
\tilde \sigma(N,M_H^2,\mu^2) = \int_0^1 d\tau_H~\tau_H^{N-1}~\sigma
(\tau_H,M_H^2,\mu^2) = \tilde g_0^2(N+1)~\tilde{\hat\sigma}^{\!\!
~_{\scriptstyle 0}}_{gg} (N,M_H^2,\mu^2,\epsilon) \, .
\eeq
The bare correction factor $\rho_0$ may be expanded perturbatively,
\beq
\rho_0 \left(z,\frac{M_H^2}{\mu^2},\alpha_s(\mu),\epsilon \right) =
\sum_{n=0}^{\infty}~\left( \frac{\alpha_s(\mu)}{\pi} \right)^n \rho_0^{(n)}
\left(z,\frac{M_H^2}{\mu^2},\epsilon \right) \, .
\label{eq:rho0exp}
\eeq
The first two [unrenormalized] coefficients are known from the explicit NLO
calculation \cite{higgsqcd,hgg,dawson,gghsm}, see eq.~(\ref{eq:gghqcd}):
\begin{eqnarray}
\rho_0^{(0)} \left(z,\frac{M_H^2}{\mu^2},\epsilon \right) & = & \delta(1-z) \\
\rho_0^{(1)} \left(z,\frac{M_H^2}{\mu^2},\epsilon \right) & = &
\left(\frac{\mu^2}{M_H^2}\right)^\ep \left\{ -3\frac{z^\epsilon}{\epsilon}
\left[\frac{1+z^4+(1-z)^4}
{(1-z)^{1+2\epsilon}}\right]_+ \right. \nonumber \\
&+& \left. \delta(1-z)\left(\frac{11}{2\epsilon}
 + \frac{203}{12} + \pi^2\right)
-\frac{11}{2} z^\epsilon (1-z)^{3-2\epsilon} \right\}
\label{rhonlo}
\end{eqnarray}
where we have absorbed trivial constants into a redefinition of the scale,
$\mu^2 \to \mu^2 \exp[\gamma_E - \log (4\pi)]$.

The starting point for the resummation is provided by the Sudakov evolution
equation \cite{sudevol}
\beq
M_H^2 \frac{d}{dM_H^2}
\rho_0\!\left(z,\frac{M_H^2}{\mu^2},\alpha_s(\mu),\epsilon \right)\!  = \!
\int^1_z \frac{dz'}{z'} W_0\!\left(z',\frac{M_H^2}{\mu^2},
\alpha_s(\mu),\epsilon\right) \rho_0\!\left(\frac{z}{z'},\frac{M_H^2}{\mu^2},
\alpha_s(\mu),\epsilon\right) \, ,
\label{eq:sudevol}
\eeq
which follows from the basic factorization theorems for partonic cross
sections into soft, collinear and hard parts at the boundaries of the phase
space \cite{factheo}. The solution for the moments of eq.~(\ref{eq:sudevol}) is
given by
\bea
\tilde \rho_0 \left(N,\frac{M_H^2}{\mu^2},\alpha_s(\mu),\ep \right)
& = & \exp\left[ \int_0^{M_H^2/\mu^2}
 {d \lambda \over \lambda} 
{\tilde W}_0 \biggl(N,\lambda,\alpha_s(\mu),\ep\biggr)
\right] \nonumber \\
& = & \exp\left[ \int_0^1 dz z^{N-1} \int_0^{M_H^2/\mu^2}
 {d \lambda \over \lambda} 
W_0 \biggl(z,\lambda,\alpha_s(\mu),\ep\biggr)
\right]
\label{eq:sudsolv}
\eea
where we have imposed the boundary condition
\beq
\rho_0\left(z,\frac{M_H^2}{\mu^2}=0,\alpha_s(\mu),\epsilon\right) =
\delta(1-z) \, ,
\eeq
which is valid in $n$ dimensions. The bare evolution kernel $W_0(z,\lambda,
\alpha_s(\mu),\eps)$ can be evaluated perturbatively. After renormalizing the
strong coupling $\alpha_s$ and the gluon densities in the $\msb$ scheme all
singularities cancel, and the finite renormalized correction
factor reads as \cite{gghresum}
\bea
\rho\left(N,\frac{M_H^2}{\mu^2},\alpha_s(\mu)\right) \!\!\!\! &=&
\!\!\!\! \exp\left[-6\int_0^1 dz\frac{z^{N-1}-1}{1-z}
\int_{(1-z)^2\frac{M_H^2}{\mu^2}}^1\frac{d\lambda}{\lambda}
\frac{\alpha_s(\lambda\mu^2)}{\pi} \right] \nonumber\\
&\times & \!\!\!\!  \exp\left\{\frac{\alpha_s(M_H^2)}{\pi} \left[\pi^2+203/12
-11/2\, \log \left(\frac{M_H^2}{\mu^2}\right) \right] \right.
\nonumber \\
&& \quad\quad\left. -\frac{11}{8}\frac{\alpha_s^2(M_H^2)}{\pi^2}\beta_0\log^2
\left( \frac{M_H^2}{\mu^2}\right) \right\} \nonumber \\
&\times & \!\!\!\! \exp\left[-6\!\int_0^1\!dz(2z-z^2+z^3)\!
\int_{(1-z)^2\frac{M_H^2}{\mu^2}}^1\frac{d\lambda}{\lambda}
\frac{\alpha_s(\lambda \mu^2)}{\pi} \right]
\nonumber \\
&\times & \!\!\!\! \exp\left[+12\int_0^1dz~z^{N-1}
\int_{(1-z)^2\frac{M_H^2}{\mu^2}}^1\frac{d\lambda}{\lambda}
\frac{\alpha_s(\lambda \mu^2)}{\pi} \right] \, ,
\label{eq:rhoresum}
\eea
with $\beta_0 = (33-2 N_F)/6$.
It should be noted that in the last exponential we have kept terms of ${\cal
O}(\log^i N/N)~(i\ge 1)$ in the moments of the correction factor, which are
not covered by the basic factorization theorems near the soft and collinear
edges of phase space.  On the other hand at NLO they turn out to originate
from collinear gluon radiation and are thus universal, so that they can be
included in the resummation\footnote{Their inclusion in the Drell--Yan process
and deep-inelastic scattering yields the correct coefficients of the
$\log^3N/N$ terms and those $\log^2N/N$ terms, which are related to the strong
coupling constant, at NNLO, which supports the consistency of
their resummation.  However, a rigorous proof has not been worked out so far.}.
In order to define the resummed correction factor we have to perform a
regularization of the singularity at $\lambda\sim \Lambda_{\rm QCD}$, which is
related to an infrared renormalon. Nevertheless, the perturbative expansion
is well defined. The NLO and NNLO results for $\mu = M$ read \cite{gghresum}
\bea
\!\!\!\!\!\!
\rho^{(1)} \left(z,\frac{M_H^2}{\mu^2} \right)\!\!\!\! & = &\!\!\!\!
12 {\cal D}_1(z) - 24{\cal E}_1(z) - 6 {\cal D}_0(z) L_\mu + \pi^2 \delta(1-z)
\label{eq:rhonlo} \\ \nonumber \\
\rho^{(2)} \left(z,\frac{M_H^2}{\mu^2} \right) & = &
3\left\{
      24 {\cal D}_3(z) +
      (-2 \beta_0 - 36 L_\mu ) {\cal D}_2(z)
    + ( - 24 \zeta_2  +2 \beta_0 L_\mu  +
            12 L_\mu^2) {\cal D}_1(z) \right. \nonumber \\
&+&
      (48 \zeta_3 + 12 \zeta_2 L_\mu -
            \frac{1}{2} \beta_0 L_\mu^2){\cal D}_0(z)
       - 48 {\cal E}_3(z)
\nonumber \\
&+&
       (4\beta_0 + 24 + 72 L_\mu ){\cal E}_2(z)
  + (48 \zeta_2-4\beta_0 L_\mu - 24 L_\mu - 24 L_\mu^2){\cal E}_1(z)
\nonumber \\
&+& \left.
    (18 \zeta_2^2 - 36 \zeta_4 - \frac{2909}{432}\beta_0
               + \zeta_2 \beta_0 L_\mu
          - 24 \zeta_3L_\mu -
              6 \zeta_2 L_\mu^2 ) \delta(1-z)  \right\} \, ,
\label{rhonnlo}
\eea
where we use the notation
\beq
{\cal D}_i(z) = \left[\frac{\log^i(1-z)}{1-z}\right]_+ \, ,
\quad \quad \quad
{\cal E}_i(z) = \log^i(1-z) \, ,
\quad \quad \quad
 L_\mu = \log \left( \frac{\mu^2}{M_H^2} \right) \, .
\eeq
The novel contributions of ${\cal O}(\log^i N/N)$ appear as the non-infrared
functions ${\cal E}_i(z)$. They are of significant importance for processes at
the LHC and therefore have to be included to gain a reliable approximation by
means of soft gluon resummation.

\begin{figure}[hbt]

\vspace*{0.5cm}
\hspace*{-1.0cm}
\begin{turn}{-90}%
\epsfxsize=11cm \epsfbox{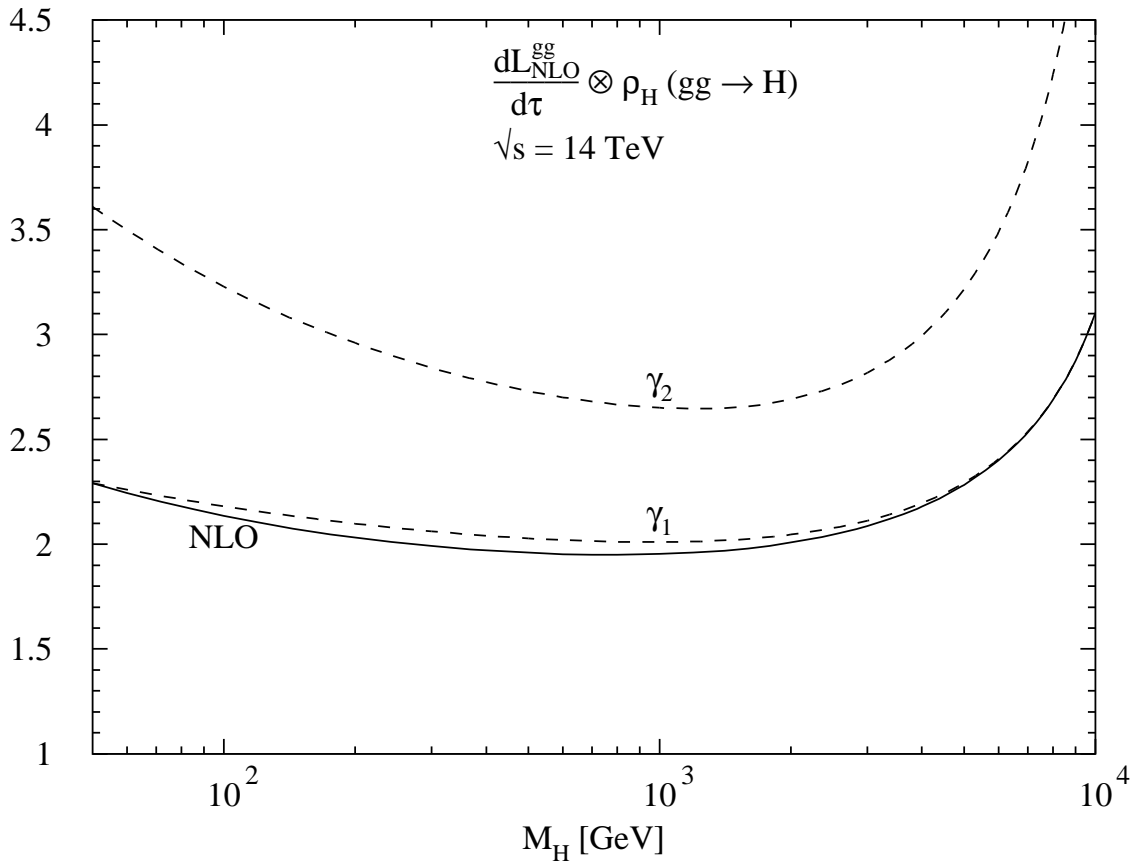}
\end{turn}
\vspace*{-0.1cm}

\caption[]{\label{fg:gghrho} \it Exact and approximate two- and three-loop
correction factor convoluted with NLO gluon densities in the heavy top quark
limit for the SM Higgs boson. The CTEQ4M parton densities have been adopted
with $\alpha_s(M_Z)=0.116$ at NLO.}
\end{figure}
The convolution of the correction factor with NLO gluon densities and strong
coupling is presented in Fig.~\ref{fg:gghrho} as a function of the Higgs mass
at the LHC. The solid line corresponds to the exact NLO result and the lower
dashed line to the NLO expansion of the resummed correction factor. It can be
inferred from this figure that the soft gluon approximation reproduces the
exact result within $\sim 5\%$ at NLO. The upper dashed line shows the NNLO
expansion of the resummed correction factor. From the analogous analysis of
the Drell--Yan process at NNLO we gain confidence that the NNLO expansion of the
resummed result reliably approximates the exact NNLO correction \cite{gghresum}.
Fig.~\ref{fg:gghrho} demonstrates that the correction factor amounts to about
2--2.3 at NLO and 2.7--3.5 at NNLO in the phenomenologically relevant Higgs
mass range $M_H \lsim 1$ TeV. However, in order to evaluate the size of the
QCD corrections, each order of the perturbative expansion has to be computed
with the strong coupling and parton densities of the {\it same} order, i.e.~LO
cross section with LO quantities, NLO cross section with NLO quantities and
NNLO cross
section with NNLO quantities. This consistent $K$ factor amounts to about
1.5--1.9 at NLO and is thus about 50--60\% smaller than the result in
Fig.~\ref{fg:gghrho}. Therefore a reliable prediction of the gluon-fusion cross
section
at NNLO requires the convolution with NNLO parton densities, which are not yet
available. Thus it is impossible to predict the Higgs production cross section
with NNLO accuracy until NNLO structure functions will be accessible.

\begin{figure}[hbtp]

\vspace*{0.3cm}
\hspace*{-0.0cm}
\begin{turn}{-90}%
\epsfxsize=9cm \epsfbox{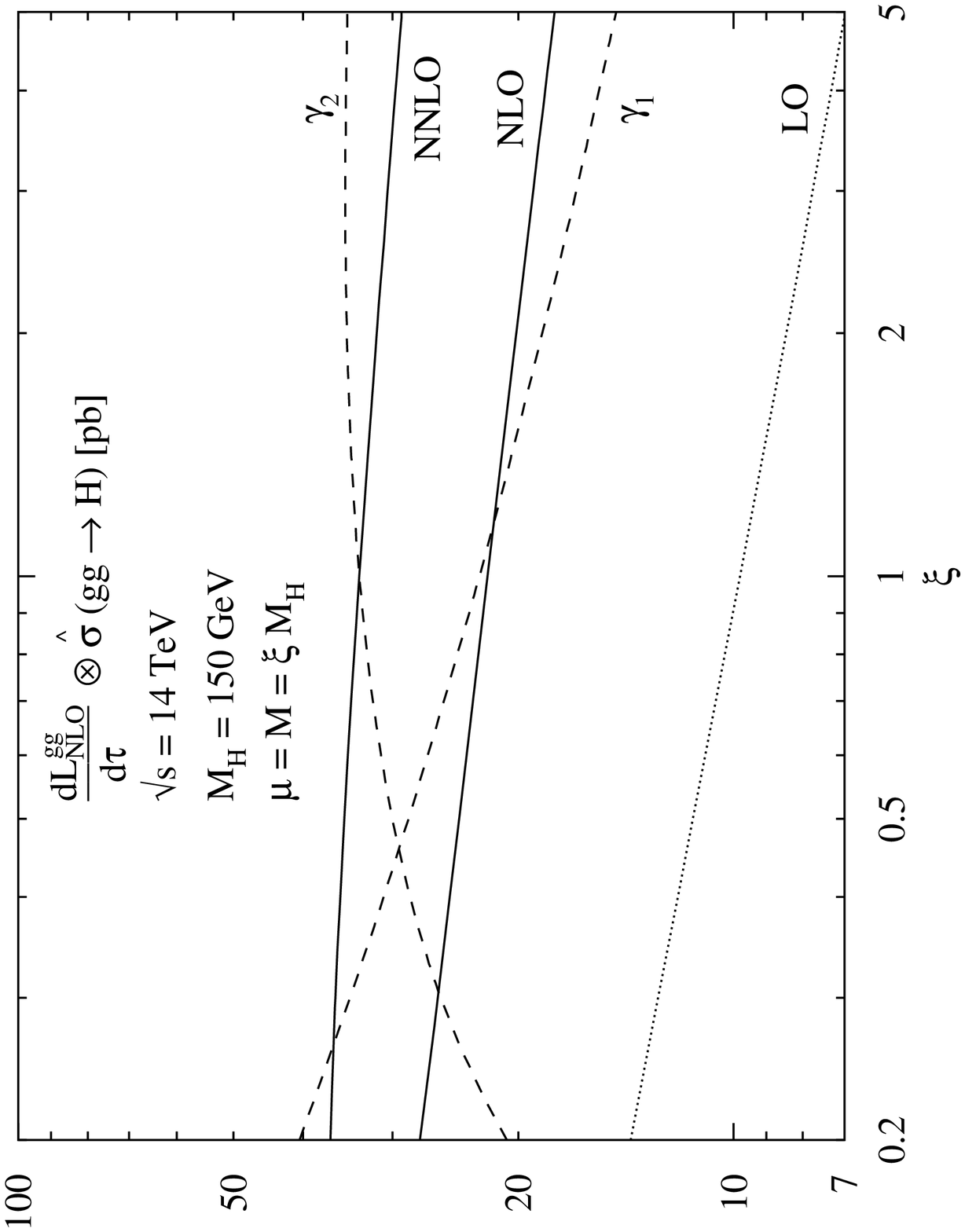}
\end{turn}
\vspace*{0.4cm}

\hspace*{-0.0cm}
\begin{turn}{-90}%
\epsfxsize=9cm \epsfbox{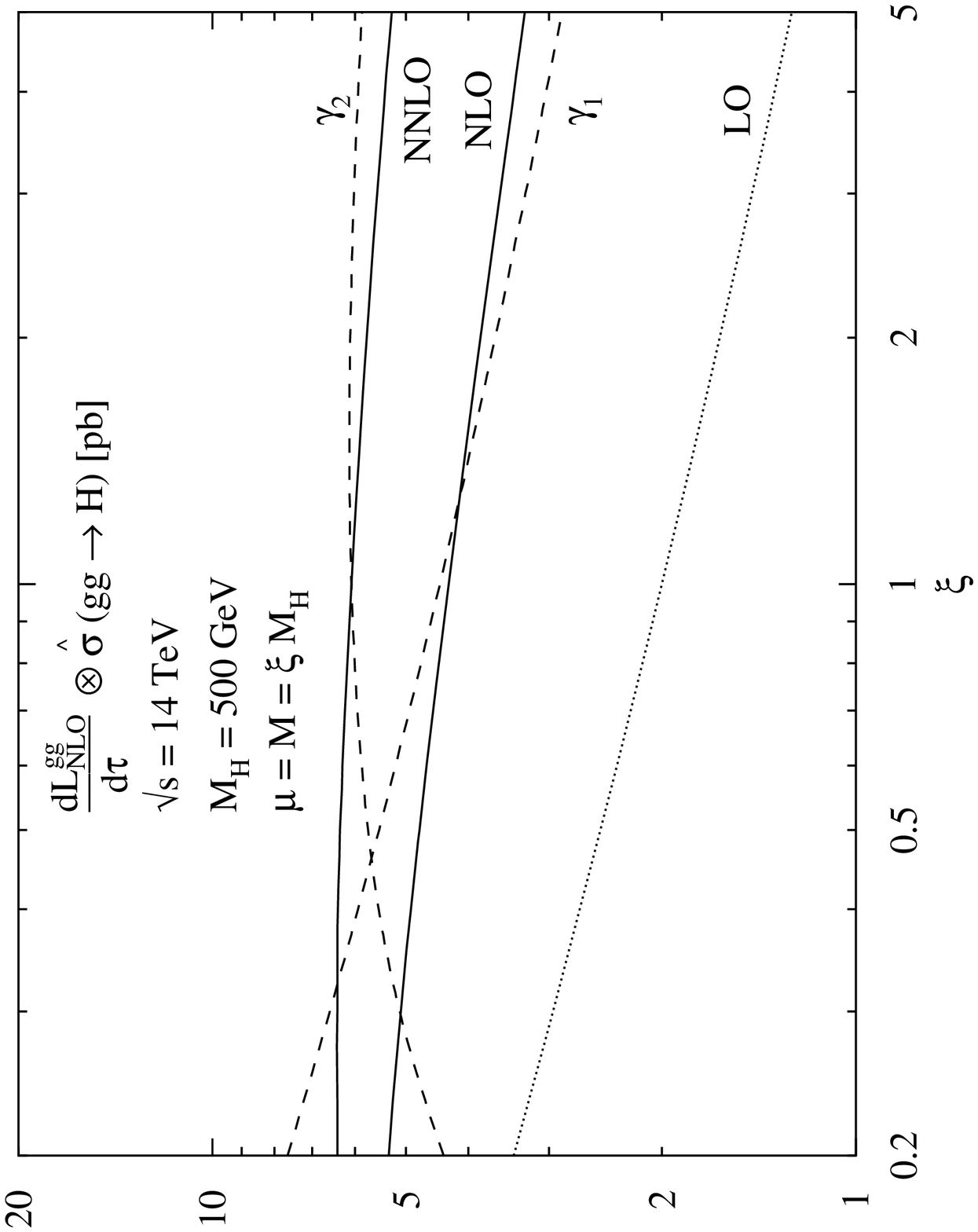}
\end{turn}
\vspace*{-0.1cm}

\caption[]{\label{fg:gghresscal} \it Scale dependence of the Higgs production
cross section as a function of the common renormalization and factorization
scale in units of the Higgs mass for two values of $M_H=150,
500$ GeV. All orders of the cross section are evaluated with NLO parton
densities [CTEQ4M] and strong coupling constant [$\alpha_s (M_Z)=0.116$].}
\end{figure}
The scale dependence of the gluon-fusion cross section [neglecting $gq$ and
$q\bar q$ initial states] is presented in Fig.~\ref{fg:gghresscal} as a
function of the scale in units of the Higgs mass, $\xi = \mu/M_H$. All orders
of the perturbative expansion are evaluated with NLO parton densities and
strong coupling, so that the LO and NNLO curves do {\it not} correspond to
physically consistent values. The dotted line represents the LO and the
lower full line the exact NLO scale dependence. The two dashed curves
correspond to the NLO and the NNLO expansions of the resummed cross section.
The upper solid line shows the full NNLO scale dependence, which has been
obtained from the exact NLO result by means of renormalization group methods
\cite{gghresum}.
This curve has been identified with the approximate NNLO expansion at $\xi=1$.
Fig.~\ref{fg:gghresscal} supports the validity of the resummed expression
within a reasonable accuracy for physically relevant scale choices $1/2 \lsim
\xi \lsim 2$. Moreover, the upper solid line clearly indicates that the NNLO
scale dependence develops a broad maximum around the natural scale $\mu \sim
M_H$ for large Higgs masses and thus a significant theoretical improvement.

\paragraph{Electroweak corrections.}
The electroweak corrections to the gluon-fusion cross section have been
computed in two different limits. The leading top mass corrections of
${\cal O}(G_F M_t^2)$ coincide
with the corrections to the gluonic decay mode of eq.~(\ref{eq:hggelw}) and are
thus small \cite{abdelgambino}.
For large Higgs masses the electroweak corrections of ${\cal O}(G_F M_H^2)$
have been evaluated by means of the equivalence theorem \cite{gghmh}. They
enhance the cross section by about 10--20\% for large Higgs masses.

\subsubsection{Vector-boson fusion: $qq\to qqV^*V^* \to qqH$}
\begin{figure}[hbt]
\begin{center}
\setlength{\unitlength}{1pt}
\begin{picture}(120,110)(0,0)

\ArrowLine(0,0)(50,0)
\ArrowLine(50,0)(100,0)
\ArrowLine(0,100)(50,100)
\ArrowLine(50,100)(100,100)
\Photon(50,0)(50,50){3}{5}
\Photon(50,50)(50,100){3}{5}
\DashLine(50,50)(100,50){5}
\put(105,46){$H$}
\put(-15,-2){$q$}
\put(-15,98){$q$}
\put(55,21){$W,Z$}
\put(55,71){$W,Z$}

\end{picture}  \\
\setlength{\unitlength}{1pt}
\caption[ ]{\label{fg:vvhlodia} \it Diagram contributing to $qq \to qqV^*V^*
\to qqH$ at lowest order.}
\end{center}
\end{figure}
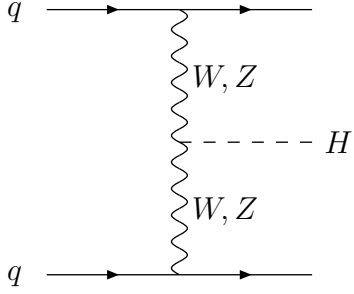
\noindent
The second important Higgs production channel at the LHC is the
vector-boson-fusion mechanism [see Fig.~\ref{fg:vvhlodia}], which will be
competitive with the dominant gluon-fusion mechanism for large Higgs masses
$M_H\sim 1$ TeV \cite{vvh,vvhqcd}. For intermediate Higgs masses
the vector-boson-fusion cross section is about one order of magnitude smaller
than the gluon one. The leading order partonic vector-boson-fusion
cross section \cite{vvh} can be cast into the form [$V = W,Z$]:
\bea
d\sigma_{LO} & = & \frac{1}{8} \frac{\sqrt{2}G_F^3M_V^8 q_1^2 q_2^2}
{[q_1^2-M_V^2]^2 [q_2^2-M_V^2]^2} \nonumber \\
& & \left\{
F_1(x_1,M^2) F_1(x_2,M^2) \left[ 2+\frac{(q_1 q_2)^2}{q_1^2q_2^2} \right]
\right. \nonumber \\
&&+\frac{F_1(x_1,M^2)F_2(x_2,M^2)}{P_2q_2}\left[\frac{(P_2q_2)^2}{q_2^2}-
M_P^2+\frac{1}{q_1^2}\left(P_2q_1-\frac{P_2q_2}{q_2^2}q_1q_2\right)^2 \right]
\nonumber \\
&&+\frac{F_2(x_1,M^2)F_1(x_2,M^2)}{P_1q_1}\left[\frac{(P_1q_1)^2}{q_1^2}-
M_P^2+\frac{1}{q_2^2}\left(P_1q_2-\frac{P_1q_1}{q_1^2}q_1q_2\right)^2 \right]
\nonumber \\
& & +\frac{F_2(x_1,M^2)F_2(x_2,M^2)}{(P_1q_1)(P_2q_2)}\left[P_1P_2 -
\frac{(P_1q_1)(P_2q_1)}{q_1^2} - \frac{(P_2q_2)(P_1q_2)}{q_2^2} \right.
\nonumber \\
& & \left. \hspace*{4.5cm}+\frac{(P_1q_1)(P_2q_2)(q_1q_2)}{q_1^2q_2^2}\right]^2
\nonumber \\
& &\left. +\frac{F_3(x_1,M^2)F_3(x_2,M^2)}{2(P_1q_1)(P_2q_2)}\left[
(P_1P_2)(q_1q_2) - (P_1q_2)(P_2q_1) \right] \right\} dx_1 dx_2
\frac{dP\!S_3}{\hat s}
\label{eq:vvhlo}
\eea
where $dP\!S_3$ denotes the three-particle phase space of the final-state
particles, $M_P$ the proton mass, $P_{1,2}$ the proton momenta and $q_{1,2}$ the
momenta of the virtual vector bosons $V^*$.
The functions $F_i(x,M^2)~(i=1,2,3)$ are the usual structure functions from
deep-inelastic scattering processes at the factorization scale $M$:
\bea
F_1(x,M^2) & = & \sum_q (v_q^2+a_q^2) [q(x,M^2) + \bar q(x,M^2)] \nonumber \\
F_2(x,M^2) & = & 2x \sum_q (v_q^2+a_q^2) [q(x,M^2) + \bar q(x,M^2)] \nonumber \\
F_3(x,M^2) & = & 4 \sum_q v_qa_q [-q(x,M^2) + \bar q(x,M^2)]
\label{eq:stfu}
\eea
where $v_q\, (a_q)$ are the (axial) vector couplings of the quarks $q$
to the vector bosons $V$: $v_q = -a_q = \sqrt{2}$ for $V=W$ and $v_q = 2I_{3q}
- 4e_q \sin^2\theta_W$, $a_q = 2 I_{3q}$ for $V=Z$. $I_{3q}$ is the third weak
isospin component and $e_q$ the electric charge of the quark $q$.

\begin{figure}[hbt]

\vspace*{0.5cm}
\hspace*{-0.5cm}
\begin{turn}{-90}%
\epsfxsize=11cm \epsfbox{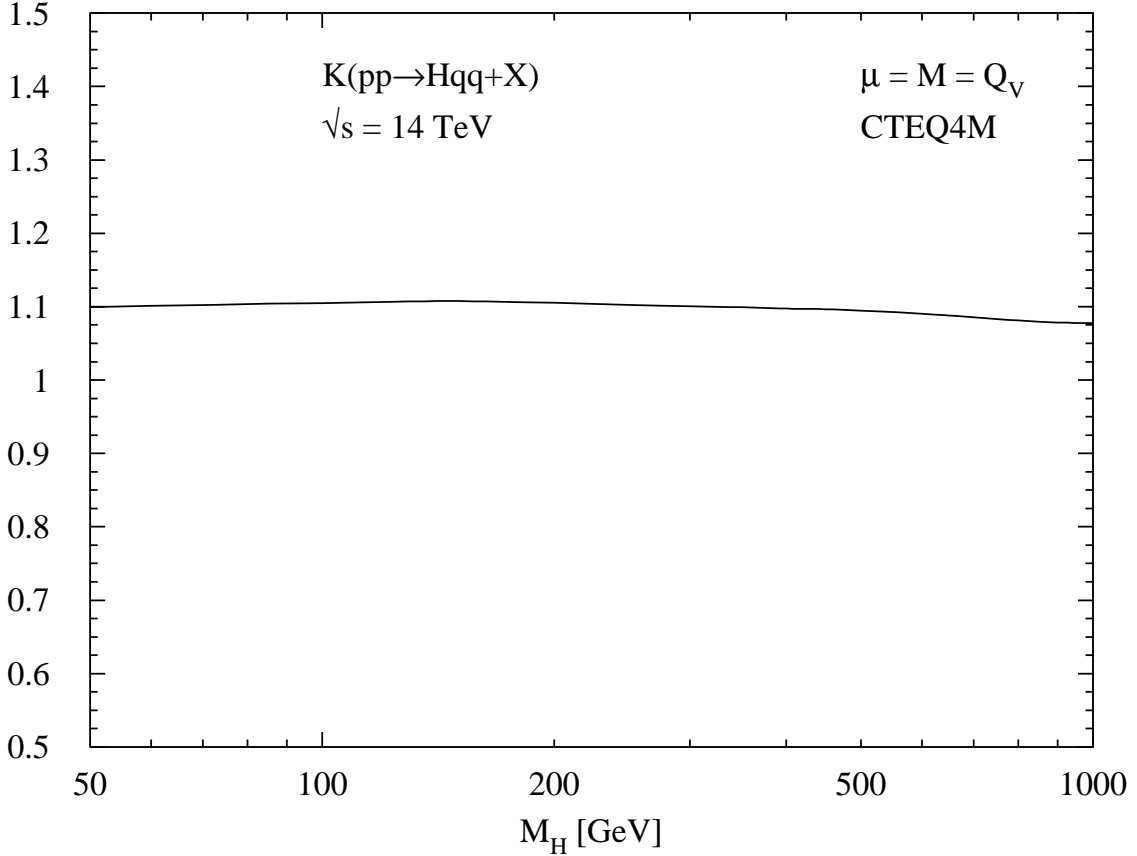}
\end{turn}
\vspace*{0.0cm}

\caption[]{\label{fg:vvhqcd} \it  K factor of the QCD corrections to $VV\to H$
as a function of the SM Higgs mass. The CTEQ4M parton densities have been
adopted, and the running strong coupling constant has been normalized to
$\alpha_s (M_Z)=0.116$ at NLO.}
\end{figure}

In the past the QCD corrections have been calculated within the structure
function approach \cite{vvhqcd}. Since, at lowest order, the
proton remnants are color singlets, no color will be exchanged between the
first and the second incoming (outgoing) quark line and hence the QCD
corrections only consist of the well-known corrections to the structure
functions $F_i(x,M^2)~(i=1,2,3)$. The final result for the QCD-corrected
cross section leads to the replacements
\bea
F_i(x,M^2) & \to & F_i(x,M^2) + \Delta F_i(x,M^2,Q^2) \hspace*{1cm} (i=1,2,3)
\nonumber \\
\Delta F_1(x,M^2,Q^2) & = & \frac{\alpha_s(\mu)}{\pi}\sum_q (v_q^2+a_q^2)
\int_x^1 \frac{dy}{y} \left\{ \frac{2}{3} [q(y,M^2) + \bar q(y,M^2)]
\right. \nonumber \\
& &
\left[ -\frac{3}{4} P_{qq}(z) \log \frac{M^2z}{Q^2} + (1+z^2) {\cal D}_1(z)
- \frac{3}{2} {\cal D}_0(z) \right. \nonumber \\
& & \left. \hspace*{6cm} + 3 - \left(
\frac{9}{2} + \frac{\pi^2}{3} \right) \delta(1-z) \right]
\nonumber \\
& & \left. + \frac{1}{4} g(y,M^2) \left[ -2 P_{qg}(z) \log \frac{M^2z}{Q^2(1-z)}
- 1 \right] \right\} \\
\Delta F_2(x,M^2,Q^2) & = & 2x\frac{\alpha_s(\mu)}{\pi}\sum_q (v_q^2+a_q^2)
\int_x^1 \frac{dy}{y} \left\{ \frac{2}{3} [q(y,M^2) + \bar q(y,M^2)]
\right. \nonumber \\
& &
\left[ -\frac{3}{4} P_{qq}(z) \log \frac{M^2z}{Q^2} + (1+z^2) {\cal D}_1(z)
- \frac{3}{2} {\cal D}_0(z) \right. \nonumber \\
& & \left. \hspace*{3.0cm} + 3 + 2z - \left(
\frac{9}{2} + \frac{\pi^2}{3} \right) \delta(1-z) \right]
\nonumber \\
& & \left. + \frac{1}{4} g(y,M^2) \left[ -2P_{qg}(z) \log \frac{M^2z}{Q^2(1-z)}
+ 8z(1-z) - 1 \right] \right\} \\
\Delta F_3(x,M^2,Q^2) & = & \frac{\alpha_s(\mu)}{\pi} \sum_q 4 v_q a_q
\int_x^1 \frac{dy}{y} \left\{ \frac{2}{3} [-q(y,M^2) + \bar q(y,M^2)]
\right. \nonumber \\
& &
\left[ -\frac{3}{4} P_{qq}(z) \log \frac{M^2z}{Q^2} + (1+z^2) {\cal D}_1(z)
- \frac{3}{2} {\cal D}_0(z) \right. \nonumber \\
& & \left. \left. \hspace*{3cm} + 2 + z - \left(
\frac{9}{2} + \frac{\pi^2}{3} \right) \delta(1-z) \right] \right\} \, ,
\eea
where $z=x/y$ and
the functions $P_{qq}, P_{qg}$ denote the well-known Altarelli--Parisi splitting
functions, which are given by \cite{apsplit}
\bea
P_{qq}(z) & = & \frac{4}{3} \left\{ 2{\cal D}_0(z)-1-z+\frac{3}{2}\delta(1-z)
\right\} \nonumber \\
P_{qg}(z) & = & \frac{1}{2} \left\{ z^2 + (1-z)^2 \right\} \, .
\eea
The physical scale $Q$ is given by $Q^2 = -q_i^2$ for $x=x_i~(i=1,2)$.
These expressions have to be inserted in eq.~(\ref{eq:vvhlo}) and the full
result expanded up to NLO. The typical renormalization and factorization scales
are fixed by the vector-boson momentum transfer $\mu=M=Q$. The $K$ factor,
defined as $K = \sigma_{NLO}/\sigma_{LO}$,
is presented in Fig.~\ref{fg:vvhqcd} as a function of the Higgs mass. The size 
of the QCD corrections amounts to about 8--10\% and is thus small
\cite{vvhqcd}.

\subsubsection{Higgs-strahlung: $q\bar q\to V^* \to VH$}
\begin{figure}[hbt]
\begin{center}
\setlength{\unitlength}{1pt}
\begin{picture}(160,120)(0,-10)

\ArrowLine(0,100)(50,50)
\ArrowLine(50,50)(0,0)
\Photon(50,50)(100,50){3}{5}
\Photon(100,50)(150,100){3}{6}
\DashLine(100,50)(150,0){5}
\put(155,-4){$H$}
\put(-15,-2){$\bar q$}
\put(-15,98){$q$}
\put(65,65){$W,Z$}
\put(155,96){$W,Z$}

\end{picture}  \\
\setlength{\unitlength}{1pt}
\caption[ ]{\label{fg:vhvlodia} \it Diagram contributing to $q\bar q \to V^*
\to VH$ at lowest order.}
\end{center}
\end{figure}
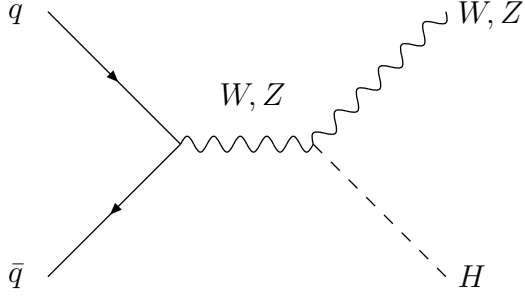
\noindent
The Higgs-strahlung mechanism $q\bar q\to V^* \to VH~(V=W,Z)$ [see
Fig.~\ref{fg:vhvlodia}] may be important in the intermediate Higgs
mass range due to the possibility to tag the associated vector boson. Its
cross section is about one to two orders of magnitude smaller than the
gluon-fusion
cross section for Higgs masses $M_H\lsim 200$ GeV. The lowest-order partonic
cross section can be expressed as \cite{vhv}
\beq
\hat\sigma_{LO}(q\bar q\to VH)=\frac{G_F^2 M_V^4}{288\pi Q^2}(v_q^2 + a_q^2)
\sqrt{\lambda(M_V^2,M_H^2;Q^2)} \frac{\lambda(M_V^2,M_H^2;Q^2) + 12 M_V^2
/ Q^2}{(1-M_V^2/Q^2)^2} \, ,
\label{eq:vhvpart}
\eeq
where $\lambda(x,y;z) = (1-x/z-y/z)^2-4xy/z^2$ denotes the usual two-body
phase-space factor and $v_q\, (a_q)$ are the (axial) vector couplings of the
quarks $q$
to the vector bosons $V$, which have been defined after eq.~(\ref{eq:stfu}).
The partonic c.m.~energy squared $\hat s$ coincides at
lowest order with the invariant mass $Q^2 = M^2_{VH}$ of the
Higgs--vector-boson pair squared, $\hat s=Q^2$. The hadronic cross section
can be obtained from convoluting eq.~(\ref{eq:vhvpart}) with the corresponding
(anti)quark densities of the protons:
\beq
\sigma_{LO}(pp \to q\bar q\to VH) = \int_{\tau_0}^1 d\tau \sum_q
\frac{d{\cal L}^{q\bar q}}{d\tau} \hat\sigma_{LO}(Q^2=\tau s) \, ,
\eeq
with $\tau_0 = (M_H+M_V)^2/s$ and $s$ the total hadronic c.m.~energy
squared.

\begin{figure}[hbt]

\vspace*{0.5cm}
\hspace*{-0.0cm}
\begin{turn}{-90}%
\epsfxsize=11cm \epsfbox{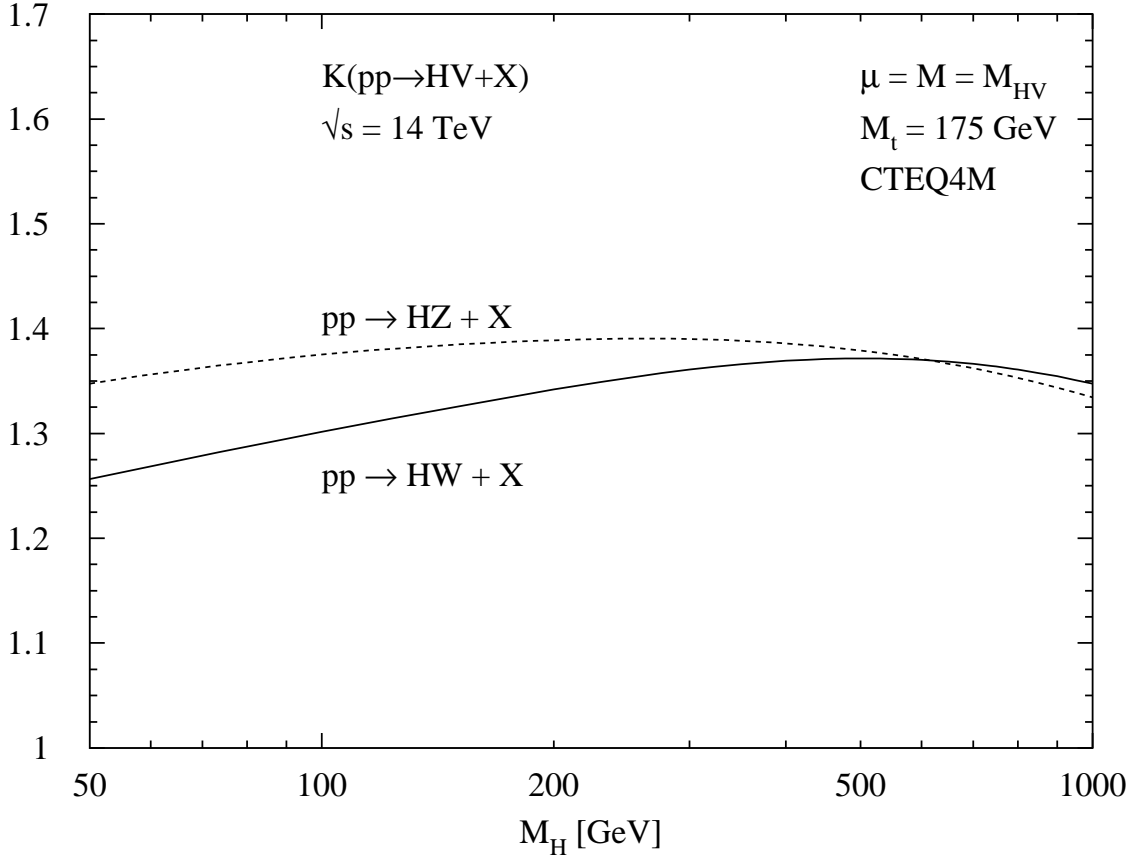}
\end{turn}
\vspace*{-0.0cm}

\caption[]{\label{fg:vhvqcd} \it K factor of the QCD corrections to $V^*\to HV$
as a function of the SM Higgs mass. The CTEQ4M parton densities have been
adopted, and the running strong coupling constant has been normalized to
$\alpha_s (M_Z)=0.116$ at NLO. The solid line corresponds to $W$ bremsstrahlung
and the dashed to $Z$ bremsstrahlung.}
\end{figure}

The QCD corrections are identical to the corresponding corrections to the
Drell--Yan process. They modify the lowest order cross section in the following
way \cite{vhvqcd}:
\bea
\sigma(pp\to VH) & = & \sigma_{LO} + \Delta\sigma_{q\bar q} + \Delta\sigma_{qg}
\nonumber \\
\Delta\sigma_{q\bar q} & = & \frac{\alpha_s(\mu)}{\pi} \int_{\tau_0}^1 d\tau
\sum_q \frac{d{\cal L}^{q\bar q}}{d\tau} \int_{\tau_0/\tau}^1 dz~\hat
\sigma_{LO}(Q^2 = \tau z s)~\omega_{q\bar q}(z) \nonumber \\
\Delta\sigma_{qg} & = & \frac{\alpha_s(\mu)}{\pi} \int_{\tau_0}^1 d\tau
\sum_{q,\bar q} \frac{d{\cal L}^{qg}}{d\tau} \int_{\tau_0/\tau}^1 dz~\hat
\sigma_{LO}(Q^2 = \tau z s)~\omega_{qg}(z)
\eea
with the coefficient functions
\bea
\omega_{q\bar q}(z) & = & -P_{qq}(z) \log \frac{M^2}{\tau s}
+ \frac{4}{3}\left\{ 2[\zeta_2-2]\delta(1-z) + 4{\cal D}_1(z) - 2(1+z)\log(1-z)
\phantom{\frac{M^2}{s}}\!\!\!\!\!\!\!\!\!\!
\right\} \nonumber \\
\omega_{qg}(z) & = & -\frac{1}{2} P_{qg}(z) \log \left(
\frac{M^2}{(1-z)^2 \tau s} \right) + \frac{1}{8}\left\{ 1+6z-7z^2 \right\} \, ,
\eea
where $M$ denotes the factorization and $\mu$ the renormalization scale.
The natural scale of the process is given by the invariant mass of the
Higgs--vector-boson pair in the final state, $\mu=M=Q$. The $K$ factors, defined
as $K=\sigma_{NLO}/\sigma_{LO}$,
are shown in Fig.~\ref{fg:vhvqcd} as a function of the Higgs mass.
The size of the QCD corrections is of about 25--40\% and they are thus of
moderate magnitude \cite{vhvqcd}.

\subsubsection{Higgs bremsstrahlung off top quarks}
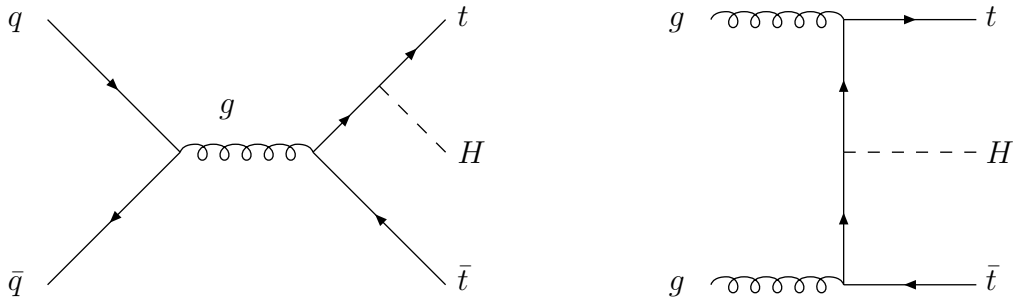
\begin{figure}[hbt]
\begin{center}
\setlength{\unitlength}{1pt}
\begin{picture}(360,120)(0,-10)

\ArrowLine(0,100)(50,50)
\ArrowLine(50,50)(0,0)
\Gluon(50,50)(100,50){3}{5}
\ArrowLine(100,50)(125,75)
\ArrowLine(125,75)(150,100)
\ArrowLine(150,0)(100,50)
\DashLine(125,75)(150,50){5}
\put(155,46){$H$}
\put(-15,98){$q$}
\put(-15,-2){$\bar q$}
\put(65,65){$g$}
\put(155,98){$t$}
\put(155,-2){$\bar t$}

\Gluon(250,0)(300,0){3}{5}
\Gluon(250,100)(300,100){3}{5}
\ArrowLine(350,0)(300,0)
\ArrowLine(300,0)(300,50)
\ArrowLine(300,50)(300,100)
\ArrowLine(300,100)(350,100)
\DashLine(300,50)(350,50){5}
\put(355,46){$H$}
\put(235,98){$g$}
\put(235,-2){$g$}
\put(355,98){$t$}
\put(355,-2){$\bar t$}

\end{picture}  \\
\setlength{\unitlength}{1pt}
\caption[ ]{\label{fg:httlodia} \it Typical diagrams contributing to
$q\bar q/gg \to Ht\bar t$ at lowest order.}
\end{center}
\end{figure}
\noindent
In the intermediate mass range the cross section of the associated production
of the Higgs boson with a $t\bar t$ pair can reach values similar to those of
the Higgs-strahlung process. It may thus provide an additional possibility
to find
a Higgs boson with mass $M_H\lsim 130$ GeV by tagging the additional $t\bar t$
pair and the rare photonic decay mode $H\to \gamma\gamma$ \cite{htt}.
This process takes
place through gluon--gluon and $q\bar q$ initial states at lowest order [see
Fig.~\ref{fg:httlodia}]. The result for the lowest order cross section is too
involved to be presented here. We have recalculated the cross section and
found numerical agreement with Refs.~\cite{htt,KMS}.

At the LHC the gluon--gluon channel dominates due to the enhanced gluon
structure function analogous to the leading Higgs production mechanism via
gluon fusion. The QCD corrections to the $Ht\bar t$ production are still
unknown.  They require the evaluation of several one-loop five-point functions
for the virtual corrections and real contributions involving four particles in
the final state, where three of them [$H,t,\bar t$] are massive.

\subsubsection{Cross sections for Higgs boson production at the LHC}
\begin{figure}[hbt]

\vspace*{0.5cm}
\hspace*{-0.5cm}
\begin{turn}{-90}%
\epsfxsize=11cm \epsfbox{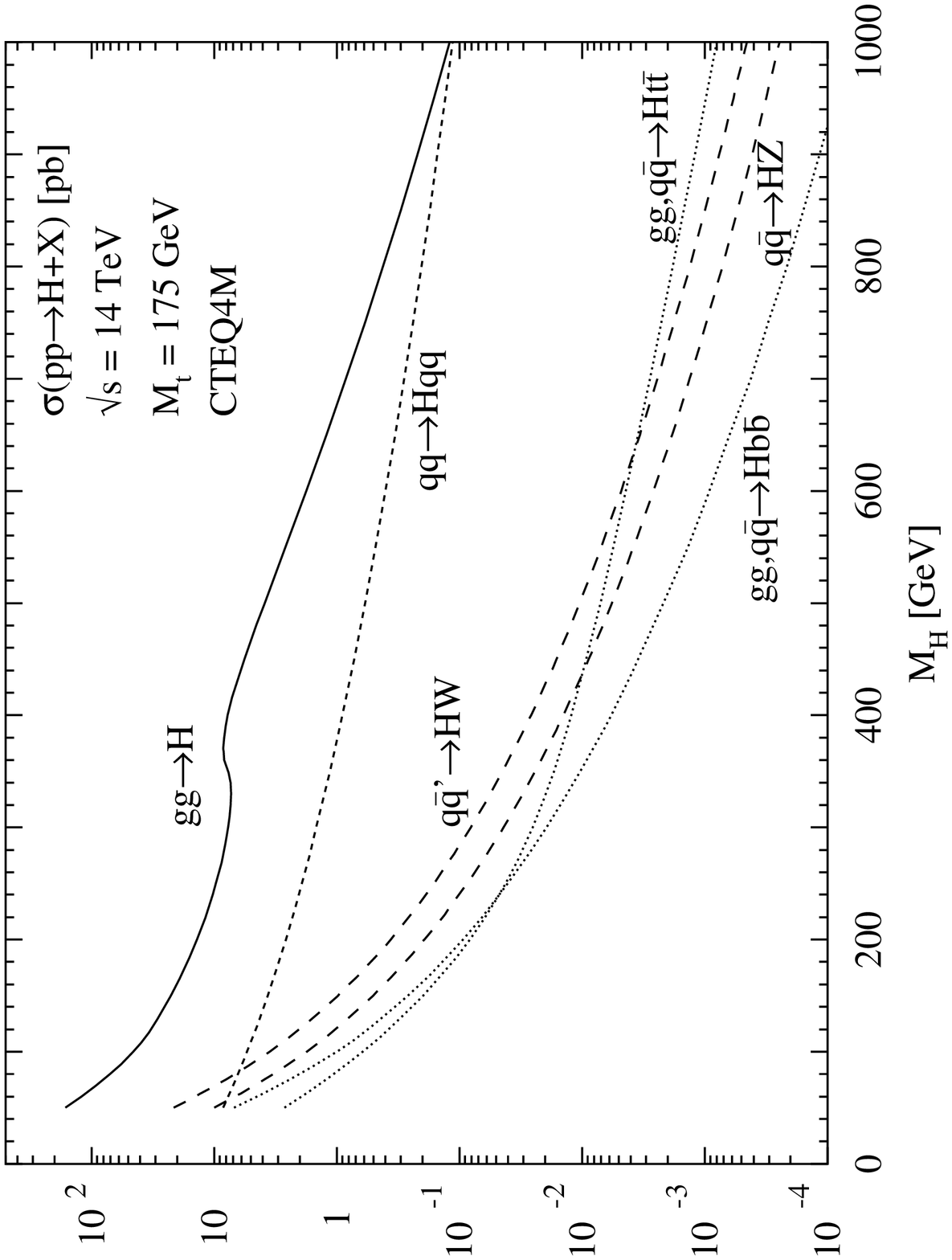}
\end{turn}
\vspace*{0.0cm}

\caption[]{\label{fg:prohiggs} \it Higgs production cross sections at the LHC
[$\sqrt{s}=14$ TeV] for the various production mechanisms as a function of the
Higgs mass. The full QCD-corrected results for the gluon fusion $gg
\to H$, vector boson fusion $qq\to VVqq \to Hqq$, vector boson bremsstrahlung
$q\bar q \to V^* \to HV$ and associated production $gg,q\bar q \to Ht\bar t,
Hb\bar b$ are shown.
The QCD corrections to the last process are unknown and thus not included.}
\end{figure}
The results for the cross sections of the various Higgs production mechanisms
at the LHC are presented in Fig.~\ref{fg:prohiggs}, which is an update of
Ref.~\cite{KMS}, as a function of the Higgs
mass. The total c.m.~energy has been chosen as $\sqrt{s} = 14$ TeV, the
CTEQ4M parton densities have been adopted with $\alpha_s(M_Z)=0.116$, and the
top and bottom masses have been set to $M_t=175$ GeV and $M_b=5$ GeV. For the
cross section of $Ht\bar t$ and $Hb\bar b$ we have used the leading order
CTEQ4L parton
densities due to the fact that the NLO QCD corrections are unknown. Thus the
consistent evaluation of this cross section requires LO parton densities and
strong coupling. The latter is normalized as $\alpha_s(M_Z) = 0.132$ at lowest
order. The gluon-fusion cross section provides the dominant production cross
section for the entire Higgs mass region up to $M_H\sim 1$ TeV. Only for Higgs
masses $M_H\gsim 800$~GeV the $VV$-fusion mechanism $qq\to qqH$ becomes
competitive and deviates from the gluon-fusion cross section by less than a
factor 2 for $M_H\gsim 800$ GeV.
In the intermediate mass range the gluon-fusion cross section
is at least one order of magnitude larger than all other Higgs production
mechanisms. At the lower end of the Higgs mass range $M_H\lsim 100$ GeV the
associated production channels of $H+V, H+t\bar t$ yield sizeable cross sections
of about one order below the gluon-fusion process and can thus allow for an
additional possibility to find the Higgs particle.

The search for the standard Higgs boson will be different within three major
mass ranges, the lower mass range $M_H\lsim 140$ GeV and the higher one, 140
GeV $\!\! \lsim M_H\lsim 800$ GeV, and the very high mass region
$M_H\gsim 800$~GeV.
\vspace*{0.3cm}

\noindent
\underline{$M_H\lsim 140$ GeV}
\vspace*{0.3cm}

\noindent
In the lower mass range the standard Higgs particle dominantly decays into
$b\bar b$ pairs. Because of the overwhelming QCD background the signal will be
very difficult to extract. Only excellent $b$ tagging, which may be provided
by excellent $\mu$-vertex detectors, might allow a sufficient rejection of
the QCD background \cite{btag}, although this task seems to be very difficult
\cite{antibtag}. The associated production of the Higgs boson with a $t\bar t$
pair or a $W$ boson may increase the significance of the $H\to b\bar b$ decay
due to the additional isolated leptons from $t$ and $W$ decays, but the rates
will be considerably smaller than single Higgs production via gluon fusion
\cite{htt,vhv}.

Studies for the detection of the $H\to \tau^+ \tau^-$ decay mode have
also been performed. Again because of the overwhelming backgrounds from
$t\bar t$ and Drell--Yan $\tau^+ \tau^-$ pair production, this possibility has
been found to be hopeless for the Standard Model Higgs boson \cite{htau}. The
branching
ratio into off-shell $Z$ pairs $H\to Z^* Z^*$ is too small to allow for a
detection of four-lepton final states \cite{hzz4l}.

The only promising channel for the detection of the Higgs boson with masses
$M_H\lsim 140$ GeV is provided by the rare $H\to \gamma \gamma$ decay mode
\cite{atlascms} with
a branching ratio of ${\cal O}(10^{-3})$. For an LHC luminosity of $\int
{\cal L} = 10^5 pb^{-1}$ the cross section times branching ratio for $pp\to
H (\to \gamma \gamma) + X$ yields ${\cal O}(0.5$--$1 \times 10^{4})$ events in
the mass range $80$ GeV $\lsim M_H \lsim 140$~GeV. In order to reject the large
backgrounds from the $\gamma \gamma$ continuum production and the two-photon
decay mode of the neutral pions $\pi^0 \to \gamma \gamma$, the detection of the
rare photonic decay mode requires excellent energy and geometric resolution
of the photon detectors \cite{atlascms}. Moreover, a necessary rejection factor
of $10^4$ for jets faking photons in the detector seems to be feasible
\cite{atlascms}. Thus the LHC studies
conclude that the rare photonic decay mode will be a possibility to find the
standard Higgs particle in the lower mass range.
\vspace*{0.3cm}

\noindent
\underline{140 GeV $\lsim M_H\lsim 800$ GeV}
\vspace*{0.3cm}

\noindent
Above the $ZZ$ threshold, on-shell $H \to ZZ \to 4 l^\pm$ decays of the Higgs
particle provide a very clean signature with small SM backgrounds \cite{hzz4l}.
The two
pairs of electrons or muons of this `gold-plated' decay channel carry invariant
masses equal to the $Z$ boson mass, thus allowing for very stringent cuts
against background processes.
Below the $ZZ$ threshold, off-shell $H\to ZZ^* \to 4 l^\pm$ decays, where one
of the $Z$ bosons is on-shell, yield clean signatures with rather small
SM backgrounds \cite{hzz4l}. However, in the mass range 155~GeV $\lsim M_H
\lsim 180$ GeV,
where the $ZZ$ branching ratio drops down to values of about 2\%, the number
of events at the LHC allows for a discovery of the Higgs boson only, if the
maximal luminosity will be reached \cite{atlascms}. On the other hand the
dominant $WW$ decay mode
of the Higgs boson leads to $l^+l^- \nu \bar \nu$ final states with strong
spin correlations of the visible charged lepton pair. A recent analysis has
shown that the Higgs particle can easily be detected within a few days in
this mass range \cite{dreiditt}.
\vspace*{0.3cm}

\noindent
\underline{$M_H\gsim 800$ GeV}
\vspace*{0.3cm}

\noindent
For large Higgs masses the total Higgs decay width exceeds 100 GeV and reaches
a value of about 600 GeV for $M_H = 1$ TeV. Thus the Higgs resonance peaks in
the 4-lepton final states become broad and, owing to the decreasing number of
events with growing Higgs mass, the `gold-plated' signal $H \to ZZ \to 4l^\pm$
will no longer be visible. In order to extend the Higgs search to masses
beyond 1 TeV, the decay modes $H \to ZZ,WW \to l^+ l^- \nu \bar \nu$ will be
the only possible signatures. The present status of the studies is not fully
conclusive, but promising \cite{atlascms}.

\begin{figure}[hbtp]

\vspace*{-1.0cm}
\hspace*{0.5cm}
\epsfxsize=15cm \epsfbox{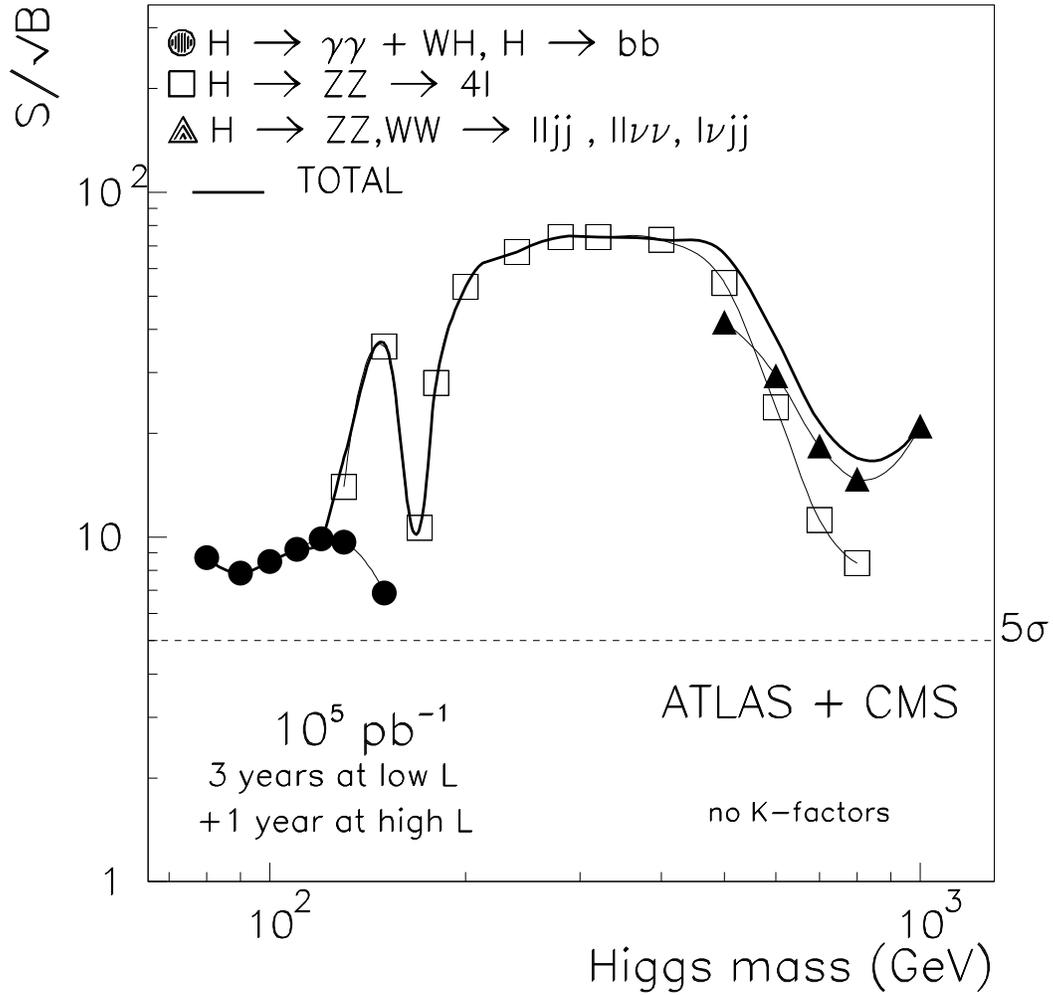}
\vspace*{0.0cm}

\caption[]{\label{fg:lhcsmhiggs} \it Expected significance of the SM Higgs
boson search at the LHC as a function of the Higgs boson mass after reaching
the anticipated integrated luminosity $\int {\cal L} = 10^5 pb^{-1}$ and
combining the experimental data of both LHC experiments, ATLAS and CMS.
Produced from Refs.~\cite{atlascms} -- courtesy of F.\ Gianotti.}
\end{figure}
Fig.~\ref{fg:lhcsmhiggs} shows the expected signal significance at the LHC as
a function of the SM Higgs mass after using the full experimental data samples
of both experiments, ATLAS and CMS. It is apparent that after reaching the
full integrated luminosity the SM Higgs signal may be extracted in the whole
relevant mass region \cite{atlascms}.

\section{Minimal Supersymmetric Extension of the Standard Model}
The couplings of the MSSM Higgs bosons to MSSM particles grow with the
MSSM particle masses, if these are generated by the Higgs mechanism. Thus the
MSSM Higgs bosons predominantly couple to heavy quarks and gauge bosons.
However, for large values of $\tb$ the couplings to down-type quarks are
enhanced, so that the coupling to bottom quarks may be much larger than to top
quarks. Moreover, the Higgs boson interaction with the intermediate gauge
bosons is always reduced with respect to the SM. The decays into heavy
particles will be dominant, if they are kinematically allowed.
The analysis includes the complete radiative corrections to the MSSM Higgs
sector due to top/bottom quark and squark loops within the effective potential
approach, as discussed in the introduction. Next-to-leading order QCD
corrections and the full mixing in the stop and sbottom sectors are
incorporated. The corresponding formulae are based on the works of
Ref.~\cite{mssmrad2}.
As for the SM case,
the decay widths and branching ratios of the MSSM Higgs bosons are evaluated
by means of the FORTRAN program HDECAY \cite{hdecay}.

\subsection{Decay Modes}

\subsubsection{Decays into lepton and heavy quark pairs}
At lowest order the leptonic decay width of neutral MSSM Higgs
boson\footnote{In the following we denote the different types of neutral Higgs
particles by $\Phi = h,H,A$.} decays is given by \cite{prohiggs,hll}
\begin{equation}
\Gamma [\Phi \to l^+ l^-] = \frac{G_F M_\Phi }{4\sqrt{2}\pi}
(g_l^\Phi)^2 m_l^2 \beta^p \, ,
\end{equation}
where $g_l^\Phi$ denotes the corresponding MSSM coupling, presented in Table
\ref{tb:hcoup}, $\beta = (1-4m_l^2/M_\Phi^2)^{1/2}$ the velocity of the
final-state leptons and $p=3\, (1)$ the exponent for scalar (pseudoscalar) Higgs
particles. The $\tau$ pair
decays play a significant r\^ole, with a branching ratio of up to about 10\%.
Muon decays can develop branching ratios of a few $10^{-4}$. All other leptonic
decay modes are phenomenologically irrelevant.

The analogous expression for the leptonic decays of the charged Higgs reads as
\begin{equation}
\Gamma [H^{+} \to  \nu{\overline{l}}] = \frac{G_F M_{H^\pm}}{4\sqrt{2}\pi}
m_l^2 \mbox{tg}^2 \beta \left(1-\frac{m_l^2}{M_{H^\pm}^2} \right)^3 \, .
\label{eq:hcnl}
\end{equation}
The decay mode into $\tau^+ \nu_\tau$ reaches branching ratios of about 100\%
below the $tb$ threshold and the muonic one ranges at a few $10^{-4}$. All
other leptonic decay channels of the charged Higgs bosons are unimportant.

For large Higgs masses [$M_\Phi \gg M_Q^2$] the QCD-corrected decay
widths of the MSSM Higgs particles into quarks can be obtained from evaluating
the analogous diagrams as presented in Fig.~\ref{fg:hqqdia}, where the Standard
Model Higgs particle $H$ has to be substituted by the corresponding MSSM Higgs
boson $\Phi$ \citer{drees,chet}:
\begin{equation}
\Gamma [\Phi \, \ra \, Q{\overline{Q}}] =
\frac{3G_F M_\Phi }{4\sqrt{2}\pi} \overline{m}_Q^2(M_\Phi) (g_Q^\Phi)^2
\left[ \Delta_{\rm QCD} + \Delta_t^\Phi \right] \, .
\end{equation}
Neglecting regular quark mass effects, the QCD corrections $\Delta_{\rm QCD}$
are presented in eq.~(\ref{eq:hqq}) and the top quark induced contributions
read as \cite{chet}
\begin{eqnarray}
\Delta_t^{h/H} & =& \frac{g_t^{h/H}}{g_Q^{h/H}}~\left(\frac{\alpha_s
(M_{h/H})}{\pi}
\right)^2 \left[ 1.57 - \frac{2}{3} \log \frac{M_{h/H}^2}{M_t^2}
+ \frac{1}{9} \log^2 \frac{\overline{m}_Q^2 (M_{h/H})}{M_{h/H}^2}\right]\non \\
\Delta_t^A & = & \frac{g_t^A}{g_Q^A}~\left(\frac{\alpha_s (M_A)}{\pi} \right)^2
\left[ 3.83 - \log \frac{M_A^2}{M_t^2} + \frac{1}{6} \log^2
\frac{\overline{m}_Q^2 (M_A)}{M_A^2} \right] \non
\end{eqnarray}
Analogous to the Standard Model case the large logarithmic contributions of
the QCD corrections are absorbed in the running $\overline{\rm MS}$ quark mass
$\overline{m}_Q(M_\Phi)$ at the scale of the corresponding Higgs mass $M_\Phi$.
In the large Higgs mass regimes the QCD corrections reduce the $b\bar b$
($c\bar c$) decay widths by about 50 (75)\% due to the large logarithmic
contributions.

The heavy quark decay width of the charged Higgs boson reads, in the large
Higgs mass regime $M_{H^\pm} \gg M_U + M_D$, as \cite{hud1,hud}
\begin{equation}
\Gamma [\,H^{+} \ra \, U{\overline{D}}\,] =
\frac{3 G_F M_{H^\pm}}{4\sqrt{2}\pi} \, \left| V_{UD} \right|^2 \,
\left[ \overline{m}_U^2(M_{H^\pm}) (g_U^{A})^2 + \overline{m}_D^2(M_{H^\pm})
 (g_D^{A})^2 \right] \Delta_{\rm QCD}
\label{eq:hcud}
\end{equation}
[Eq.~(\ref{eq:hcud}) is valid if either the first or the
second term is dominant.] The relative couplings $g_{Q}^A$ have been
collected in Table \ref{tb:hcoup} and the coefficient $V_{UD}$ denotes the
CKM matrix element of the transition of $D$ to $U$ quarks. The QCD correction
factor $\Delta_{\rm QCD}$ is given in eq.~(\ref{eq:hqq}), where large
logarithmic terms are again absorbed in the running $\overline{\rm MS}$ masses
$\overline{m}_{U,D} (M_{H^\pm})$ at the scale of the charged Higgs mass
$M_{H^\pm}$. In the large Higgs mass regimes, the QCD corrections reduce the
$c\bar b$ and $c\bar s$ decay widths by about 50--75\%, because of the large
logarithmic contributions.

In the threshold regions mass effects play a significant r\^ole. The partial
decay widths of the neutral Higgs bosons $\Phi=h,H$ and $A$ into
heavy quark pairs, in terms of the quark {\it pole} mass $M_Q$, can be cast
into the form \cite{drees}
\begin{eqnarray}
\Gamma [\Phi \ra Q\bar{Q}]= \frac{3 G_F M_\Phi}{4 \sqrt{2} \pi} \,
(g_Q^\Phi)^2 \, M_Q^2 \, \beta^p \, \left[ 1 +\frac{4}{3} \frac{\alpha_s}{\pi}
\Delta^{\Phi} \right] \, ,
\label{eq:phiqqmass}
\end{eqnarray}
where $\beta = (1-4M_Q^2/M_\Phi^2)^{1/2}$ denotes the velocity of the
final-state quarks and $p=3\, (1)$ the exponent for scalar (pseudoscalar) Higgs
bosons. To next-to-leading order, the QCD correction factor is given by
eq.~(\ref{eq:dqcdmass}) for the scalar Higgs particles $h,H$, while for
the CP-odd Higgs boson $A$ they read correspondingly as \cite{drees}
\begin{eqnarray}
\Delta^{A} = \frac{1}{\beta}A(\beta) + \frac{1}{16\beta}(19+2\beta^2+
3 \beta^4)\log \frac{1+\beta}{1-\beta} +\frac{3}{8}(7 -\beta^2) \, ,
\end{eqnarray}
with the function $A(\beta)$ defined after eq.~(\ref{eq:dqcdmass}). The QCD
corrections in the $t\bar t$ threshold region are moderate, apart from a
Coulomb singularity, which is regularized by taking into account the finite
top quark decay width.

The partial decay width of the charged Higgs particles into heavy quarks may be
written as \cite{hud}
\begin{eqnarray}
\Gamma [H^+\rightarrow U\bar{D}\,] &=& \frac{3 G_F M_{H^\pm}}{4\sqrt{2}\pi}
|V_{UD}|^2 \, \lambda^{1/2} \, \left\{ (1-\mu_U -\mu_D) \left[ \frac{M_U^2}
{ {\rm tg}^2
\beta } \left( 1+ \frac{4}{3} \frac{\alpha_s}{\pi} \Delta_{UD}^+ \right)
\right. \right. \label{eq:h+udmass} \\
&& \left. \left. +M_D^2 {\rm tg}^2 \beta \left( 1+ \frac{4}{3} \frac{\alpha_s}
{\pi} \Delta_{DU}^+ \right) \right]
-4M_UM_D \sqrt{\mu_U \mu_D} \left( 1+ \frac{4}{3}
\frac{\alpha_s}{\pi} \Delta_{UD}^- \right) \right\} \non
\end{eqnarray}
where $\mu_i=M_i^2/M_{H^\pm}^2$, and $\lambda=(1-\mu_U-\mu_D)^2-4\mu_U\mu_D$
denotes the usual two-body phase-space function;
the quark masses $M_{U,D}$ are the {\it pole} masses. The QCD factors
$\Delta_{ij}^\pm~~(i,j=U,D)$ are given by
\cite{hud}
\begin{eqnarray}
\Delta_{ij}^{+} &=&  \frac{9}{4} + \frac{ 3-2\mu_i+2\mu_j}{4} \log
\frac{\mu_i}{\mu_j} + \frac{ (\frac{3}{2}-\mu_i-\mu_j) \lambda+5 \mu_i
\mu_j}{2 \lambda^{1/2} (1-\mu_i -\mu_j)} \log x_i x_j  +B_{ij} \non \\
\Delta_{ij}^{-} &=&  3 + \frac{ \mu_j-\mu_i}{2} \log \frac{\mu_i}{\mu_j}
+ \frac{ \lambda +2(1-\mu_i-\mu_j)} { 2 \lambda^{1/2} } \log x_i x_j +B_{ij}
\end{eqnarray}
with the scaling variables $x_i= 2\mu_i/[1-\mu_i-\mu_j+\lambda^{1/2}]$ and the
generic function
\begin{eqnarray*}
B_{ij} &=& \frac{1-\mu_i-\mu_j} { \lambda^{1/2} } \left[ 4{\rm Li_{2}}(x_i
x_j)- 2{\rm Li_{2}}(-x_i) -2{\rm Li_{2}}(-x_j) +2 \log x_i x_j \log (1-x_ix_j)
\right. \non \\
&& \left. \hspace*{2.2cm} - \log x_i \log (1+x_i) - \log x_j \log (1+x_j)
\right] \non \\
&& - 4 \left[ \log (1-x_i x_j)+ \frac{x_i x_j}{1-x_i x_j} \log x_i x_j
\right]\non
\\ && +\frac{ \lambda^{1/2}+\mu_i-\mu_j } {\lambda^{1/2} } \left[
\log (1+x_i) -\frac{x_i}{1+x_i} \log x_i \, \right] \non \\
&& +\frac{\lambda^{1/2}-\mu_i+\mu_j} { \lambda^{1/2} } \left[
\log (1+x_j) -\frac{x_j}{1+x_j} \log x_j \right] \, .
\end{eqnarray*}
The transition from the threshold region, involving mass effects, to the
renormalization-group-improved large Higgs mass regime is provided by a
smooth linear interpolation analogous to the SM case in all heavy quark decay
modes.

The full MSSM electroweak  and SUSY-QCD corrections to the fermionic decay
modes have been computed \cite{dabel}. They turn out to be moderate, less than
about 10\%. Only for large values of $\tb > 10$ do the gluino corrections reach
values of 20 to 50\%, if the relevant squark masses are less than $\sim 300$
GeV. The electroweak and SUSY-QCD corrections are neglected in this analysis.

Below the $t\bar t$ threshold, heavy neutral Higgs boson decays
into off-shell top quarks are sizeable, thus modifying the profile of these
Higgs particles significantly in this region. The dominant below-threshold
contributions can be obtained from the SM expression eq.~(\ref{eq:httdalitz})
\cite{1OFF}
\begin{equation}
\frac{d\Gamma}{dx_1 dx_2} (H\to tt^*\to Wtb) = 
(g_t^H)^2 \frac{d\Gamma}{dx_1 dx_2} (H_{SM} \to tt^*\to Wtb) \, .
\label{eq:mssmhttdalitz}
\end{equation}
The corresponding dominant below-threshold contributions of the pseudoscalar
Higgs particle are given by \cite{1OFF}
\begin{equation}
\frac{d\Gamma}{dx_1 dx_2} (A\to tt^*\to Wtb) = \frac{3G_F^2}{32\pi^3} M_t^2
M_A^3 (g_t^A)^2 \frac{\Gamma_0}{y_1^2 + \gamma_t \kappa_t} \, ,
\label{eq:mssmattdalitz}
\end{equation}
with the reduced energies $x_{1,2}=2E_{t,b}/M_A$, the scaling variables
$y_{1,2} = 1-x_{1,2}$, $\kappa_i = M_i^2/M_A^2$ and the reduced decay
widths of the virtual particles $\gamma_i=\Gamma_i^2/M_A^2$. The squared
amplitude may be written as \cite{1OFF}
\begin{equation}
\Gamma_0 = y_1^2(1-y_1-y_2+\kappa_W-\kappa_t) + 2\kappa_W(y_1y_2-\kappa_W)
-\kappa_t(y_1y_2-2y_1-\kappa_W-\kappa_t) \, .
\end{equation} 
The differential decay widths of
eqs.~(\ref{eq:mssmhttdalitz}), (\ref{eq:mssmattdalitz}) have be integrated
over the $x_1, x_2$ region, bounded by eq.~(\ref{eq:dalitzbound}).
In these formulae $W$ and charged Higgs boson exchange contributions are
neglected, because they are suppressed with respect to the off-shell top
quark
contribution to $Wtb$ final states. However, for the sake of completeness they
are included in the analysis. Their explicit expressions can be found in
\cite{1OFF}. The transition from below to above the threshold is provided by a
smooth cubic interpolation. Below-threshold decays yield a $t\bar t$ branching
ratio at the per cent level already for heavy scalar and pseudoscalar Higgs
masses $M_{H,A}\sim 300$ GeV.

Below the $t\bar b$ threshold off-shell
decays $H^+\to t^*\bar b \to b\bar b W^+$ are important. For $M_{H^\pm} < M_t
+ M_b - \Gamma_t$ their expression can be cast into the form \cite{1OFF}
\begin{eqnarray}
\Gamma (H^+\to t^*\bar b \to Wb\bar b) & = &
\frac{3G_F^2M_t^4}{64\pi^3\mbox{tg}^2\beta} M_{H^\pm}
\left\{\frac{\kappa_W^2}{\kappa_t^3}(4\kappa_W\kappa_t + 3\kappa_t - 4\kappa_W)
\log \frac{\kappa_W(\kappa_t-1)}{\kappa_t-\kappa_W} \right.  \nonumber \\
& & +(3\kappa_t^2 - 4\kappa_t - 3\kappa_W^2 + 1)
\log\frac{\kappa_t-1}{\kappa_t-\kappa_W} - \frac{5}{2} \label{eq:pttoff} \\
& & \left. +\frac{1-\kappa_W}{\kappa_t^2} (3\kappa_t^3 - \kappa_t\kappa_W
- 2\kappa_t \kappa_W^2 + 4\kappa_W^2) + \kappa_W\left(4-\frac{3}{2}\kappa_W
\right) \right\} \nonumber
\end{eqnarray}
with the scaling variables $\kappa_i = M_i^2/M_{H^\pm}^2~~(i=t,W)$. The $b$
mass has been neglected in eq.~(\ref{eq:pttoff}), but it has been taken into
account in the present analysis by performing a numerical integration of the
corresponding Dalitz plot density, given in \cite{1OFF}.
The off-shell branching ratio can reach the per cent level for charged Higgs
masses above about 100~GeV for small $\tb$, which is significantly below the
$t\bar b$ threshold $M_{H^\pm}\sim 180$ GeV.

\subsubsection{Gluonic decay modes}
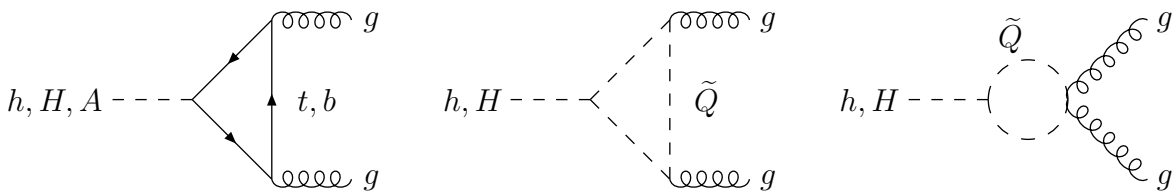
\begin{figure}[hbt]
\begin{center}
\setlength{\unitlength}{1pt}
\begin{picture}(400,100)(-20,0)

\Gluon(60,20)(90,20){-3}{4}
\Gluon(60,80)(90,80){3}{4}
\ArrowLine(60,20)(60,80)
\ArrowLine(60,80)(30,50)
\ArrowLine(30,50)(60,20)
\DashLine(0,50)(30,50){5}
\put(-40,46){$h,H,A$}
\put(70,46){$t,b$}
\put(95,18){$g$}
\put(95,78){$g$}

\Gluon(210,20)(240,20){-3}{4}
\Gluon(210,80)(240,80){3}{4}
\DashLine(210,80)(210,20){5}
\DashLine(210,20)(180,50){5}
\DashLine(180,50)(210,80){5}
\DashLine(150,50)(180,50){5}
\put(125,46){$h,H$}
\put(220,46){$\widetilde{Q}$}
\put(245,18){$g$}
\put(245,78){$g$}

\DashLine(300,50)(330,50){5}
\DashCArc(345,50)(15,0,180){5}
\DashCArc(345,50)(15,180,360){5}
\Gluon(360,50)(390,80){3}{5}
\Gluon(360,50)(390,20){-3}{5}
\put(275,46){$h,H$}
\put(335,70){$\widetilde{Q}$}
\put(395,18){$g$}
\put(395,78){$g$}

\end{picture}  \\
\setlength{\unitlength}{1pt}
\caption[ ]{\label{fg:mssmhgglodia} \it Typical diagrams contributing to
$\Phi \to gg$ at lowest order.}
\end{center}
\end{figure}
\begin{figure}[hbt]
\vspace*{0.5cm}
\hspace*{-0.5cm}
\begin{turn}{-90}%
\epsfxsize=11cm \epsfbox{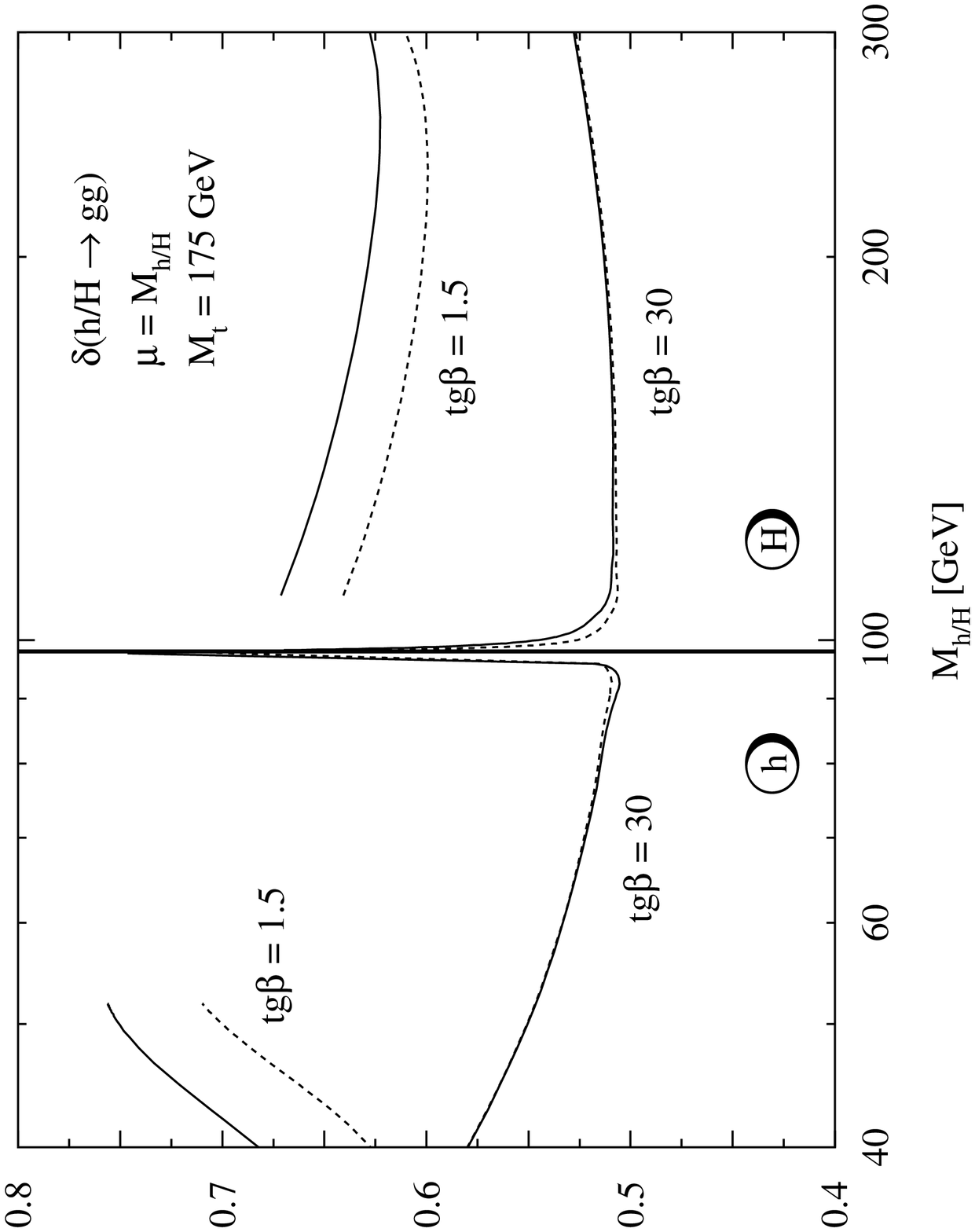}
\end{turn}
\vspace*{0.0cm}
\caption[ ]{\label{fg:mssmhggqcd} \it Size of the QCD correction factor for
$h/H\to gg$, defined as $\Gamma = \Gamma_{LO} (1+\delta)$, as a function of
the corresponding Higgs mass for two values of $\tb=1.5,30$. The full lines
include the full mass dependence on the top and bottom masses and, in addition,
the squark contributions in the heavy-squark limit. The dashed curves
correspond to the omission of the squark contributions.}
\end{figure}

\noindent
Since the $b$ quark couplings to the Higgs bosons may be strongly enhanced for
large $\tb$ and the $t$ quark couplings suppressed in the MSSM [see
Fig.~\ref{fg:mssmcoup}], $b$ loops can contribute significantly to the $gg$
coupling so that the approximation $M_{Q}^{2} \gg M_{H}^{2}$ can in general no
longer be applied. The leading order width for $h,H \ra gg$ is generated by
quark and squark loops, the latter contributing significantly for squark
masses below about 400~GeV \cite{SQCD}. The contributing diagrams are depicted
in Fig.~\ref{fg:mssmhgglodia}. The partial decay widths are given by
\cite{hunter,higgsqcd,SQCD}
\begin{eqnarray}
\Gamma_{LO} (h/H\rightarrow gg) & = & \frac{G_F \alpha_s^2M_{h/H}^3}
{36\sqrt{2}\pi^3} \left| \sum_Q g_Q^{h/H} A^{h/H}_Q(\tau_Q) +
\sum_{\widetilde{Q}} g_{\widetilde{Q}}^{h/H} A_{\widetilde{Q}}^{h/H}
(\tau_{\widetilde{Q}}) \right|^2 \label{eq:mssmhgg} \\
A_Q^{h/H} (\tau) & = & \frac{3}{2}\tau [1+(1-\tau) f(\tau)] \nonumber \\
A_{\widetilde Q}^{h/H} (\tau) & = & -\frac{3}{4}\tau [1-\tau f(\tau)]
\nonumber \\
\Gamma [\,h/H\,\ra \, gg(g),q{\overline{q}}g\,] &=&
\Gamma_{LO}\,\left[\alpha_{s}^{(N_{F})}(M_{h/H}
)\right]\,\left\{1+E^{N_{F}}\frac{\alpha_{s}^
{(N_{F})}(M_{h/H})}{\pi}\ \right\} \label{eq:mssmhggqcd} \\
E^{N_{F}}& \to &\frac{95}{4}-\frac{7}{6}N_{F} + \frac{17}{6} \Re e \left\{
\frac{\sum_{\widetilde{Q}} g_{\widetilde{Q}}^{h/H} A_{\widetilde{Q}}^{h/H}
(\tau_{\widetilde{Q}})}
{\sum_Q g_Q^{h/H} A^{h/H}_Q(\tau_Q)} \right\} \nonumber \\
& & \hspace*{6cm} \mbox{for $M_{h/H}^2 \ll 4 M_{Q,\widetilde{Q}}^2$} \nonumber
\end{eqnarray}
with $\tau_i = 4M_i^2/M_{h/H}^2~~(i=Q, \widetilde{Q})$.
The function $f(\tau)$ is defined in eq.~(\ref{eq:ftau}) and the MSSM couplings
$g^{h/H}_Q$ can be found in Table \ref{tb:hcoup}. The squark couplings
$g^{h/H}_{\widetilde{Q}}$ are summarized in Table \ref{tb:hsqcoup}.
The amplitudes approach constant values in the limit of large loop particle
masses:
\begin{eqnarray*}
A_Q^{h/H} (\tau) & \to & 1 \hspace*{1cm} \mbox{for $M_{h/H}^2 \ll 4M_Q^2$} \\
A_{\widetilde{Q}}^{h/H} (\tau) & \to & \frac{1}{4} \hspace*{1cm} \mbox{for
$M_{h/H}^2 \ll 4M_{\widetilde{Q}}^2$} \, .
\end{eqnarray*}
\begin{figure}[hbt]
\vspace*{-1.4cm}
\hspace*{2.15cm}
\epsfxsize=11cm \epsfbox{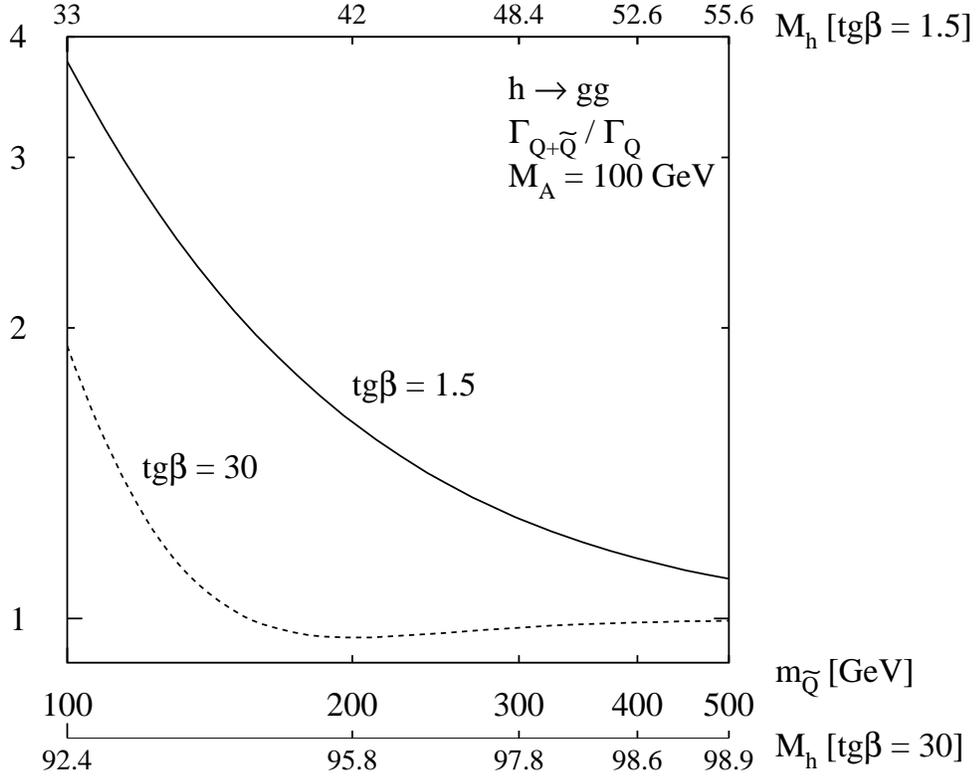}
\vspace*{-3.1cm}
\caption[ ]{\label{fg:mssmhggqsq} \it Ratio of the QCD-corrected decay width
$\Gamma(h\to gg)$ with and without squark loops for two values of $\tb = 1.5,
30$ as a function of the common squark mass $M_{\widetilde{Q}}$. The
pseudoscalar has been identified with $M_A=100$ GeV. The secondary axes show
the corresponding values of the light scalar Higgs mass.}
\end{figure}
The squark loop contributions are significant for squark masses
$M_{\widetilde{Q}}\lsim 400$ GeV and negligible above \cite{SQCD}. This can be
inferred
from Fig.~\ref{fg:mssmhggqsq}, where the ratio of the gluonic decay width with
and without the squark contributions is shown as a function of the squark mass
$M_{\widetilde{Q}}$ for two values of $\tb=1.5,30$. The QCD corrections to the
squark contribution are
only known in the heavy squark mass limit.  The relative QCD corrections are
presented in Fig.~\ref{fg:mssmhggqcd} as a function of the corresponding Higgs
mass for two representative values of $\tb = 1.5,30$.  The solid lines include
the top and bottom quark as well as squark contributions [for
$M_{\widetilde{Q}}=200$ GeV] and the dashed lines only the quark contributions.
The comparison of the solid and dashed curves implies that the
squark loop contributions cause a small effect on the relative QCD
corrections, so that a reasonable approximation within about 10\% to the
gluonic decay width can be obtained by multiplying the full lowest order
expression with the relative QCD corrections including only quark loops.

In complete analogy to the quark contributions the heavy squark loop correction
can be obtained by means of the extension of the previously described
low-energy theorem to scalar squark particles \cite{SQCD}. The effective NLO
Lagrangian for the squark part is given, according to eq.~(\ref{eq:hggleff}), by
\begin{equation}
{\cal L}_{eff} = \frac{1}{4} \frac{\beta_{\widetilde{Q}}(\alpha_s)/\alpha_s}{1+
\widetilde{\gamma}_m(\alpha_s)} G^{a\mu\nu} G^a_{\mu\nu} \frac{H}{v}
\end{equation}
where $\beta_{\widetilde{Q}}(\alpha_s) = \alpha_s^2/(12\pi) [1+11
\alpha_s/(2\pi)]$ denotes the heavy squark contribution to the QCD $\beta$
function \cite{betaqedsq} and $\widetilde{\gamma}_m (\alpha_s) =
4\alpha_s/(3\pi)$ the anomalous squark mass dimension \cite{anommasssq}.
Up to NLO the effective coupling is described
by \cite{SQCD}
\begin{equation}
{\cal L}_{eff} = \frac{\alpha_s}{48\pi} G^{a\mu\nu} G^a_{\mu\nu} \frac{H}{v}
\left[ 1 + \frac{25}{6} \frac{\alpha_s}{\pi} \right] \, .
\end{equation}
Thus the only difference to the quark loops in the heavy loop mass limit arises
in the virtual corrections. This leads to the additional last term of
eq.~(\ref{eq:mssmhggqcd}).

It turns out {\it a posteriori} that the heavy quark limit
$M_{h/H}^2 \ll 4M_Q^2$ is an excellent approximation for the QCD corrections
within a maximal deviation of about 10\% in the parameter ranges where this
decay mode is relevant.
\begin{table}[hbt]
\renewcommand{\arraystretch}{2.0}
\begin{center}
\begin{tabular}{|lc||cc|} \hline
\multicolumn{2}{|c||}{$\Phi$} & $H^\pm$ & $\tilde \chi^\pm_i$ \\
\hline \hline
SM~ & $H$ & 0 & 0 \\ \hline
MSSM~ & $h$ & $\frac{M_W^2}{M_{H^\pm}^2} \left[ \sin(\beta-\alpha)+\frac{\cos
2\beta \sin (\beta+\alpha)}{2\cos^2 \theta_W} \right]$
& $2\frac{M_W}{M_{\tilde \chi^\pm_i}}(S_{ii}\cos\alpha - Q_{ii}\sin\alpha)$ \\
& $H$ & $\frac{M_W^2}{M_{H^\pm}^2} \left[ \cos(\beta-\alpha)-\frac{\cos
2\beta \cos (\beta+\alpha)}{2\cos^2 \theta_W} \right]$
& $2\frac{M_W}{M_{\tilde \chi^\pm_i}}(S_{ii}\sin\alpha + Q_{ii}\cos\alpha)$ \\
& $A$ & 0 & $2\frac{M_W}{M_{\tilde \chi^\pm_i}}(-S_{ii}\cos\beta - Q_{ii}
\sin\beta)$ \\[0.2cm] \hline
\end{tabular} \\[0.3cm]

\begin{tabular}{|lc||c|} \hline
\multicolumn{2}{|c||}{$\Phi$} & $\tilde f_{L,R}$ \\
\hline \hline
SM~ & $H$ & 0 \\ \hline
MSSM~ & $h$ & $\frac{M_f^2}{M_{\tilde f}^2} g_f^h \mp
\frac{M_Z^2}{M_{\tilde f}^2} (I_3^f - e_f \sin^2\theta_W) \sin
(\alpha+\beta)$ \\
& $H$ & $\frac{M_f^2}{M_{\tilde f}^2} g_f^H \pm
\frac{M_Z^2}{M_{\tilde f}^2} (I_3^f - e_f \sin^2\theta_W) \cos
(\alpha+\beta)$ \\
& $A$ & 0 \\ \hline
\end{tabular}
\renewcommand{\arraystretch}{1.2}
\caption[]{\label{tb:hsqcoup}
\it MSSM Higgs couplings to charged
Higgs bosons, charginos and sfermions relative to SM couplings.
$Q_{ii}$ and $S_{ii}$ $(i=1,2)$ are related to the mixing angles between
the charginos $\tilde \chi^\pm_1$ and $\tilde \chi^\pm_2$,
see Refs.~\cite{hunter,mssmbase}.}
\end{center}
\end{table}

\begin{figure}[hbt]
\vspace*{0.5cm}
\hspace*{-0.5cm}
\begin{turn}{-90}%
\epsfxsize=11cm \epsfbox{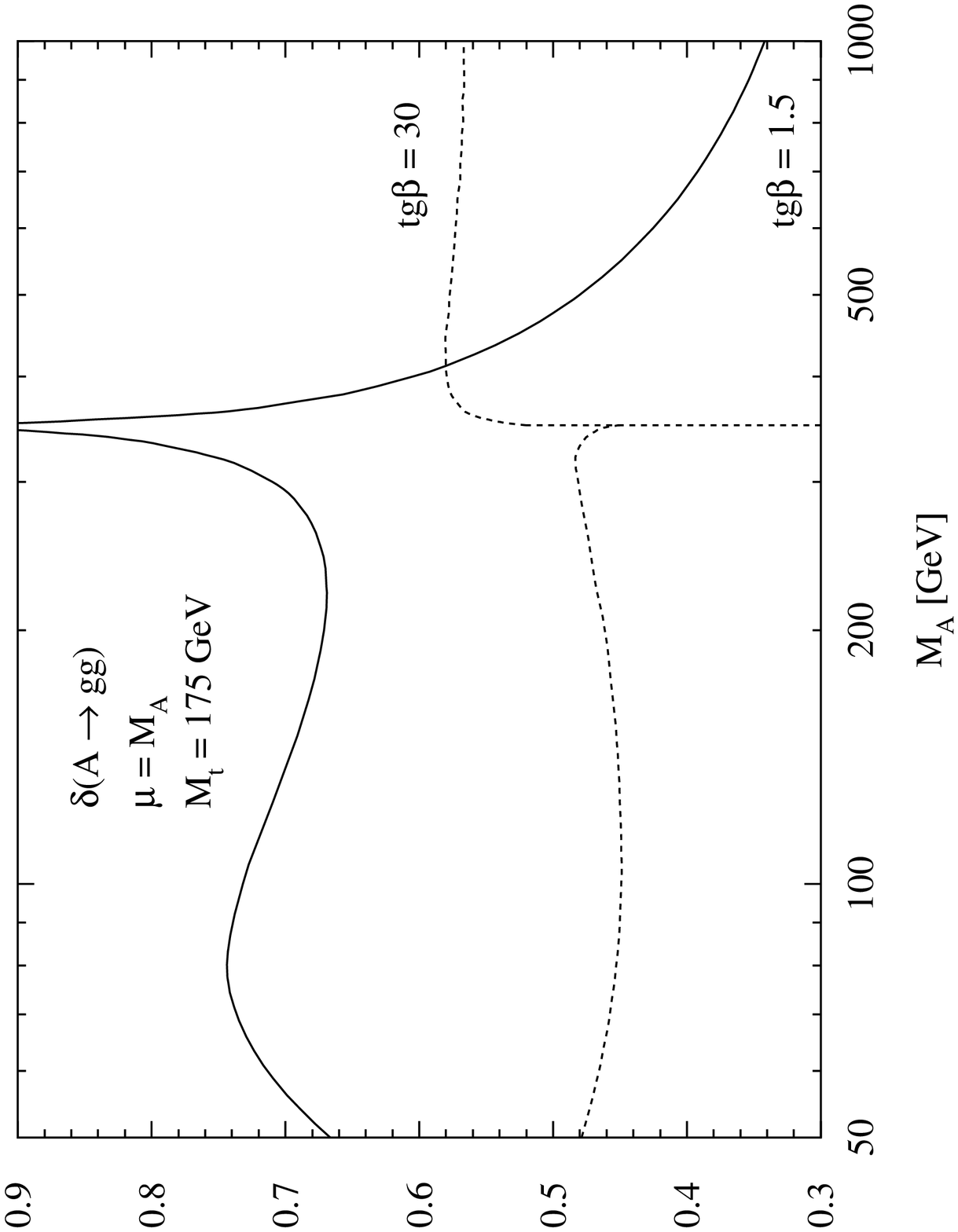}
\end{turn}
\vspace*{0.0cm}
\caption[ ]{\label{fg:aggqcd} \it Size of the QCD correction factor for
$A\to gg$, defined as $\Gamma = \Gamma_{LO} (1+\delta)$, as a function of
the pseudoscalar Higgs mass for two values of $\tb=1.5,30$.}
\end{figure}

For the pseudoscalar Higgs decays only quark loops are contributing, and we
find \cite{higgsqcd}
\begin{eqnarray}
\Gamma_{LO}\,[A\ra gg]&=&\frac{G_{F}\,
\alpha_{s}^{2}\,M_{A}^{3}}{16\,\sqrt{2}\,
\pi^{3}} \left| \sum_{Q} g_Q^A A_Q^A(\tau_Q) \right|^2 \label{eq:mssmagg} \\
A_Q^A (\tau) & = & \tau f(\tau) \nonumber \\
\Gamma [\,A\,\ra \, gg(g),q{\overline{q}}g\,] &=&
\Gamma_{LO}\,\left[\alpha_{s}^{(N_{F})}(M_A
)\right]\,\left\{1+E^{N_{F}}\frac{\alpha_{s}^
{(N_{F})}(M_A)}{\pi}\ \right\} \\
E^{N_{F}}& \to &\frac{97}{4}-\frac{7}{6}N_{F} \hspace*{1cm}
\mbox{for $M_A^2 \ll 4M_Q^2$}
\nonumber
\end{eqnarray}
with $\tau_Q = 4M_Q^2/M_A^2$.
The MSSM couplings $g_Q^A$ can be found in Table \ref{tb:hcoup}.  For large
quark masses the quark amplitude approaches unity.  In order to get a
consistent result for the two-loop QCD corrections, the pseudoscalar
$\gamma_5$ coupling has been regularized in the 't~Hooft--Veltman scheme
\cite{thoovel}, which
requires an additional finite renormalization of the $AQ\bar Q$ vertex
\cite{higgsqcd,gghsusy}.  The
relative QCD corrections are presented in Fig.~\ref{fg:aggqcd} as a function of
the pseudoscalar Higgs mass $M_A$ for two values of $\tb =1.5,30$.
The heavy quark limit $M_A^2 \ll 4M_Q^2$ provides a reasonable
approximation in the MSSM parameter range where this decay mode is
significant. At the threshold $M_A=2M_t$, the QCD corrections develop a
Coulomb singularity, which will be regularized by including the finite top
decay width \cite{agagathresh}.

The heavy quark limit can also be obtained by means of a low-energy theorem.
The starting point is the ABJ anomaly in the divergence of the axial vector
current \cite{ABJ},
\begin{equation}
\partial^\mu j_\mu^5 = 2M_Q \bar Q i\gamma_5 Q + \frac{\alpha_s}{2\pi}
G^{a\mu\nu} \widetilde{G}^a_{\mu\nu}
\label{eq:abj}
\end{equation}
with $\widetilde{G}^a_{\mu\nu} = \frac{1}{2} \epsilon_{\mu\nu\alpha\beta}
G^{a\alpha\beta}$
denoting the dual field strength tensor. Since, according to the
Sutherland--Veltman paradox \cite{suthvel}, the matrix element $\langle gg|
\partial^\mu
j_\mu^5 | 0 \rangle$ vanishes for zero momentum transfer, the matrix element
$\langle gg| M_Q \bar Q i\gamma_5 Q | 0 \rangle$ of the Higgs source can be
related to the ABJ anomaly in eq.~(\ref{eq:abj}). Thanks to the Adler--Bardeen
theorem, the ABJ anomaly is not modified by radiative corrections at vanishing
momentum transfer \cite{ABJ}, so that the effective Lagrangian
\begin{equation}
{\cal L}_{eff} = g_Q^A\frac{\alpha_s}{4\pi}G^{a\mu\nu}\widetilde{G}^a_{\mu\nu}
\frac{A}{v}
\end{equation}
is valid to all orders of perturbation theory. In order to calculate the full
QCD corrections to the $gg$ decay width, this effective coupling has to be
inserted in the effective diagrams analogous to those of
Fig.~\ref{fg:hgglimdia}. The
final result agrees with the explicit expansion of the two-loop diagrams in
terms of the heavy quark mass.

In analogy to the SM case the bottom and charm final states from gluon
splitting may be added to the corresponding $b\bar b$ and $c\bar c$ decay
modes so that the number of light flavors has to be chosen as $N_F=3$ in the
scalar and pseudoscalar decays into gluons \cite{QCD}.

\subsubsection{Decays into photon pairs}
\begin{figure}[hbt]
\begin{center}
\setlength{\unitlength}{1pt}
\begin{picture}(400,100)(-20,0)

\Photon(60,20)(90,20){-3}{4}
\Photon(60,80)(90,80){3}{4}
\ArrowLine(60,20)(60,80)
\ArrowLine(60,80)(30,50)
\ArrowLine(30,50)(60,20)
\DashLine(0,50)(30,50){5}
\put(-40,46){$h,H,A$}
\put(70,46){$f,\tilde \chi^\pm$}
\put(95,18){$\gamma$}
\put(95,78){$\gamma$}

\Photon(210,20)(240,20){-3}{4}
\Photon(210,80)(240,80){3}{4}
\DashLine(210,80)(210,20){5}
\DashLine(210,20)(180,50){5}
\DashLine(180,50)(210,80){5}
\DashLine(150,50)(180,50){5}
\put(125,46){$h,H$}
\put(215,46){$W, H^\pm, \widetilde{f}$}
\put(245,18){$\gamma$}
\put(245,78){$\gamma$}

\DashLine(300,50)(330,50){5}
\DashCArc(345,50)(15,0,180){4}
\DashCArc(345,50)(15,180,360){4}
\Photon(360,50)(390,80){3}{5}
\Photon(360,50)(390,20){3}{5}
\put(275,46){$h,H$}
\put(325,70){$W, H^\pm, \widetilde{f}$}
\put(395,18){$\gamma$}
\put(395,78){$\gamma$}

\end{picture}  \\
\setlength{\unitlength}{1pt}
\caption[ ]{\label{fg:mssmhgagalodia} \it Typical diagrams contributing to
$\Phi \to \gamma \gamma$ at lowest order.}
\end{center}
\end{figure}
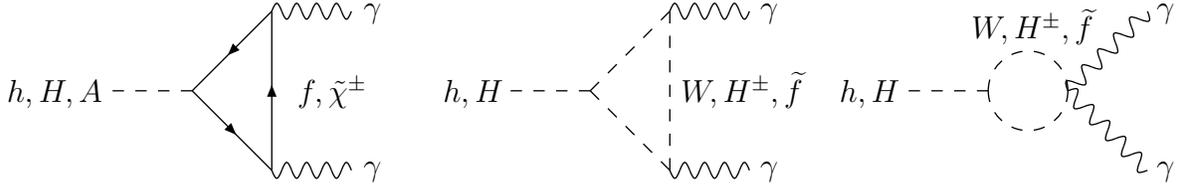
\noindent
The decays of the scalar Higgs bosons to photons are mediated by $W$ and heavy
fermion loops as in the Standard Model and, in addition, by charged Higgs,
sfermion and chargino loops; the relevant diagrams are shown in
Fig.~\ref{fg:mssmhgagalodia}. The partial decay widths \cite{hunter,higgsqcd}
are given by
\begin{eqnarray}
\!\!\!\!\!\!
\Gamma [h/H\to \gamma\gamma] & = & \frac{G_{F} \alpha^{2}M_{h/H}^{3}}
{128\sqrt{2}\pi^{3}} \left| \sum_{f} N_{cf} e_f^2 g_f^{h/H}
A_f^{h/H}(\tau_f) + g^{h/H}_W A^{h/H}_W(\tau_W)
\right. \nonumber \\
& + & \left. g_{H^\pm}^{h/H} A_{H^\pm}^{h/H}(\tau_{H^\pm})
+ \sum_{\tilde \chi^\pm} g_{\tilde \chi^\pm}^{h/H}
A_{\tilde \chi^\pm}^{h/H} (\tau_{\tilde \chi^\pm}) +
\sum_{\tilde f}N_{cf}e_{\tilde f}^2 g_{\tilde f}^{h/H} A_{\tilde f}^{h/H}
(\tau_{\tilde f})
\right|^2
\end{eqnarray}
with the form factors
\begin{eqnarray*}
A_{f,\tilde \chi^\pm}^{h/H} (\tau) & = & 2 \tau \left[ 1+(1-\tau) f(\tau)
\right] \\ \\
A_{H^\pm,\tilde f}^{h/H} (\tau) & = & - \tau \left[1-\tau f(\tau) \right] \\ \\
A_W^{h/H} (\tau) & = & -\left[ 2+3\tau+3\tau (2-\tau) f(\tau) \right] \, ,
\end{eqnarray*}
where the function $f(\tau)$ is defined in eq.~(\ref{eq:ftau}). For large loop
particle masses the form factors approach constant values,
\begin{eqnarray*}
A_{f,\tilde \chi^\pm}^{h/H} (\tau) & \to & \frac{4}{3} \hspace*{1cm}
\mbox{for $M_{h/H}^2 \ll 4M_{f,\tilde \chi^\pm}^2$} \nonumber \\
A_{H^\pm,\widetilde{f}}^{h/H} (\tau) & \to & \frac{1}{3} \hspace*{1cm}
\mbox{for $M_{h/H}^2 \ll 4M_{H^\pm,\widetilde{f}}^2$} \nonumber \\
A_W^{h/H} (\tau) & \to & - 7 \hspace*{1cm} \mbox{for $M_{h/H}^2 \ll 4M_W^2$}
\, .
\end{eqnarray*}
Sfermion loops start to be sizeable for sfermion masses $M_{\widetilde{f}} \lsim
300$ GeV. For larger sfermion masses they are negligible.

The photonic decay mode of the pseudoscalar Higgs boson is generated by heavy
charged fermion and chargino loops, see Fig.~\ref{fg:mssmhgagalodia}. The
partial decay width reads as \cite{hunter,higgsqcd}
\begin{equation}
\Gamma(A^0 \rightarrow\gamma\gamma)=
\frac{G_F\alpha^2M_A^3}{32\sqrt{2}\pi^3}
\left| \sum_f N_{cf} e_f^2 g_f^A A_f^A (\tau_f)
+ \sum_{\tilde \chi^\pm} g_{\tilde \chi\pm}^A A_{\tilde \chi^\pm}^A
(\tau_{\tilde \chi^\pm}) \right|^2 \, ,
\end{equation}
with the amplitudes
\begin{eqnarray}
A_{f, \tilde \chi^\pm}^A (\tau) & = & \tau f(\tau) \, .
\end{eqnarray}
For large loop particle masses the pseudoscalar amplitudes approach unity.

The parameters $\tau_i= 4M_i^2/M_\Phi^2~~(i=f,W,H^\pm,\tilde \chi^\pm,
\tilde f)$ are defined by the corresponding mass of the heavy loop particle
and the MSSM couplings $g^\phi_{f,W,H^\pm,\tilde \chi^\pm, \tilde f}$ are
summarized in Tables \ref{tb:hcoup} and \ref{tb:hsqcoup}.

\begin{figure}[hbtp]

\vspace*{1.0cm}
\hspace*{-5.0cm}
\begin{turn}{-90}%
\epsfxsize=17cm \epsfbox{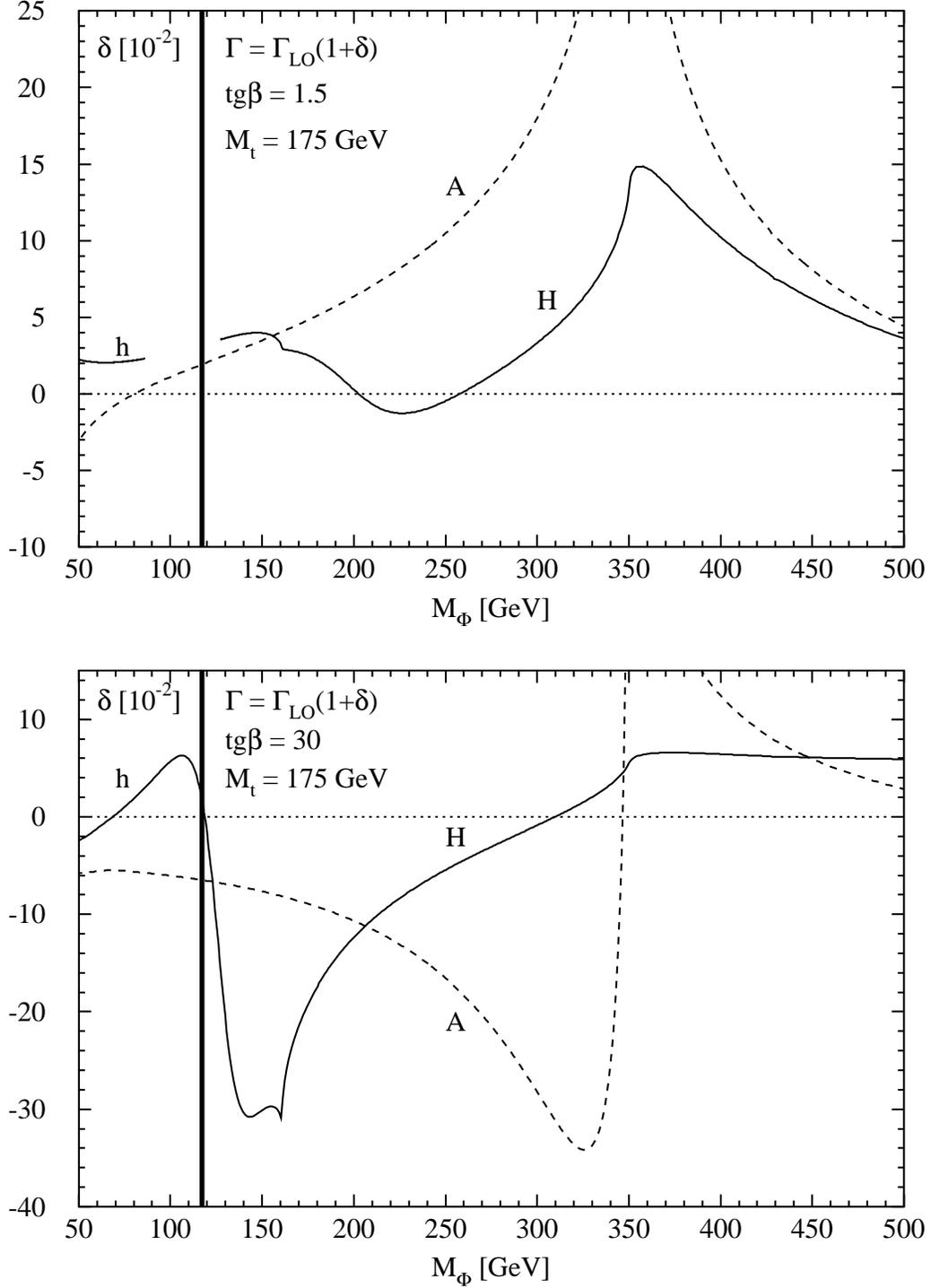}
\end{turn}
\vspace*{0.5cm}

\caption[ ]{\label{fg:mssmhgagaqcd} \it Size of the QCD correction factor for
$\Phi\to \gamma \gamma$, defined as $\Gamma = \Gamma_{LO} (1+\delta)$, as a
function of the corresponding Higgs mass for two values of $\tb=1.5,30$. The
renormalization scale of the running quark masses is identified with $\mu_Q
= M_\Phi/2$. The common squark mass has been chosen as $M_S=1$ TeV.}
\end{figure}

The QCD corrections to the quark and squark loop contributions have
been evaluated. For the $t,b$ quark loops they are known for finite quark and
Higgs masses \cite{higgsqcd,hgaga}, while in the case of squark loops only
the large squark mass limit has been computed so far \cite{hgagasq}. The QCD
corrections rescale the lowest order quark amplitudes
\cite{higgsqcd,hgaga,hgagasq},
\begin{eqnarray}
A_Q^\Phi (\tau_Q) & \to & A_Q^\Phi (\tau_Q) \left[ 1+C_\Phi(\tau_Q)
\frac{\alpha_s}{\pi} \right] \\
C_{h/H} (\tau_Q) & \to & -1 \hspace*{1cm} \mbox{for $M_{h/H}^2 \ll 4M_Q^2$}
\nonumber \\
C_A (\tau_Q) & \to & 0 \hspace*{1cm} \mbox{for $M_A^2 \ll 4M_Q^2$} \nonumber \\
\non \\
A_{\widetilde{Q}}^{h/H}(\tau_{\widetilde{Q}}) & \to & A_{\widetilde{Q}}^{h/H}
(\tau_{\widetilde{Q}}) \left[ 1+\widetilde{C}_{h/H}(\tau_{\widetilde{Q}})
\frac{\alpha_s}{\pi} \right] \\
\widetilde{C}_{h/H} (\tau_{\widetilde{Q}}) & \to & \frac{8}{3} \hspace*{1cm}
\mbox{for $M_{h/H}^2 \ll 4M_{\widetilde{Q}}^2$}
\nonumber
\end{eqnarray}
The QCD corrections to the $\gamma\gamma$ decay width are plotted in
Fig.~\ref{fg:mssmhgagaqcd} for two values of $\tb=1.5,30$ in the case of heavy
charginos and sfermions. They are defined in
terms of the running quark masses in the same way as the SM photonic decay
width. The QCD radiative corrections are moderate in the intermediate mass range
\cite{higgsqcd,hgaga}, where this decay mode will be important, and therefore
neglected in the analysis. Owing to the narrow-width approximation of the
virtual quarks, the QCD corrections to the pseudoscalar decay width exhibit a
Coulomb singularity at the $t\bar t$ threshold, which is regularized by taking
into account the finite top quark decay width \cite{agagathresh}.

The QCD corrections to the quark loops in the heavy quark limit can be obtained
by means of the low-energy theorems for scalar as well as pseudoscalar Higgs
particles, which have been discussed before. The result for the scalar Higgs
bosons agrees with the SM result of eq.~(\ref{eq:hgagaqcd}), and the QCD
corrections to the pseudoscalar decay mode vanish in this
limit due to the Adler--Bardeen theorem. In complete analogy to the gluonic
decay mode, the effective Lagrangian can be derived from the ABJ anomaly and
is given to all orders of perturbation theory by \cite{higgsqcd}
\begin{equation}
{\cal L}_{eff} = g_Q^A e_Q^2 \frac{3\alpha}{4\pi}F^{\mu\nu}
\widetilde{F}_{\mu\nu} \frac{A}{v} \, .
\end{equation}
Since there are no effective diagrams generated by light
particle interactions that contribute to the photonic decay width at
next-to-leading order, the QCD corrections to the pseudoscalar decay width
vanish, in agreement with the explicit expansion of the massive two-loop
result.

Completely analogous the QCD corrections to the squark loops for the scalar
Higgs particles in the heavy squark limit can be obtained by the extension of
the scalar low-energy theorem to the scalar squarks. Their coupling to photons
at NLO can be described by the effective Lagrangian \cite{hgagasq}
\begin{equation}
{\cal L}_{eff} = g_{\widetilde{Q}}^H \frac{e_{\widetilde{Q}}^2}{4}
\frac{\beta^{\widetilde{Q}}_\alpha/\alpha}{1+ \widetilde{\gamma}_m(\alpha_s)}
F^{\mu\nu} F_{\mu\nu} \frac{H}{v}
\end{equation}
where $\beta^{\widetilde{Q}}_\alpha = \alpha/(2\pi) [1+4
\alpha_s/\pi]$ denotes the heavy squark contribution to the QED $\beta$
function \cite{betaqedsq} and $\widetilde{\gamma}_m (\alpha_s) = 4\alpha_s/
(3\pi)$ the anomalous squark mass dimension \cite{anommasssq}. Up to NLO the
effective coupling reads as \cite{hgagasq}
\begin{equation}
{\cal L}_{eff} = g_{\widetilde{Q}}^H e_{\widetilde{Q}}^2 \frac{\alpha}{8\pi}
F^{\mu\nu} F_{\mu\nu}
\frac{H}{v} \left[ 1 + \frac{8}{3} \frac{\alpha_s}{\pi} \right] \, .
\end{equation}
This correction is small and thus neglected in the present analysis.

\subsubsection{Decays into $Z$ boson and photon}
\begin{figure}[hbt]
\begin{center}
\setlength{\unitlength}{1pt}
\begin{picture}(400,100)(0,0)

\Photon(60,20)(90,20){-3}{4}
\Photon(60,80)(90,80){3}{4}
\ArrowLine(60,20)(60,80)
\ArrowLine(60,80)(30,50)
\ArrowLine(30,50)(60,20)
\DashLine(0,50)(30,50){5}
\put(-40,46){$h,H,A$}
\put(65,46){$f,\tilde \chi^\pm$}
\put(95,78){$Z$}
\put(95,18){$\gamma$}

\Photon(210,20)(240,20){-3}{4}
\Photon(210,80)(240,80){3}{4}
\DashLine(210,80)(210,20){5}
\DashLine(210,20)(180,50){5}
\DashLine(180,50)(210,80){5}
\DashLine(150,50)(180,50){5}
\put(125,46){$h,H$}
\put(220,46){$W, H^\pm, \widetilde{f}$}
\put(245,78){$Z$}
\put(245,18){$\gamma$}

\DashLine(300,50)(330,50){5}
\DashCArc(345,50)(15,0,180){4}
\DashCArc(345,50)(15,180,360){4}
\Photon(360,50)(390,80){3}{5}
\Photon(360,50)(390,20){3}{5}
\put(275,46){$h,H$}
\put(325,70){$W, H^\pm, \widetilde{f}$}
\put(395,78){$Z$}
\put(395,18){$\gamma$}

\end{picture}  \\
\setlength{\unitlength}{1pt}
\caption[ ]{\label{fg:mssmhzgalodia} \it Typical diagrams contributing to
$\Phi \to Z \gamma$ at lowest order.}
\end{center}
\end{figure}
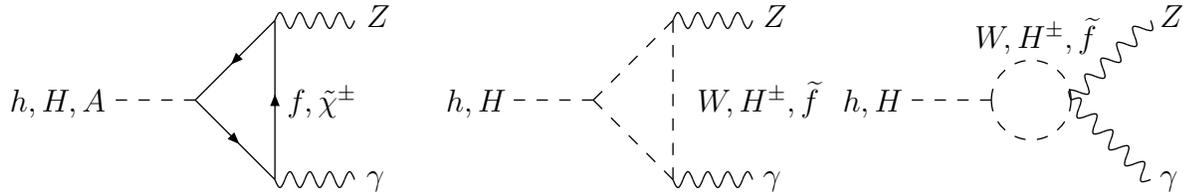
\noindent
The decays of the scalar Higgs bosons into $Z$ boson and photon are mediated
by $W$ and heavy
fermion loops as in the Standard Model and, in addition, by charged Higgs,
sfermion and chargino loops; the contributing diagrams are shown in
Fig.~\ref{fg:mssmhzgalodia}. The partial decay widths read as
\cite{higgsqcd,hzgamssm}
\begin{eqnarray}
\Gamma\, [h/H\to Z\gamma] & = & \frac{G^2_{F}M_W^2\, \alpha\,M_{h/H}^{3}}
{64\,\pi^{4}} \left(1-\frac{M_Z^2}{M_{h/H}^2} \right)^3 \left|\sum_{f}g_f^{h/H}
A_f^{h/H}(\tau_f,\lambda_f)
\right. \nonumber \\
& & \left. + g^{h/H}_W A^{h/H}_W(\tau_W,\lambda_W)
+ g_{H^\pm}^{h/H} A_{H^\pm}^{h/H}(\tau_{H^\pm},\lambda_{H^\pm})
\right. \nonumber \\
& & \left. + \sum_{\tilde \chi^\pm_i, \tilde \chi^\mp_j}
g_{\tilde \chi^\pm_i \tilde \chi^\mp_j}^{h/H}
g_{\tilde \chi^\mp_i \tilde \chi^\pm_j}^Z
A_{\tilde \chi^\pm_i \tilde \chi^\mp_j}^{h/H}
+ \sum_{\tilde f_i, \tilde f_j} g_{\tilde f_i \tilde f_j}^{h/H}
g_{\tilde f_i \tilde f_j}^Z A_{\tilde f_i \tilde f_j}^{h/H} \right|^2 \, ,
\end{eqnarray}
with the form factors $A_f^{h/H},A_W^{h/H}$ given in eq.~(\ref{eq:hzgaform}),
and
\begin{equation}
A_{H^\pm}^{h/H} (\tau,\lambda) = \frac{\cos 2\theta_W}{\cos\theta_W}
I_1(\tau,\lambda) \, ,
\end{equation}
where the function $I_1(\tau,\lambda)$ is defined after eq.~(\ref{eq:hzgaform}).

The $Z \gamma$ decay mode of the pseudoscalar Higgs boson is generated by heavy
charged fermion and chargino loops, see Fig.~\ref{fg:mssmhzgalodia}. The
partial decay width is given by \cite{hzgamssm}
\begin{equation}
\Gamma(A \rightarrow Z\gamma)=
\frac{G^2_F M_W^2 \alpha M_A^3}{16\pi^4} \left(1-\frac{M_Z^2}{M_A^2} \right)^3
\left| \sum_f g_f^A A_f^A (\tau_f,\lambda_f)
+ \sum_{\tilde \chi^\pm_i, \tilde \chi^\mp_j}
g_{\tilde \chi^\pm_i \tilde \chi^\mp_j}^A
g_{\tilde \chi^\mp_i \tilde \chi^\pm_j}^Z
A_{\tilde \chi^\pm_i \tilde \chi^\pm_j}^A \right|^2 \, ,
\end{equation}
with the fermion amplitudes
\begin{equation}
A_f^A (\tau,\lambda) = 2 N_{cf} \frac{e_f (I_{3f} - 2e_f\sin^2\theta_W )}
{\cos\theta_W}~I_2(\tau,\lambda) \, .
\end{equation}
The contributions of charginos and sfermions involve mixing terms. Their
analytical expressions can be found in \cite{hzgamssm}.
For large loop particle masses and small $Z$ mass, the form factors approach
the photonic amplitudes {\it modulo} couplings.
The parameters $\tau_i= 4M_i^2/M_\Phi^2, \lambda_i= 4M_i^2/M_Z^2~~(i=f,W,H^\pm,
\tilde \chi^\pm,\tilde f)$ are defined by the corresponding mass of the
heavy loop particle and the non-mixing MSSM
couplings $g^\phi_{f,W,H^\pm,\tilde\chi^\pm,\tilde f}$ are summarized in Tables
\ref{tb:hcoup} and \ref{tb:hsqcoup}, while the mixing and $Z$ boson
couplings $g_i^Z$ can be found in \cite{hunter}.

The branching ratios of the $Z\gamma$ decay modes range at a level of up to a
few $10^{-4}$ in the intermediate mass ranges of the Higgs bosons and are thus
phenomenologically unimportant in the MSSM.

\subsubsection{Decays into intermediate gauge bosons}
The partial widths of the scalar MSSM Higgs bosons into $W$ and $Z$ boson
pairs can be obtained from the SM Higgs decay widths by rescaling with the
corresponding MSSM couplings $g_V^{h/H}$, which are listed in Table
\ref{tb:hcoup}:
\begin{equation}
\Gamma(h/H \to V^{(*)}V^{(*)}) = (g^{h/H}_V)^2 \Gamma(H_{SM} \to
V^{(*)}V^{(*)}) \, .
\end{equation}
They are strongly decreased by kinematic suppression and reduced MSSM
couplings, and thus do not play a dominant r\^ole as in the SM case.
Nevertheless the $WW,ZZ$ branching ratios can reach values of ${\cal O}(10\%)$
for the heavy scalar Higgs boson $H$ for small $\tb$. Off-shell $WW,ZZ$ decays
can pick up several per cent of the light scalar Higgs decays at the upper end
of its mass range. The pseudoscalar Higgs particle does not couple to $W$ and
$Z$ bosons at tree level.

\subsubsection{Decays into Higgs particles}
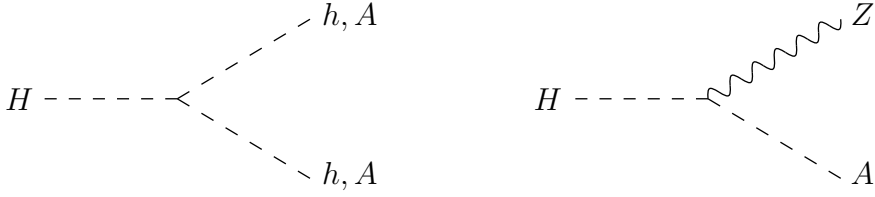
\begin{figure}[hbt]
\begin{center}
\setlength{\unitlength}{1pt}
\begin{picture}(350,100)(0,0)

\DashLine(0,50)(50,50){5}
\DashLine(50,50)(100,80){5}
\DashLine(50,50)(100,20){5}
\put(-15,46){$H$}
\put(105,18){$h,A$}
\put(105,78){$h,A$}

\DashLine(200,50)(250,50){5}
\Photon(250,50)(300,80){3}{6}
\DashLine(250,50)(300,20){5}
\put(185,46){$H$}
\put(305,18){$A$}
\put(305,78){$Z$}

\end{picture}  \\
\setlength{\unitlength}{1pt}
\caption[ ]{\label{fg:haadia} \it Typical diagrams contributing to Higgs decays
with Higgs bosons in the final state.}
\end{center}
\end{figure}
The heavy scalar Higgs particle can decay into pairs of light scalar as well
as pseudoscalar Higgs bosons, see Fig.~\ref{fg:haadia}. The partial decay widths
are given by \cite{hunter}
\begin{eqnarray}
\Gamma(H \to hh) & = & \lambda_{Hhh}^2 \frac{G_F M_Z^4}{16\sqrt{2}\pi M_H}
\sqrt{1-4\frac{M_h^2}{M_H^2}} \\
\Gamma(H \to AA) & = & \lambda_{HAA}^2 \frac{G_F M_Z^4}{16\sqrt{2}\pi M_H}
\sqrt{1-4\frac{M_A^2}{M_H^2}}
\end{eqnarray}
The self-couplings $\lambda_{Hhh}$ and $\lambda_{HAA}$ can be derived from
the effective Higgs potential \cite{mssmrad2}.
The decay mode into pseudoscalar particles is restricted to small regions of
the MSSM parameter space, where the pseudoscalar mass $M_A$ is small. The
decay into light scalar bosons is dominant for small $\tb$
below the $t\bar t$ threshold.

The contributions of final states
containing off-shell scalar or pseudoscalar Higgs bosons may be significant and
are thus included in the analysis. Their expressions read as \cite{1OFF}
\begin{eqnarray}
\Gamma(H\to \phi\phi^*) & = & \lambda_{H\phi\phi}^2
g_{\phi bb}^2 m_b^2 \frac{3G_F^2M_Z^4}{16\pi^3M_H}
\left\{ (\kappa_\phi -1) \left(2-\frac{1}{2}\log
\kappa_\phi \right) \right. \nonumber \\
& & \left. + \frac{1-5\kappa_\phi}{\sqrt{4\kappa_\phi-1}} \left( \arctan
\frac{2\kappa_\phi-1}{\sqrt{4\kappa_\phi-1}} - \arctan
\frac{1}{\sqrt{4\kappa_\phi-1}} \right) \right\} \, .
\end{eqnarray}
where $\kappa_\phi = M_\phi^2 / M_H^2$.
They slightly enhance the regions, where the $hh,AA$ decay modes of the heavy
scalar Higgs boson $H$ are sizeable.

Moreover, Higgs bosons can decay into a gauge and a Higgs boson, see
Fig.~\ref{fg:haadia}. The various partial widths can be expressed as
\begin{eqnarray}
\Gamma (H\to AZ) & = & \lambda_{HAZ}^2 \frac{G_F M_Z^4}{8\sqrt{2}\pi M_H}
\sqrt{\lambda(M_A^2,M_Z^2;M_H^2)} \lambda(M_A^2,M_H^2;M_Z^2) \\
\Gamma (H\to H^\pm W^\mp) & = & \lambda_{HH^+W}^2 \frac{G_F M_W^4}{8\sqrt{2}\pi
M_H} \sqrt{\lambda(M_{H^\pm}^2,M_W^2;M_H^2)} \lambda(M_{H^\pm}^2,M_H^2;M_W^2)
\\
\Gamma (A\to hZ) & = & \lambda_{hAZ}^2 \frac{G_F M_Z^4}{8\sqrt{2}\pi M_A}
\sqrt{\lambda(M_h^2,M_Z^2;M_A^2)} \lambda(M_h^2,M_A^2;M_Z^2) \\
\Gamma (H^+\to h W^+) & = & \lambda_{hH^+W}^2 \frac{G_F M_W^4}{8\sqrt{2}\pi
M_{H^\pm}} \sqrt{\lambda(M_h^2,M_W^2;M_{H^\pm}^2)}
\lambda(M_h^2,M_{H^\pm}^2;M_W^2) \, ,
\end{eqnarray}
where the couplings $\lambda^2_{ijk}$ can be determined from the effective
Higgs potential \cite{mssmrad2}. The functions $\lambda(x,y;z) = (1-x/z-y/z)^2
- 4xy/z^2$ denote the usual two-body phase-space factors.
The branching ratios of these decay modes may be sizeable in
specific regions of the MSSM parameter space.

Below-threshold decays into
a Higgs particle and an off-shell gauge boson turn out to be very important
for the heavy Higgs bosons of the MSSM. The individual contributions are given
by \cite{1OFF}
\begin{eqnarray}
\Gamma (H\to AZ^*) & = & \lambda_{HAZ}^2 \delta'_Z \frac{9G_F^2M_Z^4M_H}{8\pi^3}
G_{AZ} \\
\Gamma (H\to H^\pm W^{\mp*}) & = & \lambda_{HH^\pm W}^2 \frac{9G_F^2M_W^4M_H}
{8\pi^3} G_{H^\pm W} \\
\Gamma (A\to hZ^*) & = & \lambda_{hAZ}^2 \delta'_Z \frac{9G_F^2M_Z^4M_A}{8\pi^3}
G_{hZ} \\
\Gamma (H^+\to h W^{+*}) & = & \lambda_{hH^\pm W}^2 \frac{9G_F^2M_W^4M_{H^\pm}}
{8\pi^3} G_{h W} \\
\Gamma (H^+\to A W^{+*}) & = & \frac{9G_F^2M_W^4M_{H^\pm}}{8\pi^3} G_{A W} \, .
\end{eqnarray}
The generic functions $G_{ij}$ can be written as
\begin{eqnarray}
G_{ij} & = & \frac{1}{4} \left\{ 2(-1+\kappa_j-\kappa_i)\sqrt{\lambda_{ij}}
\left[ \frac{\pi}{2} + \arctan \left(\frac{\kappa_j (1-\kappa_j+\kappa_i) -
\lambda_{ij}}{(1-\kappa_i) \sqrt{\lambda_{ij}}} \right) \right] \right. \\
& & \left. + (\lambda_{ij}-2\kappa_i) \log \kappa_i + \frac{1}{3} (1-\kappa_i)
\left[ 5(1+\kappa_i) - 4\kappa_j - \frac{2}{\kappa_j} \lambda_{ij} \right]
\right\}
\end{eqnarray}
using the parameters
\begin{equation}
\lambda_{ij} = -1+2\kappa_i+2\kappa_j-(\kappa_i-\kappa_j)^2, \hspace{2cm}
\kappa_i = \frac{M_i^2}{M_\phi^2} \, .
\end{equation}
The coefficient $\delta'_Z$ is defined after eq.~(\ref{eq:hvvp}).
Off-shell $hZ^*$ decays are important for the pseudoscalar Higgs boson for
masses above about 130 GeV for small $\tb$ \cite{1OFF}. The decay modes
$H^\pm \to hW^*, AW^*$ reach branching ratios of several tens of per cent and
lead to a significant reduction of the dominant branching ratio into $\tau\nu$
final states to a level of 60--70\% for small $\tb$ \cite{1OFF}. 

\subsubsection{Total decay widths and branching ratios of non-SUSY particle
               decays}
\begin{figure}[hbtp]

\vspace*{0.5cm}
\hspace*{-0.0cm}
\begin{turn}{-90}%
\epsfxsize=9.5cm \epsfbox{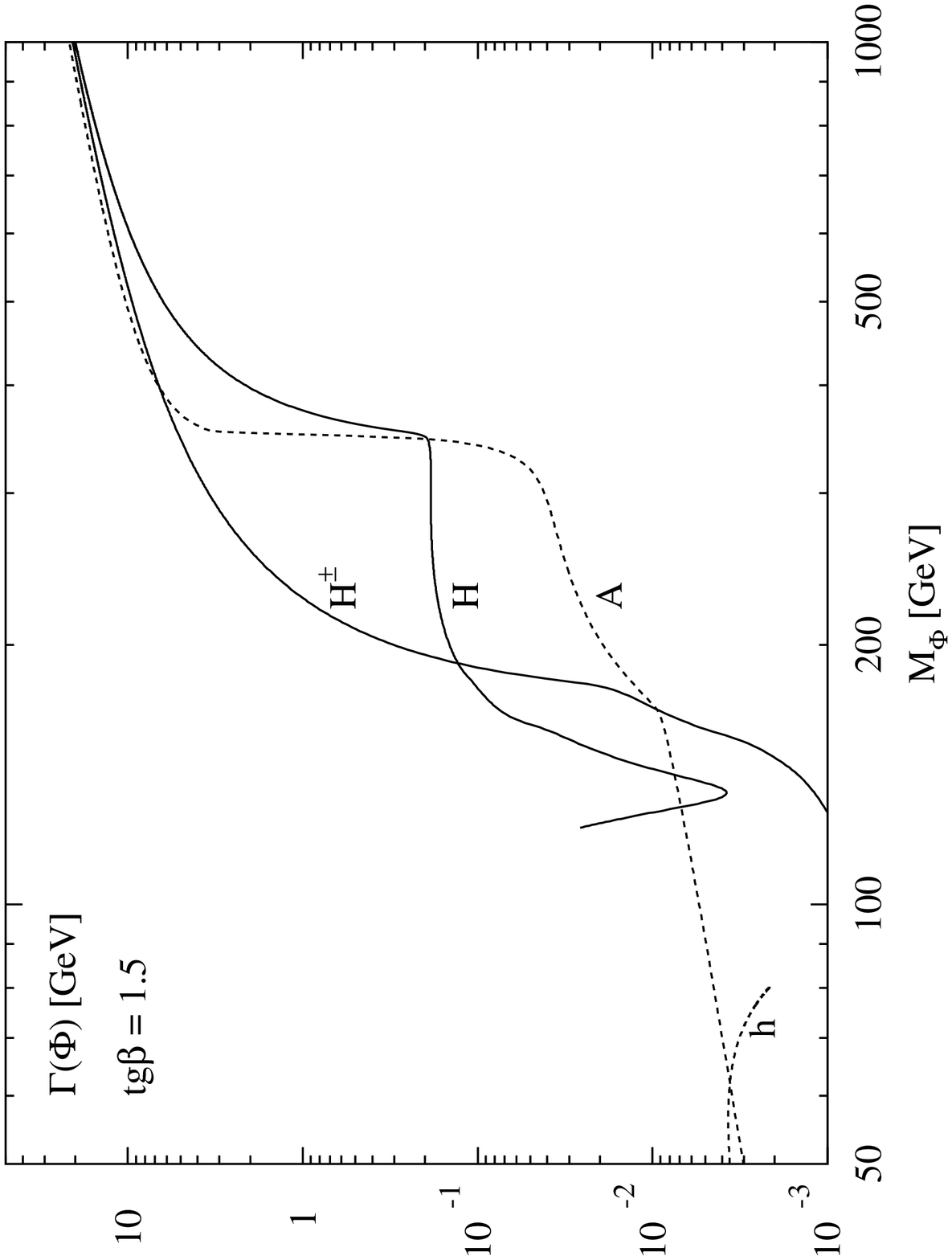}
\end{turn}

\vspace*{0.5cm}
\hspace*{-0.0cm}
\begin{turn}{-90}%
\epsfxsize=9.5cm \epsfbox{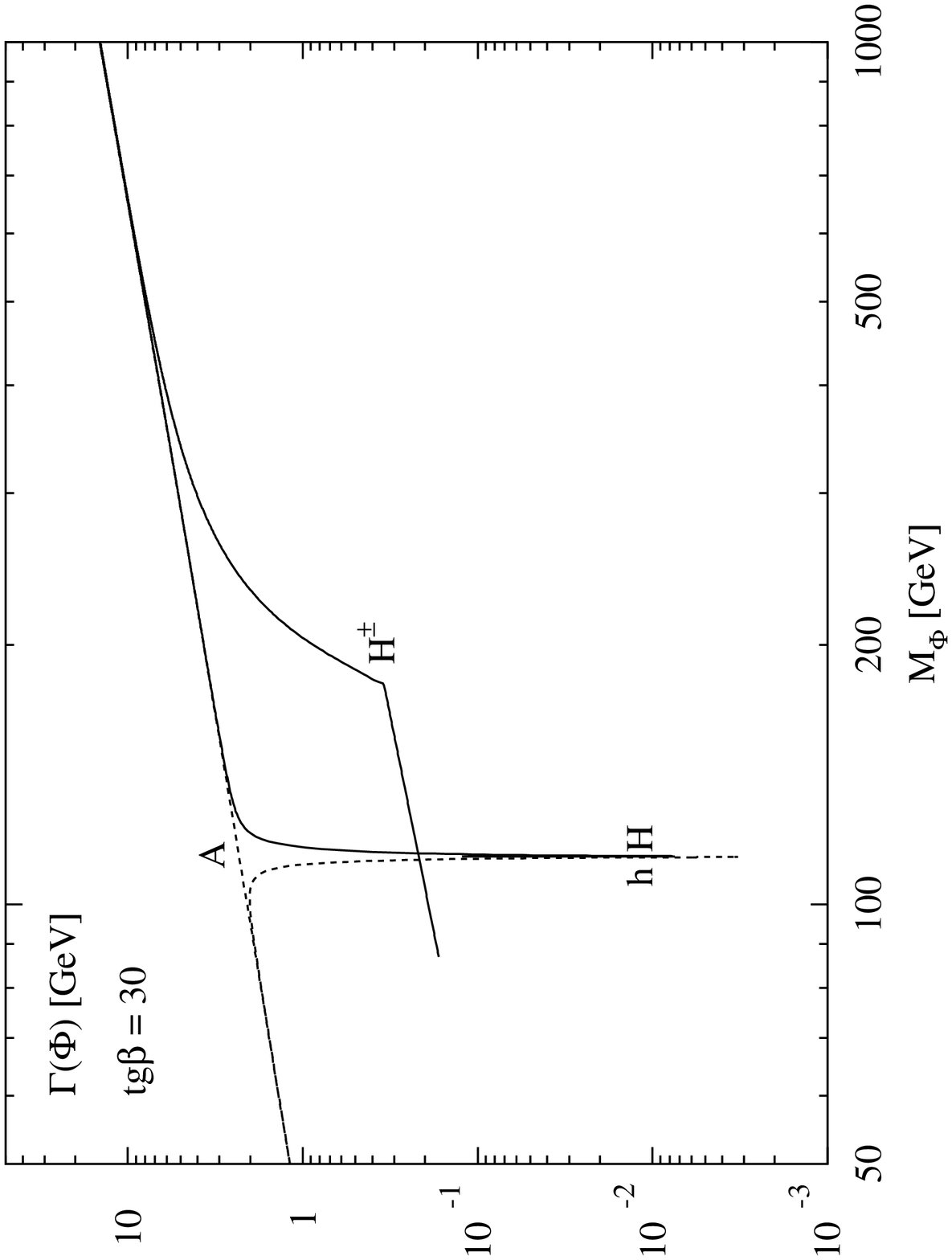}
\end{turn}
\vspace*{-0.0cm}

\caption[]{\label{fg:mssmwtot} \it Total decay widths of the MSSM Higgs bosons
$h,H,A,H^\pm$ for non-SUSY decay modes as a function of their masses for two
values of $\tb=1.5,30$ and vanishing mixing. The common squark mass has been
taken to be $M_S=1$ TeV.}
\end{figure}
\begin{figure}[hbtp]

\vspace*{-2.5cm}
\hspace*{-4.5cm}
\begin{turn}{-90}%
\epsfxsize=16cm \epsfbox{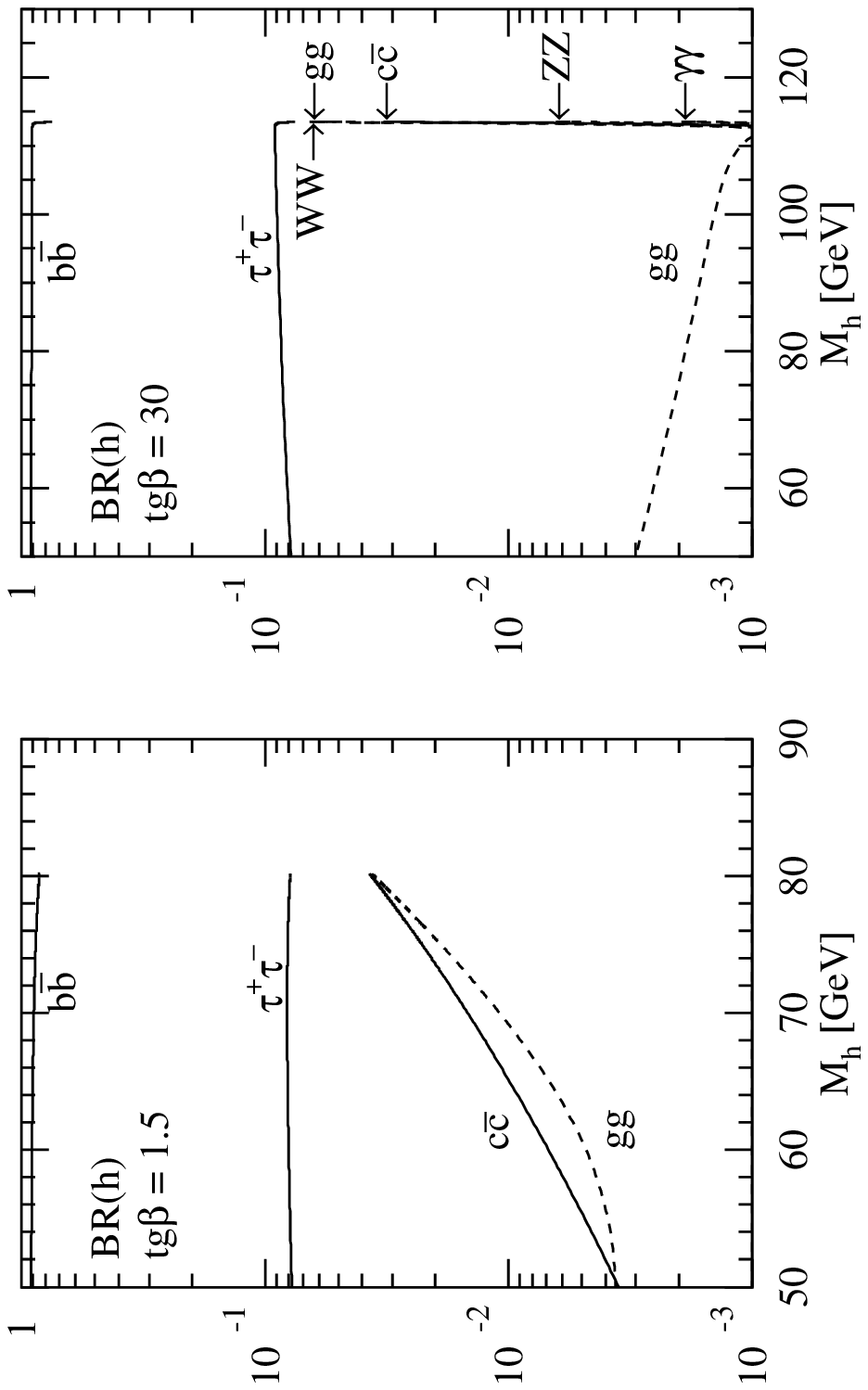}
\end{turn}
\vspace*{-4.2cm}

\centerline{\bf Fig.~\ref{fg:mssmbr}a}
 
\vspace*{-2.5cm}
\hspace*{-4.5cm}
\begin{turn}{-90}%
\epsfxsize=16cm \epsfbox{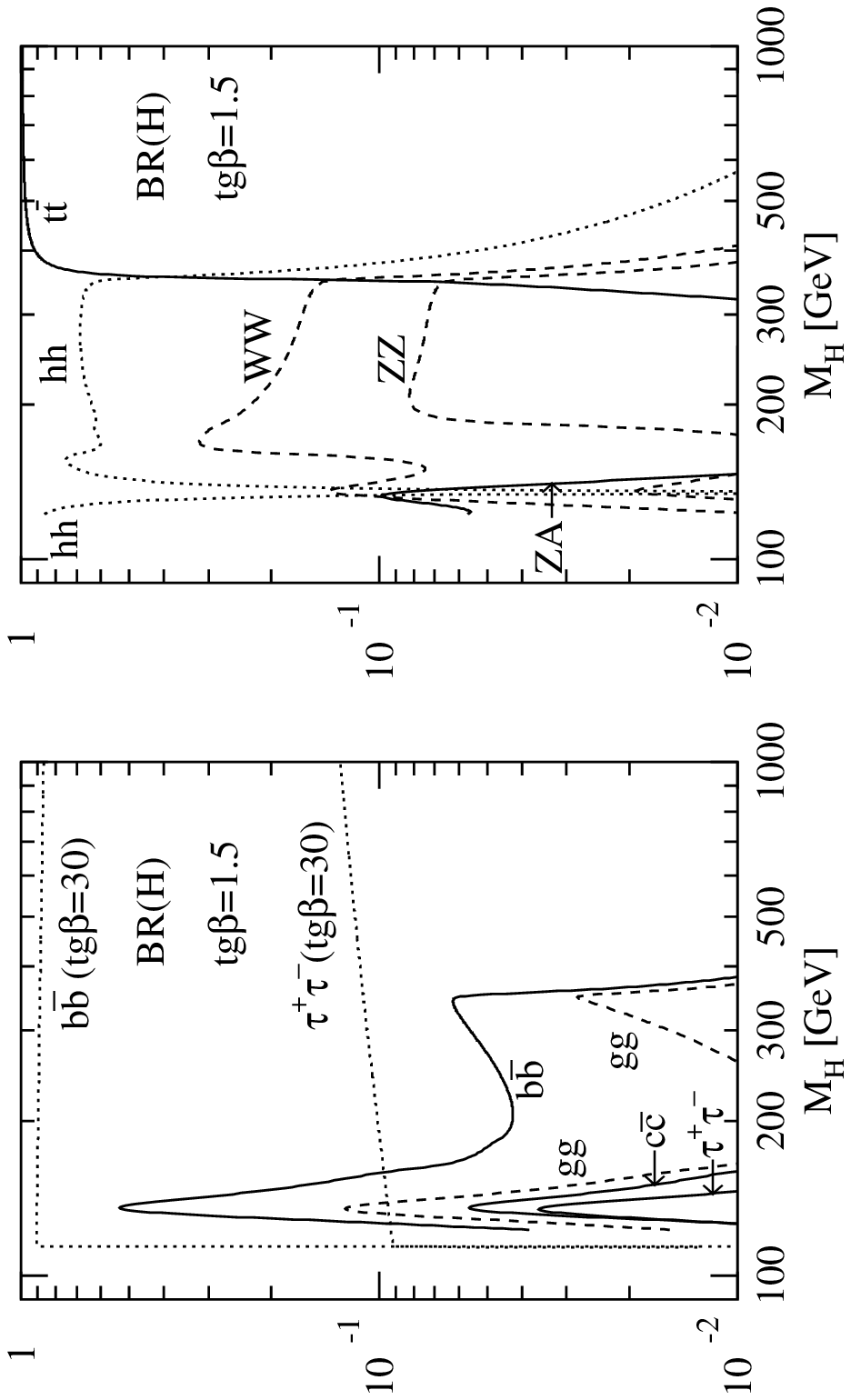}
\end{turn}
\vspace*{-4.2cm}
 
\centerline{\bf Fig.~\ref{fg:mssmbr}b}
 
\caption[]{\label{fg:mssmbr} \it Branching ratios of the MSSM Higgs bosons $h
(a), H (b), A (c), H^\pm (d)$ for non-SUSY decay modes as a function of their
masses for two values of $\tb=1.5, 30$ and vanishing mixing. The common squark
mass has been chosen as $M_S=1$ TeV.}
\end{figure}
\addtocounter{figure}{-1}
\begin{figure}[hbtp]

\vspace*{-2.5cm}
\hspace*{-4.5cm}
\begin{turn}{-90}%
\epsfxsize=16cm \epsfbox{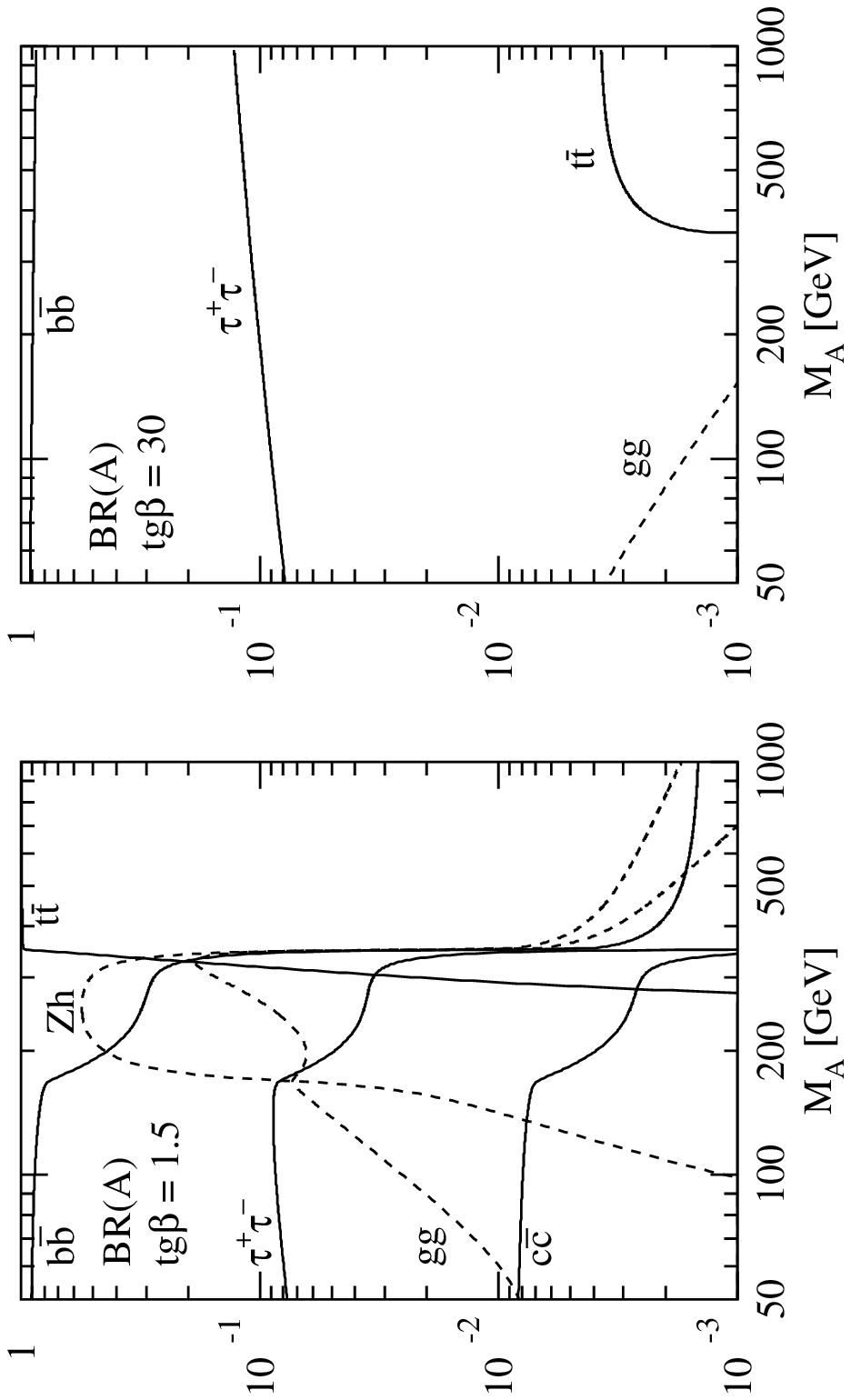}
\end{turn}
\vspace*{-4.2cm}

\centerline{\bf Fig.~\ref{fg:mssmbr}c}
 
\vspace*{-2.5cm}
\hspace*{-4.5cm}
\begin{turn}{-90}%
\epsfxsize=16cm \epsfbox{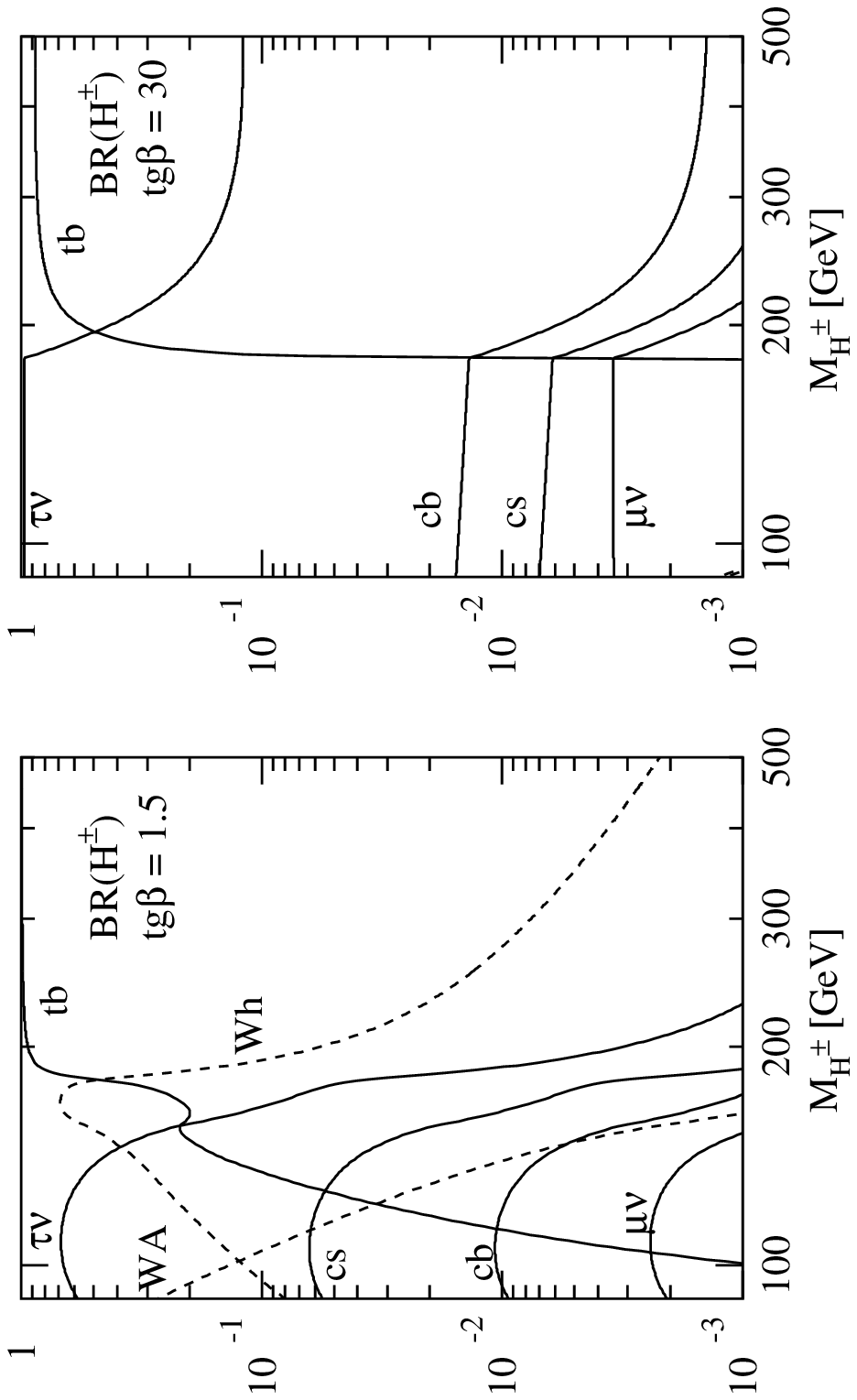}
\end{turn}
\vspace*{-4.2cm}

\centerline{\bf Fig.~\ref{fg:mssmbr}d}
 
\caption[]{\it Continued.}
\end{figure}
Fig.~\ref{fg:mssmwtot} presents the total decay widths and Fig.~\ref{fg:mssmbr}
the branching ratios of the various Higgs decay modes
into non-SUSY particles, i.e.~SM and Higgs particles,
as a function of the corresponding Higgs masses for two representative values
of $\tb = 1.5,30$. Since the Higgs self-interactions are
determined by the gauge couplings, the total decay widths of all MSSM Higgs
bosons do not exceed about 30 GeV, so that these states will appear
as rather narrow resonances. The small decay widths are a direct consequence of
the absence of quadratic divergences in the MSSM Higgs sector and the solution
of the hierarchy problem.

For the light scalar Higgs boson $h$ the $b\bar b$ decays dominate, with a
branching ratio of up to about 90\%, see Fig.~\ref{fg:mssmbr}a. The bulk of the
remaining decay modes is taken by $\tau^+ \tau^-$ decays, the branching ratio
of which ranges at about 8--9\%. At the
upper bound of the light Higgs boson mass all decay modes, as for the
intermediate SM Higgs particle, are important. Their branching ratios coincide
with the SM values for the corresponding SM Higgs mass, in accordance with the
condition that in the decoupling regime the light scalar Higgs particle behaves
as the SM Higgs boson.

Fig.~\ref{fg:mssmbr}b shows that for large $\tb$ the heavy scalar Higgs boson
$H$ predominantly decays into
$b\bar b$ final states with a branching ratio of about 90\%, and to a lesser
extent into $\tau^+\tau^-$ pairs with a branching ratio of about 10\%. All
other decay modes are unimportant for large $\tb$. In contrast, the
heavy scalar
Higgs particle exhibits a very rich spectrum of decay modes for small $\tb$.
For $\tb=1.5$ the $hh$ decay mode plays the dominant r\^ole below the $t\bar t$
threshold with a branching ratio of up to 90\%. Only in the vicinity of
$M_H\sim 130$ GeV does this decay mode drop down, because the trilinear
self-coupling $\lambda_{Hhh}$ changes sign and crosses zero. This is the only
range where the $b\bar b$ decay channels provides the dominant contribution,
but it falls off very quickly above and below this Higgs mass. Moreover,
$WW$ decays are sizeable with a branching ratio of about 10--30\% below the
$t\bar t$ threshold, while the $ZZ$ decays reach values of less than 8\%.
Above the $t\bar t$ threshold, $t\bar t$ decays are overwhelming
and their branching ratio amounts to up to 98\%.

From Fig.~\ref{fg:mssmbr}c it can be inferred that for large $\tb$ the
pseudoscalar Higgs particle $A$ only decays into $b\bar b$ [BR $\sim$ 90\%] and
$\tau^+ \tau^-$ pairs [BR $\sim$ 10\%]. All other decay channels are suppressed
and thus unimportant. Contrary to that at small $\tb$ the $b\bar b$ decay mode
dominates only below the $Zh$ threshold with a branching ratio $\sim$ 80--90\%.
The branching ratio of $\tau^+ \tau^-$ decays ranges at about 8--9\% in this
mass regime. Above the $Zh$ threshold, the $Zh$ decay channel plays the dominant
r\^ole and its branching ratio can reach about 50\% below the $t\bar t$
threshold. It should be noted that already below the $Zh$ threshold off-shell
$Z^* h$ decays are sizeable and thus important. In addition the $gg$ decay
channel grows rapidly from 2\% up to about 20\% at the $t\bar t$ threshold.
Above this threshold $t\bar t$ decays overwhelm with
a branching ratio of nearly 100\%.

Fig.~\ref{fg:mssmbr}d shows that below the $t\bar b$ threshold charged Higgs
$H^+ \to \tau^+ \nu_\tau$ decays provide the dominant contribution. Owing to the
sizeable below-threshold decays into $W^*h$ and $W^*A$, the branching ratio of
the $\tau \nu_\tau$ decays does not
exceed 70\% for small $\tb$, but amounts to about 99\% for large $\tb$. Above
the $t\bar b$ threshold, $H^\pm \to t\bar b$ is dominant. For small $\tb$
its branching ratio reaches about 99\%, whereas for large $\tb$ it does not
exceed about 80\% due to a still sizeable contribution of $\tau^+\nu_\tau$
decays. For small $\tb$ a long off-shell tail below the $t\bar b$ threshold
arises from off-shell $H^+ \to t^* \bar b$ decays. Just below the $t\bar b$
threshold $Wh$ decays can be dominant for small $\tb$ within a very restricted
charged Higgs mass range. For small charged Higgs masses the off-shell decays
into $W^*h$ and $W^*A$ can acquire branching ratios of more than 10\% for small
$\tb$. Below the $Wh$ threshold $cs$ and $cb$ decays reach branching ratios
of a few per cent.

\subsubsection{Decays into SUSY particles}
\paragraph{Chargino/neutralino masses and couplings.}
The chargino/neutralino masses and couplings to the MSSM Higgs bosons are fixed
by the Higgs mass parameter $\mu$ and the SU(2) gaugino mass parameter $M_2$.
The mass matrix of the charginos is given by \cite{mssmbase}
\begin{equation}
{\cal M}_{\chi^\pm} = \left[ \begin{array}{cc}
M_2 & \sqrt{2} M_W \sin\beta \\
\sqrt{2} M_W \cos\beta & \mu
\end{array} \right]
\end{equation}
This can be diagonalized by two mixing matrices $U,V$, yielding the masses of
the physical $\chi^\pm_{1,2}$ states:
\begin{eqnarray}
M_{\chi^\pm_{1,2}} & = & \frac{1}{\sqrt{2}} \left\{ M_2^2 + \mu^2 + 2M_W^2
\right. \nonumber \\
& & \left. \mp \sqrt{(M_2^2-\mu^2)^2 + 4M_W^4\cos^2 2\beta + 4M_W^2 (M_2^2 +
\mu^2 + 2M_2\mu\sin 2\beta)} \right\}^{1/2}
\end{eqnarray}
If either $\mu$ or $M_2$ is large, one chargino corresponds to a pure gaugino
state and the other to a pure higgsino state. The Higgs couplings to charginos
\cite{DSUSY,DSUSY1} can be expressed as [$k = 1,2,3,4$ correspond to
$H,h,A,H^\pm$]
\begin{equation}
H_k \to \chi_i^+ \chi_j^-: \hspace*{1cm} F_{ijk} = \frac{1}{\sqrt{2}} [e_k
V_{i1}U_{j2} - d_k V_{i2}U_{j1}] \, ,
\label{eq:hccoup}
\end{equation}
where the coefficients $e_k$ and $d_k$ are defined to be
\begin{equation}
\begin{array}{lllclllclll}
e_1 & = & \cos\alpha & , & e_2 & = & \sin\alpha & , & e_3 & = &-\sin\beta \\
d_1 & = &-\sin\alpha & , & d_2 & = & \cos\alpha & , & e_3 & = & \cos\beta \, .
\end{array}
\label{eq:ekdk}
\end{equation}

The mass matrix of the four neutralinos depends in addition on the U(1) gaugino
mass parameter $M_1$, which is constrained by SUGRA models to be $M_1 =
\frac{5}{3} \tan\theta_W M_2$. In the bino-wino-higgsino basis, it has the form
\cite{mssmbase}
\begin{equation}
{\cal M}_{\chi^0} = \left[ \begin{array}{cccc}
M_1 & 0   & -M_Z \sin\theta_W \cos\beta &  M_Z \sin\theta_W \sin\beta \\
0   & M_2 &  M_Z \cos\theta_W \cos\beta & -M_Z \cos\theta_W \sin\beta \\
-M_Z \sin\theta_W \cos\beta &  M_Z \cos\theta_W \cos\beta & 0    & -\mu \\
 M_Z \sin\theta_W \sin\beta & -M_Z \cos\theta_W \sin\beta & -\mu & 0    \\
\end{array} \right]
\end{equation}
which can be diagonalized by a single mixing matrix $Z$. The final results are
too involved to be presented here. They can be found in \cite{DSUSY}. If either
$\mu$ or $M_2$ is large, two neutralinos are pure gaugino states and the other
two pure higgsino states.  The Higgs couplings to neutralino pairs
\cite{DSUSY,DSUSY1} can be written as [$k = 1,2,3,4$ correspond to
$H,h,A,H^\pm$]
\begin{equation}
H_k \to \chi_i^0 \chi_j^0: \hspace*{1cm} F_{ijk} = \frac{1}{2} (Z_{j2} -
\tan\theta_W Z_{j1})(e_k Z_{i3} + d_k Z_{i4}) + (i\leftrightarrow j)
\label{eq:hncoup}
\end{equation}
with the coefficients $e_k, d_k$ defined in eq.~(\ref{eq:ekdk}).

The charged Higgs couplings to chargino--neutralino pairs are fixed to be
\cite{DSUSY}
\begin{eqnarray}
H^\pm \to \chi_i^\pm \chi_j^0: \hspace*{1cm} F_{ij4} & = & \cos\beta \left[
V_{i1}Z_{j4} + \frac{1}{\sqrt{2}} V_{i2} (Z_{j2} + \tan\theta_W Z_{j1}) \right]
\nonumber \\
F_{ji4} & = & \sin\beta \left[ U_{i1}Z_{j3} - \frac{1}{\sqrt{2}} U_{i2}
(Z_{j2} + \tan\theta_W Z_{j1}) \right]
\label{eq:chcncoup}
\end{eqnarray}

\paragraph{Sfermion masses and couplings.}
The scalar partners $\tilde f_{L,R}$ of the left- and right-handed fermion
components mix with each other. The mass eigenstates $\tilde f_{1,2}$ of the
sfermions $\tilde f$ are related to the current eigenstates $\tilde f_{L,R}$
by mixing angles $\theta_f$,
\begin{eqnarray}
\tilde f_1 & = & \tilde f_L \cos\theta_f + \tilde f_R \sin \theta_f \nonumber \\
\tilde f_2 & = & -\tilde f_L\sin\theta_f + \tilde f_R \cos \theta_f \, ,
\label{eq:sfmix}
\end{eqnarray}
which are proportional to the masses of the ordinary fermions. Thus mixing
effects are only important for the third-generation sfermions $\tilde t, \tilde
b, \tilde \tau$, the mass matrix of which is given by \cite{mssmbase}
\begin{equation}
{\cal M}_{\tilde f} = \left[ \begin{array}{cc}
M_{\tilde f_L}^2 + M_f^2 & M_f (A_f-\mu r_f) \\
M_f (A_f-\mu r_f) & M_{\tilde f_R}^2 + M_f^2
\end{array} \right] \, ,
\end{equation}
with the parameters $r_b = r_\tau = 1/r_t = \tb$. The parameters $A_f$ denote
the Yukawa mixing parameters of the soft supersymmetry breaking part of the
Lagrangian. Consequently the mixing angles acquire the form
\begin{equation}
\sin 2\theta_f = \frac{2M_f (A_f-\mu r_f)}{M_{\tilde f_1}^2 - M_{\tilde f_2}^2}
~~~,~~~
\cos 2\theta_f = \frac{M_{\tilde f_L}^2 - M_{\tilde f_R}^2}{M_{\tilde f_1}^2
- M_{\tilde f_2}^2}
\end{equation}
and the masses of the squark eigenstates are given by
\begin{equation}
M_{\tilde f_{1,2}}^2 = M_f^2 + \frac{1}{2}\left[ M_{\tilde f_L}^2 +
M_{\tilde f_R}^2 \mp \sqrt{(M_{\tilde f_L}^2 - M_{\tilde f_R}^2)^2 + 4M_f^2
(A_f - \mu r_f)^2} \right] \, .
\end{equation}
The neutral Higgs couplings to sfermions read as \cite{DSUSY}
\begin{eqnarray}
g_{\tilde f_L \tilde f_L}^\Phi & = & M_f^2 g_1^\Phi + M_Z^2 (I_{3f}
- e_f\sin^2\theta_W) g_2^\Phi \nonumber \\
g_{\tilde f_R \tilde f_R}^\Phi & = & M_f^2 g_1^\Phi + M_Z^2 e_f\sin^2\theta_W
g_2^\Phi \nonumber \\
g_{\tilde f_L \tilde f_R}^\Phi & = & -\frac{M_f}{2} (\mu g_3^\Phi
- A_f g_4^\Phi) \, ,
\label{eq:hsfcouprl}
\end{eqnarray}
with the couplings $g_i^\Phi$ listed in Table \ref{tb:hsfcoup}.
\begin{table}[hbt]
\renewcommand{\arraystretch}{1.5}
\begin{center}
\begin{tabular}{|l|c||c|c|c|c|} \hline
$\tilde f$ & $\Phi$ & $g^\Phi_1$ & $g^\Phi_2$ & $g^\Phi_3$ & $g^\Phi_4$ \\
\hline \hline
& $h$ & $\cos\alpha/\sin\beta$ & $-\sin(\alpha+\beta)$ &
$-\sin\alpha/\sin\beta$ & $\cos\alpha/\sin\beta$ \\
$\tilde u$ & $H$ & $\sin\alpha/\sin\beta$ & $\cos(\alpha+\beta)$ &
$\cos\alpha/\sin\beta$ & $\sin\alpha/\sin\beta$ \\
& $A$ & 0 & 0 & 1 & $-1/\tb$ \\ \hline
& $h$ & $-\sin\alpha/\cos\beta$ & $-\sin(\alpha+\beta)$ &
$\cos\alpha/\cos\beta$ & $-\sin\alpha/\cos\beta$ \\
$\tilde d$ & $H$ & $\cos\alpha/\cos\beta$ & $\cos(\alpha+\beta)$ &
$\sin\alpha/\cos\beta$ & $\cos\alpha/\cos\beta$ \\
& $A$ & 0 & 0 & 1 & $-\tb$ \\ \hline
\end{tabular} 
\renewcommand{\arraystretch}{1.2}
\caption[]{\label{tb:hsfcoup}
\it Coefficients of the neutral MSSM Higgs couplings to sfermion pairs.}
\end{center}
\end{table}
The charged Higgs couplings to sfermion pairs \cite{DSUSY} can be expressed
as [$\alpha,\beta = L,R$]
\begin{equation}
g_{\tilde u_\alpha \tilde d_\beta}^{H^\pm} = -\frac{1}{\sqrt{2}} [g_1^{\alpha
\beta} + M_W^2 g_2^{\alpha\beta} ] \, ,
\label{eq:chsfcouprl}
\end{equation}
with the coefficients $g_{1,2}^{\alpha\beta}$ summarized in Table \ref{tb:chsfcoup}.
\begin{table}[hbt]
\renewcommand{\arraystretch}{1.5}
\begin{center}
\begin{tabular}{|l||c|c|c|c|} \hline
$i$ & $g^{LL}_i$ & $g^{RR}_i$ & $g^{LR}_i$ & $g^{RL}_i$\\ \hline \hline
1 & $M_u^2/\tb+M_d^2 \tb$ & $M_u M_d (\tb + 1/\tb)$ & $M_d (\mu+A_d\tb)$ &
$M_u (\mu+A_u/\tb)$ \\
2 & $-\sin 2\beta$ & 0 & 0 & 0 \\ \hline
\end{tabular} 
\renewcommand{\arraystretch}{1.2}
\caption[]{\label{tb:chsfcoup}
\it Coefficients of the charged MSSM Higgs couplings to sfermion pairs.}
\end{center}
\end{table}

\begin{figure}[hbt]

\vspace*{-2.5cm}
\hspace*{-4.5cm}
\begin{turn}{-90}%
\epsfxsize=16cm \epsfbox{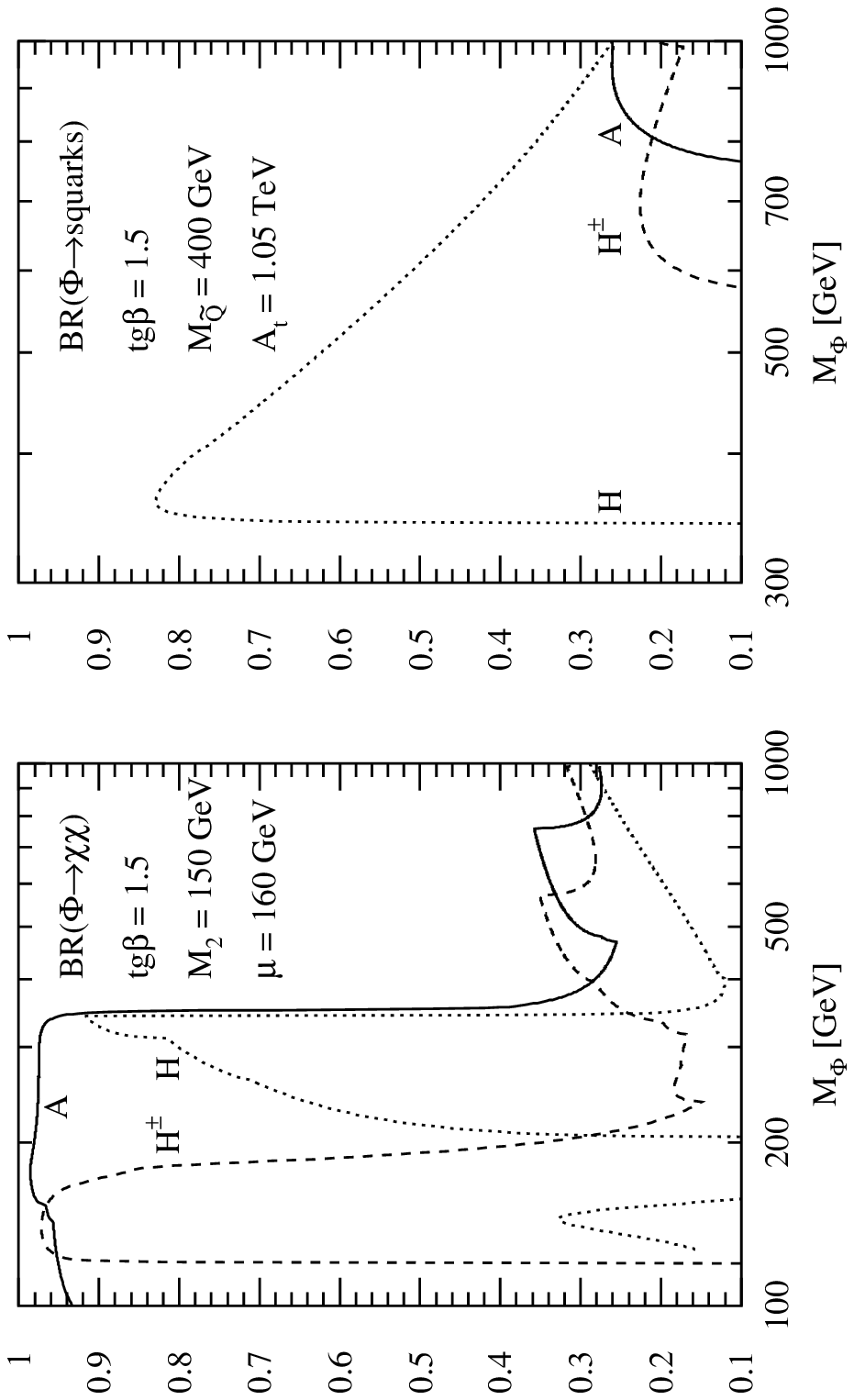}
\end{turn}
\vspace*{-4.2cm}
 
\caption[]{\label{fg:hcharneutsq} \it Branching ratios of the MSSM Higgs boson
$H,A,H^\pm$ decays into charginos/neutralinos and squarks as a function of their
masses for $\tb=1.5$. The mixing parameters have been chosen as $\mu=160$ GeV,
$A_t=1.05$ TeV, $A_b=0$ and the squark masses of the first two generations as
$M_{\widetilde{Q}}=400$ GeV. The gaugino mass parameter has been set to
$M_2=150$ GeV.}
\end{figure}

\paragraph{Decays into charginos and neutralinos.}
The decay widths of the MSSM Higgs particles $H_k$ [$k = 1,2,3,4$ correspond
to $H,h,A,H^\pm$] into neutralino and chargino pairs can be cast into the form
\cite{DSUSY,DSUSY1}
\begin{eqnarray}
\Gamma (H_k \to \chi_i \chi_j) & = & \frac{G_F M_W^2}{2\sqrt{2}\pi}\frac{M_{H_k}
\sqrt{\lambda_{ij,k}}}{1+\delta_{ij}} \left[ (F_{ijk}^2 + F_{jik}^2) \left( 1
- \frac{M_{\chi_i}^2}{M_{H_k}^2} - \frac{M_{\chi_j}^2}{M_{H_k}^2} \right)
\right. \nonumber \\
& & \left. -4\eta_k \epsilon_i \epsilon_j F_{ijk} F_{jik} \frac{M_{\chi_i}
M_{\chi_j}}{M_{H_k}^2} \right] \, ,
\end{eqnarray}
where $\eta_{1,2,4}=+1,~\eta_3 = -1$ and $\delta_{ij}=0$ unless the final state
consists of two identical (Majorana) neutralinos, in which case $\delta_{ii}=1$;
$\epsilon_i = \pm 1$ stands for the sign of the $i$'th eigenvalue of the
neutralino mass matrix, which can be positive or negative. For charginos these
parameters are always equal to unity. The symbols $\lambda_{ij,k}$ denote the
usual two-body phase-space functions
\begin{equation}
\lambda_{ij,k} = \left( 1 - \frac{M_i^2}{M_k^2} - \frac{M_j^2}{M_k^2} \right)^2
- 4 \frac{M_i^2 M_j^2}{M_k^4} \, .
\label{eq:kaellen}
\end{equation}
If chargino/neutralino decays are kinematically allowed, which may be the case
for the heavy MSSM Higgs particles $H,A,H^\pm$, their branching ratios can
reach values up to about 100\% below the corresponding top quark thresholds.
They can thus be dominant, jeopardizing the Higgs search at the LHC due to the
invisibility of a significant fraction of these decay modes \cite{DSUSY}. A
typical example
of the total sum of chargino/neutralino decay branching ratios is shown in
Fig.~\ref{fg:hcharneutsq} for the heavy Higgs bosons. Even above the
corresponding top quark thresholds the chargino/neutralino branching ratios
will be sizeable. For large Higgs masses they reach common values of about
20--80\%: in the asymptotic regime $M_{H_k} \gg M_\chi$, the total sum of decay
widths into charginos and neutralinos acquires the simple form
\cite{DSUSY,DSUSY1}
\begin{equation}
\Gamma\left(H_k \to \sum_{i,j}\chi_i\chi_j\right) =
\frac{3G_F M_W^2}{4\sqrt{2}\pi}M_{H_k}
\left(1+\frac{1}{3} \tan^2\theta_W \right)
\end{equation}
for all three Higgs bosons $H,A,H^\pm$, which is independent of any MSSM
parameter [$\tb,\mu,A_{t,b},M_2$]. Normalized to the total width, which is
dominated by $t\bar t, b\bar b$ $(t\bar b)$ decay modes for the neutral
(charged) Higgs particles the branching ratio of chargino/neutralino decays
will exceed a level of about 20\% even for small and large $\tb$. In some part
of the MSSM parameter space, invisible light scalar Higgs boson decays into the
lightest neutralino $h\to \chi_1^0 \chi_1^0$ will be possible and their
branching ratio can exceed 50\% \cite{DSUSY,DSUSY1}.

\paragraph{Decays into sleptons and squarks.}
The sfermionic decay widths of the MSSM Higgs bosons $H_k$ [$k = 1,2,3,4$
correspond to $H,h,A,H^\pm$ and $i,j = 1,2$] can be written as \cite{DSUSY}
\begin{equation}
\Gamma(H_k\to\tilde f_i\overline{\tilde f}_j)=\frac{3G_F}{2\sqrt{2}\pi M_{H_k}}
\sqrt{\lambda_{\tilde f_i \tilde f_j, H_k}}(g^{H_k}_{\tilde f_i \tilde f_j})^2
\, .
\end{equation}
The physical MSSM couplings $g^{H_k}_{\tilde f_i \tilde f_j}$ can be obtained
from the couplings presented in eqs.~(\ref{eq:hsfcouprl}) and
(\ref{eq:chsfcouprl}) by means of the mixing relations in eq.~(\ref{eq:sfmix}).
The symbol $\lambda_{ij,k}$ denotes the usual two-body phase-space factor of
eq.~(\ref{eq:kaellen}).

In the limit of massless fermions, which is a valid approximation for the first
two generations, the pseudoscalar Higgs bosons $A$ do not decay into sfermions
due to the suppression of sfermion mixing by the fermion mass. In the decoupling
regime, where the Higgs masses $M_{H,H^\pm}$ are large, the decay widths of the
heavy scalar and charged Higgs particles into sfermions are proportional to
\cite{DSUSY}
\begin{equation}
\Gamma(H,H^\pm\to\tilde f\overline{\tilde f})\propto
\frac{G_F M_W^4}{M_{H,H^\pm}} \sin^2 2\beta \, .
\end{equation}
Thus they are only important for small $\tb \sim 1$. However, they are
suppressed by an inverse power of the large Higgs masses, rendering unimportant
the sfermion decays of the first two generations.

Decay widths into third-generation sfermions [$\tilde t,\tilde b,\tilde \tau$]
can be much larger, thanks to the significantly larger fermion masses. For
instance,
in the asymptotic regime the heavy scalar Higgs decay into stop pairs of the
same helicity is proportional to \cite{DSUSY}
\begin{equation}
\Gamma(H\to\tilde t\overline{\tilde t})\propto
\frac{G_F M_t^4}{M_H\mbox{tg$^2\beta$}} \, ,
\end{equation}
which will be enhanced by large coefficients compared to the
first/second-generation squarks for small $\tb$. At large $\tb$ sbottom decays
will be
significant. Moreover, for large Higgs masses the decay widths of heavy neutral
CP-even and CP-odd Higgs particles into stop pairs of different helicity will
be proportional to \cite{DSUSY}
\begin{equation}
\Gamma (H,A\to\tilde t\overline{\tilde t})\propto\frac{G_F M_t^2}{M_{H,A}}
\left[\mu + \frac{A_t}{\tb} \right]^2
\end{equation}
and hence will be of the same order of magnitude as standard fermion and
char\-gi\-no/neu\-tra\-li\-no decay widths. In summary, if third-generation
sfermion decays are kinematically allowed, they have to be taken into account.
An extreme example for the total branching ratios of decays into squarks is
depicted in Fig.~\ref{fg:hcharneutsq}, where they can reach values of $\sim 80\%$
for the heavy scalar Higgs boson $H$.

Very recently the SUSY-QCD corrections to the stop and sbottom decays of the
MSSM Higgs bosons have been calculated \cite{hsqqcd}. They reach about
30\%, especially in the threshold regions. They are not included in the
present analysis.

\subsection{Neutral Higgs Boson Production at the LHC}

\subsubsection{Gluon fusion: $gg\to \Phi~~[\Phi=h,H,A]$}
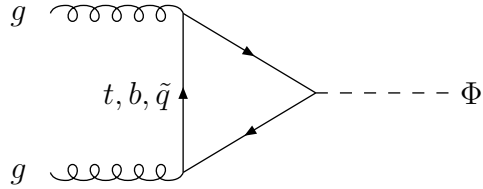
\begin{figure}[hbt]
\begin{center}
\setlength{\unitlength}{1pt}
\begin{picture}(180,100)(0,0)

\Gluon(0,20)(50,20){-3}{5}
\Gluon(0,80)(50,80){3}{5}
\ArrowLine(50,20)(50,80)
\ArrowLine(50,80)(100,50)
\ArrowLine(100,50)(50,20)
\DashLine(100,50)(150,50){5}
\put(155,46){$\Phi$}
\put(20,46){$t,b,\tilde q$}
\put(-15,18){$g$}
\put(-15,78){$g$}

\end{picture}  \\
\setlength{\unitlength}{1pt}
\caption[ ]{\label{fg:mssmgghlodia} \it Typical diagram contributing to
$gg\to \Phi$ at lowest order.}
\end{center}
\end{figure}

The gluon-fusion mechanism \cite{glufus}
\begin{displaymath}
pp \to gg \to \Phi
\end{displaymath}
dominates the neutral Higgs boson production at the LHC in the
phenomenologically relevant Higgs
mass ranges for small and moderate values of $\tb$. Only for large $\tb$ can the
associated $\Phi b\bar b$ production channel develop a larger cross section
due to the enhanced Higgs couplings to bottom quarks \cite{htt,att}. Analogous
to the gluonic
decay modes, the gluon coupling to the neutral Higgs bosons in the MSSM is
built up by loops involving top and bottom quarks as well as squarks, see
Fig.~\ref{fg:mssmgghlodia}.

\begin{figure}[hbt]

\vspace*{-1.4cm}
\hspace*{3.1cm}
\epsfxsize=10cm \epsfbox{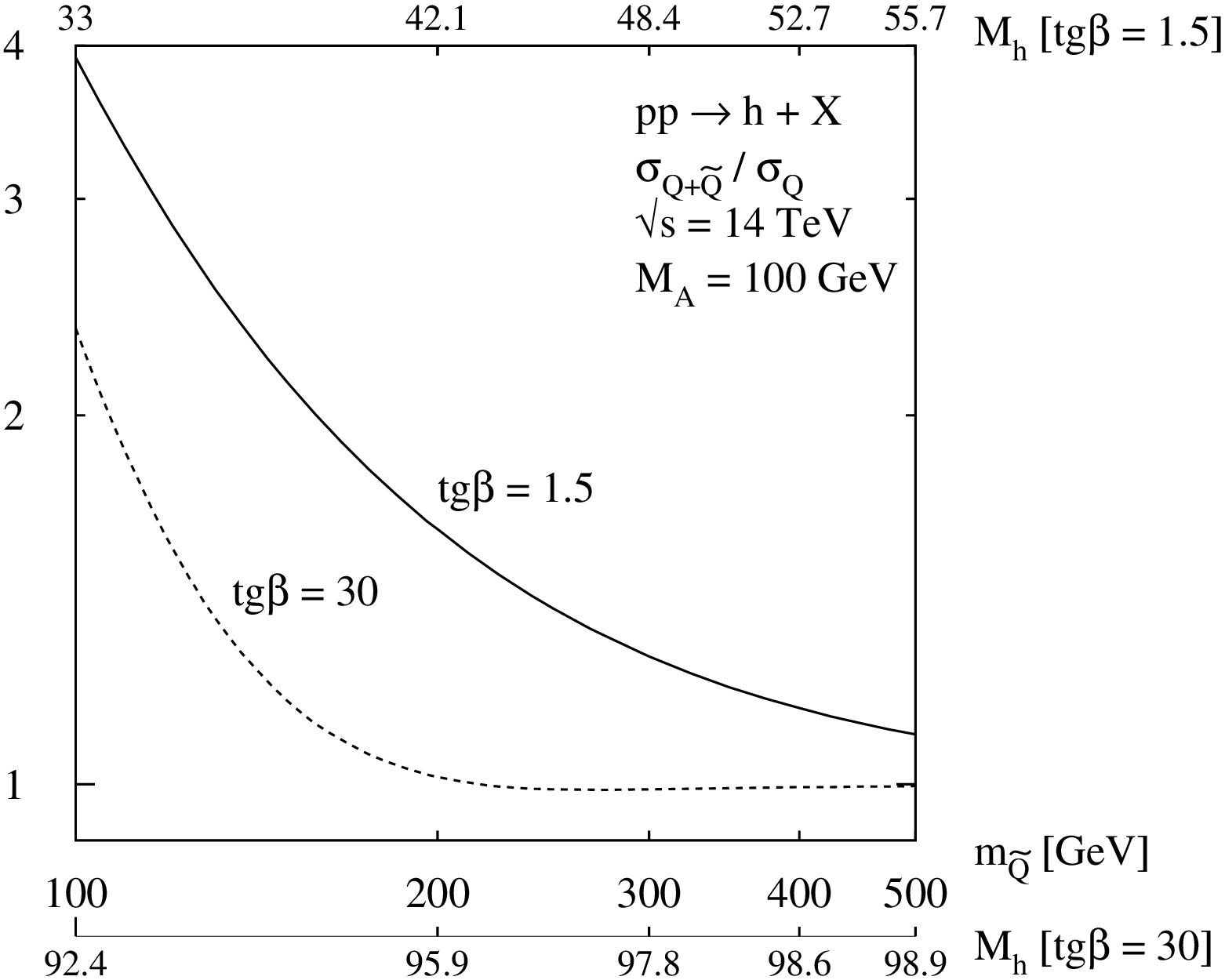}
\vspace*{-3.0cm}
 
\caption[]{\label{fg:squarkeffect} \it Ratio of the QCD-corrected cross section
$\sigma(pp\to h + X)$ with and without squark loops as a function of the
common squark mass $M_{\widetilde{Q}}$ for two values of $\tb=1.5, 30$, and
for $M_A=100$ GeV. The secondary axes present the corresponding light scalar
Higgs mass $M_h$. The top and bottom masses have been chosen as $M_t=175$ GeV,
$M_b=5$ GeV, and the cross sections are convoluted with CTEQ4M parton densities
using $\alpha_s (M_Z)=0.116$ as the normalization of the NLO strong coupling
constant.}
\end{figure}

The partonic cross sections can be obtained from the
gluonic widths of the Higgs bosons at lowest order \cite{higgsqcd,SQCD}:
\begin{eqnarray}
\hat\sigma^\Phi_{LO} (gg\to \Phi) & = & \sigma^\Phi_0 \delta
(1 - z) \label{eq:mssmgghlo} \\
\sigma^\Phi_0 & = & \frac{\pi^2}{8M_\Phi^3}\Gamma_{LO}(\Phi\to gg) \nonumber \\
\sigma^{h/H}_0 & = & \frac{G_{F}\alpha_{s}^{2}(\mu)}{288 \sqrt{2}\pi} \
\left| \sum_{Q} g_Q^{h/H} A_Q^{h/H} (\tau_{Q})
+ \sum_{\widetilde{Q}} g_{\widetilde{Q}}^{h/H} A_{\widetilde{Q}}^{h/H}
(\tau_{\widetilde{Q}}) \right|^{2} \nonumber \\
\sigma^A_0 & = & \frac{G_{F}\alpha_{s}^{2}(\mu)}{128 \sqrt{2}\pi} \
\left| \sum_{Q} g_Q^A A_Q^A (\tau_{Q}) \right|^{2} \nonumber
\end{eqnarray}
where the scaling variables are defined as $z=M_\Phi^2/\hat s$, $\tau_i
=4M_i^2/M_\Phi^2~~(i=Q,\widetilde{Q})$, and $\hat{s}$ denotes the
partonic c.m.~energy squared. The amplitudes $A_{Q,\widetilde{Q}}^\Phi
(\tau_{Q,\widetilde{Q}})$ are defined in eqs.~(\ref{eq:mssmhgg}),
(\ref{eq:mssmagg}), and the MSSM couplings $g_Q^\Phi,g^\Phi_{\widetilde{Q}}$ can
be found in Tables \ref{tb:hcoup} and \ref{tb:hsqcoup}.
In the narrow-width approximation the hadronic cross sections are given by
\begin{equation}
\sigma_{LO}(pp\to \Phi) = \sigma^\Phi_0 \tau_\Phi \frac{d{\cal L}^{gg}}
{d\tau_\Phi}
\end{equation}
with the gluon luminosity defined in eq.~(\ref{eq:gglum})
and the scaling variables $\tau_\Phi = M^2_\Phi/s$ where $s$ specifies the
total hadronic c.m.~energy squared. For small $\tb$ the top loop contribution
is dominant, while for large values of $\tb$ the bottom quark contribution
is strongly enhanced. If the squark masses are less than $\sim 400$ GeV, their
contribution is significant, and for squark masses beyond $\sim 500$ GeV they
can safely be neglected \cite{SQCD}. This is demonstrated in
Fig.~\ref{fg:squarkeffect},
where the ratio of the cross section with and without the squark contribution
is presented as a function of the corresponding scalar Higgs mass. In the
phenomenological mass range the squark loops may enhance the cross section by
up to a factor 2.

\paragraph{QCD corrections.}
In the past the two-loop QCD corrections to the gluon-fusion cross
section were calculated \cite{higgsqcd,gghsusy}. In complete analogy to
the SM case they consist
of virtual corrections to the basic $gg\to \Phi$ process and real corrections
due to the associated production of the Higgs bosons with massless partons,
\begin{displaymath}
gg \rightarrow \Phi g \hspace{0.5cm} \mbox{and} \hspace{0.5cm}
gq \rightarrow \Phi q,~q\overline{q} \rightarrow \Phi g\, .
\end{displaymath}
Thus the contributions to the final result for the hadronic cross section can
be classified as
\begin{equation}
\sigma(pp \rightarrow \Phi +X) = \sigma^\Phi_{0} \left[ 1+ C^\Phi
\frac{\alpha_{s}}{\pi} \right] \tau_\Phi \frac{d{\cal L}^{gg}}{d\tau_\Phi} +
\Delta\sigma^\Phi_{gg} + \Delta\sigma^\Phi_{gq} + \Delta\sigma^\Phi_{q\bar{q}}
\, .
\label{eq:mssmgghqcd5}
\end{equation}
The analytic expressions for arbitrary Higgs boson and quark masses are rather
involved and can be found in \cite{higgsqcd}. As in the SM case the
(s)quark-loop masses have been identified with the pole masses $M_Q\,
(M_{\widetilde{Q}})$, while
the QCD coupling is defined in the $\overline{\rm MS}$ scheme. We have adopted
the $\overline{\rm MS}$ factorization scheme for the NLO parton densities. 
The axial $\gamma_5$ coupling has been regularized in the 't~Hooft--Veltman
scheme \cite{thoovel}, which preserves the chiral symmetry in the massless
quark limit and
fulfills the non-renormalization theorem of the ABJ anomaly at vanishing
momentum transfer \cite{ABJ}.

The coefficients $C^\Phi(\tau_Q,\tau_{\widetilde{Q}})$ split into the infrared
$\pi^2$ term, a logarithmic term including the renormalization scale $\mu$, and
finite (s)quark mass-dependent pieces $c^\Phi(\tau_Q,\tau_{\widetilde{Q}})$:
\begin{equation}
C^\Phi(\tau_Q,\tau_{\widetilde{Q}}) = \pi^{2}+ c^\Phi(\tau_Q,
\tau_{\widetilde{Q}}) +
\frac{33-2N_{F}}{6} \log \frac{\mu^{2}}{M_\Phi^{2}} \, .
\label{eq:mssmCvirt}
\end{equation}
The terms $c^\Phi (\tau_Q)$ originating from quark loops have been reduced
analytically to one-dimensional Feynman-parameter integrals, which were
evaluated numerically \cite{higgsqcd,gghsusy}. The QCD corrections to the squark
contributions are only known in the heavy-squark limit \cite{SQCD},
which however provides a reasonable approximation to
the $K$ factor due to the dominance of soft and collinear gluon radiation for
heavy particle loops in the gluon-fusion process.

The remaining contributions of eq.~(\ref{eq:mssmgghqcd5}) can be cast into the
form \cite{higgsqcd,gghsusy}
\begin{eqnarray}
\Delta \sigma^\Phi_{gg} & = & \int_{\tau_\Phi}^{1} d\tau \frac{d{\cal
L}^{gg}}{d\tau} \times \frac{\alpha_{s}}{\pi} \sigma^\Phi_{0} \left\{ - z
P_{gg} (z) \log \frac{M^{2}}{\hat{s}} + d^\Phi_{gg} (z,\tau_Q,
\tau_{\widetilde{Q}}) \right. \non \\
& & \left. \hspace{3.7cm} + 12 \left[ \left(\frac{\log
(1-z)}{1-z} \right)_+ - z[2-z(1-z)] \log (1-z) \right] \right\} \non \\ \non \\
\Delta \sigma^\Phi_{gq} & = & \int_{\tau_\Phi}^{1} d\tau \sum_{q,
\bar{q}} \frac{d{\cal L} ^{gq}}{d\tau} \times \frac{\alpha_{s}}{\pi}
\sigma^\Phi_{0}\left\{ -\frac{z}{2} P_{gq}(z) \log\frac{M^{2}}{\hat{s}(1-z)^2}
+ d^\Phi_{gq} (z,\tau_Q,\tau_{\widetilde{Q}}) \right\}
\non \\ \non \\
\Delta \sigma^\Phi_{q\bar{q}} & = & \int_{\tau_\Phi}^{1} d\tau
\sum_{q} \frac{d{\cal L}^{q\bar{q}}}{d\tau} \times \frac{\alpha_{s}}{\pi}
\sigma^\Phi_{0}~d^\Phi_{q\bar q} (z,\tau_Q,\tau_{\widetilde{Q}}) \, ,
\label{eq:mssmgghqcd}
\end{eqnarray}
with $z = \tau_\Phi / \tau = M_\Phi^2/\hat s$. $P_{gg}$ and $P_{gq}$ are the
standard Altarelli--Parisi splitting functions defined in
eq.~(\ref{eq:APKernel}).
The coefficients $d^\Phi_{gg}, d^\Phi_{gq}$ and $d^\Phi_{q \ov{q}}$ have been
reduced to one-dimensional integrals for the quark loops, which can be
evaluated numerically \cite{higgsqcd,gghsusy} for arbitrary quark masses. They
can be calculated analytically in the heavy- and light-quark limits.

\begin{figure}[hbt]

\vspace*{1.0cm}
\hspace*{0.15cm}
\begin{turn}{-90}%
\epsfxsize=10cm \epsfbox{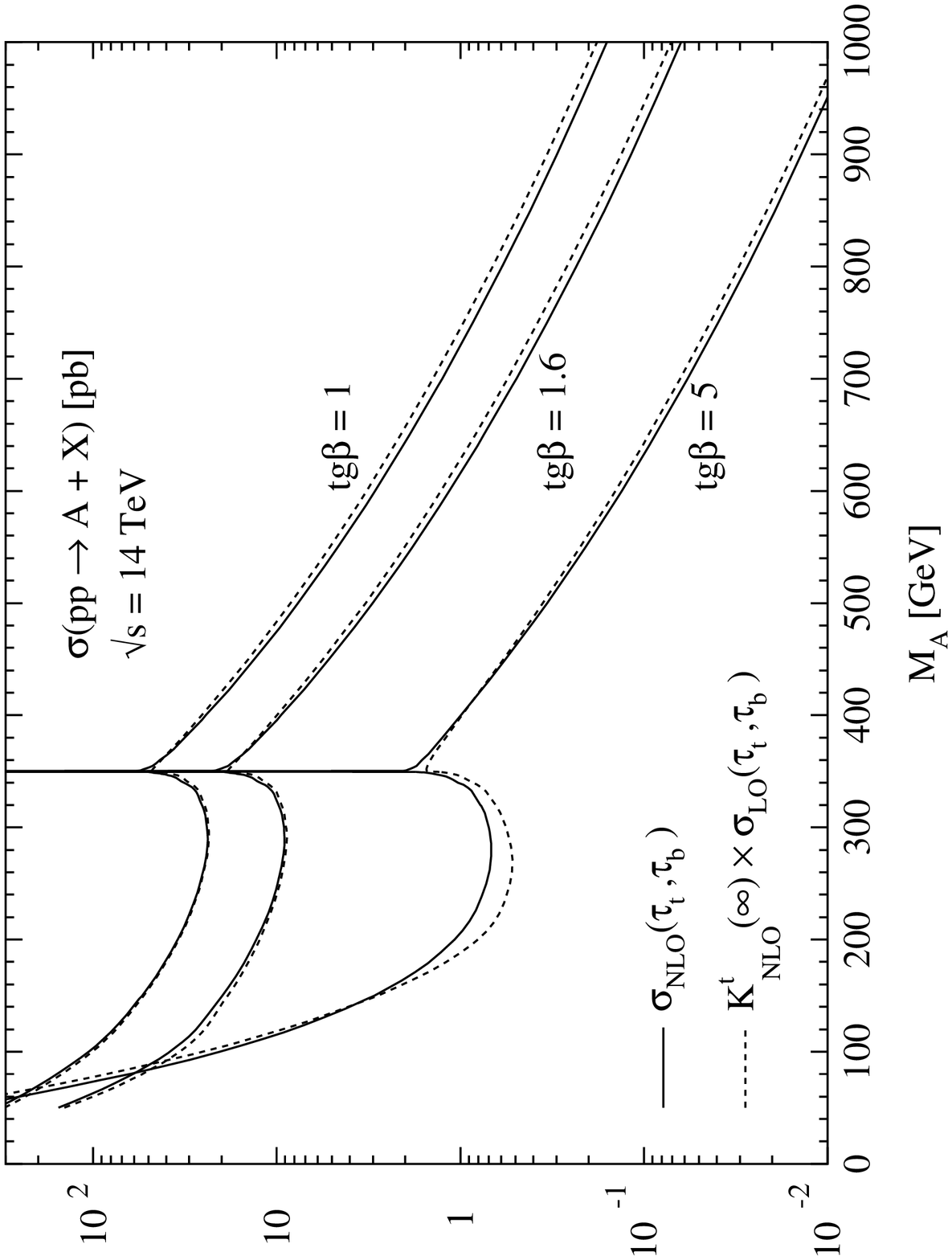}
\end{turn}
\vspace*{-0.0cm}
 
\caption[]{\label{fg:alimit} \it Comparison of the exact and approximate
NLO cross section $\sigma(pp\to A+X)$ at the LHC with c.m.~energy $\sqrt{s}=14$
TeV. The solid lines show the exact cross sections including the full $t,b$
quark mass dependence and the dashed lines correspond to the heavy-quark
approximation of the K factor. The renormalization and factorization
scales have been identified with the Higgs mass, $\mu=M=M_A$ and the CTEQ4M
parton densities with NLO strong coupling [$\alpha_s(M_Z)=0.116$] have been
adopted. The top mass has been chosen as $M_t=175$ GeV, the bottom mass as
$M_b=5$ GeV and the common squark mass as $M_S=1$ TeV.}
\end{figure}

In the heavy-quark limit the quark contributions to the coefficients
$c^\Phi(\tau_Q)$ and $d^\Phi_{ij} (z,\tau_Q)$ reduce to the same expressions as
in the SM case of eq.~(\ref{eq:gghqcdlim}) for the scalar Higgs particles $h,H$.
For the pseudoscalar Higgs boson only the coefficient $c^A(\tau_Q)$
differs from the scalar case,
\begin{equation}
\tau_Q = 4M_Q^2/M_A^2 \gg 1\,: \hspace{3cm}
c^A(\tau_Q) \to 6 \, .
\hspace{5.5cm}
\end{equation}
In fact, the leading terms in the heavy-quark limit provide a reliable
approximation for small $\tb$ up to Higgs masses of $\sim 1$ TeV as can be
inferred from Fig.~\ref{fg:alimit}, which shows the exact pseudoscalar cross 
sections
(solid lines) as a function of the pseudoscalar Higgs mass for three values of
$\tb$ and the approximation obtained by multiplying the full massive
leading-order cross section with the $K$ factor obtained in the heavy-quark
limit. A maximal deviation $\sim 25\%$ for $\tb\lsim 5$ occurs in the
intermediate mass range.
The squark contribution in the heavy-squark limit coincides with the
heavy-quark case apart from the virtual piece \cite{SQCD},
\begin{equation}
\tau_{\widetilde{Q}} = 4M_{\widetilde{Q}}^2/M_{h/H}^2 \gg 1\,: \hspace{3cm}
c^{h/H}(\tau_{\widetilde{Q}}) \to \frac{25}{3} \, .
\hspace{4.5cm}
\end{equation}
The QCD corrections to the squark loops have been evaluated for degenerate
squark masses, so that no mixing effects occur, and for heavy gluinos, such
that their contributions are suppressed. In this case there are no squark loop
effects in pseudoscalar Higgs production.

In the opposite limit, where the Higgs mass is much larger than the quark mass,
the analytic results coincide with the SM expressions for both the scalar and
pseudoscalar Higgs particles \cite{higgsqcd}. This coincidence reflects
the restoration of the chiral symmetry in the massless quark limit.

\begin{figure}[hbtp]

\vspace*{1.4cm}
\hspace*{-5.2cm}
\begin{turn}{-90}%
\epsfxsize=16.5cm \epsfbox{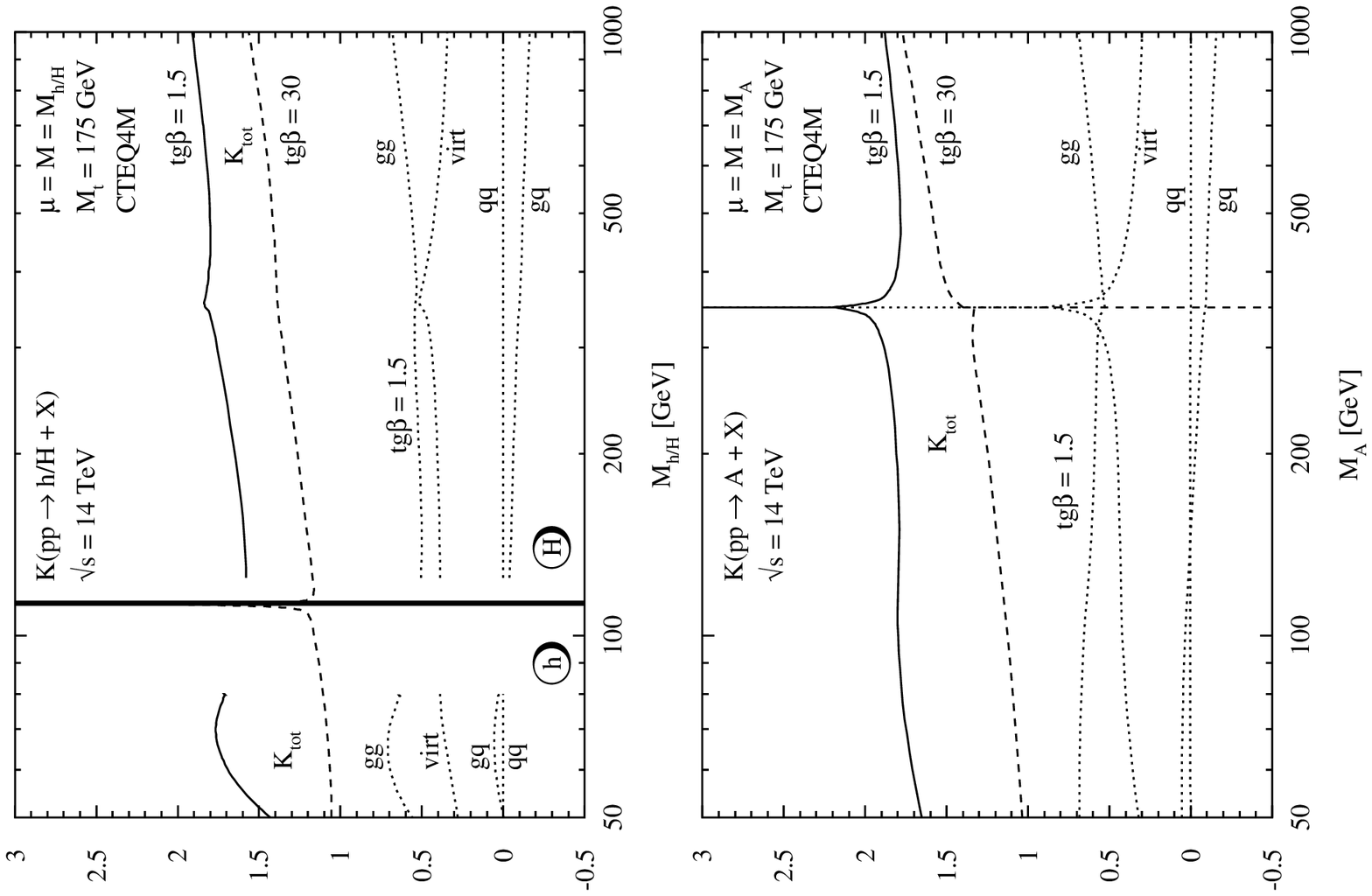}
\end{turn}
\vspace*{0.7cm}

\caption[]{\label{fg:mssmgghk} \it K factors of the QCD-corrected gluon-fusion
cross section $\sigma(pp\to\Phi+X)$ at the LHC with c.m.~energy $\sqrt{s}=14$
TeV. The dashed lines show the individual contributions of the four terms of
the QCD corrections given in eq.~(\ref{eq:mssmgghqcd5}). The renormalization and
factorization scales have been identified with the corresponding Higgs mass,
$\mu=M=M_\Phi$, and the CTEQ4M parton densities have been adopted.}
\end{figure}
The $K$ factors $K_{tot} = \sigma_{NLO}/\sigma_{LO}$ are presented for LHC
energies in Fig.~\ref{fg:mssmgghk} as a function of the corresponding Higgs
boson mass.  Both the renormalization and the factorization scales have been
identified with the Higgs masses $\mu = M = M_\Phi$.  The variation of the
$K$ factors with the Higgs masses is mainly caused by the MSSM couplings apart
from the threshold region, where in the pseudoscalar case a Coulomb
singularity emerges in analogy to the gluonic and photonic decay modes
\cite{higgsqcd,gghsusy}. The
corrections are positive and large, increasing the MSSM Higgs production cross
sections at the LHC by up to about 100\%.

\begin{figure}[hbt]

\vspace*{-2.0cm}
\hspace*{2.15cm}
\epsfxsize=11cm \epsfbox{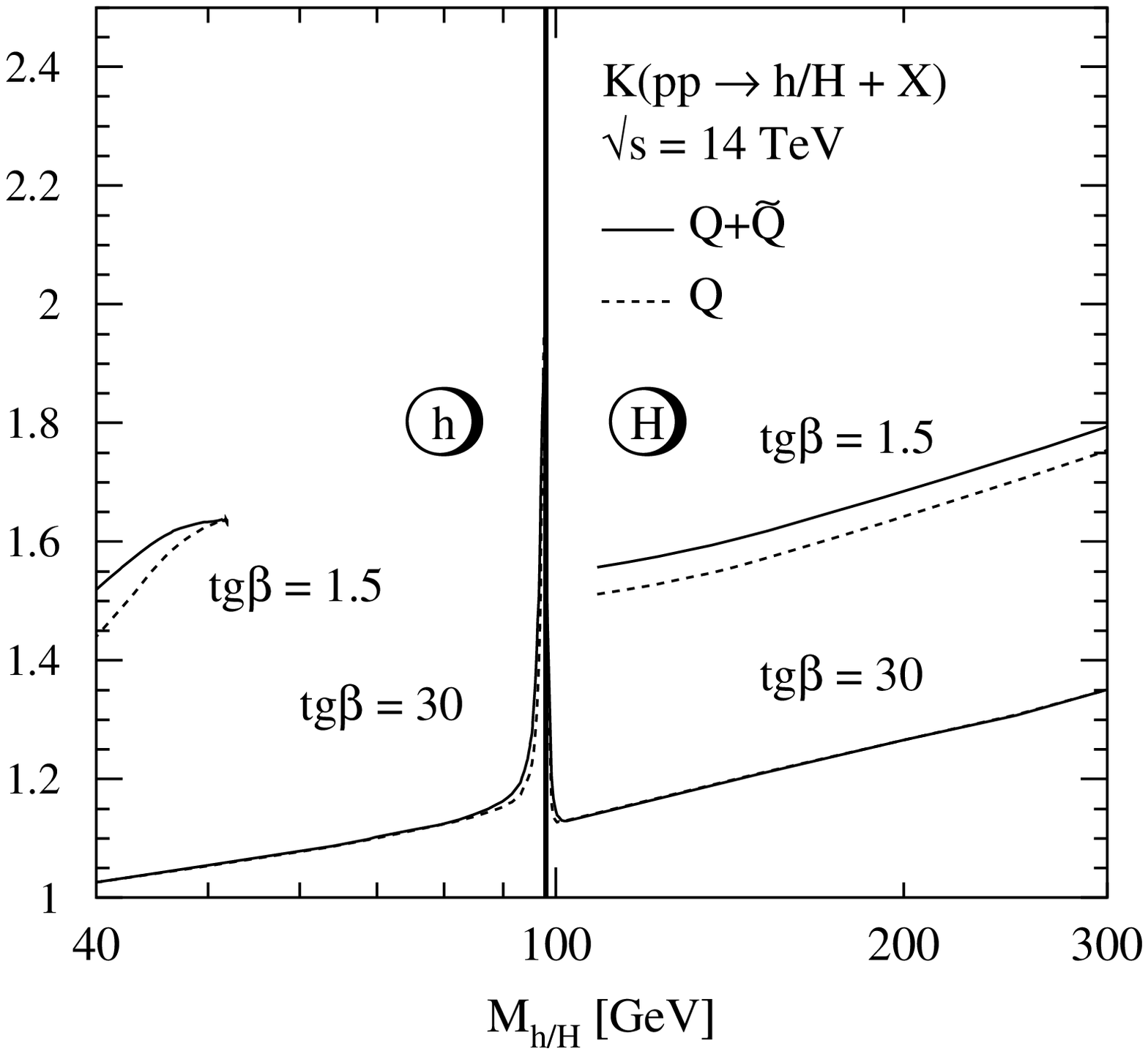}
\vspace*{-3.1cm}

\caption[]{\label{fg:kqksq} \it K factors of the cross sections $\sigma(pp\to
h/H + X)$ with [solid lines] and without [dashed lines] squark loops as a
function of the corresponding scalar Higgs mass for two values of $\tb=1.5,30$.
The common squark mass has been chosen as $M_{\widetilde{Q}}=200$ GeV. The top
and bottom masses have been set to $M_t=175$ GeV, $M_b=5$ GeV, and the NLO
cross sections are convoluted with CTEQ4M parton densities using $\alpha_s
(M_Z)=0.116$ as the normalization of the NLO strong coupling constant. The LO
cross sections are evaluated with CTEQ4L parton densities with the LO strong
coupling $\alpha_s(M_Z)=0.132$.}
\end{figure}
The effect of the squark loops on the scalar
Higgs $K$ factors is presented in Fig.~\ref{fg:kqksq}, which shows the
$K$ factors of scalar Higgs boson production with and without squark loops as
a function of the corresponding Higgs mass. It is clearly visible that the
squark loops hardly change the $K$ factors, making the $K$ factors from the
pure quark contributions an excellent approximation within maximal
deviations of about 10\%.

Theoretical uncertainties in the prediction of the Higgs cross section
originate from two sources, the dependence of the cross section on different
parametrizations of the parton densities and the unknown
NNLO corrections. For representative sets of recent
parton distributions \cite{cteq4,stfu}, we find a variation of about $\pm
10\%$ of the cross section for Higgs masses larger than $\sim 100$ GeV
analogous to the SM case.  The uncertainty due to the gluon density will be
smaller in the near future when the deep-inelastic electron/positron--nucleon
scattering experiments at HERA will have reached the anticipated level of
accuracy.

\begin{figure}[hbtp]

\vspace*{1.7cm}
\hspace*{-6.0cm}
\begin{turn}{-90}%
\epsfxsize=18cm \epsfbox{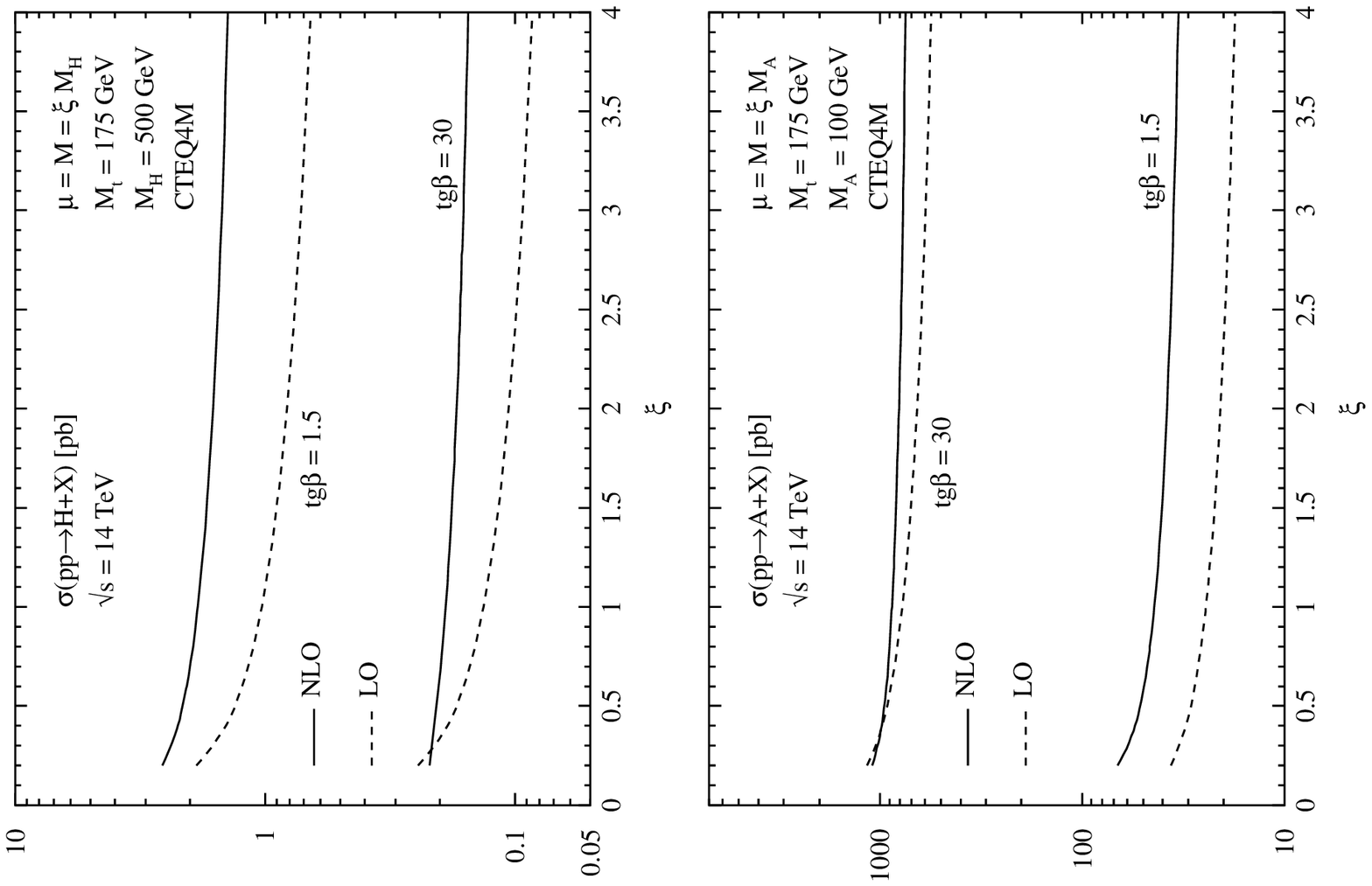}
\end{turn}
\vspace*{0.5cm}

\caption[]{\label{fg:mssmgghscale} \it The renormalization and factorization
scale dependence of the Higgs production cross section at lowest and
next-to-leading order for two different Higgs bosons $H,A$ with masses $M_H =
500$ GeV and $M_A=100$ GeV and two values of $\tb=1.5,30$.}
\end{figure}
The [unphysical] variation of the cross sections with the
renormalization and factorization scales is reduced by
including the NLO corrections. This is
shown in Fig.~\ref{fg:mssmgghscale} for the heavy scalar and pseudoscalar Higgs
particles with masses $M_H = 500$ GeV and $M_A=100$ GeV. The
renormalization/factorization
scale $\mu = M$ is varied in units of the Higgs mass $\mu = \xi M_\Phi$.
The remaining uncertainties due to the scale
dependence appear to be less than about 15\%.

\paragraph{Soft gluon resummation.}
Recently soft and collinear gluon radiation effects for the total gluon-fusion
cross section have been resummed \cite{gghresum}. In complete analogy to the
SM case, the
perturbative expansion of the resummed result leads to a quantitative
approximation of the three-loop NNLO corrections of the partonic cross section
in the heavy top mass limit, which approximates the full massive NLO result
with a reliable precision  for small and medium values of $\tb$ [see
Fig.~\ref{fg:alimit}]. Owing to the low-energy theorems discussed before [see
the gluonic decay modes $\Phi\to gg$], the unrenormalized partonic cross section
factorizes in the same way as the SM cross section. The scalar factors
$\kappa^{h/H}$ coincide with the SM values of eq.~(\ref{eq:kappa})
and the pseudoscalar factor is equal to unity, because of the
non-renormalization of the ABJ anomaly \cite{ABJ},
\beq
\kappa^A = 1 \, .
\label{eq:akappa}
\eeq
The resummation of soft and collinear gluon effects proceeds along the same
lines as in the SM case. The final results for the scalar correction factors
$\rho^{h/H}$ are identical to the SM result eq.~(\ref{eq:rhoresum}), and the
pseudoscalar correction factor can be cast into the form \cite{gghresum}
\beq
\rho^A\left(N,\frac{M_A^2}{\mu^2},\alpha_s(\mu)\right) =
\rho^{h/H}\left(N,\frac{M_A^2}{\mu^2},\alpha_s(\mu)\right) \times
\exp\left\{6 \frac{\alpha_s(M_A^2)}{\pi} \right\} \, .
\label{eq:arhoresum}
\eeq
The perturbative expansions at NLO and NNLO \cite{gghresum} read as [for
$\mu = M$]
\bea
\rho_A^{(1)} \left(z,\frac{M_A^2}{\mu^2} \right) & = &
\rho_{h/H}^{(1)} \left(z,\frac{M_A^2}{\mu^2} \right) + 6 \delta(1-z)
\label{eq:arhonlo} \\ \nonumber \\
\rho_A^{(2)} \left(z,\frac{M_A^2}{\mu^2} \right) & = &
\rho_{h/H}^{(2)} \left(z,\frac{M_A^2}{\mu^2} \right)
+ 3\left\{ 24 {\cal D}_1(z) - 12 L_\mu {\cal D}_0(z) - 48 {\cal E}_1(z)
\right. \nonumber \\
& & \left. + ( 12 \zeta_2 + 6 + \beta_0 L_\mu) \delta(1-z) \right\}
\label{arhonnlo}
\eea
where we use the same notation as in the SM case.

\begin{figure}[hbt]

\vspace*{0.5cm}
\hspace*{-1.0cm}
\begin{turn}{-90}%
\epsfxsize=11cm \epsfbox{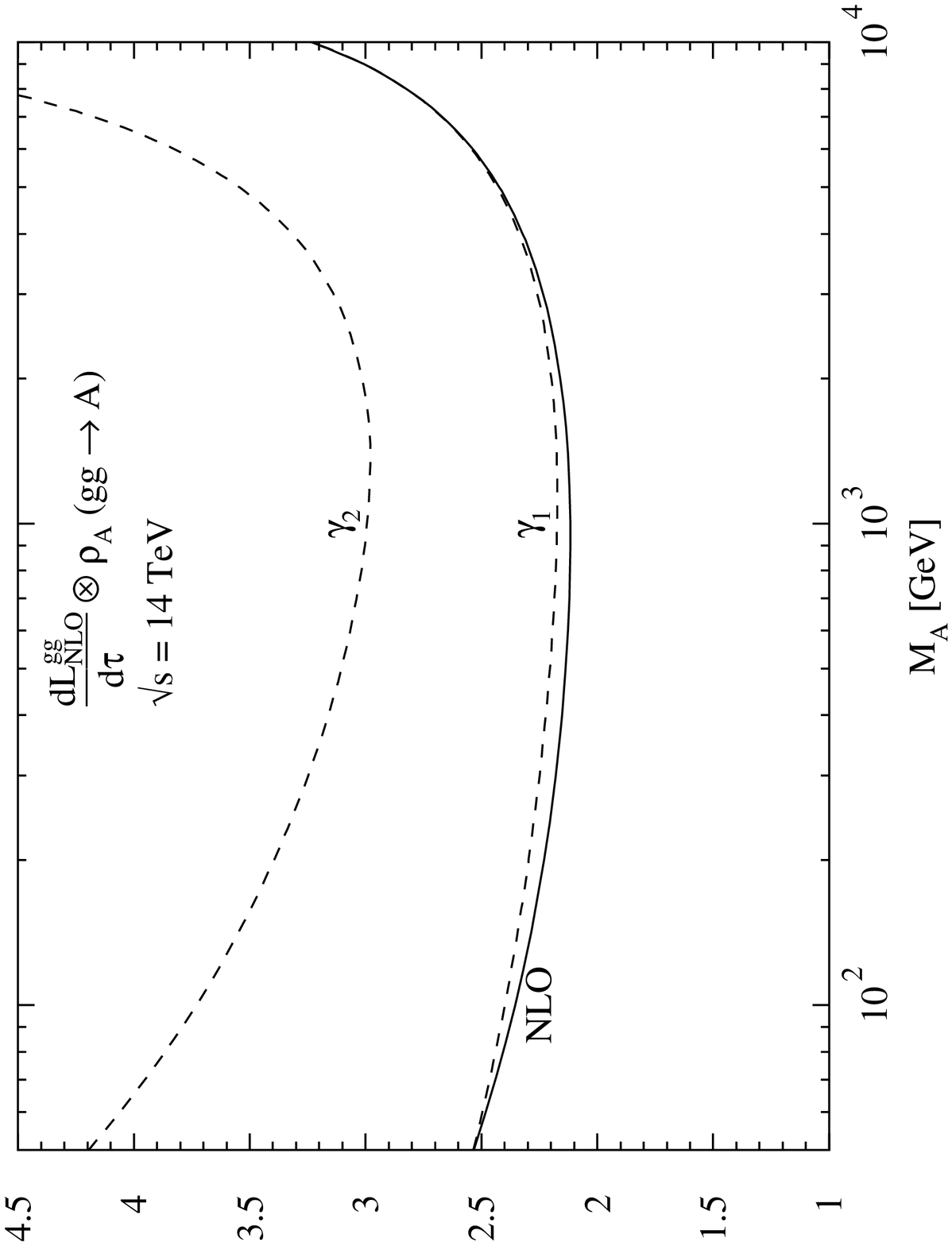}
\end{turn}
\vspace*{-0.1cm}

\caption[]{\label{fg:ggarho} \it Exact and approximate two- and three-loop
correction factor convoluted with NLO gluon densities in the heavy top quark
limit for the pseudoscalar MSSM Higgs boson. The CTEQ4M parton densities have
been adopted with $\alpha_s(M_Z)=0.116$ at NLO.}
\end{figure}
The convolution of the scalar correction factors with NLO gluon densities and
strong coupling coincides with the SM case in Fig.~\ref{fg:gghrho}, while the
pseudoscalar case is presented in Fig.~\ref{fg:ggarho} as a function of the
pseudoscalar Higgs mass
at the LHC. The solid line corresponds to the exact NLO result and the lower
dashed line to the NLO expansion of the resummed correction factor. It can be
inferred from this figure that the soft gluon approximation reproduces the
exact result within $\sim 5\%$ at NLO. The upper dashed line shows the NNLO
expansion of the resummed correction factor. Fig.~\ref{fg:ggarho} demonstrates
that the correction factor amounts to about 2.2--2.5 at NLO and 3.2--4.1
at NNLO in the phenomenologically relevant Higgs
mass range $M_A \lsim 1$ TeV. However, in order to evaluate the size of the
QCD corrections, each order of the perturbative expansion has to be integrated
with the strong coupling and parton densities of the {\it same} order, i.e.~LO
cross section with LO quantities, NLO cross section with NLO quantities and
NNLO cross
section with NNLO quantities. This consistent $K$ factor amounts to about
1.5--2.0 at NLO and is thus about 50--60\% smaller than the result in
Fig.~\ref{fg:ggarho}. A reliable prediction of the gluon-fusion cross section
at NNLO requires the convolution with NNLO parton densities, which are not yet
available. It is thus impossible to predict the Higgs production cross sections
with NNLO accuracy until NNLO structure functions are accessible.

The scale dependence at NNLO develops a similar picture as in the SM case. For
large Higgs masses a
broad maximum appears near the natural scale $\mu=M=M_\Phi$ indicating an
important theoretical improvement in the prediction of the Higgs production
cross section \cite{gghresum}.

\subsubsection{Vector boson fusion: $qq\to qqV^*V^* \to qqh/qqH$}
\begin{figure}[hbt]
\begin{center}
\setlength{\unitlength}{1pt}
\begin{picture}(120,120)(0,0)

\ArrowLine(0,0)(50,0)
\ArrowLine(50,0)(100,0)
\ArrowLine(0,100)(50,100)
\ArrowLine(50,100)(100,100)
\Photon(50,0)(50,50){3}{5}
\Photon(50,50)(50,100){3}{5}
\DashLine(50,50)(100,50){5}
\put(105,46){$h,H$}
\put(-15,-2){$q$}
\put(-15,98){$q$}
\put(55,21){$W,Z$}
\put(55,71){$W,Z$}

\end{picture}  \\
\setlength{\unitlength}{1pt}
\caption[ ]{\label{fg:mssmvvhlodia} \it Diagram contributing to $qq \to qqV^*V^*
\to qqh/qqH$ at lowest order.}
\end{center}
\end{figure}
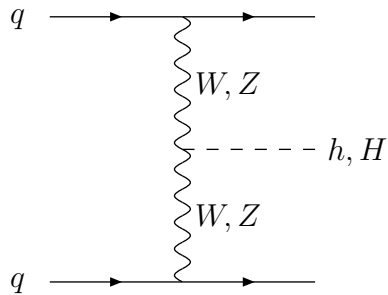
\noindent
Due to the absence of vector boson couplings to pseudoscalar Higgs particles
$A$, only the scalar Higgs bosons $h,H$ can be produced via the
vector-boson-fusion mechanism at tree level [see Fig.~\ref{fg:mssmvvhlodia}].
However, these processes are
suppressed with respect to the SM cross section due to the MSSM couplings
[$g_V^{h/H} = \sin (\alpha - \beta)/\cos (\alpha - \beta)$],
\beq
\sigma(pp\to qq\to qqh/qqH) = \left(g_V^{h/H} \right)^2 \sigma
(pp\to qq\to qqH_{SM}) \, .
\eeq
It turns out that the vector-boson-fusion mechanism is unimportant in the MSSM,
because for large heavy scalar Higgs masses $M_H$, the MSSM couplings $g_V^H$
are very small.
The relative QCD corrections are the same as for the SM Higgs particle
[Fig.~\ref{fg:vvhqcd}] and thus small \cite{vvhqcd}.

\subsubsection{Higgs-strahlung: $q\bar q\to V^* \to Vh/VH$}
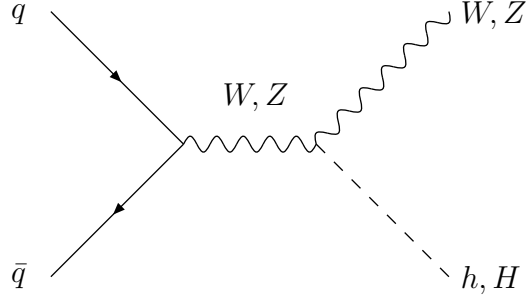
\begin{figure}[hbt]
\begin{center}
\setlength{\unitlength}{1pt}
\begin{picture}(160,120)(0,-10)

\ArrowLine(0,100)(50,50)
\ArrowLine(50,50)(0,0)
\Photon(50,50)(100,50){3}{5}
\Photon(100,50)(150,100){3}{6}
\DashLine(100,50)(150,0){5}
\put(155,-4){$h,H$}
\put(-15,-2){$\bar q$}
\put(-15,98){$q$}
\put(65,65){$W,Z$}
\put(155,96){$W,Z$}

\end{picture}  \\
\setlength{\unitlength}{1pt}
\caption[ ]{\label{fg:mssmvhvlodia} \it Diagram contributing to $q\bar q \to
V^* \to Vh/VH$ at lowest order.}
\end{center}
\end{figure}
\noindent
For the same reasons as in the vector-boson-fusion mechanism case, the
Higgs-strah\-lung off $W,Z$ bosons, $q\bar q\to V^* \to Vh/VH~(V=W,Z)$ [see
Fig.~\ref{fg:mssmvhvlodia}], is unimportant for the scalar MSSM Higgs particles
$h,H$. The cross sections can be easily related to the SM cross sections,
\beq
\sigma(pp \to Vh/VH) = \left(g_V^{h/H} \right)^2
\sigma(pp \to VH_{SM}) \, .
\eeq
Pseudoscalar couplings to intermediate vector bosons are absent so that
pseudoscalar Higgs particles cannot be produced at tree level in this channel.
The relative QCD corrections are the same as in the SM case, see
Fig.~\ref{fg:vhvqcd}, and thus of moderate size \cite{vhvqcd}.

\subsubsection{Higgs bremsstrahlung off top and bottom quarks}
\begin{figure}[hbt]
\begin{center}
\setlength{\unitlength}{1pt}
\begin{picture}(360,120)(0,-10)

\ArrowLine(0,100)(50,50)
\ArrowLine(50,50)(0,0)
\Gluon(50,50)(100,50){3}{5}
\ArrowLine(100,50)(125,75)
\ArrowLine(125,75)(150,100)
\ArrowLine(150,0)(100,50)
\DashLine(125,75)(150,50){5}
\put(155,46){$\Phi$}
\put(-15,98){$q$}
\put(-15,-2){$\bar q$}
\put(65,65){$g$}
\put(155,98){$t/b$}
\put(155,-2){$\bar t/\bar b$}

\Gluon(250,0)(300,0){3}{5}
\Gluon(250,100)(300,100){3}{5}
\ArrowLine(350,0)(300,0)
\ArrowLine(300,0)(300,50)
\ArrowLine(300,50)(300,100)
\ArrowLine(300,100)(350,100)
\DashLine(300,50)(350,50){5}
\put(355,46){$\Phi$}
\put(235,98){$g$}
\put(235,-2){$g$}
\put(355,98){$t/b$}
\put(355,-2){$\bar t/\bar b$}

\end{picture}  \\
\setlength{\unitlength}{1pt}
\caption[ ]{\label{fg:mssmhqqlodia} \it Typical diagrams contributing to
$q\bar q/gg \to \Phi Q\bar Q~~(Q=t,b)$ at lowest order.}
\end{center}
\end{figure}
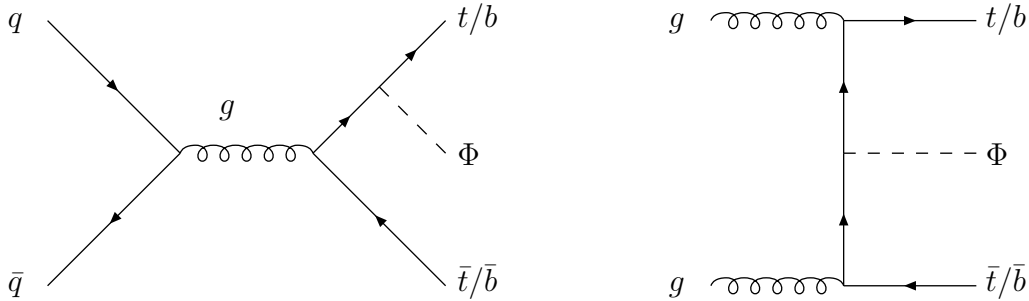
\noindent
The scalar Higgs cross sections for Higgs bremsstrahlung off heavy quarks $Q$
can simply be related to the SM case:
\beq
\sigma (pp\to hQ\bar Q/HQ\bar Q) = \left( g_Q^{h/H} \right)^2
\sigma (pp\to H_{SM}Q\bar Q)
\eeq
The expressions for the pseudoscalar Higgs boson \cite{att} are similarly
involved as the scalar case and will not be presented here.

The top quark coupling to MSSM Higgs bosons is suppressed with respect to the SM
for $\tb >1$. Therefore Higgs bremsstrahlung off top quarks $pp\to\Phi t\bar t$
is less important for MSSM Higgs particles. On the other hand Higgs
bremsstrahlung
off bottom quarks $pp \to Hb\bar b$ will be the dominant Higgs production
channel for large $\tb$ due to the strongly enhanced bottom quark Yukawa
couplings \cite{htt}. The QCD corrections to $HQ\bar Q$ production are
still unknown.

\subsubsection{Cross sections for Higgs boson production at the LHC}
\begin{figure}[hbtp]

\vspace*{0.3cm}
\hspace*{1.0cm}
\begin{turn}{-90}%
\epsfxsize=8.5cm \epsfbox{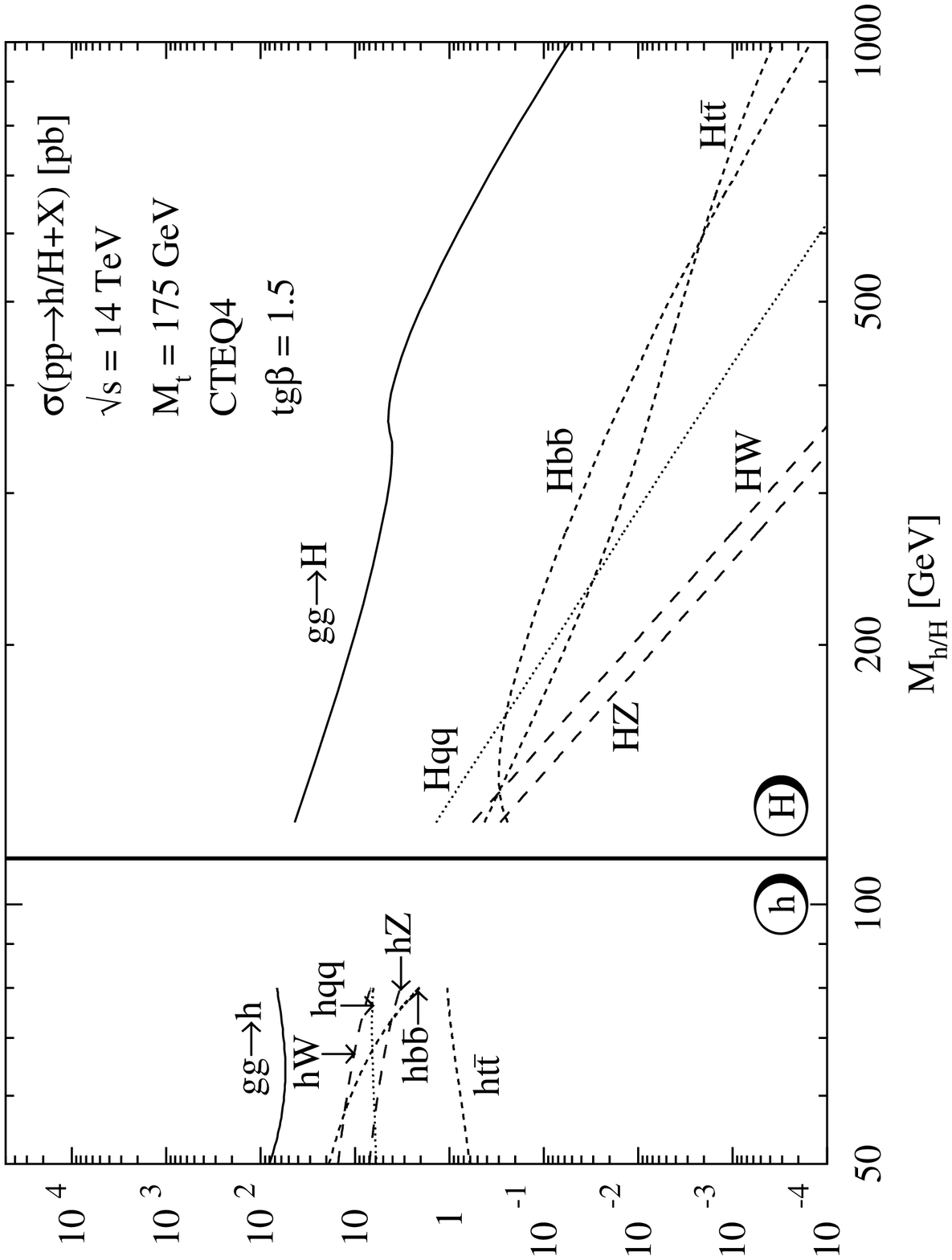}
\end{turn}
\vspace*{0.3cm}

\centerline{\bf Fig.~\ref{fg:mssmprohiggs}a}

\vspace*{0.2cm}
\hspace*{1.0cm}
\begin{turn}{-90}%
\epsfxsize=8.5cm \epsfbox{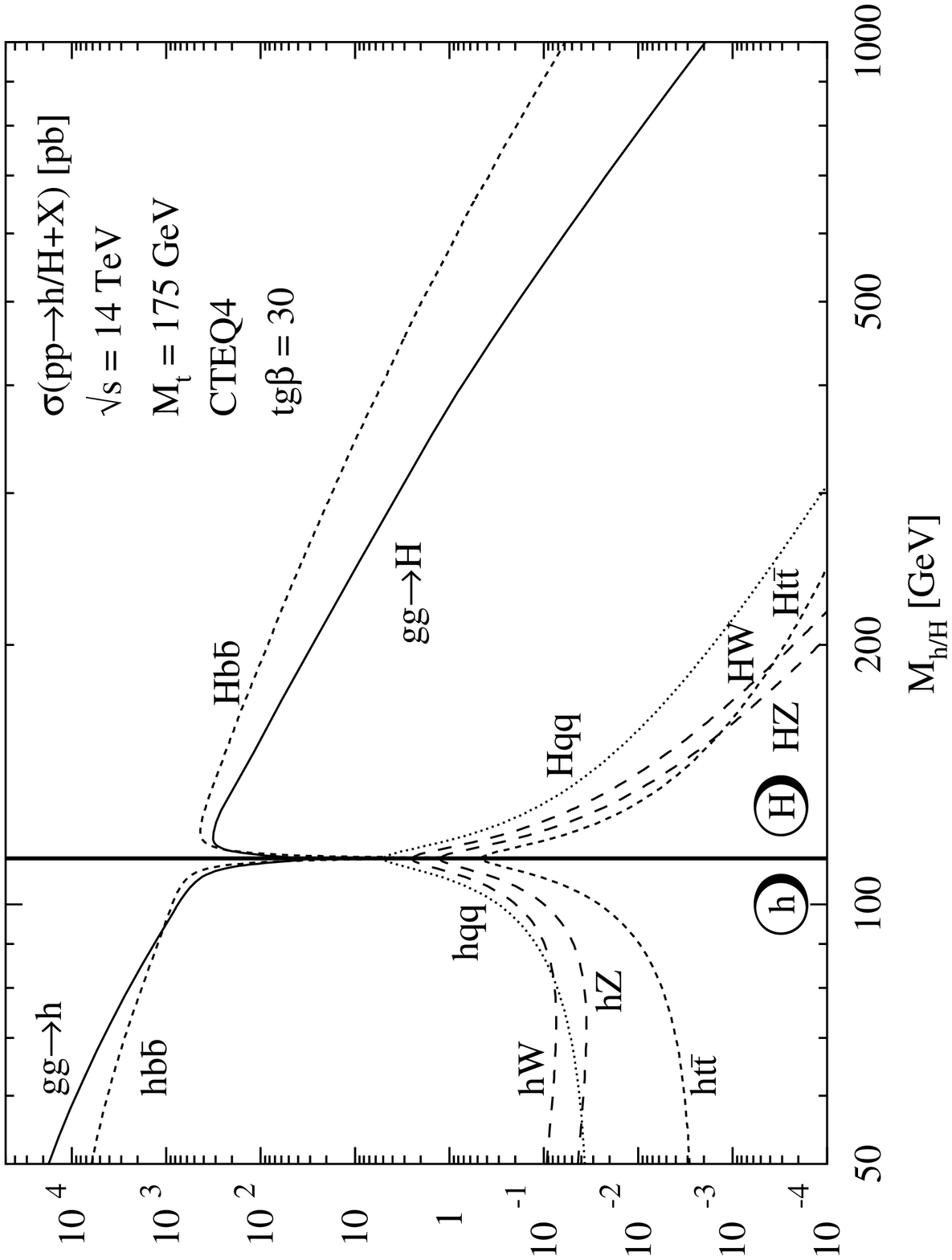}
\end{turn}
\vspace*{0.3cm}

\centerline{\bf Fig.~\ref{fg:mssmprohiggs}b}

\caption[]{\label{fg:mssmprohiggs} \it Neutral MSSM Higgs production cross
sections at the LHC [$\sqrt{s}=14$ TeV] for gluon fusion $gg\to \Phi$,
vector-boson fusion $qq\to qqVV \to qqh/
qqH$, vector-boson bremsstrahlung $q\bar q\to V^* \to hV/HV$ and the associated
production $gg,q\bar q \to \Phi b\bar b/ \Phi t\bar t$ including all known
QCD corrections. (a) $h,H$ production for $\tb=1.5$, (b) $h,H$ production for
$\tb=30$, (c) $A$ production for $\tb=1.5$, (d) $A$ production for $\tb=30$.}
\end{figure}
\addtocounter{figure}{-1}
\begin{figure}[hbtp]

\vspace*{0.5cm}
\hspace*{0.5cm}
\begin{turn}{-90}%
\epsfxsize=9.0cm \epsfbox{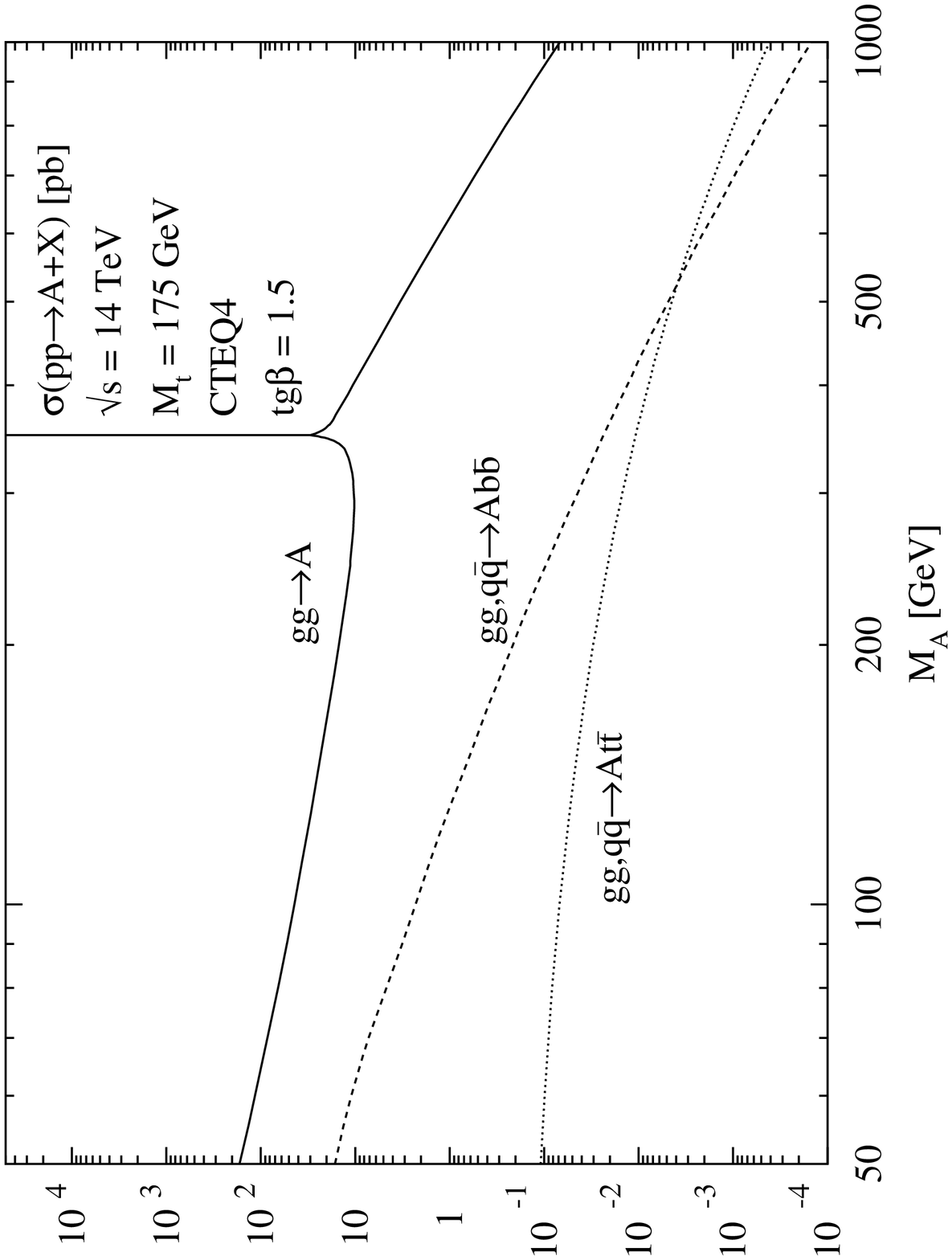}
\end{turn}
\vspace*{0.5cm}

\centerline{\bf Fig.~\ref{fg:mssmprohiggs}c}

\vspace*{0.5cm}
\hspace*{0.5cm}
\begin{turn}{-90}%
\epsfxsize=9.0cm \epsfbox{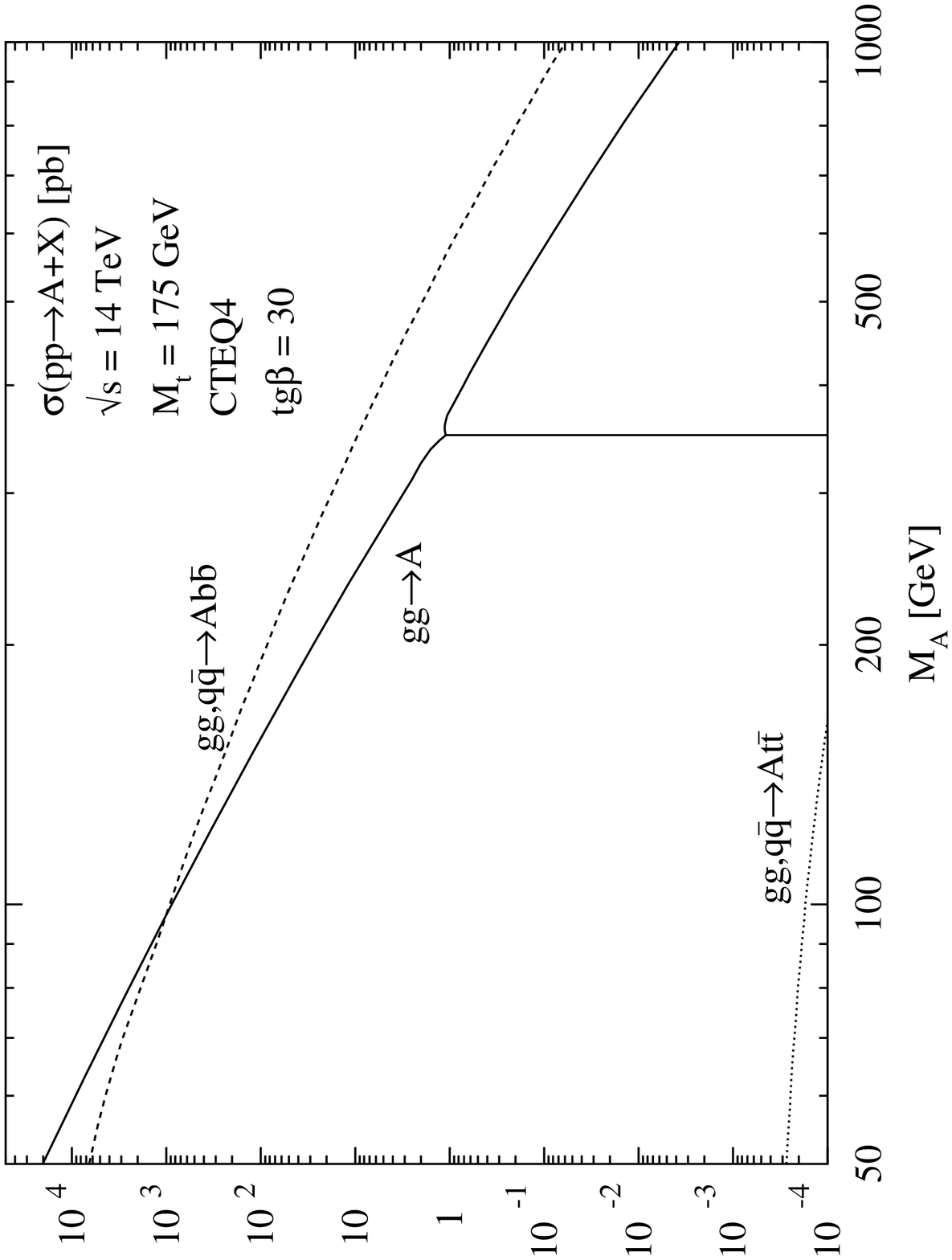}
\end{turn}
\vspace*{0.5cm}

\centerline{\bf Fig.~\ref{fg:mssmprohiggs}d}

\caption[]{\it Continued.}
\end{figure}

Previous studies of MSSM Higgs boson production at the LHC \cite{mssmpro} were
based on lowest-order cross sections or included a part of the QCD corrections.
We have updated these analyses by including all known QCD corrections to the
production processes and the two-loop corrections to the MSSM Higgs sector, thus
rendering the results more accurate and reliable than in the previous studies.

The cross sections of the various MSSM Higgs production mechanisms at the LHC
are shown in Figs.~\ref{fg:mssmprohiggs}a--d for two representative values of
$\tb = 1.5, 30$ as a function of the corresponding Higgs mass.
The total c.m.~energy has been chosen as $\sqrt{s} = 14$ TeV, the CTEQ4M
parton densities have been adopted with $\alpha_s(M_Z)=0.116$, and the top and
bottom masses have been set to $M_t=175$ GeV and $M_b=5$ GeV. For the
Higgs bremsstrahlung off $t,b$ quarks, $pp\to\Phi Q\bar Q +X$, we have used the
leading order CTEQ4L parton densities, because the NLO QCD corrections are
unknown. Thus the consistent evaluation of this cross section requires LO
parton densities and strong coupling. The latter is normalized as $\alpha_s
(M_Z) = 0.132$ at lowest order. For small and moderate values of $\tb\lsim 10$
the gluon-fusion cross section provides the dominant production cross section
for the entire Higgs mass region up to $M_\Phi\sim 1$ TeV. However, for large
$\tb$, Higgs bremsstrahlung off bottom quarks, $pp\to\Phi b\bar b+X$, dominates
over the gluon-fusion mechanism through the strongly enhanced bottom Yukawa
couplings.

\begin{figure}[hbtp]

\vspace*{-5.0cm}
\hspace*{-3.5cm}
\epsfxsize=20cm \epsfysize=28cm \epsfbox{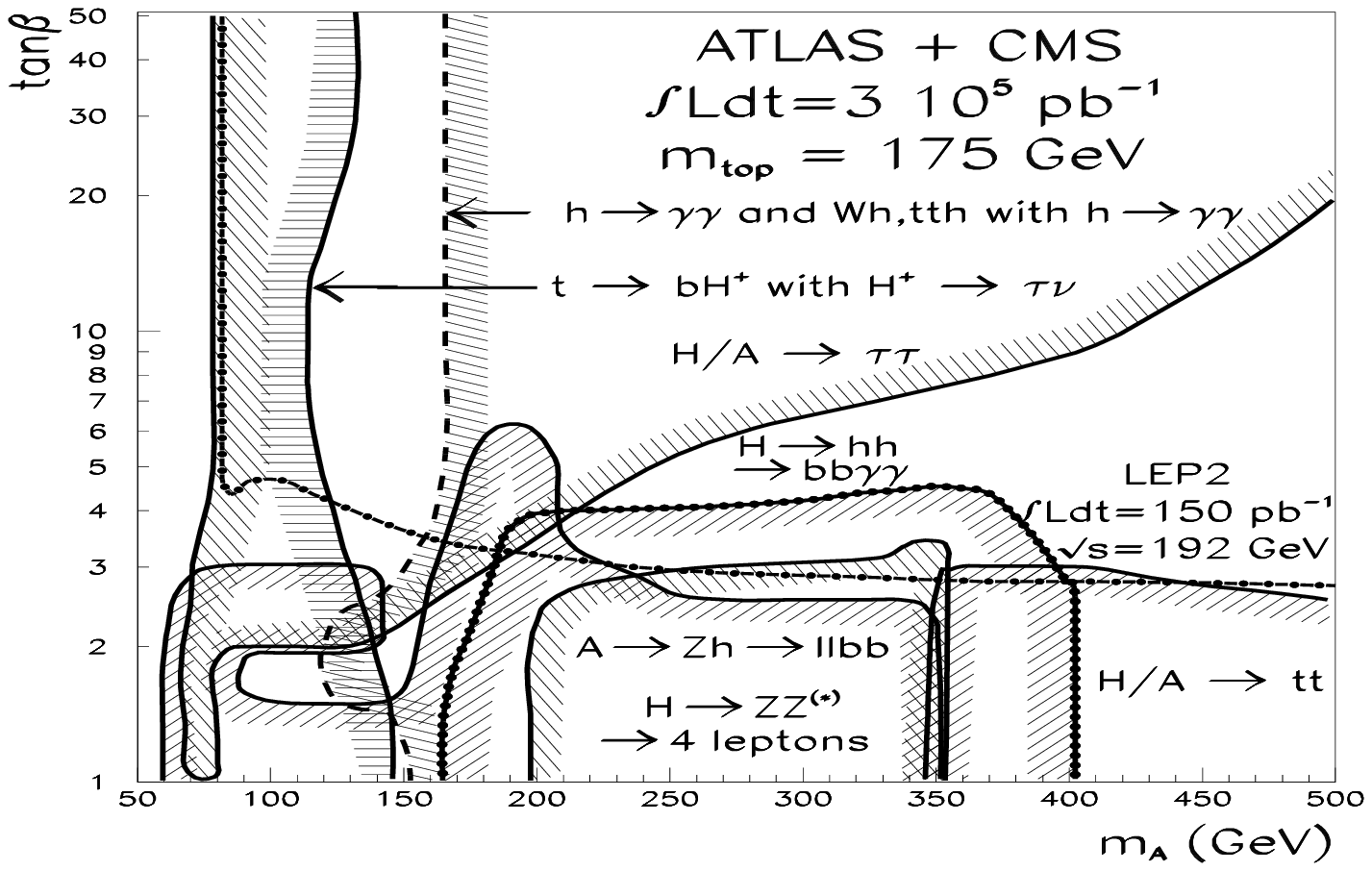}
\vspace*{-6.5cm}

\caption[]{\label{fg:atlascms} \it MSSM parameter space including the contours
of the various Higgs decay modes, which will be visible at the LHC after
reaching the anticipated integrated luminosity $\int {\cal L} dt = 3\times
10^5\, pb^{-1}$ and combining the experimental data of both LHC experiments,
ATLAS and CMS [taken from Ref.~\cite{richter}].}
\end{figure}

The MSSM Higgs search at the LHC will be more involved than the SM Higgs
search. The basic features can be summarized as follows.
\begin{description}
\item[(i)~] For large pseudoscalar Higgs masses $M_A \gsim 200$ GeV the light
scalar Higgs boson $h$ can only be found via its photonic decay mode $h\to
\gamma \gamma$. In a significant part of this MSSM parameter region, especially
for moderate values of $\tb$, no other MSSM Higgs particle can be discovered.
Because of the decoupling limit for large $M_A$ the MSSM cannot be distinguished
from the SM in this mass range.

\item[(ii)~] For small values of $\tb \lsim 3$ and pseudoscalar Higgs masses
between about 200 and 350 GeV, the heavy scalar Higgs boson can be searched
for in the `gold-plated' channel $H \to ZZ \to 4l^\pm$. Otherwise this
`gold-plated' signal does not play any r\^ole in the MSSM.
However, the detectable MSSM parameter region hardly exceeds the anticipated
exclusion limits of the LEP2 experiments.

\item[(iii)] For large and moderate values of $\tb \gsim 3$ the decays $H,A
\to \tau^+\tau^-$
become visible at the LHC. Thus this decay mode plays a significant r\^ole for
the MSSM in contrast to the SM. Moreover, it will also be detectable for
small values of $\tb \gsim 1$--$2$ and $M_A\lsim 200$ GeV.

\item[(iv)] For $\tb \lsim 4$ and 150 GeV $\!\!\lsim M_A \lsim 400$ GeV the
heavy scalar Higgs
particle can be detected via its decay mode $H\to hh\to b\bar b\gamma\gamma$.
However, the MSSM parameter range for this signature is very limited.

\item[(v)~] For $\tb\lsim 3$--$5$ and 50 GeV $\!\!\lsim M_A\lsim 350$ GeV the
pseudoscalar decay mode $A\to Zh \to l^+ l^- b\bar b$ will be visible, but
hardly exceeds the exclusion limits from LEP2.

\item[(vi)] For pseudoscalar Higgs masses $M_A\lsim 100$ GeV charged Higgs
bosons, produced from top quark decays $t\to H^+ b$, can be discovered via its
decay mode $H^+ \to \tau^+ \bar\nu_\tau$.
\end{description}
The final picture exhibits a difficult region for the MSSM Higgs search at the
LHC. For $\tb \sim 5$ and $M_A \sim 150$ GeV the full
luminosity and the full data sample of both the ATLAS and CMS experiments at the
LHC, are needed to cover the problematic parameter region \cite{richter}, see
Fig.~\ref{fg:atlascms}. On the other hand, if no excess of Higgs events
above the SM background processes beyond 2 standard deviations will be found,
the MSSM Higgs bosons can easily be excluded at 95\% C.L.

\section{Summary}
In this review the decay widths and branching ratios of SM and MSSM Higgs
bosons have been updated. All relevant higher order corrections, which are
dominated by QCD corrections, have been taken into account.
We have thus presented the branching ratios and decay widths of SM and MSSM
Higgs particles with the best available theoretical accuracy.

At the LHC the SM Higgs particle will be produced predominantly by gluon fusion
$gg\to H$, followed by vector-boson fusion $VV\to H$ ($V=W,Z$) and, to a lesser
extent, Higgs-strahlung off vector bosons, $V^*\to VH$, and top quarks,
$gg/q\bar q\to t\bar t H$. The cross sections of these production channels
have been updated by including all known QCD corrections, which are important
in particular for the dominant gluon-fusion mechanism. Thus the final
results of this review may serve as a benchmark of the theoretical predictions
for SM Higgs boson production at the LHC.

For Higgs masses $M_H\gsim 140$ GeV the SM Higgs search at the LHC will proceed
via the `gold-plated' $H\to ZZ^{(*)}\to 4l^\pm$ decay mode with small SM
backgrounds. The extraction of four charged lepton signals at the LHC will probe
Higgs masses up to about 800 GeV. In the Higgs mass range 155 GeV$\lsim M_H
\lsim 180$ GeV the SM Higgs boson can also easily be found via its decay
channel $H\to WW\to l^+l^- \nu\bar \nu$, by means of the specific angular
correlations among the charged leptons. Moreover, the decay channel $H\to
l^+l^-\nu \bar \nu$ may allow for an extension of the Higgs search up to Higgs
masses beyond 1 TeV. For $M_W\lsim M_H\lsim 140$~GeV, the only promising decay
mode seems to be provided by the photonic decay $H\to\gamma\gamma$, the
detection of which, however, requires excellent energetic and geometric
resolutions of the detectors in order to suppress the large QCD backgrounds.
Higgs-strahlung off vector bosons or top quarks, with the Higgs decaying into
a photon pair, may allow a further reduction of the background, but
unfortunately the signal rates are small. In the case of excellent
$b$-tagging the dominant $b\bar b$ decay mode of the Higgs might be detectable,
if the Higgs particle is produced in association with a $W$ boson or
$t\bar t$ pair.

In the MSSM the neutral Higgs bosons will mainly be produced via gluon fusion
$gg\to \Phi$. However, through the enhanced $b$ quark couplings, Higgs
bremsstrahlung
off $b$ quarks, $gg/q\bar q\to b\bar b \Phi$, will dominate for large $\tb$.
All other Higgs production mechanisms, i.e.\ vector-boson fusion and
Higgs-strahlung off vector bosons or $t\bar t$ pairs, will be less important
than in the SM.

The `gold-plated' $H\to ZZ\to 4l^\pm$ signal does not play an important r\^ole
in the MSSM. On the other hand the $\tau$ lepton pair decays
$H,A\to\tau^+\tau^-$
will be visible for large values of $\tb$. The light scalar Higgs particle will
only be visible via its photonic decay mode $h\to\gamma\gamma$, the branching
ratio of which will be smaller than in the SM because of the suppressed SUSY
couplings. Moreover, the decay modes $H\to hh\to b\bar b\gamma\gamma$, $H/A\to
t\bar t$ and $A\to Zh\to l^+l^- b\bar b$ will be detectable in very restricted
regions of the MSSM parameter space. Finally charged Higgs particles may be
looked for in the top quark decays $t\to H^+ b$ at the lower end of the charged
Higgs mass range. In the search for the MSSM Higgs particles at the LHC, the
maximal anticipated integrated luminosity will be needed, especially to cover
the difficult region around $M_A\sim 150$ GeV and $\tb \sim 5$.

\vspace*{1cm}

\noindent
{\large \bf Acknowledgements} \\[0.5cm]
First I would like to thank G.\ Kramer and P.M.\ Zerwas for the fruitful
collaboration on QCD and Higgs physics. Thanks also go to other collaborators:
A.\ Djouadi, S.\ Dawson, J.\ Kalinowski, B.\ Kniehl, M.~Kr\"amer, E.\ Laenen
and D.\ Graudenz. Moreover, I am grateful to G.\ Kramer for encouraging me to
complete this work. I thank F.\ Gianotti for providing me with
Fig.~\ref{fg:lhcsmhiggs}. Finally I am also grateful to P.M.\ Zerwas,
A.\ Djouadi, S.\ Dawson and A.\ Kataev for reading the manuscript and their
valuable comments.

\end{document}